\documentclass[11pt,reqno]{amsart}    

\usepackage[margin=37mm]{geometry}

\usepackage{amssymb,amsmath,amsfonts,amsthm, amscd, mathrsfs, helvet, mathtools}
\usepackage{arydshln}
\usepackage{bm, stmaryrd}
\usepackage{setspace}
\usepackage{caption}
\usepackage{graphicx,verbatim}
\usepackage[usenames, dvipsnames]{color}
\usepackage{color}
\usepackage[mathcal]{euscript}
\usepackage{multirow}
\usepackage{frcursive}
\usepackage[T1]{fontenc}

\usepackage{tikz}    
\usetikzlibrary{arrows,matrix,snakes}
\usetikzlibrary{calc} 
\usetikzlibrary{decorations.pathmorphing,decorations.markings,shapes.geometric,shapes.misc}    

\theoremstyle{plain}
\newtheorem{theorem}{Theorem}[section]
\newtheorem*{theorem*}{Theorem}
\newtheorem{proposition}[theorem]{Proposition}
\newtheorem*{proposition*}{Proposition}

\newtheorem{lemma}[theorem]{Lemma}
\newtheorem{corollary}[theorem]{Corollary}

\theoremstyle{remark}

\newtheorem{examplex}[theorem]{Example}
\newtheorem{remarkx}[theorem]{Remark}
\newenvironment{example}
  {\pushQED{\qed}\examplex}
  {\popQED\endexamplex}
\newenvironment{remark}
  {\pushQED{\qed}\remarkx}
  {\popQED\endremarkx}

\definecolor{brightPurple}{rgb}{.47,0,1}

\definecolor{brightRed}{rgb}{1,0,0}

\definecolor{brightBlue}{rgb}{0,0,1}

\definecolor{myGreen}{rgb}{0,0.7,0}

\newcommand{\mathfrc}[1]{\mbox{\small\cursive\slshape#1}}
\def\r{\mathfrc{r}}

\newcommand{\tensor}[1]{{\mathfrak{#1}}}

\newcommand{\co}[1]{\check{ #1}}

\DeclareMathOperator{\res}{res}
\DeclareMathOperator{\id}{id}
\DeclareMathOperator{\ad}{ad}

\DeclareMathOperator{\End}{End}
\DeclareMathOperator{\Aut}{Aut}
\DeclareMathOperator{\bAut}{\overline{Aut}}

\DeclareMathOperator{\Conn}{Conn}
\DeclareMathOperator{\proj}{\pi}
\DeclareMathOperator{\hotimes}{\hat\otimes}
\DeclareMathOperator{\hgt}{ht}
\DeclareMathOperator{\rk}{rk}

\DeclareSymbolFont{widetriangleaccent}{OMX}{yhex}{m}{n}
\DeclareMathAccent{\widetriangle}{\mathord}{widetriangleaccent}{"E6}

\newcommand{\longhookrightarrow}{\lhook\joinrel\relbar\joinrel\rightarrow}
\newcommand{\longtwoheadrightarrow}{\relbar\joinrel\twoheadrightarrow}

\def\Loop{\mathcal{L}}
\def\cent{\mathsf K}
\def\cocent{\mathsf D}
\def\s{\mathsf s}
\def\t{\mathsf t}
\def\z{{\bm z}}
\def\Z{{\bm{\mathcal Z}}}
\def\lvl{\ell}
\def\rep{\varrho}
\def\a{\mathfrak{a}}
\def\b{\mathfrak{b}}
\def\c{\mathfrak{c}}

\def\g{\mathfrak{g}}
\def\Fil{\mathsf F}
\def\cFil{\overline{\mathsf F}}
\def\F{\mathfrak F}
\def\hg{\widehat{\mathfrak{g}}}
\def\tg{\widetilde{\mathfrak{g}}}
\def\bg{\overline{\mathfrak{g}}}
\def\Lg{L\tg}
\def\Og{\Omega\tg}
\def\h{\mathfrak{h}}

\def\m{\mathfrak{m}} 

\def\p{\mathfrak{p}} 
\def\q{\mathfrak{q}}

\def\VV{\mathbb{V}}
\def\ZZ{\mathbb{Z}}

\def\ha{\mbox{\small $\frac{1}{2}$}}
\def\qa{\mbox{\small $\frac{1}{4}$}}

\newcommand{\SimTo}{%
\xrightarrow{\raisebox{-.3em}{\tiny $\sim$}}
}

\newcommand{\ms}[1]{{\mathsf #1}}
\newcommand{\bb}[1]{[\kern-.1em[ #1 ]\kern-.1em]}
\newcommand{\lau}[1]{(\kern-.2em( #1 )\kern-.2em)}

\newcommand{\lbf}[1]{\langle\kern-.2em\langle #1 \rangle\kern-.2em\rangle}
\newcommand{\biglbf}[1]{\big\langle \kern-.25em \big\langle #1 \big\rangle \kern-.25em \big\rangle}

\newcommand{\rbf}[1]{(\kern-.2em( #1 )\kern-.2em)}
\newcommand{\bigrbf}[1]{\big( \kern-.25em \big( #1 \big) \kern-.25em \big)}

\newcommand{\wh}[1]{\widehat #1}
\newcommand{\wt}[1]{\widetilde #1}

\newcommand{\ul}[1]{[#1]}

\def\A{\mathcal{A}}

\def\CC{\mathbb{C}}
\def\CP{\mathbb{P}^1}
\def\RR{\mathbb{R}}
\def\H{\mathcal{H}}

\def\K{\mathcal{K}}

\def\P{\mathcal{P}}
\def\R{\mathcal{R}}
\def\D{\mathcal{D}}
\def\sO{\mathscr{O}}
\def\sK{\mathscr{K}}

\def\Tak{\mathcal{T}}
\def\L{\mathcal{L}}

\def\1{\tensor{1}}
\def\2{\tensor{2}}
\def\3{\tensor{3}}
\def\4{\tensor{4}}

\newlength{\dhatheight}
\newcommand{\doublewh}[1]{%
    \settoheight{\dhatheight}{\ensuremath{\widehat{#1}}}%
    \addtolength{\dhatheight}{-0.35ex}%
    \widehat{\vphantom{\rule{1pt}{\dhatheight}}%
    \smash{\widehat{#1}}}}

\numberwithin{equation}{section}

\begin{document}

\title[On integrable field theories as dihedral affine Gaudin models]{On integrable field theories as\\[1mm]
dihedral affine Gaudin models}

\author{Beno\^{\i}t Vicedo}
\address{Department of Mathematics, University of York, York YO10 5DD, U.K.}
\email{benoit.vicedo@gmail.com}

\begin{abstract}
We introduce the notion of a \emph{classical dihedral affine Gaudin model}, associated with an untwisted affine Kac-Moody algebra $\tg$ equipped with an action of the dihedral group $D_{2T}$, $T \geq 1$ through (anti-)linear automorphisms. We show that a very broad family of classical integrable field theories can be recast as examples of such classical dihedral affine Gaudin models. Among these are the principal chiral model on an arbitrary real Lie group $G_0$ and the $\ZZ_T$-graded coset $\sigma$-model on any coset of $G_0$ defined in terms of an order $T$ automorphism of its complexification. Most of the multi-parameter integrable deformations of these $\sigma$-models recently constructed in the literature provide further examples. The common feature shared by all these integrable field theories, which makes it possible to reformulate them as classical dihedral affine Gaudin models, is the fact that they are non-ultralocal. In particular, we also obtain affine Toda field theory in its lesser-known non-ultralocal formulation as another example of this construction.

We propose that the interpretation of a given classical non-ultralocal integrable field theory as a classical dihedral affine Gaudin model provides a natural setting within which to address its quantisation. At the same time, it may also furnish a general framework for understanding the massive ODE/IM correspondence since the known examples of integrable field theories for which such a correspondence has been formulated can all be viewed as dihedral affine Gaudin models.
\end{abstract}

\input{epsf}

\maketitle
\setcounter{tocdepth}{1}
\tableofcontents

\section{Motivation and introduction}

The ODE/IM correspondence describes a striking and rather unexpected relation between the theory of linear Ordinary Differential Equations in the complex plane on the one hand, and that of quantum Integrable Models on the other. Concretely, the first instance of such a correspondence was formulated by V. Bazhanov, S. Lukyanov and A. Zamolodchikov for quantum KdV theory in the series of seminal papers \cite{Bazhanov:1994ft} -- \cite{Bazhanov:2003ni}, building on from the original insight of P. Dorey and R. Tateo in \cite{Dorey:1998pt}. These works culminated in the remarkable conjecture of \cite{Bazhanov:2003ni} stating that the joint spectrum of the quantum KdV Hamiltonians on the level $L \in \ZZ_{\geq 0}$ subspace of an irreducible module over the Virasoro algebra is in bijection with the set of certain one-dimensional Schr\"odinger operators $-\partial^2_z + V_L(z)$ with `monster' potentials $V_L(z)$ of a given form. The justification for this conjecture comes from the central observation that the functional relations and analytic properties characterising the eigenvalues of the $Q$-operators of quantum KdV theory \cite{Bazhanov:1996dr, Bazhanov:1998dq}
on a given joint eigenvector coincide with those satisfied by certain connection coefficients of the associated one-dimensional Schr\"odinger equation. These ideas were soon extended to other massless integrable field theories associated with higher rank Lie algebras of classical type, see \emph{e.g.} \cite{Dorey:1999pv,Dorey:2000ma,Bazhanov:2001xm,Dorey:2006an} and the review \cite{Dorey:2007zx}.

Despite the variety of examples of the ODE/IM correspondence, its mathematical underpinning remained elusive for a number of years. This problem was addressed by B. Feigin and E. Frenkel in \cite{Feigin:2007mr} where they argued that the ODE/IM correspondence for quantum $\hg$-KdV theory could be understood as originating from an affine analogue of the geometric Langlands correspondence.
To explain this connection we make a brief digression on Gaudin models, which provide a realisation of the global geometric Langlands correspondence for rational curves over the complex numbers.

\medskip

Let $\g$ be a finite-dimensional complex semisimple Lie algebra. The Gaudin model, or $\g$-Gaudin model to emphasise the dependence on $\g$, is a quantum integrable spin-chain with long-range interactions \cite{Gaudin}. If we let $N \in \ZZ_{\geq 1}$ denote the number of sites then the algebra of observables of the model is the $N$-fold tensor product $U(\g)^{\otimes N}$ of the universal enveloping algebra $U(\g)$ of $\g$. The quadratic Gaudin Hamiltonians are elements of $U(\g)^{\otimes N}$ given by
\begin{equation} \label{Gaudin Ham intro}
H_i \coloneqq \sum_{\substack{j=1\\ j \neq i}}^N \frac{I^{a(i)} I_a^{(j)}}{z_i - z_j}
\end{equation}
where the $z_i$, $i = 1, \ldots, N$ are arbitrary distinct complex numbers, $\{ I^a \}$ and $\{ I_a \}$ are dual bases of $\g$ with respect to a fixed non-degenerate invariant bilinear form $\langle \cdot, \cdot \rangle$ on $\g$, and $\ms x^{(i)}$ is the element of $U(\g)^{\otimes N}$ with $\ms x \in \g$ in the $i^{\rm th}$ tensor factor and $1$'s in every other factor. The quantum integrability of the model is characterised by the existence of a large commutative subalgebra $\mathscr Z_{\z}(\g) \subset U(\g)^{\otimes N}$ with $\z \coloneqq \{ z_i \}_{i=1}^N \cup \{ \infty \}$, known as the \emph{Gaudin algebra}, containing in particular the quadratic Gaudin Hamiltonians.

Let $M_i$, $i = 1, \ldots, N$ be $\g$-modules. One is interested in finding the joint spectrum of $\mathscr Z_{\z}(\g)$ on the spin-chain $\bigotimes_{i=1}^N M_i$.
Note that a joint eigenvalue of the Gaudin algebra defines a homomorphism $\mathscr Z_{\z}(\g) \to \CC$ sending each element of $\mathscr Z_{\z}(\g)$ to its eigenvalue. The joint spectrum can therefore be described as a subset of the maximal spectrum of the commutative algebra $\mathscr Z_\z(\g)$, \emph{i.e.} the set of all homomorphisms $\mathscr Z_\z(\g) \to \CC$. It was shown by E. Frenkel in \cite[Theorem 2.7(1)]{Frenkel2} that the maximal spectrum of the Gaudin algebra $\mathscr Z_\z(\g)$ is isomorphic to a certain subquotient of the space of $\null^L \g$-valued connections on $\CP$, known as \emph{$\null^L \g$-opers}, with regular singularities in the set $\z$, where $\null^L \g$ denotes the Langlands dual of the Lie algebra $\g$. In other words, each joint eigenvalue of the Gaudin algebra $\mathscr Z_\z(\g)$ on the given spin-chain $\bigotimes_{i=1}^N M_i$ will be described by such an $\null^L \g$-oper. In fact, when all the $\g$-modules $M_i$ are finite-dimensional irreducibles $V_{\lambda_i}$ of highest weights $\lambda_i \in \h^\ast$, it was Conjectured in \cite{Frenkel2} and proved recently in \cite{Rybnikov:2016} that for each integral dominant weight $\lambda_\infty \in \h^\ast$, the joint spectrum of $\mathscr Z_\z(\g)$ on the subspace of weight $\lambda_\infty$ singular vectors in $\bigotimes_{i=1}^N V_{\lambda_i}$ is in bijection with the subspace of such $\null^L \g$-opers with residue at the points $z_i$ and infinity given by the shifted Weyl orbits of the weights $\lambda_i$ and $\lambda_\infty$ respectively, and with trivial monodromy representation.

The description of the maximal spectrum of the Gaudin algebra $\mathscr Z_\z(\g)$ in terms of $\null^L \g$-opers also generalises to the case of Gaudin models with irregular singularities; see \cite{FFT, FFRy}. Another possible generalisation of Gaudin models is given by \emph{cyclotomic} Gaudin models, introduced in \cite{Vicedo:2014zza, ViY2} and more recently \cite{ViY3} for the case with irregular singularities. A similar description of the corresponding cyclotomic Gaudin algebra of \cite{Vicedo:2014zza} was recently conjectured in \cite{Lacroix:2016mpg} in terms of \emph{cyclotomic $\null^L \g$-opers}, \emph{i.e.} $\null^L \g$-opers equivariant under an action of the cyclic group.
In fact, these descriptions of the various Gaudin algebras in terms of global $\null^L \g$-opers on $\CP$ follow (conjecturally in the cyclotomic case) from the `local' version proved by B. Feigin and E. Frenkel in their seminal paper \cite{FeiginFrenkel} (see also \cite{Frenkel3, Langlands}) which states that the space of singular vectors in the vacuum Verma module $\VV_0^{\rm crit}(\g)$ at the critical level over the untwisted affine Kac-Moody algebra $\hg$, which naturally forms a commutative algebra, is isomorphic to the algebra of functions on the space of $\null^L \g$-opers on the formal disc.

\medskip

The apparent similarity between the description of the joint spectrum of the Gaudin algebra on any given spin-chain in terms of certain $\null^L \g$-opers and the statement of the ODE/IM correspondence for quantum KdV theory is more than just a coincidence. Indeed, as argued in \cite{Feigin:2007mr}, quantum $\hg$-KdV theory can be regarded as a generalised Gaudin model associated with the untwisted affine Kac-Moody algebra $\hg$, or $\hg$-Gaudin model for short, with a regular singularity at the origin and an irregular singularity of the mildest possible form at infinity. Unfortunately, much less is know at present about Gaudin models associated with general Kac-Moody algebras; see however \cite{MV, Frenkel}. In particular, there is currently no known analogue of the Feigin-Frenkel isomorphism for describing the space of singular vectors in the suitably completed vacuum Verma module over the double affine, or toroidal, Lie algebra $\doublewh{\g}$. It is not even clear what the critical level should be in this setting. Nevertheless, the notion of an \emph{affine oper}, or \emph{$\hg$-oper}, on $\CP$ can certainly be defined \cite{Frenkel} and so it is tempting to speculate that the description of the spectrum of the $\g$-Gaudin Hamiltonians in terms of $\null^L \g$-opers persists when $\g$ is replaced by an affine Kac-Moody algebra.

In this spirit, the explicit form of the $\null^L \hg$-opers which ought to describe the joint spectrum of the quantum $\hg$-KdV Hamiltonians on certain irreducible modules over the $W$-algebra associated with $\g$ was conjectured in \cite{Feigin:2007mr}, by using as a finite-dimensional analogy a certain description of the finite $W$-algebra for a regular nilpotent element in terms of $\null^L \g$-opers. Remarkably, when $\hg = \widehat{\mathfrak{sl}}_2$ so that also $\null^L \hg = \widehat{\mathfrak{sl}}_2$, these $\widehat{\mathfrak{sl}}_2$-opers were shown to coincide exactly, after a simple change of coordinate on $\CP$, with the one-dimensional Schr\"odinger operators written down in \cite{Bazhanov:2003ni}. This result not only confirms the idea that the ODE/IM correspondence can be thought of as a particular instance of the geometric Langlands correspondence but also provides strong evidence in support of the general claim that the joint spectrum of the higher Hamiltonians of an affine Gaudin model can be described in terms of affine opers for the Langlands dual affine Kac-Moody algebra.

Another approach to testing the proposed link between the joint spectrum of the quantum $\hg$-KdV Hamiltonians and $\null^L \hg$-opers of the prescribed form is to follow the same strategy originally used to establish the ODE/IM correspondence for quantum KdV theory. Specifically, one should compare the functional relations and analytic properties of the joint eigenvalues of the $Q$-operators of quantum $\hg$-KdV theory on joint eigenvectors in the irreducibles over the $W$-algebra associated with $\g$, with those satisfied by the connection coefficients of the associated $\null^L \hg$-opers. This programme was initiated in \cite{Sun:2012xw} and was further developed very recently in \cite{Masoero:2015lga, Masoero:2015rcz} where some remarkable functional relations, referred to as the \emph{$Q\wt Q$-system}, were obtained for certain generalised spectral determinants of the ODE associated with the $\null^L \hg$-opers of \cite{Feigin:2007mr} corresponding to highest weight states in representations of the $W$-algebra. Even more recently in \cite{Frenkel:2016gxg}, the very same $Q \wt Q$-system was shown to arise as relations in the Grothendieck ring $K_0(\mathcal O)$ of the category $\mathcal O$ of representations of the Borel subalgebra of the quantum affine algebra $U_q(\hg)$ for an untwisted affine Kac-Moody algebra $\hg$. Analogous relations were also conjectured to hold when $\hg$ is a twisted affine Kac-Moody algebra. Since non-local quantum $\hg$-KdV Hamiltonians can be associated with elements of $K_0(\mathcal O)$ by the construction of \cite{Bazhanov:1996dr, Bazhanov:1998dq, Bazhanov:2001xm}, the joint spectrum of these Hamiltonians also satisfy the $Q \wt Q$-system, thereby providing further evidence in favour of the ODE/IM correspondence for quantum $\hg$-KdV theory.

\medskip

The recent developments towards formulating and ultimately proving the ODE/IM correspondence for quantum $\hg$-KdV theory, which we briefly recalled above, can be summarised in the following commutative diagram
\begin{equation} \label{gKdV picture}
\begin{tikzpicture}
  \node (a) at (0,0) [align=center, draw, rounded corners] {$\hg$-Gaudin\\ model};
  \node (b) at (-2.5,0) [left, align=center, draw, rounded corners]
    {
      quantum\\ $\hg$-KdV theory
    };
  \node (c) at (2.5,0) [right, align=center, draw, rounded corners]
    {
      $\null^L \hg$-opers
    };
  \node (d) at (0,-1.5) [align=center, draw, rounded corners] {$Q \wt Q$-system};
  
  \draw[->, shorten >=3pt, shorten <=3pt] (b) -- node[above] {\cite{Feigin:2007mr}} (a);
  \draw[<-, dashed, shorten >=3pt, shorten <=3pt] (c) -- node[above] {\cite{Feigin:2007mr}} (a);
  \draw[->, shorten >=5pt, shorten <=3pt] (b) -- node[below left] {\cite{Frenkel:2016gxg}} (d);
  \draw[->, shorten >=6pt, shorten <=4pt] (c) -- node[below right] {\cite{Masoero:2015lga, Masoero:2015rcz}} (d);
\end{tikzpicture}
\end{equation}
The top line of this diagram, corresponding to the work \cite{Feigin:2007mr}, consisted of two steps. The first was to reinterpret quantum $\hg$-KdV theory as a particular affine $\hg$-Gaudin model. The second, which we represent by a dashed arrow to emphasise its conjectural status, was to make use of the existing description of the spectrum of $\g$-Gaudin models in terms of $\null^L \g$-opers as an analogy. The big open problem here is to establish the affine counterpart of the latter statement to put the second step on a firm mathematical footing. Indeed, this would promote the top line in the above diagram to a proof of the ODE/IM correspondence for quantum $\hg$-KdV theory. While the top line is still partly conjectural, the bottom part of the diagram provides a solid link between both sides of the `KdV-oper' correspondence of \cite{Feigin:2007mr} through the common $Q \wt Q$-system.

\medskip

Until relatively recently, the study of the ODE/IM correspondence had been limited to describing integrable structures in conformal field theories only. This left open the important question of whether similar ideas could be used to describe the spectrum of \emph{massive} quantum integrable field theories as well. The first example of such a \emph{massive} ODE/IM correspondence was put forward by S. Lukyanov and A. Zamolodchikov for quantum sine-Gordon and sinh-Gordon theories in their pioneering paper \cite{Lukyanov:2010rn}. Specifically, they showed that the functional relations and analytic properties characterising the vacuum eigenvalues of the $Q$-operators of quantum sine/sinh-Gordon theory were the same as those satisfied by certain connection coefficients of the auxiliary linear problem of the classical modified sinh-Gordon equation for a suitably chosen classical solution. Subsequently, various higher rank generalisations of this massive ODE/IM correspondence for quantum affine $\tg$-Toda field theories were also conjectured, when $\tg$ is of type $A$ for rank $3$ in \cite{Dorey:2012bx} and for general rank $n$ in \cite{Adamopoulou:2014fca}, and more recently for a general untwisted affine Kac-Moody algebra $\tg$ in \cite{Ito:2013aea, Ito:2015nla} as well as examples of twisted type in \cite{Ito:2016qzt}. Another important quantum integrable field theory for which a massive ODE/IM correspondence has been conjectured in \cite{Lukyanov:2013wra}, and further studied in \cite{Bazhanov:2013cua, Bazhanov:2014joa}, is the Fateev model \cite{Fateev:1996ea}. It can be viewed as a two-parameter deformation of the $SU_2$ principal chiral model and as such it is equivalent \cite{Hoare:2014pna} to the so called $SU_2$ bi-Yang-Baxter $\sigma$-model \cite{Klimcik:2008eq}. Here as well the correspondence is a conjectured link between the spectrum of the Fateev model on the one hand, and solutions of the classical modified sinh-Gordon equation on the other.

A noteworthy feature of the massive ODE/IM correspondence for quantum affine $\tg$-Toda field theory in the non-simply-laced case is the appearance of the Langlands dual $\null^L \tg$ of the affine Kac-Moody algebra $\tg$ on the ODE side. This strongly suggests that the geometric Langlands correspondence may also underly the \emph{massive} ODE/IM correspondence. One of the aims of the present paper is to make the first step towards generalising the above picture in \eqref{gKdV picture} for quantum $\hg$-KdV theory to massive quantum integrable field theories. In fact, as in the case of $\hg$-(m)KdV theory, one typically starts from a description of the \emph{classical} integrable field theory. Therefore, in a first instance, one is faced with the initial problem of how to quantise the given classical integrable field theory. We will argue that both problems are in fact closely related.

\medskip

The most effective approach for quantising a given classical integrable field theory and establishing its quantum integrability is the quantum inverse scattering method \cite{FT, KS}, whose mathematical underpinning gave rise to the theory of quantum affine algebras. In particular, it can be used to obtain functional equations such as Baxter's $TQ$-relation and the $Q\widetilde{Q}$-system, all of which follow from corresponding relations in the Grothendieck ring of category $\mathcal O$. Unfortunately, the quantum inverse scattering method is well known to apply only under the restrictive assumption that the classical integrable field theory one starts with is \emph{ultralocal}. Since the main focus of the present paper is to address the problem of quantising classical integrable field theories which violate this condition, we begin by briefly recalling why this condition is necessary in the standard quantum inverse scattering method.

The starting point of the classical inverse scattering method, as crystalised by A. Reiman and M. Semenov-Tian-Shansky in \cite{RS, STS1}, is to identify the phase space of the given classical integrable field theory with a coadjoint orbit in the smooth dual $\hat{\mathfrak G}_1^\ast$ of a hyperplane $\hat{\mathfrak{G}}_1$ in a certain central extension of the double loop algebra $\mathfrak G$. The latter consists of smooth maps from the circle $S^1$ to the loop algebra $\g(\!( z )\!)$, or possibly its twist by some finite-order automorphism of $\g$. The smooth dual is defined relative to a certain bilinear form on $\mathfrak G$ given in terms of a model dependent rational function $\varphi(z)$, called the \emph{twist function}, as
\begin{equation} \label{bilinear form intro}
(\!(\mathscr X, \mathscr Y)\!)_\varphi \coloneqq \int_{S^1} d\theta \res_z \big\langle \mathscr X(\theta, z), \mathscr Y(\theta, z) \big\rangle \varphi(z) dz
\end{equation}
for $\mathscr X, \mathscr Y \in \mathfrak{G}$. The Poisson bracket on $\hat{\mathfrak G}_1^\ast$ is the Kostant-Kirillov $\R$-bracket associated with some solution $\R \in \End \mathfrak{G}$ of the modified classical Yang-Baxter equation. Given any pair of differentiable functionals $f$ and $g$ on $\hat{\mathfrak G}_1^\ast$, their Poisson bracket at a generic point $(\mathscr L, 1) \in \hat{\mathfrak G}_1^\ast$, where $\mathscr L \in \mathfrak{G}$, may be written as
\begin{equation} \label{PB Lax intro}
\{ f, g \}\big( (\mathscr L, 1) \big) = \big(\!\big( df_{\mathscr L}, (\ad \mathscr L \circ \R + \R^\ast \circ \ad \mathscr L - (\R + \R^\ast) \partial_\theta) \cdot dg_{\mathscr L} \big)\!\big)_\varphi
\end{equation}
where $df_{\mathscr L} \in \mathfrak{G}$ denotes the Fr\'echet derivative of $f$ at $(\mathscr L, 1)$ and $\R^\ast$ is the adjoint operator of $\R$ with respect to \eqref{bilinear form intro}. In this language the theory is said to be \emph{ultralocal} if $\R^\ast = - \R$, otherwise it is \emph{non-ultralocal}.

It is well known that the integrals of motion of the classical integrable field theory can be obtained from spectral invariants of the monodromy $M_{\mathscr L}$ of the differential operator $\partial_\theta + \mathscr L$, which is valued in the loop group $G(\!( z )\!)$. Given any smooth functional $\phi$ on $G(\!( z )\!)$ which is central, the Fr\'echet derivative of the functional $\phi^M : \mathscr L \mapsto \phi(M_{\mathscr L})$ on $\hat{\mathfrak{G}}_1^\ast$ defines an element of $\mathfrak{G}$, and hence its Poisson bracket \eqref{PB Lax intro} with any other smooth functional $f$ on $\hat{\mathfrak{G}}_1^\ast$ is well defined. In particular, the involution property of the integrals of motion is established by showing that for any pair of central functionals $\phi$ and $\psi$ on $G(\!( z )\!)$ we have $\{ \phi^M, \psi^M \} = 0$ \cite{STS2}. If a functional $\phi$ on $G(\!( z )\!)$ is not central, however, then the Fr\'echet derivative $d \phi^M$ will in general exhibit a jump discontinuity at the base point of $M_{\mathscr L}$. In the case of an ultralocal theory where the $\partial_\theta$-term in \eqref{PB Lax intro} is absent, the bracket naturally extends to such functionals $f$ and $g$ with discontinuous Fr\'echet derivatives. One can then evaluate $\{ \phi^M, \psi^M \}$ for arbitrary smooth functionals $\phi$ and $\psi$ \cite{STS2}, yielding the celebrated Sklyanin bracket on $G(\!( z )\!)$. The quantisation of the latter then serves as a starting point for the quantum inverse scattering method. By contrast, in the non-ultralocal case the bracket $\{ \phi^M, \psi^M \}$ is clearly ill-defined for arbitrary smooth functionals $\phi$ and $\psi$. This issue has precluded the direct application of the quantum inverse scattering method to a wide range of important integrable field theories due to their non-ultralocal nature.

\medskip

Although generalisations of the quantum inverse scattering method capable of also accommodating non-ultralocal systems do exist, see for instance \cite{Freidel-Maillet, Hlavaty:1994vh, SemenovTianShansky:1995ha}, these remain applicable only to a very restricted class of non-ultralocal systems. Faced with this limitation, the common strategy for quantising a given non-ultralocal system is to attempt to `ultralocalise' it by different means.
These include modifying the classical field theory itself by altering its twist function, see \emph{e.g.} \cite{Faddeev:1985qu} (and also \cite{SemenovTianShansky:1995ha, Delduc:2012qb}), finding a suitable gauge transformation which will bring it to an ultralocal form, see \emph{e.g.} \cite{Bazhanov:1994ft, Ridout:2011wx}, or possibly by finding a dual description of the theory which would be ultralocal.
Yet such attempts at `curing' a classical integrable field theory of its non-ultralocality ultimately work only in a limited number of cases. Let us mention also some alternative approaches to dealing with the problem of non-ultralocality which have been put forward recently in \cite{Melikyan:2016gkd} for the Alday-Arutyunov-Frolov model and very recently in \cite{Schmidtt:2017ngw} for the $\lambda$-deformation of the $AdS_5 \times S^5$ superstring.

A possible way around the problem of dealing with the monodromy in a non-ultralocal field theory, at least for describing local integrals of motion, was first proposed in \cite{Evans:1999mj} on the example of the principal chiral model associated with any Lie algebra $\g$ of classical type. These ideas were subsequently developed for other models in \cite{Evans:2000hx, Evans:2000qx, Evans:2005zd, Evans:2001sz}, and more recently in \cite{Lacroix:2017isl}. Specifically, it was shown in \cite{Evans:1999mj} that local integrals of motion in involution could be constructed directly from the Lax matrix, without needing to use the monodromy matrix. A striking property of this family of local charges is that their degrees precisely match the pattern of exponents of the untwisted affine Kac-Moody algebra $\tg$ associated with $\g$. This is very reminiscent of what happens in the $\g$-Gaudin model, where the degrees of the homogeneous integrals of motion also correspond to the pattern of exponents of $\g$. Indeed, this parallel also served as a source of inspiration for the present work.

In fact, classical $\hg$-(m)KdV theory is one of those distinguished classical integrable field theories which admits both an ultralocal and a non-ultralocal formulation, related to one another through a gauge transformation. Its ultralocal description serves as the starting point in the approach of \cite{Bazhanov:1994ft, Bazhanov:1996dr, Bazhanov:1998dq, Bazhanov:2001xm} for quantising the theory using the quantum inverse scattering method. As recalled above, within this approach non-local quantum $\hg$-(m)KdV Hamiltonians can be associated with elements of $K_0(\mathcal O)$ and the $Q \wt Q$-system is obtained from corresponding relations in $K_0(\mathcal O)$ established in \cite{Frenkel:2016gxg}. In other words, the bottom left arrow of the diagram in \eqref{gKdV picture} starts from the quantisation of $\hg$-(m)KdV theory in its ultralocal formulation. By contrast, treating classical $\hg$-KdV as a non-ultralocal theory enables one to regard it as a classical affine Gaudin model\footnote{Let us note here that it will also follow from \S\ref{sec: Toda FT}, where we discuss affine $\tg$-Toda field theory, that $\hg$-mKdV theory can be regarded as a classical \emph{cyclotomic} affine Gaudin model.}. The proposal of \cite{Feigin:2007mr} to then quantise $\hg$-KdV theory by viewing it as a classical affine Gaudin model led to the conjectural description of its quantum spectrum in terms of affine $\null^L \hg$-opers, corresponding to the top line of the diagram in \eqref{gKdV picture}. Since the ultralocal and non-ultralocal formulations of classical $\hg$-KdV theory are related by a gauge transformation we expect that their respective quantisations should agree. In this setting, the fact that the work \cite{Masoero:2015lga, Masoero:2015rcz} makes the diagram in \eqref{gKdV picture} commutative can be seen as evidence of this.

\medskip

The goal of the present paper is to initiate a program for quantising non-ultralocal classical integrable field theories and propose a framework within which to understand the massive ODE/IM correspondence for such models. Specifically, we introduce the notion of a classical \emph{dihedral} (or \emph{real cyclotomic}) affine $\tg$-Gaudin model associated with an arbitrary untwisted affine Kac-Moody algebra $\tg$. We then show that classical dihedral $\tg$-Gaudin models describe a general class of classical non-ultralocal integrable field theories, namely those whose Poisson bracket is as in \eqref{PB Lax intro} with $\R$-matrix given by the standard solution of the classical Yang-Baxter equation on the (twisted) double loop algebra $\mathfrak G$. We illustrate this relation between classical dihedral $\tg$-Gaudin models and non-ultralocal classical integrable field theories on a wide variety of examples, listed in Table \ref{table intro}, including the principal chiral model on a real semisimple Lie group $G_0$ and the $\ZZ_T$-graded coset $\sigma$-models for any $T \in \ZZ_{\geq 2}$ as well as some of their various multi-parameter deformations introduced in recent years \cite{Klimcik, Klimcik:2008eq, Delduc:2013fga, Sfetsos:2013wia, Hollowood:2014rla, Delduc:2014uaa}. Replacing the semisimple Lie algebra $\g$ by the Grassmann envelope of a Lie superalgebra, the present formalism also describes $\ZZ_T$-graded supercoset $\sigma$-models \cite{Young:2005jv, Magro, Vicedo:2009sn, Ke:2011zzb} and various deformations recently constructed \cite{Delduc:2013qra, Delduc:2014kha, Hollowood:2014qma}. It is interesting to note, in particular, that the examples of integrable field theories for which a massive ODE/IM correspondence has been formulated can all be recast as classical dihedral affine Gaudin models. Our proposal is therefore that the problem of quantising non-ultralocal integrable field theories and that of formulating an ODE/IM correspondence for such models can both be addressed within the context of quantisation of dihedral (affine) Gaudin models.

\medskip

\begin{center}
\begin{tabular}{|c|c|c|}
\hline
Non-ultralocal field theory & Divisor $\D$ & $\sigma \in \Aut \tg$\\
\hline
Principal chiral model (PCM) & $2 \cdot 0 + 2 \cdot \infty$ & \multirow{5}{*}{$\id$}\\
PCM with WZ-term & $2 \cdot k + 2 \cdot \infty$, $k \in \RR^\times$ & \\
Yang-Baxter (YB) $\sigma$-model & $(i \eta) + 2 \cdot \infty$, $\eta \in \RR_{>0}$ & \\
YB $\sigma$-model with WZ-term & $(k + i A) + 2 \cdot \infty$, $k \in \RR^\times$, $A \in \RR_{>0}$ & \\
bi-Yang-Baxter $\sigma$-model & $e^{i \vartheta} + e^{i(\psi + \pi)} + \infty$, $\vartheta, \psi \in \, ]0, \frac{\pi}{2}[$ & \\
\hline
$\ZZ_T$-graded coset $\sigma$-model & $2 \cdot 1 + \infty$ & \multirow{3}{*}{$\begin{array}{c} |\sigma| = T, \\ \; T \in \ZZ_{\geq 2} \end{array}$}\\
$q$-deformation ($q \in \RR$) & $e^{i \vartheta} + \infty$, $\vartheta \in \, ] 0, \frac{\pi}{T} [$ & \\
$q$-deformation ($|q| = 1$) & $p + p^{-1} + \infty$, $p \in \, ]0, 1[$ & \\
\hdashline
Affine Toda field theory & $2 \cdot 0 + 2 \cdot \infty$ & $\sigma$ Coxeter\\
\hline
\end{tabular}
\captionof{table}{Examples of dihedral affine Gaudin models associated with an untwisted affine Kac-Moody algebra $\tg$.}
\label{table intro}
\end{center}

\medskip

To end this introduction we motivate the definition of classical dihedral $\tg$-Gaudin models by considering the simpler case where $\tg$ is replaced by a finite-dimensional complex semisimple Lie algebra $\g$. The datum for a (classical) $\g$-Gaudin model with irregular singularities can be described by a divisor $\D$ on $\CP$, \emph{i.e.} a formal sum of a finite subset of points $\z = \{ z_i \}_{i=1}^N \cup \{ \infty \}$ on $\CP$ weighted by positive integers $n_x \in \ZZ_{\geq 1}$ for each $x \in \z$. We further restrict attention in this introduction to the case where $n_x = 1$ for all $x \in \z$ for simplicity. The algebra of observables of the classical $\g$-Gaudin model is then given by the $N$-fold tensor product $S(\g)^{\otimes N}$ of the symmetric algebra $S(\g)$ on $\g$. The classical quadratic Hamiltonians $H^{cl}_i$, $i = 1, \ldots, N$ of the model are given by the same expressions as the quantum Hamiltonians $H_i$ in \eqref{Gaudin Ham intro} but regarded as elements of $S(\g)^{\otimes N}$. They can be obtained from the \emph{Lax matrix} $L(z)$, defined by the expression
\begin{equation*}
L(z) dz = \sum_{j=1}^N \frac{I_a dz}{z - z_j} \otimes I^{a (j)},
\end{equation*}
as the spectral invariants $H^{cl}_i = \res_{z_i} \langle L(z), L(z) \rangle dz$ where the inner product is taken over the first tensor factor, \emph{i.e.} the auxiliary space.

Now let $\sigma \in \Aut \g$ be an automorphism of $\g$ whose order divides $T \in \ZZ_{\geq 1}$ and pick a primitive $T^{\rm th}$-root of unity $\omega^{-1} \in \CC^\times$. These both induce actions of the cyclic group $\Gamma \coloneqq \ZZ_T$ on $\g$ and $\CP$, respectively. The quadratic Gaudin Hamiltonians of a classical \emph{cyclotomic} $\g$-Gaudin model are similarly obtained from the same spectral invariants but using the Lax matrix defined by
\begin{equation*}
L(z) dz = \frac{1}{T} \sum_{j=1}^N \sum_{\alpha \in \Gamma} \hat \alpha \bigg( \frac{I_a dz}{z - z_j} \bigg) \otimes I^{a(j)},
\end{equation*}
where $\hat \alpha$ denotes the action of $\alpha \in \Gamma$ on $\g$-valued meromorphic differentials defined by combining the action on $\g$ with the pullback on differentials over $\CP$. This Lax matrix has the $\Gamma$-equivariance property $\sigma L(z) = \omega L(\omega z)$, where $\sigma$ acts on the auxiliary space.

If we are also given an anti-linear automorphism $\tau \in \bAut \g$ of $\g$ which preserves the eigenspaces of $\sigma$ then we obtain an action of the dihedral group $\Pi \coloneqq D_{2T}$ of order $2T$ on $\g$. Promoting also the action of $\Gamma$ on $\CP$ to an action of $\Pi$ by adding complex conjugation $z \mapsto \bar z$, we can define the Lax matrix of the classical \emph{dihedral} $\g$-Gaudin model by an expression similar to the above but replacing the sum over $\Gamma$ by a sum over $\Pi$. Specifically, we should now take the tensor product over $\RR$ rather than $\CC$ and use dual basis elements of the realification of $\g$ so that we set
\begin{equation} \label{Lax matrix intro}
L(z) dz = \frac{1}{2T} \sum_{j=1}^N \sum_{\alpha \in \Pi} \Bigg( \hat \alpha \bigg( \frac{I_a dz}{z - z_j} \bigg) \otimes I^{a (j)} + \hat \alpha \bigg( \frac{- i I_a dz}{z - z_j} \bigg) \otimes i I^{a (j)} \Bigg).
\end{equation}
By construction, this Lax matrix is $\Pi$-equivariant in the sense that $\sigma L(z) = \omega L(\omega z)$ and $\tau L(z) = L(\bar z)$ where $\sigma$ and $\tau$ both act on the auxiliary space.

In order to describe integrable field theories on the circle one should replace $\g$ in the above discussion by an untwisted affine Kac-Moody algebra $\tg$. Concretely this means replacing the dual bases $\{ I^a \}$ and $\{ I_a \}$ of $\g$ in the expression for the Lax matrix \eqref{Lax matrix intro} by dual bases $\{ I^{\wt a} \}$ and $\{ I_{\wt a} \}$ of $\tg$ and working in a suitable completion of the tensor product. We will demonstrate that in this affine setting the Lax matrix \eqref{Lax matrix intro}, or its generalisation to other divisors $\D$, reproduces the Lax matrices and twist functions of all the integrable field theories in Table \ref{table intro}.

Let us finally note that reductions of general $2$-dimensional integrable field theories of Zakharov-Shabat-Mikhailov type by discrete symmetry groups, such as cyclic or dihedral groups, were introduced in the important paper \cite{Mikhailov}. The purpose of the present work is to show that many integrable field theories of interest fall, in fact, within a very restricted subclass of reduced Zakharov-Shabat-Mikhailov type field theories, namely that of affine Gaudin models with dihedral symmetry. Indeed, given an integrable field theory which can be formulated within the Zakharov-Shabat-Mikhailov scheme (often in different gauge-inequivalent ways), it is by no means guaranteed that it can also be regarded as an affine Gaudin model. In particular, an affine Gaudin formulation is typically not preserved by formal gauge transformations of the zero curvature representation. In \S\ref{sec: examples of NUL} we give explicit examples of well known zero curvature representations of integrable field theories which are not of Gaudin type. Yet, as outlined above, having an affine Gaudin formulation of a given classical non-ultralocal integrable field theory puts the problem of its quantisation on a firm mathematical footing and should provide a framework for establishing an ODE/IM correspondence for it. In the present context, the reduction by a dihedral group is expected to play an important role in extensions of this correspondence to the massive case.

\medskip

The plan of the article is as follows. We begin in \S\ref{sec: KM} by recalling some basic results about (anti-)linear automorphisms on finite-dimensional Lie algebras and affine Kac-Moody algebras. In \S\ref{sec: Double loop} we construct a direct sum of Takiff algebras for $\tg$ attached to the finite subset $\z \subset \CP$ as a quotient of a direct sum of loop algebras of $\tg$, and describe its dual space in terms of certain $\tg$-valued meromorphic differentials on $\CP$. The main section is \S\ref{sec: affine Gaudin} where we define classical dihedral $\tg$-Gaudin models and establish their relation to a general family of non-ultralocal integrable field theories. The Lax matrix is defined as the canonical element of the dual pair constructed in \S\ref{sec: Double loop}. Finally, \S\ref{sec: examples of NUL} is devoted to a detailed construction of important non-ultralocal integrable field theories as dihedral $\tg$-Gaudin models. We collect in an appendix some facts about dual pairs and our conventions on tensor index notation.

\subsubsection*{Acknowledgements}

The author thanks Sylvain Lacroix for a careful reading of the draft and many useful comments and suggestions.

\section{Real affine Kac-Moody algebras} \label{sec: KM}

Let $T \in \ZZ_{\geq 1}$. We denote the dihedral group of order $2T$ by
\begin{equation*}
\Pi \coloneqq D_{2T} = \langle \s, \t \,|\, \s^T = \t^2 = (\s \, \t)^2 = 1 \rangle.
\end{equation*}
Let $\Gamma \coloneqq \langle \s \rangle \subset \Pi$ be the cyclic subgroup of order $T$, which is normal in $\Pi$. We refer to elements of $\Gamma$ as \emph{orientation preserving} and to elements of the coset $\Gamma \t$ as \emph{orientation reversing}.

Given a complex Lie algebra $\a$, we let $\bAut \a$ denote the group of all linear and anti-linear automorphisms of $\a$. The subgroup $\Aut \a$ of linear automorphisms is normal of index $2$. We denote by $\bAut_- \a$ the subset in $\bAut \a$ of all anti-linear automorphisms of $\a$ so that
\begin{equation*}
\bAut \a = \Aut \a \sqcup \bAut_- \a.
\end{equation*}
For any $\chi \in \bAut_- \a$ we can identify $\bAut_- \a$ with the coset $\chi \Aut \a$.

\subsection{Finite-dimensional Lie algebras} \label{sec: fin dim g}

Let $\g$ be a finite-dimensional complex Lie algebra and $\sigma \in \Aut \g$ be a linear automorphism whose order divides $T$, \emph{i.e.} such that $\sigma^T = \id$. Fix a primitive $T^{\rm th}$-root of unity $\omega$ and let
\begin{equation} \label{g eigen decomp}
\g = \bigoplus_{j=0}^{T-1} \g_{(j), \CC}
\end{equation}
be the decomposition of $\g$ into the eigenspaces $\g_{(j), \CC} \coloneqq \{ \ms x \in \g \,|\, \sigma \ms x = \omega^j \ms x \}$ of $\sigma$.

Let $\tau \in \bAut_- \g$ be an anti-linear involutive automorphism of $\g$, namely such that $\tau^2 = \id$, and let $\g_0 \coloneqq \{ \ms x \in \g \,|\, \tau \ms x = \ms x \}$ denote the corresponding real form of $\g$. Its complexification $\g_0 \otimes_\RR \CC$ is naturally isomorphic to $\g$.

We shall assume that each of the eigenspaces $\g_{(j), \CC}$ for $j \in \ZZ_T$ is $\tau$-stable, \emph{i.e.}
\begin{equation} \label{tau fix gj assume}
\tau \g_{(j), \CC} = \g_{(j), \CC}.
\end{equation}
It follows from this, and using the property $\bar \omega = \omega^{-1}$, that $(\sigma \circ \tau)^2 = \id$.
We use the property \eqref{tau fix gj assume} to define the real subspaces $\g_{(j)} \coloneqq \g_{(j), \CC} \cap \g_0$ for each $j \in \ZZ_T$ so that
\begin{equation} \label{g0 eigen decomp}
\g_0 = \bigoplus_{j=0}^{T-1} \g_{(j)}
\end{equation}
Note that $\sigma$ preserves the real subspace $\g_{(j)}$ only if $2j = 0$ in $\ZZ_T$. Indeed, given any $\ms x \in \g_{(j)}$ we have $\sigma \ms x = \omega^j \ms x$ and $\tau \ms x = \ms x$ but $\tau(\sigma \ms x) = \sigma^{-1} (\tau \ms x) = \omega^{-j} \ms x = \omega^{-2j} \sigma \ms x$.

Note that $\sigma^k \circ \tau \in \bAut_- \g$ defines an anti-linear involutive automorphism of $\g$ for each $k \in \ZZ_T$. Introduce the corresponding real forms of $\g$ by
\begin{equation} \label{real forms gk}
\g_k \coloneqq \g^{\sigma^k \circ \tau} = \{ \ms x \in \g \,|\, \sigma^k \tau \ms x = \ms x \}.
\end{equation}
The notation reflects the fact that the case $k=0$ gives back the original real form $\g_0$.
We note that for each $k \in \ZZ_T$, the anti-linear involutive automorphism $\sigma^k \circ \tau$ clearly also preserves the eigenspace $\g_{(j), \CC}$ for each $j \in \ZZ_T$.
For any $p \in \ZZ_T$ the anti-linear map $\omega^{- k p} \sigma^k \circ \tau$ is also an involution (but in general not an automorphism). We shall make use of the corresponding real subspaces
\begin{equation} \label{real forms gkp}
\g_{k, p} \coloneqq \g^{\omega^{- k p} \sigma^k \circ \tau} = \{ \ms x \in \g \,|\, \sigma^k \tau \ms x = \omega^{kp} \ms x \}.
\end{equation}
In this notation we have $\g_k = \g_{k, 0}$. We shall also use the notation $\g_{k, p}$ for any $p \in \ZZ$, which will be understood to mean $\g_{k, p \, \text{mod}\, T}$.

By virtue of the relations $\sigma^T = \tau^2 = (\sigma \circ \tau)^2 = \id$ satisfied by the automorphisms $\sigma$ and $\tau$, we have an action of the dihedral group $\Pi$ on the complex Lie algebra $\g$ by linear and anti-linear automorphisms. That is, we have a group homomorphism
\begin{equation} \label{hom r g}
r : \Pi \longhookrightarrow \bAut \g, \qquad \alpha \longmapsto r_\alpha
\end{equation}
defined by $r_\s \coloneqq \sigma$, $r_\t \coloneqq \tau$.

Suppose, moreover, that $\g_0$ is equipped with a non-degenerate invariant symmetric bilinear form
\begin{equation} \label{ip on g0}
\langle \cdot, \cdot \rangle : \g_0 \times \g_0 \longrightarrow \RR.
\end{equation}
We extend it to a non-degenerate symmetric invariant bilinear form on $\g$ as
\begin{equation} \label{ip on g}
\begin{split}
\langle \cdot, \cdot \rangle : \g \times \g &\longrightarrow \CC\\
( \ms x \otimes u, \ms y \otimes v ) &\longmapsto \langle \ms x, \ms y \rangle uv,
\end{split}
\end{equation}
using the canonical isomorphism between $\g$ and $\g_0 \otimes_\RR \CC$.
The bilinear form \eqref{ip on g} has the property that $\langle \tau \ms x, \tau \ms y \rangle = \overline{\langle \ms x, \ms y \rangle}$ for any $\ms x, \ms y \in \g$. We assume it is also $\sigma$-invariant, in other words that $\langle \sigma \ms x, \sigma \ms y \rangle = \langle \ms x, \ms y \rangle$ for any $\ms x, \ms y \in \g$.

In the following lemma we make use of the notion of dual pair recalled in \S\ref{app: dual pairs}.

\begin{lemma} \label{lem: gj and gkp}
For each $j \in \ZZ_T$, the triple $(\g_{(-j)}, \g_{(j)}, \langle \cdot, \cdot \rangle|_{\g_{(-j)} \times \g_{(j)}})$ is a dual pair. In particular, the restriction of \eqref{ip on g} to $\g_{(0)}$ is non-degenerate.

For any $k, p \in \ZZ_T$, the triple $(\g_{k, -p}, \g_{k, p}, \langle \cdot, \cdot \rangle|_{\g_{k, -p} \times \g_{k, p}})$ is a dual pair. In particular, \eqref{ip on g} restricts to a non-degenerate invariant symmetric bilinear form on the real form $\g_k$ for each $k \in \ZZ_T$.
\begin{proof}
Let $j \in \ZZ_T$. Using the properties of the bilinear form \eqref{ip on g} with respect to $\tau$ and $\sigma$ we see that it restricts to a bilinear form $\g_{(-j)} \times \g_{(j)} \to \RR$. Now let $\ms y \in \g_{(j)}$ be non-zero. It follows from the non-degeneracy of \eqref{ip on g} that there exists an $\ms x \in \g$ such that $\langle \ms x, \ms y \rangle = 1$. Next, using the $\sigma$-invariance we have $\langle \omega^{k j} \sigma^k \ms x, \ms y \rangle = 1$ for all $k \in \ZZ_T$. We then also have $\langle \omega^{- k j} \sigma^{-k} \tau \ms x, \ms y \rangle = 1$ for all $k \in \ZZ_T$. Putting this together we obtain $\langle \ms z, \ms y \rangle = 1$ where
\begin{equation*}
\ms z = \frac{1}{2T} \sum_{k = 0}^{T-1} (\omega^{k j} \sigma^k \ms x + \omega^{-k j} \sigma^{-k} \tau \ms x).
\end{equation*}
But clearly we have $\sigma \ms z = \omega^{-j} \ms z$ and $\tau \ms z = \ms z$. Hence $\ms z \in \g_{(-j)}$, from which we conclude that $\g_{(-j)} \times \g_{(j)} \to \RR$ is non-degenerate on the right. The proof of the non-degeneracy on the left is similar.

Let $k, p \in \ZZ_T$. For all $\ms x, \ms y \in \g$ we have
\begin{equation} \label{ip prop omega tau}
\langle \omega^{kp} \sigma^k \tau \ms x, \omega^{-kp} \sigma^k \tau \ms y \rangle = \overline{\langle \ms x, \ms y \rangle}.
\end{equation}
Hence \eqref{ip on g} restricts to a bilinear form $\g_{k, -p} \times \g_{k, p} \to \RR$. To show it is non-degenerate on the right, note that for any non-zero $\ms y \in \g_{k, p}$ there is an $\ms x \in \g$ such that $\langle \ms x, \ms y \rangle = 1$, by the non-degeneracy of \eqref{ip on g}. It follows using \eqref{ip prop omega tau} that $\langle \omega^{kp} \sigma^k \tau \ms x, \ms y \rangle = 1$ and hence $\langle \ha ( \ms x + \omega^{kp} \sigma^k \tau \ms x ), \ms y \rangle = 1$. But we have $\ha ( \ms x + \omega^{kp} \sigma^k \tau \ms x ) \in \g_{k, -p}$ which proves the result. The non-degeneracy on the left is shown in a similar way.
\end{proof}
\end{lemma}

\subsubsection{Canonical element} \label{sec: can elem g}

For each $k, p \in \ZZ_T$ we fix a basis
\begin{subequations} \label{g kp bases}
\begin{equation} \label{g kp basis}
I^a_{k, p}, \qquad a = 1, \ldots, \dim \g
\end{equation}
of the real subspace $\g_{k, p}$, and we let
\begin{equation} \label{g kp dual basis}
I_{a; k, -p}, \qquad a = 1, \ldots, \dim \g
\end{equation}
\end{subequations}
denote the dual basis of $\g_{k, -p}$ with respect to \eqref{ip on g}, \emph{i.e.} such that $\big\langle I^a_{k, p}, I_{b; k, -p} \big\rangle = \delta^a_b$ for every $a, b = 1, \ldots, \dim \g$. In the case $p=0$ we obtain dual bases $I^a_k \coloneqq I^a_{k,0}$ and $I_{a; k} \coloneqq I_{a; k,0}$ of $\g_k$, for each $k \in \ZZ_T$. For the real form $\g_0$ we will denote these dual bases simply as $I^a \coloneqq I^a_0$ and $I_a \coloneqq I_{a; 0}$. Note that any basis of the real subspace $\g_{k, p}$ for any $k, p \in \ZZ_T$ also forms a basis over $\CC$ for the complexification $\g$. A basis for the realification $\g_\RR$ is then given, for instance, by
\begin{equation*}
I^a, \qquad i I^a, \qquad a = 1, \ldots, \dim \g.
\end{equation*}
We shall also fix a basis
\begin{subequations} \label{g j bases}
\begin{equation} \label{g j basis}
I^{(j, \alpha)}, \qquad \alpha=1, \ldots, \dim \g_{(j), \CC}
\end{equation}
of $\g_{(j)}$ and let
\begin{equation} \label{g j dual basis}
I_{(-j, \alpha)}, \qquad \alpha=1, \ldots, \dim \g_{(j), \CC}
\end{equation}
\end{subequations}
denote the dual basis of $\g_{(-j)}$ so that $\langle I^{(j, \alpha)}, I_{(-j, \beta)} \rangle = \delta^\alpha_\beta$ for all $\alpha, \beta = 1, \ldots, \dim \g_{(j), \CC}$.

The canonical element of $\g$ is defined as
\begin{equation*}
C \coloneqq I_a \otimes I^a \in \g \otimes \g
\end{equation*}
where sums over repeated Lie algebra indices, here $a$ from $1$ to $\dim \g$, will always be implicit.
Using the decomposition \eqref{g eigen decomp} of $\g$, for each $j \in \ZZ_T$ we also introduce the canonical element of the dual pair $\big( \g_{(-j),\CC}, \g_{(j),\CC}, \langle \cdot, \cdot \rangle|_{\g_{(-j),\CC} \times \g_{(j),\CC}} \big)$ as
\begin{equation*}
C^{(j)} \coloneqq I_{(-j, \alpha)} \otimes I^{(j, \alpha)} \in \g_{(-j), \CC} \otimes \g_{(j), \CC}.
\end{equation*}
As above, repeated indices labelling dual bases of $\g_{(-j), \CC}$ and $\g_{(j), \CC}$, here $\alpha$, will always be implicitly summed over.

\subsection{Affine Kac-Moody algebras} \label{sec: aff KM}
Let $\Loop \g \coloneqq \g \otimes \CC[t, t^{-1}]$ be the polynomial loop algebra associated with $\g$. For any $\ms x \in \g$ and $n \in \ZZ$ we define $\ms x_n \coloneqq \ms x \otimes t^n \in \Loop \g$. The Lie bracket in $\Loop \g$ is defined by letting $(\ms x_m, \ms y_n) \mapsto [\ms x, \ms y]_{m+n}$ and then extending to a bilinear map $\Loop \g \times \Loop \g \to \Loop \g$ by linearity. Similarly, define a non-degenerate invariant symmetric bilinear form on $\Loop \g$ by $(\ms x_m, \ms y_n) \mapsto \langle \ms x, \ms y \rangle \delta_{m+n, 0}$.

The (untwisted) affine Kac-Moody algebra associated with $\g$ is defined as the vector space direct sum
\begin{equation} \label{tg def}
\tg \coloneqq \Loop \g \oplus \CC \cent \oplus \CC \cocent
\end{equation}
endowed with the Lie bracket $[\cdot, \cdot] : \tg \times \tg \to \tg$ defined by
\begin{equation} \label{bracket on tg}
\begin{split}
[ \ms x_m &+ \alpha \cent + x \cocent, \ms y_n + \beta \cent + y \cocent ]\\
&\qquad\qquad\coloneqq [\ms x, \ms y]_{m+n} + n x \, \ms y_n - m y \, \ms x_m + m \delta_{m+n, 0} \langle \ms x, \ms y \rangle \cent
\end{split}
\end{equation}
for any $\ms x, \ms y \in \g$, $m, n \in \ZZ$ and $\alpha, \beta, x, y \in \CC$.
It is equipped with a non-degenerate invariant symmetric bilinear form
\begin{subequations} \label{ip on tg}
\begin{equation}
(\cdot | \cdot ) : \tg \times \tg \longrightarrow \CC,
\end{equation}
defined for any $\ms x, \ms y \in \g$, $m, n \in \ZZ$ and $\alpha, \beta, x, y \in \CC$ by
\begin{equation}
( \ms x_m + \alpha \cent + x \cocent | \ms y_n + \beta \cent + y \cocent ) \coloneqq \langle \ms x, \ms y \rangle \delta_{m+n, 0} + \alpha y + \beta x.
\end{equation}
\end{subequations}
We shall also make use of the subalgebra $\hg \subset \tg$ and the quotient $\bg \subset \tg$ defined as
\begin{equation} \label{hg bg def}
\hg \coloneqq \Loop \g \oplus \CC \cent, \qquad \bg \coloneqq \tg / \CC \cent.
\end{equation}
Let $\bar\rep : \tg \to \bg$ denote the canonical homomorphism with kernel $\CC \cent$. It follows from the defining relations \eqref{bracket on tg} that $\Loop \g$ is isomorphic as a Lie algebra to the subquotient $\hg / \CC \cent$. The relationships between the various Lie algebras introduced above can then be summarised in the following commutative diagram
\begin{equation*}
\begin{tikzpicture}
\matrix (m) [matrix of math nodes, row sep=2em, column sep=3em,text height=1.5ex, text depth=0.25ex]    
{
\CC \cocent & \bg & \hg / \CC \cent & \Loop \g\\
\CC \cocent & \tg & \hg\\
& \CC \cent & \CC \cent\\
};
\path[<<-] (m-1-1) edge (m-1-2);
\path[->] (m-1-3) edge node[above=-.2em]{$\sim$} (m-1-4);
\path[<<-] (m-2-1) edge (m-2-2);
\path[<-left hook] (m-1-2) edge (m-1-3);
\path[<-left hook] (m-2-2) edge (m-2-3);
\path[<<-] (m-1-2) edge node[left]{$\bar\rep$} (m-2-2);
\path[<<-] (m-1-3) edge node[left]{$\bar\rep|_{\hg}$} (m-2-3);
\path[right hook->] (m-3-2) edge (m-2-2);
\path[right hook->] (m-3-3) edge (m-2-3);
\end{tikzpicture}
\end{equation*}

\subsubsection{Action of the dihedral group} \label{sec: tg action of Pi}

Consider the automorphisms $\sigma, \tau \in \bAut \g$ of $\g$ introduced in \S\ref{sec: fin dim g}. We extend these to linear and anti-linear automorphisms of the loop algebra $\Loop \g$, respectively, which by abuse of notation we also denote $\sigma$ and $\tau$, first on homogeneous elements by letting
\begin{equation*}
\sigma(\ms x_n) \coloneqq (\sigma \ms x)_n, \qquad
\tau(\ms x_n) \coloneqq (\tau \ms x)_{-n}
\end{equation*}
for $\ms x \in \g$, $n \in \ZZ$ and then to the rest of $\Loop \g$ by (anti-)linearity. It is worth noting that $\Loop \g$ is invariant under the inversion $t \mapsto t^{-1}$ of the formal parameter. This would not be the case if we replaced the polynomial loop algebra $\Loop \g$ with its formal completion $\g \otimes \CC\lau{t}$ of $\g$-valued formal Laurent series. Indeed, the latter has the undesired feature of breaking the symmetry between positive and negative loops so that the anti-linear automorphism $\tau \in \bAut_- \g$ could not be extended to $\g \otimes \CC\lau{t}$ as above.

Letting $\sigma \cent \coloneqq \cent$, $\sigma \cocent \coloneqq \cocent$, $\tau \cent \coloneqq - \cent$ and $\tau \cocent \coloneqq - \cocent$ further extends $\sigma, \tau \in \bAut_- \Loop \g$ to (anti-)linear automorphisms of the affine Kac-Moody algebra $\tg$, which we shall also denote by $\sigma, \tau \in \bAut \tg$. By construction the latter satisfy $\sigma^T = \tau^2 = (\sigma \circ \tau)^2 = \id$ and thus define a representation of the dihedral group $\Pi$ on $\tg$. Specifically, we have a homomorphism
\begin{equation} \label{hom r tg}
r : \Pi \longhookrightarrow \bAut \tg, \qquad \alpha \longmapsto r_\alpha
\end{equation}
defined by $r_\s \coloneqq \sigma$, $r_\t \coloneqq \tau$.
Given any complex vector space $V$ equipped with a real structure, namely an anti-linear involution $\tau : V \to V$, we extend the action of $\Pi$ on $\tg$ given by \eqref{hom r tg} to the complex vector space $\tg \otimes V$ by defining
\begin{equation} \label{Pi act g otimes V}
r_\s ( \ms X \otimes v ) \coloneqq r_\s \ms X \otimes v, \qquad 
r_\t ( \ms X \otimes v ) \coloneqq r_\t \ms X \otimes \tau v.
\end{equation}
In other words, if we define an action of $\Pi$ on $V$ by letting $\s$ act trivially and $\t$ act as $\tau$ then \eqref{Pi act g otimes V} gives an action of $\Pi$ on the tensor product $\tg \otimes V$.

As in the finite-dimensional setting of \S\ref{sec: fin dim g}, for each pair $k, p \in \ZZ_T$ we consider the involutive anti-linear map $\omega^{-kp} \sigma^k \circ \tau : \tg \to \tg$ which for $p=0$ defines an automorphism of the affine Kac-Moody algebra $\tg$. We denote the corresponding real subspaces by
\begin{equation*}
\tg_{k, p} \coloneqq \tg^{\omega^{-kp} \sigma^k \circ \tau} = \{ \ms X \in \tg \,|\, \sigma^k \tau \ms X = \omega^{kp} \ms X \}.
\end{equation*}
When $p = 0$ we obtain the real forms $\tg_k \coloneqq \tg_{k, 0}$ of $\tg$ for each $k \in \ZZ_T$. We have the direct sum decomposition of real vector spaces
\begin{equation*}
\tg_\RR = \tg_{k, p} \dotplus i \, \tg_{k,p},
\end{equation*}
where $\tg_\RR$ denotes the realification of $\tg$. Denote the corresponding projections relative to this decomposition by
\begin{alignat*}{2}
\proj^+_{k, p} : \tg_\RR &\longrightarrow \tg_{k, p}, &\qquad \ms X &\longmapsto \ha (\ms X + \omega^{-k p} \sigma^k \tau \ms X),\\
\proj^-_{k, p} : \tg_\RR &\longrightarrow i \, \tg_{k, p}, &\qquad \ms X &\longmapsto \ha (\ms X - \omega^{-k p} \sigma^k \tau \ms X).
\end{alignat*}
When $p=0$ we denote these projections as $\pi^+_k : \tg_\RR \to \tg_k$ and $\pi^-_k : \tg_\RR \to i \, \tg_k$, which are given explicitly by $\ms X \mapsto \ha (\ms X \pm \sigma^k \tau \ms X)$ respectively.

We denote the eigenspaces of automorphism $\sigma \in \Aut \tg$ by $\tg_{(j), \CC} \coloneqq \{ \ms X \,|\, \sigma \ms X = \omega^j \ms X \}$ for each $j \in \ZZ_T$. By using the definition of $\sigma \in \Aut \tg$ these can be described explicitly as follows. For $j \neq 0$, $\tg_{(j), \CC}$ coincides with the subspace $\g_{(j), \CC} \otimes \CC[t, t^{-1}]$ of $\tg$ whereas $\tg_{(0), \CC}$ can be identified with the subspace $\g_{(0), \CC} \otimes \CC[t, t^{-1}] \oplus \CC \cent \oplus \CC \cocent$. We have the eigenspace decomposition
\begin{equation} \label{tg eigen decomp}
\tg = \bigoplus_{j=0}^{T-1} \tg_{(j), \CC}.
\end{equation}
It follows from the assumption \eqref{tau fix gj assume} that $\tau \in \bAut_- \tg$ preserves each of these eigenspaces, namely $\tau \tg_{(j), \CC} = \tg_{(j), \CC}$. Defining corresponding real subspaces $\tg_{(j)} \coloneqq \tg_{(j), \CC} \cap \tg_0$ we have
\begin{equation} \label{tg0 eigen decomp}
\tg_0 = \bigoplus_{j=0}^{T-1} \tg_{(j)}.
\end{equation}
We denote the projections relative to this decomposition by
\begin{equation*}
\proj_{(j)} : \tg_0 \longrightarrow \tg_{(j)}, \qquad \ms X \longmapsto \frac{1}{T} \sum_{k=0}^{T-1} \omega^{-k j} \sigma^k \ms X.
\end{equation*}
We will also need the projections $\proj_{(j)} : \tg \to \tg_{(j), \CC}$ defined by the same formulae.

The bilinear pairing \eqref{ip on tg} is both $\sigma$- and $\tau$-invariant in the sense that
\begin{equation} \label{ip on tg sigma tau}
( \sigma \ms X | \sigma \ms Y ) = ( \ms X | \ms Y ), \qquad
( \tau \ms X | \tau \ms Y ) = \overline{( \ms X | \ms Y)},
\end{equation}
for any $\ms X, \ms Y \in \tg$. The following lemma is proved using \eqref{ip on tg sigma tau} in exactly the same way as Lemma \ref{lem: gj and gkp} in the finite-dimensional case.
Recall the notion of dual pair from \S\ref{app: dual pairs}.

\begin{lemma} \label{lem: tgj and tgkp}
For each $j \in \ZZ_T$, the triple $(\tg_{(-j)}, \tg_{(j)}, (\cdot | \cdot)|_{\tg_{(-j)} \times \tg_{(j)}})$ is a dual pair. In particular, the restriction of \eqref{ip on tg} to $\tg_{(0)}$ is non-degenerate.

For any $k, p \in \ZZ_T$, the triple $(\tg_{k, -p}, \tg_{k, p}, (\cdot | \cdot)|_{\tg_{k, -p} \times \tg_{k, p}})$ is a dual pair. In particular, \eqref{ip on tg} restricts to a non-degenerate invariant symmetric bilinear form on the real form $\tg_k$ for each $k \in \ZZ_T$. \qed
\end{lemma}

\subsubsection{Canonical element} \label{sec: can elem tg}

Given any basis of $\g$, such as $I^a$, $a = 1, \ldots, \dim \g$ which was defined initially as a basis for the real form $\g_0$, we define a corresponding basis of $\tg$ consisting of $I^a_n \coloneqq I^a \otimes t^n$ for $a = 1, \ldots, \dim \g$ and $n \in \ZZ$ together with the elements $\cent$ and $\cocent$. We denote these basis elements of $\tg$ collectively as $I^{\wt a}$. The dual basis of $\tg$ with respect to \eqref{ip on tg} is then given by $I_{a, -n} \coloneqq I_a \otimes t^{-n}$ for $a = 1, \ldots, \dim \g$ and $n \in \ZZ$ together with the elements $\cocent$ and $\cent$. We denote the elements of this basis as $I_{\wt a}$. A basis of the realification $\tg_\RR$ is given by $I^{\wt a}$ and $i I^{\wt a}$.

For any anti-linear map $\tg \to \tg$ of the form $\chi \coloneqq z \, \sigma^k \circ \tau$ with $|z| = 1$ and $k \in \ZZ_T$ we have $\chi^2 = \id$ and $\chi \ms x_n = (\chi \ms x)_{-n}$ for all $\ms x \in \g$ and $n \in \ZZ$. Therefore $\ms x_n + (\chi \ms x)_{-n}$ is $\chi$-invariant. Recall the dual bases \eqref{g kp bases} of the real subspaces $\g_{k, p}$ and $\g_{k, -p}$ for each $k, p \in \ZZ_T$. We introduce the notation $I^a_{k, p, n} \coloneqq I^a_{k, p} \otimes t^n$ and $I_{a; k, p, n} \coloneqq I_{a; k, p} \otimes t^n$ for all $n \in \ZZ$. In terms of these, a basis of the real subspace $\tg_{k, p}$ of the affine Kac-Moody algebra $\tg$ is then given by
\begin{subequations} \label{tg kp bases}
\begin{equation} \label{tg kp basis}
I^a_{k, p, 0}, \qquad \mbox{\small $\frac{1}{\sqrt 2}$} (I^a_{k, p, n} + I^a_{k, p, -n}), \qquad
\mbox{\small $\frac{i}{\sqrt 2}$} (I^a_{k, p, n} - I^a_{k, p, -n}),
\end{equation}
for $n \in \ZZ_{> 0}$ and $a = 1, \ldots, \dim \g$ together with $i \omega^{- kp/2} \cent$ and $i \omega^{- kp/2} \cocent$. We denote the elements of this basis collectively as $I^{\wt a}_{k, p}$. The dual basis of the real space $\tg_{k, -p}$ is
\begin{equation} \label{tg kp dual basis}
I_{a; k, -p, 0}, \qquad \mbox{\small $\frac{1}{\sqrt 2}$} (I_{a; k, -p, n} + I_{a; k, -p, -n}), \qquad
\mbox{\small $\frac{i}{\sqrt 2}$} (I_{a; k, -p, n} - I_{a; k, -p, -n}),
\end{equation}
\end{subequations}
for $n \in \ZZ_{> 0}$ and $a = 1, \ldots, \dim \g$ together with $- i \omega^{kp/2} \cocent$ and $- i \omega^{kp/2} \cent$. We use the notation $I_{\wt a; k, -p}$ for these basis elements.

Similarly, recall the dual bases \eqref{g j bases} of $\g_{(j)}$ and $\g_{(-j)}$ for each $j \in \ZZ_T$, and introduce the notation $I^{(j, \alpha)}_n \coloneqq I^{(j, \alpha)} \otimes t^n$ and $I_{(-j, \alpha), n} \coloneqq I_{(-j, \alpha)} \otimes t^n$ for all $n \in \ZZ$. Define dual bases of $\tg_{(j)}$ and $\tg_{(-j)}$, which we denote respectively by $I^{(j, \wt \alpha)}$ and $I_{(-j, \wt \alpha)}$, as follows. For $j \neq 0$ the basis $I^{(j, \wt \alpha)}$ consists of elements
\begin{subequations} \label{tg j bases}
\begin{equation} \label{tg j basis}
I^{(j, \alpha)}_0, \qquad \mbox{\small $\frac{1}{\sqrt 2}$} (I^{(j, \alpha)}_n + I^{(j, \alpha)}_{-n}), \qquad
\mbox{\small $\frac{i}{\sqrt 2}$} (I^{(j, \alpha)}_n - I^{(j, \alpha)}_{-n}),
\end{equation}
for $n \in \ZZ_{> 0}$ and $\alpha = 1, \ldots, \dim \g_{(j), \CC}$. Its dual basis $I_{(-j, \wt \alpha)}$ in $\tg_{(-j)}$ consists of the dual elements
\begin{equation} \label{tg j dual basis}
I_{(-j, \alpha), 0}, \qquad \mbox{\small $\frac{1}{\sqrt 2}$} (I_{(-j, \alpha), n} + I_{(-j, \alpha), -n}), \qquad
\mbox{\small $\frac{i}{\sqrt 2}$} (I_{(-j, \alpha), n} - I_{(-j, \alpha), -n}),
\end{equation}
\end{subequations}
for $n \in \ZZ_{> 0}$ and $\alpha = 1, \ldots, \dim \g_{(j), \CC}$. The basis $I^{(0, \wt \alpha)}$ of $\tg_{(0)}$ comprises the same elements as in \eqref{tg j basis} with $j = 0$ together with $i \cent$ and $i \cocent$, and its dual basis $I_{(0, \wt \alpha)}$ of $\tg_{(0)}$ consists of \eqref{tg j dual basis} with $j=0$ together with $- i \cocent$ and $- i \cent$.

Consider the subspaces of $\tg$ defined by
\begin{equation} \label{Fil n tg}
\Fil_n \tg \coloneqq \g \otimes t^n \CC[t],
\end{equation}
for each $n \in \ZZ_{\geq 0}$. They define a descending $\ZZ_{\geq 0}$-filtration on $\tg$, denoted $(\Fil_n \tg)_{n \in \ZZ_{\geq 0}}$, in the sense that $\Fil_n \tg \subset \Fil_m \tg$ for all $n \geq m$ in $\ZZ_{\geq 0}$, \emph{i.e.}
\begin{equation*}
\Fil_0 \tg \supset \Fil_1 \tg \supset \Fil_2 \tg \supset \Fil_3 \tg \supset \ldots
\end{equation*}
and $\cap_n \Fil_n \tg = \{ 0 \}$. Note that \eqref{Fil n tg} defines, in fact, a descending $\ZZ_{\geq 0}$-filtration of $\tg$ as a Lie algebra since we have
\begin{equation*}
[\Fil_m \tg, \Fil_n \tg] \subset \Fil_{m+n} \tg
\end{equation*}
for any $m, n \in \ZZ_{\geq 0}$.
It induces descending $\ZZ_{\geq 0}$-filtrations on both the realification $\tg_\RR$ and real subspaces $\tg_{k, p}$, $k, p \in \ZZ_T$ given by $\Fil_n \tg_\RR \coloneqq (\Fil_n \tg)_\RR$ and $\Fil_n \tg_{k, p} \coloneqq \pi^+_{k, p} (\Fil_n \tg)$, respectively. We shall also make use of the `conjugate' descending $\ZZ_{\geq 0}$-filtration of $\tg$ as a Lie algebra defined by the subspaces
\begin{equation} \label{Fil- n tg}
\cFil_n \tg \coloneqq \g \otimes t^{-n} \CC[t^{-1}].
\end{equation}
Note that by definition of $\tau \in \bAut_- \tg$ in \S\ref{sec: tg action of Pi} we have $\cFil_n \tg = \tau(\Fil_n \tg)$.

We endow the tensor product $\tg \otimes \tg$ with a descending $\ZZ_{\geq 0}$-filtration defined by
\begin{equation} \label{Fil tg times tg def}
\Fil_n(\tg \otimes \tg) \coloneqq \Fil_n \tg \otimes \cFil_n \tg + \cFil_n \tg \otimes \Fil_n \tg,
\end{equation}
for each $n \in \ZZ_{\geq 0}$. Note that these subspaces are invariant under the action of $\Pi$ on the tensor product $\tg \otimes \tg$. We define the completed tensor product $\tg \hotimes \tg$ as the completion of $\tg \otimes \tg$ with respect to the associated linear topology, in which the subspaces \eqref{Fil tg times tg def} form a basis of fundamental open neighbourhoods of the origin.
In other words, it is given by the corresponding inverse limit
\begin{equation} \label{inv lim tg otimes tg}
\tg \hotimes \tg \coloneqq \varprojlim \tg \otimes \tg / \Fil_n (\tg \otimes \tg).
\end{equation}
Specifically, the descending $\ZZ_{\geq 0}$-filtration defined by the subspaces \eqref{Fil tg times tg def} gives rise to an inverse system $\big( (\tg \otimes \tg/ \Fil_n(\tg \otimes \tg) )_{n \in \ZZ_{\geq 0}}, (\pi^m_n)_{m \geq n \in \ZZ_{\geq 0}} \big)$ where for each $m \geq n \in \ZZ_{\geq 0}$ we have the canonical linear map
\begin{equation*}
\pi^m_n : \tg \otimes \tg/ \Fil_m(\tg \otimes \tg) \longtwoheadrightarrow \tg \otimes \tg/ \Fil_n(\tg \otimes \tg),\end{equation*}
sending $\ms X \otimes \ms Y + \Fil_m (\tg \otimes \tg)$ for $\ms X, \ms Y \in \tg$ to $\ms X \otimes \ms Y + \Fil_n (\tg \otimes \tg)$. The inverse limit \eqref{inv lim tg otimes tg} then consists of sequences $\big( v_n + \Fil_n (\tg \otimes \tg) \big)_{n \in \ZZ_{\geq 0}} \in (\tg \otimes \tg/ \Fil_n(\tg \otimes \tg) )_{n \in \ZZ_{\geq 0}}$, where $v_n \in \tg \otimes \tg$, such that $\pi^m_n \big( v_m + \Fil_m (\tg \otimes \tg) \big) = v_n + \Fil_n (\tg \otimes \tg)$ for all $m \geq n \in \ZZ_{\geq 0}$.

The canonical element of $\tg$ living in $\tg \hotimes \tg$ is then defined as
\begin{equation} \label{can elem}
\wt C \coloneqq I_{\widetilde{a}} \otimes I^{\widetilde{a}} = \cocent \otimes \cent + \cent \otimes \cocent + \sum_{n \in \mathbb{Z}} I_{a, -n} \otimes I^a_n
\end{equation}
where, as in the finite dimensional case, sums over repeated Lie algebra indices, here $\wt a$, shall always be implicit. The infinite sum over $n \in \ZZ$ is used here to represent the element of the inverse limit \eqref{inv lim tg otimes tg} given by the sequence
\begin{equation*}
\Bigg( \sum_{n = -k + 1}^{k-1} I_{a, -n} \otimes I^a_n + \Fil_k(\tg \otimes \tg) \Bigg)_{k \in \ZZ_{\geq 0}}.
\end{equation*}
Similarly, for each $j \in \ZZ_T$ we let
\begin{equation} \label{can elem j}
\wt C^{(j)} \coloneqq I_{(-j, \wt \alpha)} \otimes I^{(j, \wt \alpha)} = \delta_{(j)} (\cocent \otimes \cent + \cent \otimes \cocent) + \sum_{n \in \mathbb{Z}} I_{(-j, \alpha), -n} \otimes I^{(j, \alpha)}_n,
\end{equation}
where $\delta_{(j)}$ is the periodic Kronecker delta, equal to $1$ if $j \equiv 0 \; \text{mod}\, T$ and $0$ otherwise,
and summation over the repeated indices $\wt \alpha$ and $\alpha$ is implicit. As above, the infinite sum over $n \in \ZZ$ represents an element of the subspace $\tg_{(-j), \CC} \hotimes \tg_{(j), \CC}$ of the completion $\tg \hotimes \tg$. We have the decomposition
\begin{equation} \label{can elem decomp}
\wt C = \sum_{j=0}^{T-1} \wt C^{(j)}.
\end{equation}

The statement of the following lemma uses standard tensor index notation recalled in \S\ref{sec: tensor index}. Specifically, in the notation used there we take $\mathfrak a = \mathfrak b = \tg$ and $\mathfrak A = \CC$ so that we may drop the last tensor factor in $\mathfrak A$.
\begin{lemma} \label{lem: can elem identity}
For any $\ms X \in \tg$ we have
\begin{subequations}
\begin{equation} \label{can elem inv}
\big[ \ms X_\1 + \ms X_\2, \wt C_{\1\2} \big] = 0.
\end{equation}
Moreover, for any $i, j \in \ZZ_T$ and $\ms X \in \tg_{(j), \CC}$ we have
\begin{equation} \label{can elem inv j}
\big[ \ms X_\1, \wt C^{(i+j)}_{\1\2} \big] + \big[ \ms X_\2, \wt C^{(i)}_{\1\2} \big] = 0.
\end{equation}
\end{subequations}
\begin{proof}
Let $\ms X \in \tg$. Since $I^{\wt a}$ and $I_{\wt a}$ are bases of $\tg$ we may write $[\ms X, I^{\wt a}] = {x^{\wt a}}_{\wt c} I^{\wt c}$ and $[\ms X, I_{\wt c}] = {y_{\wt c}}^{\wt a} I_{\wt a}$ where all but finitely many of the coefficients ${x^{\wt a}}_{\wt c}, {y_{\wt c}}^{\wt a} \in \CC$ are non-zero. And by the invariance of the bilinear form \eqref{ip on tg} we have
\begin{equation*}
{x^{\wt a}}_{\wt c} = {x^{\wt a}}_{\wt e} \big( I^{\wt e} \big| I_{\wt c} \big) = \big( [\ms X, I^{\wt a}] \big| I_{\wt c} \big) = - \big( I^{\wt a} \big| [\ms X, I_{\wt c}] \big) = - {y_{\wt c}}^{\wt e} \big( I^{\wt a} \big| I_{\wt e} \big) = - {y_{\wt c}}^{\wt a}.
\end{equation*}
If follows that $[\ms X, I^{\wt a}] \otimes I_{\wt a} = {x^{\wt a}}_{\wt c} I^{\wt c} \otimes I_{\wt a} = - {y_{\wt c}}^{\wt a} I^{\wt c} \otimes I_{\wt a} = - I^{\wt c} \otimes [\ms X, I_{\wt c}]$, or equivalently $[\ms X_\1, \wt C_{\1\2}] = - [\ms X_\2, \wt C_{\1\2}]$, which proves \eqref{can elem inv}.

Now let $\ms X \in \tg_{(j), \CC}$. By the above result we have $[\ms X_\1, \wt C_{\1\2}] + [\ms X_\2, \wt C_{\1\2}] = 0$, which we can decompose using \eqref{can elem decomp} as
\begin{equation*}
\sum_{k = 0}^{T-1} [\ms X_\1, \wt C^{(k)}_{\1\2}] + \sum_{l = 0}^{T-1} [\ms X_\2, \wt C^{(l)}_{\1\2}] = 0.
\end{equation*}
Projecting this identity in the first tensor factor onto the subspace $\tg_{(-i), \CC}$ relative to the decomposition \eqref{tg eigen decomp} we deduce \eqref{can elem inv j}.
\end{proof}
\end{lemma}

\subsubsection{Connections on $S^1$} \label{sec: Conn on S1}

Recall the definition \eqref{hg bg def} of $\bg$ and the canonical map
\begin{equation} \label{rep tg to bg}
\bar\rep : \tg \longtwoheadrightarrow \bg,
\end{equation}
whose restriction to $\hg$ is the homomorphism $\bar\rep|_{\hg} : \hg \twoheadrightarrow \Loop \g$, where from now on we use the isomorphism $\hg/ \CC \cent \SimTo \Loop \g$ implicitly. In particular, we regard $\Loop \g$ as a subalgebra of $\bg$. To make contact in \S\ref{sec: affine Gaudin} with classical field theories on the circle $S^1 \coloneqq \RR/ 2 \pi \ZZ$, in this section we provide concrete realisations of the Lie algebras $\Loop \g$ and $\bg$ respectively in terms of $\g$-valued trigonometric polynomials and connections on $S^1$.

Let $\mathscr T(S^1)$ be the commutative differential $\CC$-algebra of trigonometric polynomials on $S^1$, namely functions $S^1 \to \CC$ of the form $\theta \mapsto P(e^{i \theta})$ with $P$ a Laurent polynomial.
We denote by $\partial : \mathscr T(S^1) \to \mathscr T(S^1)$ the derivation on $\mathscr T(S^1)$ which sends the function $\theta \mapsto P(e^{i \theta})$ to $\theta \mapsto i e^{i \theta} P'(e^{i\theta})$. A basis of $\mathscr T(S^1)$ is $\{ e_n \}_{n \in \ZZ}$ where
\begin{equation*}
e_n : S^1 \longrightarrow \CC, \qquad \theta \longmapsto e^{i n \theta}.
\end{equation*}
Complex conjugation provides $\mathscr T(S^1)$ with an anti-linear involution, which sends the basis element $e_n$ to $e_{-n}$. We equip $\mathscr T(S^1)$ with an action of $\Pi$ by letting $\s$ act trivially and $\t$ act by complex conjugation.

Let $\mathscr T(S^1, \g) \coloneqq \g \otimes \mathscr T(S^1)$ be the space of $\g$-valued trigonometric polynomial on $S^1$. We obtain an action of $\Pi$ on $\mathscr T(S^1, \g)$ by combining the above action on $\mathscr T(S^1)$ with that on $\g$ given in \eqref{hom r g}. A non-degenerate symmetric bilinear form on $\mathscr T(S^1, \g)$ is given by
\begin{equation} \label{ip Conn g}
( A | B )_{S^1} \coloneqq \frac{1}{2 \pi} \int_{S^1} d\theta \langle A(\theta), B(\theta) \rangle,
\end{equation}
for any $A, B \in \mathscr T(S^1, \g)$, where the bilinear form \eqref{ip on g} on $\g$ is extended to a map
\begin{equation*}
\langle \cdot, \cdot \rangle : \mathscr T(S^1, \g) \times \mathscr T(S^1, \g) \longrightarrow \mathscr T(S^1)
\end{equation*}
defined by $\langle \ms x \otimes f, \ms y \otimes g \rangle \coloneqq \langle \ms x, \ms y \rangle fg$ for any $\ms x, \ms y \in \g$ and $f, g \in \mathscr T(S^1)$.

Consider the complex vector space $\Conn_\g(S^1)$ of $\g$-valued connections on $S^1$ of the form $\lvl \partial + A$ where $\lvl \in \CC$ and $A \in \mathscr T(S^1, \g)$. We extend the action of $\Pi$ on $\mathscr T(S^1, \g)$ to $\Conn_\g(S^1)$ by letting it act trivially on the derivative $\partial$ and using (anti-)linearity. We refer to $\lvl \partial + A \in \Conn_\g(S^1)$ as an $\lvl$-connection to emphasise its dependence on the coefficient $\lvl$ of the derivative term. In particular, we may regard an element of $\mathscr T(S^1, \g)$ as defining a $0$-connection in $\Conn_\g(S^1)$. The commutator of two connections then defines a Lie bracket on $\Conn_\g(S^1)$,
\begin{equation*}
[\cdot, \cdot] : \Conn_\g(S^1) \times \Conn_\g(S^1) \longrightarrow \Conn_\g(S^1).
\end{equation*}

\begin{lemma} \label{lem: eval rep}
We have a $\Pi$-equivariant isomorphism $\bg \SimTo \Conn_\g(S^1)$, under which the bilinear form \eqref{ip Conn g} on $\mathscr T(S^1, \g)$ corresponds to that on $\Loop \g$. Its composition with \eqref{rep tg to bg} is the Lie algebra homomorphism $\rep : \tg \to \Conn_\g(S^1)$ given by
\begin{equation} \label{eval rep}
\rep(\cocent) = - i \partial, \qquad
\rep(\cent) = 0, \qquad
\rep(\ms x_n) = \ms x \otimes e_n,
\end{equation}
for any $\ms x \in \g$ and $n \in \ZZ$.
\begin{proof}
The isomorphism $\CC[t, t^{-1}] \SimTo \mathscr T(S^1)$, $t^n \mapsto e_n$ given by the change of variable $t = e^{i \theta}$ extends to a Lie algebra isomorphism $\Loop \g \SimTo \mathscr T(S^1, \g)$. It commutes with the action of the derivations $t \partial_t$ and $- i \partial$ on $\Loop \g$ and $\mathscr T(S^1, \g)$ respectively, so it further extends to an isomorphism $\bg \SimTo \Conn_\g(S^1)$ by letting $\cocent \mapsto - i \partial$.
And performing the change of variable $t = e^{i \theta}$ in the bilinear form \eqref{ip Conn g} we get
\begin{equation*}
\frac{1}{2 \pi} \int_{S^1} d\theta \big\langle \ms x \otimes P(e^{i \theta}) , \ms y \otimes Q(e^{i \theta}) \big\rangle
= \res_0 \frac{d t}{t} \big\langle \ms x \otimes P(t) , \ms y \otimes Q(t) \big\rangle
\end{equation*}
for any $\ms x, \ms y \in \g$ and Laurent polynomials $P, Q$, which is the bilinear form on $\Loop \g$.
\end{proof}
\end{lemma}

The vector space $\Conn_\g(S^1)$ is endowed with a pair of descending $\ZZ_{\geq 0}$-filtrations defined as the images of the subspaces $(\Fil_n \tg)_{n \in \ZZ_{\geq 0}}$ and $(\cFil_n \tg)_{n \in \ZZ_{\geq 0}}$ under the linear map $\rep$ from Lemma \ref{lem: eval rep}. Concretely, the subspace $\Fil_n(\Conn_\g(S^1))$ (resp. $\cFil_n(\Conn_\g(S^1))$) for $n \in \ZZ_{\geq 0}$ is spanned by $\ms x \otimes e_m$ (resp. $\ms x \otimes e_{-m}$) with $\ms x \in \g$ and $m \geq n$. Since $\rep \otimes \rep$ is continuous it extends to a linear map
\begin{equation*}
\rep \otimes \rep : \tg \hotimes \tg \longrightarrow \Conn_\g(S^1) \hotimes \Conn_\g(S^1).
\end{equation*}
We shall need the image of the canonical element $\widetilde C$ under this map. It follows from the form of $\wt C$ in \eqref{can elem} that its image in fact lies in the subspace
\begin{equation*}
\mathscr T(S^1, \g) \hotimes \mathscr T(S^1, \g) \cong_\CC \g \otimes \g \otimes \mathscr T(S^1) \hotimes \mathscr T(S^1),
\end{equation*}
where the pair of conjugate descending $\ZZ_{\geq 0}$-filtrations on $\mathscr T(S^1)$ are given by the subspaces $\Fil_n(\mathscr T(S^1))$ and $\cFil_n(\mathscr T(S^1))$ for $n \in \ZZ_{\geq 0}$ with bases $\{ e_{\pm m} \}_{m \in \ZZ_{\geq n}}$, respectively. Given any element $\kappa \in \mathscr T(S^1) \hotimes \mathscr T(S^1)$, for each $\theta \in S^1$ we can regard $\kappa(\theta, \cdot)$ as the kernel of a formal distribution on $S^1$, in the sense that it provides a well defined linear map
\begin{equation*}
\mathscr T(S^1) \longrightarrow \CC, \qquad
f \longmapsto \frac{1}{2 \pi} \int_{S^1} d\theta' \kappa(\theta, \theta') f(\theta').
\end{equation*}
In particular, the element $(\rep \otimes \rep) \wt C \in \g \otimes \g \otimes \mathscr T(S^1) \hotimes \mathscr T(S^1)$ given explicitly by
\begin{equation} \label{wt C delta}
\big( (\rep \otimes \rep) \widetilde{C} \big)(\theta, \theta') = \sum_{n \in \ZZ} I_a \otimes I^a e_{-n}(\theta) e_n(\theta') = C \, \delta_{\theta \theta'}
\end{equation}
is related to the Dirac $\delta$-distribution $\delta \coloneqq \sum_{n \in \ZZ} e_n \otimes e_{-n} \in \mathscr T(S^1) \hotimes \mathscr T(S^1)$. For any $\theta \in S^1$, the expression $\delta(\theta, \cdot)$ is the kernel of the distribution on $S^1$ sending the test function $f \in \mathscr T(S^1)$ to $f(\theta)$. Note that we use the notation $\delta_{\theta \theta'}$ instead of $\delta(\theta, \theta')$.

\section{Double loop algebras and Takiff algebras} \label{sec: Double loop}

Consider the Riemann sphere $\CP \coloneqq \CC \cup \{ \infty \}$ and fix a global coordinate $z$ on $\CC$. At each finite point $x \in \CC \subset \CP$ we have the local coordinate $\xi_x \coloneqq z - x$ and at the point $\infty \in \CP$ a local coordinate is given by $\xi_\infty \coloneqq z^{-1}$.

We denote by $\CC[t]$ the ring of polynomials in $t$, by $\CC\{ t \}$ the ring of convergent power series in $t$ and by $\CC(\!\{ t \}\!)$ the ring of convergent Laurent series in $t$, \emph{i.e.} $f(t) \in \CC(\!\{ t \}\!)$ if and only if $t^k f(t) \in \CC\{ t \}$ for some $k \in \ZZ_{\geq 0}$.

For each $x \in \CP$, we let $\sO_x \coloneqq \CC \{ \xi_x \}$ denote the local ring of germs of holomorphic functions at $x$, \emph{i.e.} the ring of convergent power series in $\xi_x$. Denote by $\m_x \coloneqq \xi_x \CC \{ \xi_x \}$ the maximal ideal of $\sO_x$ consisting of germs of holomorphic functions at $x$ which vanish at $x$. The ring $\sO_x$ has a natural descending $\ZZ_{\geq 0}$-filtration
\begin{equation*}
\sO_x = \m^0_x \supset \m_x \supset \m^2_x \supset \m^3_x \supset \ldots
\end{equation*}
where $\m^n_x \coloneqq \xi_x^n \CC\{ \xi_x \}$ for any $n \in \ZZ_{\geq 0}$. Denote by $\sK_x \coloneqq \CC (\!\{ \xi_x \}\!)$ the field of germs of meromorphic functions at $x$, \emph{i.e.} the field of convergent Laurent series in $\xi_x$. It has a natural descending $\ZZ$-filtration which by abuse of notation we also denote $\m^{\bullet}_x$, namely
\begin{equation*}
\ldots \supset \m^{-3}_x \supset \m^{-2}_x \supset \m^{-1}_x \supset \sO_x \supset \m_x \supset \m^2_x \supset \m^3_x \supset \ldots
\end{equation*}
where we extend the notation $\m^n_x$, $n \in \ZZ_{\geq 0}$ introduced above by letting $\m^k_x \coloneqq \xi_x^k \CC\{ \xi_x \}$ for any $k \in \ZZ$. Let $\p_x \coloneqq \xi_x^{-1} \CC[\xi_x^{-1}]$ denote the set of principal parts at $x$, which forms a ring without identity. If we choose to also include the constant term in the principal part then we obtain the corresponding ring $\p^0_x \coloneqq \CC[\xi_x^{-1}]$. This ring also has a natural descending $\ZZ_{\geq 0}$-filtration
\begin{equation*}
\p^0_x \supset \p_x \supset \p^2_x \supset \p^3_x \supset \ldots
\end{equation*}
where we use the notation $\p^n_x \coloneqq \xi_x^{-n} \CC[\xi_x^{-1}]$ defined for any $n \in \ZZ$.

We define an injective homomorphism
\begin{equation} \label{Mobius rep}
\mu : \Pi \longhookrightarrow \bAut \CP,
\qquad
\alpha \longmapsto \mu_\alpha
\end{equation}
of $\Pi$ into the full M\"obius group of holomorphic and anti-holomorphic automorphisms of $\CP$ by letting $\mu_\s : z \mapsto \omega z$ and $\mu_\t : z \mapsto \bar z$. The image of the cyclic subgroup $\Gamma$ consists of M\"obius transformations and the image of the coset $\Gamma \t$ consists of orientation-reversing M\"obius transformations. Given any point
$x \in \CP$ we let $\Pi_x \subset \Pi$ denote its stabilizer under the action \eqref{Mobius rep}. We will refer to $x \in \CP$ as a \emph{complex} point if $\Pi_x$ is trivial and as a \emph{real} point if $\Pi_x \cap \Gamma \t \neq \emptyset$. Specifically, the set of all real points is formed of the union $\{ \infty \} \cup \bigcup_{k \in \ZZ_T} \omega^{k/2} \RR$ since $\Pi_x = \langle \s^k \t \rangle$ for any $x \in \omega^{k/2} \RR \setminus \{ 0 \}$ with $k \in \ZZ_T$ and $\Pi_x = \Pi$ for any $x \in \{ 0, \infty \}$. Moreover, the set of complex points is the complement of the set of real points. We refer to $\{ 0, \infty \}$ as the set of \emph{fixed} points. This terminology reflects the fact that when $T \in \ZZ_{> 1}$ we have $\Pi_x = \Pi$ if and only if $x \in \{ 0, \infty \}$. However, by convention we will still refer to $0$ and $\infty$ as fixed points even when $T = 1$.

Let $N \in \ZZ_{\geq 0}$. Pick and fix a finite set $\z \coloneqq \{ z_1, \ldots, z_N, \infty \} \subset \CP$ which includes the point at infinity labelled as $z_{N+1} = \infty$. We will assume that the $\Pi$-orbits of the points in $\z$ are all disjoint, in other words $z_i \neq \mu_\alpha \, z_j$ for all $\alpha \in \Pi$ and $i \neq j$. Denote by $\z_{\rm c}$ the subset of complex points in $\z$ and by $\z_{\rm r}$ the subset of real points, so that in particular $\infty \in \z_{\rm r}$ and we have the disjoint union $\z = \z_{\rm c} \sqcup \z_{\rm r}$. We also define the subsets
\begin{equation*}
\z_{\rm f} \coloneqq \z \cap \{ 0, \infty \}, \qquad
\z'_{\rm r} \coloneqq \z_{\rm r} \setminus \{ 0, \infty \}, \qquad
\z^k_{\rm r} \coloneqq \{ x \in \z'_{\rm r} \,|\, \Pi_x = \langle \s^k \t \rangle \}
\end{equation*}
corresponding respectively to fixed points, to real non-fixed points and to real points with stabiliser $\langle \s^k \t \rangle$. The disjoint union decomposition of $\z$ may be refined as
\begin{equation*}
\z = \z_{\rm c} \sqcup \z'_{\rm r} \sqcup \z_{\rm f} = \z_{\rm c} \sqcup \bigsqcup_{k \in \ZZ_T} \z^k_{\rm r} \sqcup \z_{\rm f}.
\end{equation*}
For any finite subset $\bm x \subset \CP$ we also introduce the notation $\bar {\bm x} \coloneqq \{ x \in \CP \,|\, \bar x \in \bm x \}$.

Let $\tg$ be an affine Kac-Moody algebra and suppose it is equipped with an action of the dihedral group $\Pi$ as in \S\ref{sec: aff KM}.

\subsection{`Local' Lie algebras}

For any $x \in \CP$, we define an action of $\Pi_x$ on the field $\sK_x$ as follows. Given a germ $[f]_x \in \sK_x$ we choose a representative $f : D_x \to \CC$ on a small disc $D_x$ around $x$. We set
\begin{equation*}
\alpha . [f]_x \coloneqq \left\{ \begin{array}{ll} [f \circ \mu_\alpha^{-1}]_x \; , & \quad \text{if}\;\; \alpha \in \Pi_x \cap \Gamma\\[1mm]
[ c \circ f \circ \mu_\alpha^{-1} ]_x \; , & \quad \text{if} \;\; \alpha \in \Pi_x \cap \Gamma \t,
\end{array}
\right.
\end{equation*}
where $c : \CC \to \CC$ denotes complex conjugation and $\mu_\alpha : D_x \to D_x$ is the restriction of $\mu_\alpha \in \bAut \CP$ to the open disc $D_x$. This defines a left action of $\Pi_x$ on $\sK_x$ which also preserves its $\ZZ$-filtration, in the sense that $\alpha . \m^k_x \subset \m^k_x$ for any $k \in \ZZ$.
By combining the actions of $\Pi_x \subset \Pi$ on $\tg$ and $\sK_x$ we obtain a natural action on the tensor product $\tg \otimes \sK_x$. Explicitly, we define a homomorphism
\begin{equation} \label{Mobius rep g K}
\Pi_x \longhookrightarrow \bAut (\tg \otimes \sK_x), \qquad
\alpha \longmapsto \hat \alpha
\end{equation}
by letting, for any $\alpha \in \Pi_x$, $\ms X \in \tg$ and $[f]_x \in \sK_x$,
\begin{equation} \label{Gamma action}
\hat \alpha (\ms X \otimes [f]_x) \coloneqq r_\alpha \ms X \otimes \alpha . [f]_x.
\end{equation}
It induces a homomorphism $\Pi_x \hookrightarrow \Aut (\tg \otimes \sK_x)_\RR$ into $\RR$-linear automorphisms of the realification $(\tg \otimes \sK_x)_\RR$.

To each $x \in \z$ we attach the real Lie algebra
\begin{equation*}
\Lg_x \coloneqq \big( (\tg \otimes \sK_x)_\RR \big)^{\Pi_x}.
\end{equation*}
Explicitly, to any complex point $x \in \z_{\rm c}$ we attach the realification $\Lg_x = (\tg \otimes \sK_x)_\RR$ and to a non-fixed real point $x \in \z^k_{\rm r}$ we attach the real form $\Lg_x = (\tg \otimes \sK_x)^{\hat \s^k \circ \hat \t}$.
If $0 \in \z$ then we attach to it the $\Pi$-invariant subalgebra $\Lg_0 = (\tg \otimes \sK_0)^{\Pi}$. Similarly, to the point at infinity, which by assumption always belongs to $\z$, we also attach the $\Pi$-invariant subalgebra $\Lg_{\infty} = (\tg \otimes \sK_{\infty})^{\Pi}$. Define the direct sum of real Lie algebras
\begin{equation} \label{Gaudin symmetry alg}
\Lg_\z \coloneqq \bigoplus_{x \in \z} \Lg_x.
\end{equation}

We also introduce the Lie subalgebras $\Lg_x^+ \coloneqq ( (\tg \otimes \sO_x)_\RR )^{\Pi_x} \subset \Lg_x$ at every finite point $x \in \z \setminus \{ \infty \}$. In particular, if $0 \in \z$ then $\Lg_0^+ = (\tg \otimes \sO_0)^{\Pi}$. However, for reasons to be clarified in the next subsection, cf. Lemma \ref{lem: Gamma decomp}, at infinity we consider instead the Lie subalgebra $\Lg_{\infty}^+ \coloneqq (\tg \otimes \m_{\infty})^{\Pi}$ of $\Lg_\infty$. We set
\begin{equation} \label{gD+}
\Lg_\z^+ \coloneqq \bigoplus_{x \in \z} \Lg_x^+.
\end{equation}

There is a natural complementary subalgebra to $\Lg_{\z}^+$ in $\Lg_{\z}$ defined as follows. To every finite point $x \in \z \setminus \{ \infty \}$ we attach the Lie subalgebra $\Lg^-_x \coloneqq ( (\tg \otimes \p_x)_\RR)^{\Pi_x} \subset \Lg_x$ and similarly we define $\Lg^-_{\infty} \coloneqq (\tg \otimes \p^0_\infty)^\Pi$ for the point at infinity. Recall here that $\p^0_\infty = \CC[\xi_\infty^{-1}]$ includes the constant term. In particular, we then have the direct sum decomposition of linear spaces $\Lg_x = \Lg^+_x \dotplus \Lg^-_x$ for any $x \in \z$. We define the direct sum of Lie subalgebas
\begin{equation} \label{gD-}
\Lg_\z^- \coloneqq \bigoplus_{x \in \z} \Lg_x^-,
\end{equation}
so that we have the direct sum decomposition $\Lg_{\z} = \Lg^+_{\z} \dotplus \Lg^-_{\z}$. We will be interested in a different complement of $\Lg^+_\z$ in $\Lg_\z$ provided by Lemma \ref{lem: Gamma decomp} below.

\subsection{`Global' Lie algebras} \label{sec: global}

Given a finite subset $S \subset \CP$ such that $\z \subset S$, we denote by $R_S$ the ring of meromorphic functions on $\CP$ with poles contained in $S$. We let $R_S(\tg) \coloneqq \tg \otimes R_S$ be the corresponding Lie algebra of $\tg$-valued meromorphic functions on $\CP$ with poles in $S$. For each $x \in \z$ there is an injective homomorphism of rings $\iota_x : R_S \hookrightarrow \sK_x$ which assigns to a meromorphic function $f \in R_S$ its germ $[f]_x \in \sK_x$ at $x$. Correspondingly, there is an embedding of Lie algebras
\begin{equation} \label{R_S embed}
\iota_x : R_S(\tg) \longhookrightarrow \bigoplus_{x \in \z} \tg \otimes \sK_x
\end{equation}
which assigns to any meromorphic function $\ms X \otimes f \in R_S(\tg)$, where $\ms X \in \tg$ and $f \in R_S$, the set of its germs $\iota_x (\ms X \otimes f) \coloneqq \ms X \otimes [f]_x \in \tg \otimes \sK_x$ at the points $x \in \z$.

Consider the set
\begin{equation*}
\Pi \z \coloneqq \{ \mu_\alpha x \,|\, \alpha \in \Pi, x \in \z \}.
\end{equation*}
We define an action of $\Gamma$ on $R_{\Pi \z}$ by setting $\alpha . f \coloneqq f \circ \mu_\alpha^{-1}$ for any $f \in R_{\Pi \z}$ and $\alpha \in \Gamma$. This lifts to an action of $\Pi$ by letting $\t$ act as $\t . f \coloneqq c \circ f \circ \mu_\t \eqqcolon \bar f$ on any $f \in R_{\Pi \z}$. We therefore obtain an action of $\Pi$ on $R_{\Pi \z}(\tg) = \tg \otimes R_{\Pi \z}$,
\begin{equation} \label{Mobius rep Rg}
\Pi \longhookrightarrow \bAut R_{\Pi \z}(\tg), \qquad
\alpha \longmapsto \hat \alpha
\end{equation}
where the action of $\alpha \in \Pi$ is given explicitly by
\begin{equation} \label{Gamma action global}
\hat \alpha (\ms X \otimes f) \coloneqq r_\alpha \ms X \otimes \alpha . f,
\end{equation}
for $\ms X \in \tg$ and $f \in R_{\Pi \z}$.
Define the real Lie algebra of $\Pi$-invariants
\begin{equation} \label{twisted functions}
R^{\Pi}_{\z}(\tg) \coloneqq R_{\Pi \z}(\tg)^{\Pi}.
\end{equation}

The invariance property under the action \eqref{Gamma action global} may be equivalently rephrased as follows. Let $\bar R_{\Pi \z}$ be the ring of anti-meromorphic functions on $\CP$ with poles contained in $\Pi \z$, and define the corresponding Lie algebra $\bar R_{\Pi \z}(\tg) \coloneqq \tg \otimes \bar R_{\Pi \z}$ of $\tg$-valued anti-meromorphic functions. For any $\alpha \in \Gamma$ we extend $r_\alpha \in \Aut \tg$ to a linear map
\begin{equation*}
r_\alpha : R_{\Pi \z}(\tg) \longrightarrow R_{\Pi \z}(\tg), \qquad
r_\alpha (\ms X \otimes f) = r_\alpha \ms X \otimes f.
\end{equation*}
Similarly, for any $\alpha \in \Gamma \t$ we extend $r_\alpha \in \bAut_- \tg$ to an anti-linear map
\begin{equation*}
r_\alpha : R_{\Pi \z}(\tg) \longrightarrow \bar R_{\Pi \z}(\tg), \qquad
r_\alpha(\ms X \otimes f) = r_\alpha \ms X \otimes c \circ f.
\end{equation*}
Now for any $\alpha \in \Gamma$ we define a linear map $\mu_\alpha^\ast : R_{\Pi \z}(\tg) \to R_{\Pi \z}(\tg)$ using the pullback by $\mu_\alpha$ on the second tensor factor. On the other hand, the pullback by $\mu_\alpha$ for $\alpha \in \Gamma \t$ defines instead a linear map $\mu_\alpha^\ast : R_{\Pi \z}(\tg) \to \bar R_{\Pi \z}(\tg)$.
By combining the above, we can then describe the $\Pi$-invariant subalgebra \eqref{twisted functions} equivalently as
\begin{equation} \label{equivariance Rtg}
R^{\Pi}_{\z}(\tg) = \{ F \in R_{\Pi \z}(\tg) \,|\, r_\alpha F = \mu^\ast_\alpha F \; \text{for all} \; \alpha \in \Pi \}.
\end{equation}

Restricting the embedding \eqref{R_S embed} with $S = \Pi \z$ to the latter we obtain an embedding
\begin{equation} \label{iota map}
\iota_{\z} \coloneqq (\iota_{z_1}, \ldots, \iota_{z_N}, \iota_{\infty}) : R^\Pi_{\z}(\tg) \longhookrightarrow \Lg_{\z}.
\end{equation}
In what follows we shall often regard $R^\Pi_{\z}(\tg)$ as a subalgebra of $\Lg_{\z}$ by identifying an element $X \in R^\Pi_{\z}(\tg)$ with its image $\iota_{\z} X = (\iota_x X)_{x \in \z} \in \Lg_\z$ under the $\iota_{\z}$-map \eqref{iota map}.

\begin{lemma} \label{lem: Gamma decomp}
We have the direct sum decomposition of real vector spaces
\begin{equation} \label{Rzk compl}
\Lg_{\z} = \Lg_{\z}^+ \dotplus R^\Pi_{\z}(\tg).
\end{equation}
\begin{proof}
Let $X = (X_x)_{x \in \z} \in \Lg_\z$. For each $x \in \z$ we let $X^-_x \in \Lg^-_x$ denote the principal part of $X_x \in \Lg_x$, which we regard as an element of $R_{\Pi \z}(\tg)$ with a single pole at $x$.
Note that $X^-_{\infty}$ includes the constant term. Set
\begin{equation*}
F_X \coloneqq \sum_{\alpha \in \Pi} \sum_{x \in \z} \frac{1}{|\Pi_x|} \hat \alpha X^-_x \in R^\Pi_\z(\tg).
\end{equation*}
By construction, the principal part in $\Lg^-_x$ of the germ $\iota_x F_X$ at $x \in \z$ agrees with $X^-_x \in \Lg^-_x$. Therefore $X$ can be written uniquely as a sum of $F_X \in R^\Pi_\z(\tg)$, or rather its image $\iota_{\z} F_X \in \Lg_{\z}$ under \eqref{iota map}, and $( X_x - \iota_x F_X )_{x \in \z} \in \Lg^+_{\z}$, as required.
\end{proof}
\end{lemma}

\subsection{Dual spaces} \label{sec: dual spaces}

For any $x \in \CP$ we define an action of the stabilizer subgroup $\Pi_x \subset \Pi$ on the space of germs of meromorphic differenials $\sK_x d\xi_x$. We let $\alpha \in \Pi_x \cap \Gamma$ act on $[\varpi]_x \in \sK_x d\xi_x$ via pullback by $\mu_\alpha^{-1}$, namely $\alpha . [\varpi]_x \coloneqq \big[ (\mu_\alpha^{-1})^\ast \varpi \big]_x$, whereas if $\alpha \in \Pi_x \cap \Gamma \t$ then $\mu_\alpha$ is orientation-reversing and so we set $\alpha . [\varpi]_x \coloneqq \big[ c \circ (\mu_\alpha^{-1})^\ast \varpi \big]_x$ instead. Define an action of $\Pi_x$ on the tensor product $\tg \otimes \sK_x d\xi_x$,
\begin{equation} \label{Mobius rep g Kdz}
\Pi_x \longhookrightarrow \bAut (\tg \otimes \sK_x d\xi_x), \qquad
\alpha \longmapsto \hat \alpha
\end{equation}
given for any $\alpha \in \Pi_x$, $\ms X \in \tg$ and $[\varpi]_x \in \sK_x d\xi_x$ by
\begin{equation*}
\hat \alpha (\ms X \otimes [\varpi]_x) \coloneqq r_\alpha \ms X \otimes \alpha . [ \varpi ]_x.
\end{equation*}

To any point $x \in \z$ we attach the real subspace of $\Pi_x$-invariants in the realification $(\tg \otimes \sK_x d\xi_x)_\RR$, namely
\begin{equation*}
\Og_x \coloneqq \big( (\tg \otimes \sK_x d\xi_x)_\RR \big)^{\Pi_x}.
\end{equation*}
Define also the subspace $\Og^+_x \coloneqq ( (\tg \otimes \sO_x d\xi_x)_\RR )^{\Pi_x}$ for each $x \in \z \setminus \{ \infty \}$ and at infinity we define $\Og^+_\infty \coloneqq (\tg \otimes \m_\infty^{-1} d\xi_\infty)^\Pi$. These both have natural complementary subspaces in $\Og_x$ given respectively by $\Og^-_x \coloneqq ( (\tg \otimes \p_x d\xi_x)_\RR )^{\Pi_x}$ and $\Og^-_\infty \coloneqq (\tg \otimes \p^2_\infty d\xi_\infty)^\Pi$. We introduce the direct sums
\begin{equation*}
\Og_{\z} \coloneqq \bigoplus_{x \in \z} \Og_x, \qquad
\Og^+_{\z} \coloneqq \bigoplus_{x \in \z} \Og^+_x.
\end{equation*}

We shall also need the space of globally defined $\Pi$-invariant $\tg$-valued meromorphic differentials on $\CP$ with poles contained in the set $\Pi \z$.
Given a finite subset $S \subset \CP$ containing $\z$, we let $\Omega_S$ denote the space of meromorphic differentials on $\CP$ with poles at most in $S$. The differential $dz$, where $z$ is the global coordinate on $\CC$, provides an $R_S$-basis for $\Omega_S$ since any $\varpi \in \Omega_S$ can be written as $\varpi = f dz$ for some $f \in R_S$. In the case $S = \Pi \z$, the group $\Pi$ acts on the space $\Omega_{\Pi \z}$ by letting $\alpha \in \Gamma$ act as the pullback by the inverse of the multiplication map $\mu_\alpha : \CP \to \CP$ defined in \eqref{Mobius rep} and letting $\t$ send $\varpi = f dz$ to $\t . \varpi = \bar f dz$ where $\bar f = c \circ f \circ \mu_\t$. In particular, this allows us to define an action
\begin{subequations} \label{Mobius rep Og}
\begin{equation}
\Pi \longhookrightarrow \bAut \Omega_{\Pi \z}(\tg), \qquad
\alpha \longmapsto \hat \alpha
\end{equation}
on the tensor product $\Omega_{\Pi \z}(\tg) \coloneqq \tg \otimes \Omega_{\Pi \z}$, given explicitly by
\begin{equation} \label{Gamma action diff}
\hat \alpha (\ms X \otimes \varpi) \coloneqq r_\alpha \ms X \otimes \alpha . \varpi
\end{equation}
\end{subequations}
for any $\alpha \in \Pi$, $\ms X \in \tg$ and $\varpi \in \Omega_{\Pi \z}$. Define the real vector space of $\Pi$-invariants
\begin{equation} \label{invariant differentials}
\Omega^\Pi_{\z}(\tg) \coloneqq \Omega_{\Pi \z}(\tg)^\Pi.
\end{equation}

This subspace can alternatively be described in a similar fashion to the subalgebra of $\Pi$-invariant $\tg$-valued rational functions \eqref{equivariance Rtg}. For this we introduce the space $\bar \Omega_{\Pi \z}$ of anti-meromorphic differentials with poles contained in $\Pi \z$, of the form $\varphi = f d\bar z$ for some $f \in \bar R_{\Pi \z}$. We define also the corresponding space $\bar \Omega_{\Pi \z}(\tg) \coloneqq \tg \otimes \bar \Omega_{\Pi \z}$. The map $r_\alpha \in \bAut \tg$ extends to a linear map $\Omega_{\Pi \z}(\tg) \to \Omega_{\Pi \z}(\tg)$ for $\alpha \in \Gamma$ and to an anti-linear map $\Omega_{\Pi \z}(\tg) \to \bar \Omega_{\Pi \z}(\tg)$ for $\alpha \in \Gamma \t$ in the same way as done in \S\ref{sec: global} for $R_{\Pi \z}(\tg)$. If we also define linear maps $\mu^\ast_\alpha : \Omega_{\Pi \z}(\tg) \to \Omega_{\Pi \z}(\tg)$ (resp. $\mu^\ast_\alpha : \Omega_{\Pi \z}(\tg) \to \bar \Omega_{\Pi \z}(\tg)$) for each $\alpha \in \Gamma$ (resp. $\alpha \in \Gamma \t$), then the real vector space \eqref{invariant differentials} may be equivalently described as
\begin{equation} \label{equivariance Otg}
\Omega^{\Pi}_{\z}(\tg) = \{ \Phi \in \Omega_{\Pi \z}(\tg) \,|\, r_\alpha \Phi = \mu^\ast_\alpha \Phi \; \text{for all} \; \alpha \in \Pi \}.
\end{equation}

Just as in \eqref{iota map}, we have an injective map
\begin{equation} \label{iota map Om}
\iota_{\z} \coloneqq (\iota_{z_1}, \ldots, \iota_{z_N}, \iota_{\infty}) : \Omega^\Pi_{\z}(\tg) \longhookrightarrow \Og_{\z}
\end{equation}
which assigns to a meromorphic differential in $\Omega^\Pi_{\z}(\tg)$ the set of its germs at the points in $\z$. In what follows we will also often implicitly identify an element $\Phi \in \Omega^\Pi_{\z}(\tg)$ with its image $\iota_{\z} \Phi = (\iota_x \Phi)_{x \in \z} \in \Og_{\z}$ under \eqref{iota map Om}.

The proof of the following is completely analogous to that of Lemma \ref{lem: Gamma decomp}. See also Lemma \ref{lem: ann Lg+} below.
\begin{lemma} \label{lem: Gamma decomp Om}
We have the direct sum decomposition of real vector spaces
\begin{equation*}
\pushQED{\qed}
\Og_\z = \Og_\z^+ \dotplus \Omega^\Pi_\z(\tg). \qedhere
\popQED 
\end{equation*}
\end{lemma}

\subsubsection{Dual pairing}

Let $\Re : \CC \to \RR$, $u \mapsto \Re(u)$ and $\Im : \CC \to \RR$, $u \mapsto \Im(u)$ denote the maps which return the real and imaginary parts of a complex number, respectively.
By combining the non-degenerate bilinear form on $\tg$ with the residue pairing, we can define a bilinear form $\lbf{\cdot, \cdot} : \Og_{\z} \times \Lg_{\z} \to \RR$ as
\begin{equation} \label{gD pairing}
\lbf{\Phi, X} \coloneqq \Re \left( \sum_{x \in \z} \frac{2T}{|\Pi_x|} \res_x \, ( \Phi_x | X_x ) \right)
\end{equation}
for any $\Phi = (\Phi_x)_{x \in \z} \in \Og_\z$ and $X = (X_x)_{x \in \z} \in \Lg_\z$.

In what follows we make use of standard results on dual pairs recalled in \S\ref{app: dual pairs}.

\begin{lemma} \label{lem: main dual pair}
The triple $(\Og_{\z}, \Lg_{\z}, \lbf{\cdot, \cdot})$ is a dual pair.
\begin{proof}
To show the non-degeneracy on the left of the bilinear form \eqref{gD pairing}, let $\Phi \in \Og_\z$ be such that $\lbf{\Phi, X} = 0$ for all $X \in \Lg_\z$ and suppose, for a contradiction, that $\Phi$ is non-zero. Then $\Phi_x$ is non-zero for some $x \in \z$. We treat the three cases when $x \in \z_{\rm c}$, $x \in \z'_{\rm r}$ and $x \in \z_{\rm f}$ separately.

Suppose first that the germ $\Phi_x$ is non-zero for some $x \in \z_{\rm c}$. Let $\ms Y \otimes \xi^n_x d\xi_x$ for some $n \in \ZZ$ and non-zero $\ms Y \in \tg$ be its most singular term. We choose $X = (X_y)_{y \in \z} \in \Lg_\z$ such that $X_y = 0$ for all $y \neq x$ and $X_x = \ms X \otimes \xi^{-n-1}_x$ with $\ms X \in \tg$. Then
\begin{equation*}
0 = \lbf{\Phi, X} = 2 T \Re (\ms Y|\ms X), \qquad \text{and} \qquad 0 = \lbf{\Phi, i X} = - 2 T \Im (\ms Y|\ms X).
\end{equation*}
where the first equalities are by the assumption on $\Phi \in \Og_\z$. Hence $(\ms Y|\ms X) = 0$. Since $\ms X \in \tg$ was arbitrary, by the non-degeneracy of the bilinear form on $\tg$ it follows that $\ms Y = 0$, which is a contradiction.

Next, suppose $\Phi_x$ is non-zero for some $x \in \z'_{\rm r}$, specifically with $\Pi_x = \langle \s^k \t \rangle$, $k \in \ZZ_T$. Again let $\ms Y \otimes \xi^n_x d\xi_x$ be its most singular term where $n \in \ZZ$ and $\ms Y \in \tg_{k, n+1}$ is non-zero. Let $X = (X_y)_{y \in \z} \in \Lg_\z$ be such that $X_y = 0$ for all $y \neq x$ and $X_x = \ms X \otimes \xi^{-n-1}_x$ with $\ms X \in \tg_{k, -n-1}$. We then have
\begin{equation*}
0 = \lbf{\Phi, X} = T \Re (\ms Y|\ms X) = T (\ms Y|\ms X).
\end{equation*}
Since $\ms X \in \tg_{k, -n-1}$ was arbitrary and $( \tg_{k, -n-1}, \tg_{k, n+1}, (\cdot|\cdot))$ is a dual pair by Lemma \ref{lem: tgj and tgkp}, it follows once again that $\ms Y = 0$, which is a contradiction.

Finally, suppose that $\Phi_x$ with $x \in \z_{\rm f}$ is non-zero. Let $\ms Y \otimes \xi^n_x d\xi_x$ be its most singular term with $n \in \ZZ$ and $\ms Y \neq 0$ in $\tg_{(n+1)}$. Choose $X = (X_y)_{y \in \z} \in \Lg_\z$ such that $X_y = 0$ for all $y \neq x$ and $X_x = \ms X \otimes \xi^{-n-1}_x$ with $\ms X \in \tg_{(-n-1)}$. Then
\begin{equation*}
0 = \lbf{\Phi, X} = \Re (\ms Y|\ms X) = (\ms Y|\ms X).
\end{equation*}
Using the fact that $(\tg_{(-n-1)}, \tg_{(n+1)}, (\cdot | \cdot))$ is a dual pair, again by Lemma \ref{lem: tgj and tgkp}, and that $\ms X \in \tg_{(-n-1)}$ was arbitrary, we deduce that $\ms Y = 0$, which is a contradiction.

The proof that the bilinear form \eqref{gD pairing} is also non-degenerate on the right is completely analogous.
\end{proof}
\end{lemma}

The first part of the following proposition is a generalisation of the $\Gamma$-equivariant strong residue theorem on $\CP$ as in \cite[Lemma A.1]{Vicedo:2014zza} to the $\Pi$-equivariant case. For the non-equivariant version of the strong residue theorem on an arbitrary algebraic curve see, for instance, \cite[\S 2.3]{Takhtajan}.
\begin{proposition} \label{prop: eq res thm}
We have $R^\Pi_\z(\tg)^\perp = \Omega^\Pi_\z(\tg)$ and $\Omega^\Pi_\z(\tg)^\perp = R^\Pi_\z(\tg)$.

Also, $(\Og^+_\z)^\perp = \Lg^+_\z$ and $(\Lg^+_\z)^\perp = \Og^+_\z$.
\begin{proof}
According to Lemmas \ref{lem: Gamma decomp}, \ref{lem: Gamma decomp Om} and \ref{lem: dual pairs} it suffices to show that $\Omega^\Pi_\z(\tg) \perp R^\Pi_\z(\tg)$ and $\Og^+_\z \perp \Lg^+_\z$. We prove each of these statements in turn.

Let $\Phi \in \Omega^\Pi_{\z}(\tg)$ and $X \in R^\Pi_{\z}(\tg)$. It follows from \eqref{equivariance Rtg} and \eqref{equivariance Otg} that these have the properties $r_\alpha \Phi = \mu_\alpha^\ast \Phi$ and $r_\alpha X = \mu_\alpha^\ast X$ for any $\alpha \in \Pi$. Then by the $\sigma$-invariance of the bilinear form on $\tg$ it follows that, for any $x \in \CP$ and $\alpha = \s^n \in \Gamma$,
\begin{align*}
\res_x ( \iota_x \Phi | \iota_x X ) &= \frac{1}{2 \pi i} \int_{\mathscr C_x} (\Phi | X) = \frac{1}{2 \pi i} \int_{\mathscr C_x} (r_\alpha \Phi | r_\alpha X) = \frac{1}{2 \pi i} \int_{\mathscr C_x} (\mu_\alpha^\ast \Phi | \mu_\alpha^\ast X)\\
&= \frac{1}{2 \pi i} \int_{\mu_\alpha \mathscr C_x} (\Phi | X) = \frac{1}{2 \pi i} \int_{\mathscr C_{\mu_\alpha x}} (\Phi | X) = \res_{\omega^n x} ( \iota_{\omega^n x} \Phi | \iota_{\omega^n x} X ),
\end{align*}
where $\mathscr C_x$ is a sufficiently small counterclockwise contour around the point $x$. Likewise, by the $\tau$-invariance of the bilinear pairing \eqref{ip on tg}, in the sense of \eqref{ip on tg sigma tau}, we have
\begin{align*}
\overline{\res_x ( \iota_x \Phi | \iota_x X )} &= - \frac{1}{2 \pi i} \int_{\mathscr C_x} \overline{(\Phi | X)} = - \frac{1}{2 \pi i} \int_{\mathscr C_x} (\tau \Phi | \tau X) = - \frac{1}{2 \pi i} \int_{\mathscr C_x} (\mu_\t^\ast \Phi | \mu_\t^\ast X)\\
&= - \frac{1}{2 \pi i} \int_{\mu_\t \mathscr C_x} (\Phi | X) = \frac{1}{2 \pi i} \int_{\mathscr C_{\bar x}} (\Phi | X) = \res_{\bar x} ( \iota_{\bar x} \Phi | \iota_{\bar x} X ),
\end{align*}
noting that $\mu_\t \mathscr C_x$ is oriented clockwise so that $\mu_\t \mathscr C_x = - \mathscr C_{\bar x}$.
Hence
\begin{align*}
\lbf{\Phi, X} &= \Re \left( \sum_{x \in \z} \frac{2T}{|\Pi_x|} \res_x \, ( \iota_x \Phi | \iota_x X ) \right)\\
&= \sum_{x \in \z_{\rm c}} T \big( \res_x \, ( \iota_x \Phi | \iota_x X ) + \res_{\bar x} \, ( \iota_{\bar x} \Phi | \iota_{\bar x} X ) \big)\\
&\qquad\qquad + \sum_{x \in \z'_{\rm r}} T \res_x \, ( \iota_x \Phi | \iota_x X ) + \sum_{x \in \z_{\rm f}} \res_x \, ( \iota_x \Phi | \iota_x X )\\
&= \sum_{x \in \Pi \z} \res_x \, ( \iota_x \Phi | \iota_x X ) = 0,
\end{align*}
where the last equality is by the standard residue theorem.

Now let $\Phi \in \Og^+_\z$ and $X \in \Lg^+_\z$. At any point $x \in \z \setminus \{ \infty \}$ the germs $\Phi_x \in \Og^+_x$ and $X_x \in \Lg^+_x$ are elements of the spaces $\tg \otimes \sO_x d\xi_x$ and $\tg \otimes \sO_x$, respectively. Therefore $(\Phi_x | X_x) \in \sO_x d\xi_x$ so that $\res_x (\Phi_x | X_x) = 0$. On the other hand, at infinity the germs $\Phi_\infty$ and $X_\infty$ belong to $\tg \otimes \m^{-1}_\infty d\xi_\infty$ and $\tg \otimes \m_\infty$, respectively. So in this case as well we have $(\Phi_\infty | X_\infty) \in \sO_\infty d\xi_\infty$, and hence $\res_\infty (\Phi_\infty | X_\infty) = 0$. It therefore follows that each term in the sum of residues in \eqref{gD pairing} vanishes, and thus $\lbf{\Phi, X} = 0$.
\end{proof}
\end{proposition}

\begin{corollary} \label{cor: smooth dual}
The triple $(\Omega^\Pi_\z(\tg), \Lg^+_\z, \lbf{\cdot, \cdot})$ is a dual pair.
\begin{proof}
This is a direct application of Lemma \ref{lem: dual pairs}.
\end{proof}
\end{corollary}

\subsection{Divisors and Takiff algebras} \label{sec: Divisors}

Let $\text{Div}(\CP)$ be the free abelian group generated by the points of $\CP$. An element $\D \in \text{Div}(\CP)$ is called a \emph{divisor}, which we write as a formal sum
\begin{equation} \label{divisor def}
\D = \sum_{x \in \CP} n_x \, x
\end{equation}
with $n_x \in \ZZ$ being zero for all but finitely many $x \in \CP$. The divisor $\D$ is said to be non-negative, and we write $\D \geq 0$, if $n_x \geq 0$ for all $x \in \CP$. For any $\D, \D' \in \text{Div}(\CP)$ we write $\D \geq \D'$ if $\D - \D' \geq 0$. This defines a partial ordering on the set $\text{Div}(\CP)$. The support of $\D \in \text{Div}(\CP)$ is the finite subset $\text{supp}\, \D \coloneqq \{ x \in \CP \,|\, n_x \neq 0 \}$ 
and its degree is defined as $\deg \D \coloneqq \sum_{x \in \CP} n_x \in \ZZ$. Given a finite subset $S \subset \CP$ we let
\begin{equation*}
\text{Div}_{\geq 1}(S) \coloneqq \{ \D \in \text{Div}(\CP) \, |\, \D \geq 0 \;\text{and}\; \text{supp}\, \D = S \}
\end{equation*}
denote the subset of all non-negative divisors with support $S$.

Fix a divisor $\D \in \text{Div}_{\geq 1}(\z)$. We associate to it a non-negative divisor with support $\Pi \z$, given by
\begin{equation*}
\Pi \D \coloneqq \sum_{x \in \Pi \z} n_x \, x \in \text{Div}_{\geq 1}(\Pi \z)
\end{equation*}
where for any $x \in \Pi \z$, $n_x \in \ZZ_{\geq 1}$ is defined by noting that there is a unique $y \in \z$ such that $x \in \Pi\, y$, \emph{i.e.} $x$ and $y$ lie on the same orbit of $\Pi$, and we set $n_x \coloneqq n_y$. In what follows we assume $\D$ to be such that $\deg (\Pi \D) \geq 2$, \emph{i.e.} we have $n_\infty \geq 2$ or $|\z| \geq 2$.

Given a meromorphic differential $\varpi \in \Omega_{\Pi \z}$, its order at $x \in \CP$, denoted $\text{ord}_x \varpi$, is by definition equal to $n \in \ZZ$ if its germ at $x$ takes the form $[\varpi]_x = \sum_{k = n}^\infty a_k \xi_x^k d\xi_x$ with $a_k \in \CC$ and $a_n \neq 0$. The canonical divisor of $\varpi$ is then defined as
\begin{equation*}
(\varpi) \coloneqq \sum_{x \in \CP} (\text{ord}_x \varpi) \, x \in \text{Div}(\CP).
\end{equation*}
Let $\Omega_{\Pi \D} \coloneqq \{ \varpi \in \Omega_{\Pi \z} \,|\, (\varpi) \geq - \Pi \D \}$. For any complex vector space $V$ we introduce the notation
\begin{equation} \label{Om V notation}
\Omega_{\Pi \D}(V) \coloneqq V \otimes \Omega_{\Pi \D}.
\end{equation}
The linear space $\Omega_{\Pi \D}(\tg)$ admits a natural action of $\Pi$ defined as in \eqref{Gamma action diff}. We shall be interested in the subspace of $\Pi$-invariants
\begin{equation*}
\Omega^{\Pi}_\D(\tg) \coloneqq \Omega_{\Pi \D}(\tg)^\Pi.
\end{equation*}
This is a subspace of $\Omega^{\Pi}_{\z}(\tg)$ which, as usual, we will implicitly identify with its image in $\Og_{\z}$ under the $\iota_{\z}$-map \eqref{iota map Om}.

For each $x \in \z$ and $n \in \ZZ_{\geq 1}$ we define the ideal $\Lg_x^{+n} \coloneqq ((\tg \otimes \m_x^n)_\RR)^{\Pi_x}$ of $\Lg_x^+$. We note in particular that $\Lg_x^{+1}$ is a proper ideal in $\Lg_x^+$ for $x \neq \infty$ whereas $\Lg_{\infty}^{+1} = \Lg_{\infty}^+$. Set
\begin{equation} \label{LgD+ def}
\Lg_\D^+ \coloneqq \bigoplus_{x \in \z} \Lg_x^{+n_x},
\end{equation}
which is an ideal in the real Lie algebra $\Lg_\z^+$.

\begin{lemma} \label{lem: ann Lg+}
We have the direct sum decompositions of real vector spaces
\begin{subequations} \label{Ann decomp}
\begin{align}
\label{Ann Lg+ decomp} (\Lg^+_\D)^\perp = \Og^+_\z \dotplus \Omega^\Pi_\D(\tg),\\
\label{Ann OgD decomp} \Omega^\Pi_\D(\tg)^\perp = R^\Pi_\z(\tg) \dotplus \Lg^+_\D.
\end{align}
\end{subequations}
\begin{proof}
To show \eqref{Ann Lg+ decomp} we apply Lemma \ref{lem: dual pairs} to the dual pair from Lemma \ref{lem: main dual pair}, namely $(V, W, \lbf{\cdot, \cdot})$ with $V = \Og_\z$ and $W = \Lg_\z$. We consider the subspaces
\begin{gather*}
W_+ = \Lg^+_\D, \qquad W_- = \bigoplus_{x \in \z} \big( (\tg \otimes \p_x^{- n_x + 1})_\RR \big)^{\Pi_x},\\
V_+ = \bigoplus_{x \in \z} \big( (\tg \otimes \m_x^{-n_x} d\xi_x)_\RR \big)^{\Pi_x}, \qquad
V_- = \bigoplus_{x \in \z} \big( (\tg \otimes \p_x^{n_x + 1} d\xi_x)_\RR \big)^{\Pi_x}.
\end{gather*}
At every point $x \in \z$ we have, by construction, $(\Phi_x| X_x) \in \sO_x d\xi_x$ for any $\Phi \in V_+$ and $X \in W_+$ from which we deduce $\lbf{\Phi, X} = 0$ using the definition \eqref{gD pairing}. Similarly we have $(\Phi_x| X_x) \in \p^2_x d\xi_x$ for any $\Phi \in V_-$ and $X \in W_-$ so that once again $\lbf{\Phi, X} = 0$. This shows that $V_\pm \perp W_\pm$ so that the conditions of Lemma \ref{lem: dual pairs} hold. We therefore conclude that $W_+^\perp = V_+$, or in other words
\begin{equation*}
(\Lg^+_\D)^\perp = V_+.
\end{equation*}
It remains to show that $V_+ = \Og^+_\z \dotplus \Omega^\Pi_\D(\tg)$. By definition the vector space $V_+$ consists of all elements $\Phi = (\Phi_x)_{x \in \z} \in \Og_\z$ such that for each $x \in \z$ the germ $\Phi_x$ has a pole of order at most $n_x$ at $x$. Let $\Phi = (\Phi_x)_{x \in \z} \in V_+$. We denote by $\Phi^-_x \in \Og^-_x$ the principal part of $\Phi_x$, which we regard as an element of $\Omega_{\Pi \z}(\tg)$. If $x \neq \infty$ then $\Phi_x^-$ has a pole at $x$ and at most a simple pole at $\infty$. Recall, in particular, that the principal part $\Phi^-_{\infty}$ at infinity doesn't include the simple pole term. Define
\begin{equation*}
\varpi_\Phi \coloneqq \sum_{\alpha \in \Pi} \sum_{x \in \z} \frac{1}{|\Pi_x|} \hat \alpha \Phi^-_x,
\end{equation*}
which by construction belongs to $\Omega^\Pi_\D(\tg)$.
We can then write $\Phi \in V_+$ uniquely as the sum of $\iota_{\z} \varpi_\Phi \in \Og_{\z}$ and $( \Phi_x - \iota_x \varpi_\Phi )_{x \in \z} \in \Og^+_\z$, as required.

The second equality \eqref{Ann OgD decomp} can be deduced from \eqref{Ann Lg+ decomp} as follows. We have
\begin{equation} \label{ann Lg+ proof id}
\Lg^+_\D = \big( \Og^+_\z \dotplus \Omega^\Pi_\D(\tg) \big)^\perp = (\Og^+_\z)^\perp \cap \Omega^\Pi_\D(\tg)^\perp = \Lg^+_\z \cap \Omega^\Pi_\D(\tg)^\perp,
\end{equation}
where in the first equality we used the above which shows not only that $(\Lg^+_\D)^\perp = V_+$ but also $V_+^\perp = \Lg^+_\D$.
In the last equality we used Proposition \ref{prop: eq res thm}.
Now since we have $\Omega^\Pi_\D(\tg)^\perp \supset \Omega^\Pi_\z(\tg)^\perp = R^\Pi_\z(\tg)$, again using Proposition \ref{prop: eq res thm}, we obtain
\begin{equation*}
\Omega^\Pi_\D(\tg)^\perp = \Omega^\Pi_\D(\tg)^\perp \cap \big( \Lg^+_\z \dotplus R^\Pi_\z(\tg) \big) = \big( \Omega^\Pi_\D(\tg)^\perp \cap \Lg^+_\z \big) \dotplus R^\Pi_\z(\tg) = \Lg^+_\D \dotplus R^\Pi_\z(\tg),
\end{equation*}
where in the middle step we used Lemma \ref{lem: ABC}.
\end{proof}
\end{lemma}

\begin{proposition}
The triple $(\Omega^\Pi_\D(\tg), \Lg^+_\z \big/ \Lg^+_\D, \lbf{\cdot, \cdot})$ forms a dual pair, where the bilinear form
\begin{equation} \label{pairing OgD Lgp}
\lbf{\cdot, \cdot} : \Omega^\Pi_\D(\tg) \times \Lg^+_\z \big/ \Lg^+_\D \longrightarrow \RR
\end{equation}
is induced from the restriction $\lbf{\cdot, \cdot}|_{\Omega^\Pi_\D(\tg) \times \Lg^+_\z}$ of \eqref{gD pairing} to $\Omega^\Pi_\D(\tg) \times \Lg^+_\z$.
\begin{proof}
This is a direct application of Lemma \ref{lem: induced bilinear} to the subspaces $\Omega^\Pi_\D(\tg)$ and $\Lg^+_\z$. Note that $\Omega^\Pi_\D(\tg) \cap (\Lg^+_\z)^\perp = \Omega^\Pi_\D(\tg) \cap \Og^+_\z = \{ 0 \}$ using Proposition \ref{prop: eq res thm} followed by Lemma \ref{lem: Gamma decomp Om}. And using \eqref{ann Lg+ proof id} in the proof of Lemma \ref{lem: ann Lg+}, $\Lg^+_\z \cap \Omega^\Pi_\D(\tg)^\perp = \Lg^+_\D$.
\end{proof}
\end{proposition}

\subsubsection{Direct sum of real Takiff algebras} \label{sec: real Takiff}

The quotient Lie algebra $\Lg_{\z}^+ / \Lg_\D^+$ can be described in terms of real generalised Takiff algebras for $\tg$ as follows.

We assign to each $x \in \z$ a formal variable $\varepsilon_x$. At every finite point $x \in \z \setminus \{ \infty \}$ we consider the ring of polynomials $\CC[\varepsilon_x]$, whereas for the point at infinity we consider instead the ideal $\varepsilon_\infty \CC[\varepsilon_\infty]$ of polynomials without constant terms.
There is a natural action of $\Pi$ on $\CC[\varepsilon_x]$ for each $x \in \z \setminus \{ \infty \}$ given by
\begin{equation*}
\s . f(\varepsilon_x) = f(\omega^{-1} \varepsilon_x), \qquad
\t . f(\varepsilon_x) = \bar f(\varepsilon_x),
\end{equation*}
for any $f \in \CC[\varepsilon_x]$. Here $\bar f \in \CC[\varepsilon_x]$ denotes the complex conjugate polynomial, defined as $\bar f(\varepsilon_x) = \sum_i \bar a_i \varepsilon_x^i$ if $f(\varepsilon_x) = \sum_i a_i \varepsilon_x^i$. By contrast, we let $\Pi$ act on $\varepsilon_\infty \CC[\varepsilon_\infty]$ by
\begin{equation*}
\s . f(\varepsilon_\infty) = f(\omega \, \varepsilon_\infty), \qquad
\t . f(\varepsilon_\infty) = \bar f(\varepsilon_\infty),
\end{equation*}
for any $f \in \varepsilon_\infty \CC[\varepsilon_\infty]$, with $\bar f \in \varepsilon_\infty \CC[\varepsilon_\infty]$ denoting its complex conjugate.

Now given the above divisor $\D = \sum_{x \in \z} n_x\, x \in \text{Div}_{\geq 1}(\z)$, we can form the quotient ring $\CC[\varepsilon_x] / \varepsilon_x^{n_x} \CC[\varepsilon_x]$ of $n_x$-truncated polynomials for $x \in \z \setminus \{ \infty \}$. At infinity we have, instead, the quotient $\varepsilon_\infty \CC[\varepsilon_\infty] / \varepsilon_\infty^{n_\infty} \CC[\varepsilon_\infty]$ which forms a ring without identity. The actions of $\Pi$ on $\CC[\varepsilon_x]$ and $\varepsilon_\infty \CC[\varepsilon_\infty]$ defined above both descend to these quotients since the respective ideals $\varepsilon_x^{n_x} \CC[\varepsilon_x]$ and $\varepsilon_\infty^{n_\infty} \CC[\varepsilon_\infty]$ are invariant. By abuse of notation we denote the action of $\alpha \in \Pi$ on the class $f \in \CC[\varepsilon_x] / \varepsilon_x^{n_x} \CC[\varepsilon_x]$ also as $\alpha . f$ and similarly for the point at infinity.

At a finite point $x \in \z \setminus \{ \infty \}$ we form the Lie algebra
\begin{equation*}
\Tak^{n_x}_x \tg \coloneqq \tg \otimes \CC[\varepsilon_x] / \varepsilon_x^{n_x} \CC[\varepsilon_x],
\end{equation*}
which we shall refer to as a generalised Takiff algebra for $\tg$.
For the point at infinity we have instead the nilpotent Lie algebra
\begin{equation*}
\Tak^{n_\infty}_\infty \tg \coloneqq \tg \otimes \varepsilon_\infty \CC[\varepsilon_\infty] / \varepsilon_\infty^{n_\infty} \CC[\varepsilon_\infty].
\end{equation*}
By combining the action of $\Pi$ on $n_x$-truncated polynomials for each $x \in \z$ introduced above with that on $\tg$, we obtain an action of $\Pi$ on the Lie algebra $\Tak^{n_x}_x \tg$ for each $x \in \z$, \emph{i.e.} a homomorphism $\Pi \to \bAut (\Tak^{n_x}_x \tg)$, $\alpha \mapsto \hat \alpha$ given by
\begin{equation} \label{Pi action Tg}
\hat \alpha (\ms X \otimes f ) = r_\alpha \ms X \otimes \alpha . f,
\end{equation}
for any $\alpha \in \Pi$, $\ms X \in \tg$ and $f \in \CC[\varepsilon_x]/ \varepsilon_x^{n_x} \CC[\varepsilon_x]$ if $x \in \z \setminus \{ \infty \}$ or $f \in \varepsilon_\infty \CC[\varepsilon_\infty]/ \varepsilon_\infty^{n_\infty} \CC[\varepsilon_\infty]$ for the point at infinity. In what follows we will be interested only in the restriction $\Pi_x \to \bAut (\Tak^{n_x}_x \tg)$ of the above homomorphism to the stabiliser subgroup $\Pi_x \subset \Pi$.

To any $x \in \z$ we attach the real Lie algebra of $\Pi_x$-invariants
\begin{equation*}
\tg^{n_x x} \coloneqq \big( (\Tak^{n_x}_x \tg)_\RR \big)^{\Pi_x}.
\end{equation*}
We refer to these as real (generalised) Takiff algebras for $\tg$. Consider the direct sum of real Lie algebras
\begin{equation} \label{Takiff}
\tg^\D \coloneqq \bigoplus_{x \in \z} \tg^{n_x x}.
\end{equation}
Note that when $n_\infty = 1$, the summand in \eqref{Takiff} corresponding to the point at infinity is absent.

\begin{lemma} \label{lem: iso aGbG}
Let $\a$ be a real Lie algebra equipped with an action of a finite group $G$ by automorphisms. Let $\b$ be a $G$-invariant ideal in $\a$, \emph{i.e.} $G . \b = \b$. The quotient $\a / \b$ is then equipped with a natural action of $G$ defined as $G \times \a/\b \mapsto \a/\b$, $(g, x + \b) \mapsto g . x + \b$ and we have an isomorphism of real Lie algebras $\a^G / \b^G \cong (\a / \b)^G$.
\begin{proof}
Consider the map $\varphi : \a^G \to (\a / \b)^G$ given by $x \mapsto x + \b$. This map is well defined since $g . x = x$ for all $g \in G$ implies $g . (x + \b) = g . x + \b = x + \b$, in other words $x + \b \in (\a / \b)^G$. Next, $\varphi$ is clearly a homomorphism of Lie algebras since $\b$ is an ideal: $[\varphi(x), \varphi(y)] = [ x + \b, y + \b] = [x, y] + \b = \varphi([x, y])$. Now let $x + \b \in (\a / \b)^G$. This means that for all $g \in G$ we have $g . (x + \b) = x + \b$ or in other words $g . x - x \in \b$. Therefore we can write $x + \b = \bar x + \b$ where $\bar x \coloneqq \frac{1}{|G|} \sum_{g \in G} g . x$. But clearly $g . \bar x = \bar x$ for all $g \in G$ so that $x + \b = \varphi(\bar x)$ which shows that $\varphi$ is surjective. Finally, suppose $\varphi(x) = 0$. This means $x + \b = \b$ or in other words $x \in \b$. But since $x \in \a^G$ we have $x \in \a^G \cap \b = \b^G$. Thus $\ker \varphi = \b^G$. Since $\varphi$ is a surjective homomorphism the result follows from the first isomorphism theorem.
\end{proof}
\end{lemma}

\begin{proposition} \label{prop: Takiff dual pair}
We have an isomorphism of real Lie algebras $\Lg^+_\z / \Lg^+_\D \cong \tg^\D$. In particular, the triple $(\Omega^\Pi_\D(\tg), \tg^\D, \lbf{\cdot, \cdot})$ forms a dual pair where the bilinear form
\begin{equation} \label{pairing Og gD}
\lbf{\cdot, \cdot} : \Omega^\Pi_\D(\tg) \times \tg^\D \to \RR
\end{equation}
is induced from \eqref{pairing OgD Lgp}.
\begin{proof}
By definition \eqref{gD+} and \eqref{LgD+ def} of the Lie algebras $\Lg^+_\z$ and $\Lg^+_\D$, we have an isomorphism of real Lie algebras
\begin{equation*}
\Lg^+_\z / \Lg^+_\D \cong \bigoplus_{x \in \z} \Lg^+_x / \Lg^{+ n_x}_x.
\end{equation*}

Now for any $x \in \z \setminus \{ \infty \}$ we have an isomorphism
\begin{equation*}
\Lg^+_x / \Lg^{+ n_x}_x \cong \big( (\tg \otimes \sO_x / \m^{n_x}_x)_\RR \big)^{\Pi_x}
\end{equation*}
which follows from Lemma \ref{lem: iso aGbG}.
On the other hand, the injective homomorphism of rings $\CC[\varepsilon_x] \hookrightarrow \sO_x$ given by $\varepsilon_x \mapsto \xi_x$ induces an isomorphism $\CC[\varepsilon_x] / \varepsilon_x^{n_x} \CC[\varepsilon_x] \cong \sO_x / \m_x^{n_x}$ which commutes with the action of $\Pi_x$. Therefore $\Lg^+_x / \Lg^{+ n_x}_x \cong ((\Tak^{n_x}_x \tg)_\RR)^{\Pi_x} = \tg^{n_x x}$.
Similarly, for the point at infinity we have
\begin{equation*}
\Lg^+_\infty / \Lg^{+ n_\infty}_\infty \cong \big( (\tg \otimes \m_\infty / \m^{n_\infty}_\infty)_\RR \big)^\Pi
\cong \big( (\Tak^{n_\infty}_\infty \tg)_\RR \big)^\Pi = \tg^{n_\infty \infty},
\end{equation*}
as required. Note that the isomorphism $\varepsilon_\infty \CC[\varepsilon_\infty] / \varepsilon_\infty^{n_\infty} \CC[\varepsilon_\infty] \cong \m_\infty / \m_\infty^{n_\infty}$ commutes with the action of $\Pi$ by definition of the latter on $\varepsilon_{\infty} \CC[\varepsilon_\infty]$ and of the local coordinate $\xi_\infty$ at infinity.
\end{proof}
\end{proposition}

In the next section we shall make use of explicit bases for $\tg^\D$ and its dual $\Omega^\Pi_\D(\tg)$ from Proposition \ref{prop: Takiff dual pair}, which we now describe.

Recall the dual bases $\{ I^{\wt a} \}$ and $\{ I_{\wt a} \}$ of $\tg$ introduced in \S\ref{sec: aff KM}.
In terms of these, a basis of the Takiff algebra $\Tak^{n_x}_x \tg$ for $x \in \z \setminus \{ \infty \}$ is given by
$I^{\wt a} \otimes \varepsilon^p_x$ for $p = 0, \ldots, n_x - 1$. Here we denote the class $\varepsilon^p_x + \varepsilon^{n_x}_x \CC[\varepsilon_x] \in \CC[\varepsilon_x]/\varepsilon^{n_x}_x \CC[\varepsilon_x]$ simply by $\varepsilon^p_x$. In particular, we have $\varepsilon_x^{n_x} = 0$ for every $x \in \z$. A basis of the realification $(\Tak^{n_x}_x \tg)_\RR$ then consists of $I^{\wt a} \otimes \varepsilon^p_x$ and $i I^{\wt a} \otimes \varepsilon^p_x$ for $p = 0, \ldots, n_x - 1$. Likewise, a basis of $(\Tak^{n_\infty}_{\infty} \tg)_\RR$ is given by $I^{\wt a} \otimes \varepsilon^{q+1}_\infty$ and $i I^{\wt a} \otimes \varepsilon^{q+1}_\infty$ for $q = 0, \ldots, n_\infty - 2$.

A basis of $\tg^\D$, whose elements we denote collectively as $I^A$, is given by
\begin{subequations} \label{Takiff basis}
\begin{equation} \label{Takiff basis a}
I^{\wt a} \otimes \varepsilon_x^p, \qquad\qquad
i I^{\wt a} \otimes \varepsilon_x^p,
\end{equation}
for each $x \in \z_{\rm c}$ with $p = 0, \ldots, n_x - 1$,
\begin{equation} \label{Takiff basis b}
I^{\wt a}_{k, p} \otimes \varepsilon_x^p,
\end{equation}
for each $x \in \z^k_{\rm r}$, $k \in \ZZ_T$ with $p = 0, \ldots, n_x - 1$, and
\begin{align}
\label{Takiff basis c}
&I^{(p, \wt \alpha)} \otimes \varepsilon_0^p,\\
\label{Takiff basis d}
&I^{(-q-1, \wt \alpha)} \otimes \varepsilon_{\infty}^{q+1},
\end{align}
\end{subequations}
for $p = 0, \ldots, n_0 - 1$ and $q = 0, \ldots, n_\infty - 2$. The dual basis elements $I_A$, with respect to the pairing \eqref{pairing Og gD}, of the real vector space $\Omega^{\Pi}_\D(\tg)$ read
\begin{subequations} \label{dual Takiff basis}
\begin{equation} \label{dual Takiff basis a}
\frac{1}{2T} \sum_{\alpha \in \Pi} \hat \alpha \left( \frac{I_{\wt a}}{(z - x)^{p+1}} dz \right), \quad\qquad
\frac{1}{2T} \sum_{\alpha \in \Pi} \hat \alpha \left( \frac{- i I_{\wt a}}{(z - x)^{p+1}} dz \right),
\end{equation}
for each $x \in \z_{\rm c}$ with $p = 0, \ldots, n_x - 1$,
\begin{equation} \label{dual Takiff basis b}
\frac{1}{T} \sum_{\alpha \in \Gamma} \hat \alpha \left( \frac{I_{\wt a; k, -p}}{(z - x)^{p+1}} dz \right),
\end{equation}
for each $x \in \z^k_{\rm r}$, $k \in \ZZ_T$ with $p = 0, \ldots, n_x - 1$, and
\begin{align} \label{dual Takiff basis c}
& I_{(- p, \wt \alpha)} z^{-p-1} dz, \\
\label{dual Takiff basis d}
- \, & I_{(q+1, \wt \alpha)} z^q dz,
\end{align}
\end{subequations}
for $p = 0, \ldots, n_0 - 1$ and $q = 0, \ldots, n_\infty - 2$. Note that the factor of $1/2T$ (resp. $1/T$) in the differentials at points $x \in \z_{\rm c}$ (resp. $x \in \z^k_{\rm r}$) stems from the corresponding factor of $2T/|\Pi_x| = 2T$ (resp. $2T/|\Pi_x| = T$) in the definition of the bilinear form \eqref{gD pairing}.

\subsubsection{Complexification of $\tg^\D$}

In the next section we shall also make use of an explicit description of the complexification of $\tg^\D$ defined in Proposition \ref{prop: complexified Takiff} below. First we need the following lemma.

Given any complex Lie algebra $\a$ we denote by $\bar \a$ the complex conjugate Lie algebra, namely the realification $\a_\RR$ of $\a$ endowed with the opposite complex structure given by $i \cdot v = - i v$ for any $v \in \bar \a$.
\begin{lemma} \label{lem: Lie re com}
Let $\a$ be a complex Lie algebra, $\tau \in \bAut_- \a$ an anti-linear involutive automorphism and $\a^\tau \coloneqq \{ a \in \a \, | \, \tau a = a \}$ the corresponding real form of $\a$.

We have the following isomorphisms of complex Lie algebras:
\begin{itemize}
  \item[$(i)$]
$\psi_{\rm c} : \a \oplus \bar \a \SimTo \a \otimes_\RR \CC$, $(a, b) \mapsto \ha (a \otimes 1 - i a \otimes i + b \otimes 1 + i b \otimes i)$, under which the anti-linear involution $\a \otimes_\RR \CC \to \a \otimes_\RR \CC$, $a \otimes u \mapsto a \otimes \bar u$ corresponds to the exchange $\a \oplus \bar \a \to \a \oplus \bar \a$, $(a, b) \mapsto (b, a)$.
  \item[$(ii)$]
$\psi_{\rm r} : \a \SimTo \a^\tau \otimes_\RR \CC$,
$a \mapsto \ha (a + \tau a) \otimes 1 - \ha i (a - \tau a) \otimes i$,
under which the anti-linear involution $\a^\tau \otimes_\RR \CC \to \a^\tau \otimes_\RR \CC$, $a \otimes u \mapsto a \otimes \bar u$ corresponds to $\tau : \a \to \a$.
\end{itemize}
\begin{proof}
One checks that $\psi_{\rm c}$ is $\CC$-linear and that $\phi_{\rm c} : \a \otimes_\RR \CC \to \a \oplus \bar \a$, $a \otimes u \mapsto (u a, \bar u a)$ is its inverse. Moreover, the latter is seen to be a homomorphism of complex Lie algebras. It also follows from the form of the isomorphism $\psi_{\rm c}$ that complex conjugation on the second tensor factor in $\a \otimes_\RR \CC$ sends $(a, b)$ to $(b, a)$ in $\a \oplus \bar \a$. This proves $(i)$.

Next, the map $\psi_{\rm r}$ is $\CC$-linear since it is $\RR$-linear and $\psi_{\rm r}(i a) = i \psi_{\rm r}(a)$ for all $a \in \a$. Its inverse is given by $\phi_{\rm r} : \a^\tau \otimes_\RR \CC \to \a$, $a \otimes u \mapsto u a$. The latter is clearly a homomorphism of complex Lie algebras and the remaining claim in $(ii)$ follows from the explicit form of the isomorphism $\psi_{\rm r}$.
\end{proof}
\end{lemma}

For any $x \in \z_{\rm c}$ we let $\overline{\Tak^{n_x}_x \tg}$ denote the complex conjugate of the Lie algebra $\Tak^{n_x}_x \tg$.

\begin{proposition} \label{prop: complexified Takiff}
Let $\tg^\D_\CC \coloneqq \tg^\D \otimes_\RR \CC = \bigoplus_{x \in \z} \tg^{n_x x} \otimes_\RR \CC$ be the complexification of $\tg^\D$ and denote by $c : \tg^\D_\CC \to \tg^\D_\CC$ the complex conjugation in the second tensor factor. We have an isomorphism of complex Lie algebras
\begin{equation} \label{Takiff complex}
\psi : \bigoplus_{x \in \z_{\rm f}} (\Tak^{n_x}_x \tg)^\Gamma \oplus \bigoplus_{x \in \z'_{\rm r}} \Tak^{n_x}_x \tg \oplus \bigoplus_{x \in \z_{\rm c}} \big( \Tak^{n_x}_x \tg \oplus \overline{\Tak^{n_x}_x \tg} \big) \overset{\sim}\longrightarrow \tg^\D_\CC.
\end{equation}
More precisely, we have the following isomorphisms of complex Lie algebras:
\begin{subequations} \label{Lie re com all}
\begin{itemize}
  \item[$(i)$]
For every $x \in \z_{\rm c}$,
\begin{align} \label{Lie re com iso}
\Tak^{n_x}_x \tg \oplus \overline{\Tak^{n_x}_x \tg} &\overset{\sim}\longrightarrow \tg^{n_x x} \otimes_\RR \CC,\\
(\ms X \otimes \varepsilon_x^p, \ms Y \otimes \varepsilon_x^q) &\longmapsto \ms X^{(x)}_{\ul p} + \ms Y^{(\bar x)}_{\ul q} \coloneqq \ha \big( (\ms X \otimes \varepsilon_x^p) \otimes 1 - (i \ms X \otimes \varepsilon_x^p) \otimes i \notag\\
&\qquad\qquad\qquad\qquad\qquad + (\ms Y \otimes \varepsilon_x^q) \otimes 1 + (i \ms Y \otimes \varepsilon_x^q) \otimes i \big). \notag
\end{align}
Moreover, we have $c \big( \ms X^{(x)}_{\ul p} \big) = \ms X^{(\bar x)}_{\ul p}$ for any $x \in \z_{\rm c} \cup \bar \z_{\rm c}$.
  \item[$(ii)$]
For every $x \in \z^k_{\rm r}$ with $k \in \ZZ_T$,
\begin{align} \label{Lie re com iso 2}
\Tak^{n_x}_x \tg &\overset{\sim}\longrightarrow \tg^{n_x x} \otimes_\RR \CC,\\
\ms X \otimes \varepsilon_x^p &\longmapsto \ms X^{(x)}_{\ul p} \coloneqq ( \pi^+_{k,p} \ms X \otimes \varepsilon_x^p ) \otimes 1 - i ( \pi^-_{k,p} \ms X \otimes \varepsilon_x^p ) \otimes i. \notag
\end{align}
Moreover, $c \big( \ms X^{(x)}_{\ul p} \big) = (\omega^{-kp} \sigma^k \tau \ms X)^{(x)}_{\ul p}$. In particular, $c \big( \ms X^{(x)}_{\ul p} \big) = \ms X^{(x)}_{\ul p}$ for $\ms X \in \tg_{k, p}$.
  \item[$(iii)$]
For every $x \in \z_{\rm f}$,
\begin{align} \label{Lie re com iso 0}
(\Tak^{n_x}_x \tg)^\Gamma &\overset{\sim}\longrightarrow \tg^{n_x x} \otimes_\RR \CC,\\
\ms X \otimes \varepsilon_x^p &\longmapsto \ms X^{(x)}_{\ul p} \coloneqq ( \pi^+_0 \ms X \otimes \varepsilon_x^p ) \otimes 1 - i ( \pi^-_0 \ms X \otimes \varepsilon_x^p ) \otimes i, \notag
\end{align}
with $\ms X \in \tg_{(p), \CC}$ if $x = 0$ and $\ms X \in \tg_{(-p), \CC}$ if $x = \infty$. Moreover, $c \big( \ms X^{(x)}_{\ul p} \big) = (\tau \ms X)^{(x)}_{\ul p}$. In particular, $c \big( \ms X^{(0)}_{\ul p} \big) = \ms X^{(0)}_{\ul p}$ for $\ms X \in \tg_{(p)}$ and $c \big( \ms X^{(\infty)}_{\ul p} \big) = \ms X^{(\infty)}_{\ul p}$ for $\ms X \in \tg_{(-p)}$.
\end{itemize}
\end{subequations}
\begin{proof}
This is a direct application of Lemma \ref{lem: Lie re com}. If $x \in \z_{\rm c}$, in other words $\Pi_x = \{ 1 \}$, then we have $\tg^{n_x x} = (\Tak^{n_x}_x \tg)_\RR$, the complexification of which is isomorphic by Lemma \ref{lem: Lie re com}$(i)$ to the direct sum of complex Lie algebras
\begin{equation*}
\tg^{n_x x} \otimes_\RR \CC \cong
\Tak^{n_x}_x \tg \oplus \overline{\Tak^{n_x}_x \tg}.
\end{equation*}
The explicit form \eqref{Lie re com iso} of the isomorphism and the claim about complex conjugation in the second tensor factor both follow from Lemma \ref{lem: Lie re com}$(i)$.

On the other hand, if $x \in \z^k_{\rm r}$ for some $k \in \ZZ_T$, then $\tg^{n_x x}$ is the real form of $\Tak^{n_x}_x \tg$ with respect to the anti-linear involution $\hat \s^k \circ \hat \t$. It then follows that the complexification of $\tg^{n_x x}$ gives back the generalised Takiff algebra $\Tak^{n_x}_x \tg$, namely in this case
\begin{equation*}
\tg^{n_x x} \otimes_\RR \CC \cong
\Tak^{n_x}_x \tg,
\end{equation*}
with the explicit form \eqref{Lie re com iso 2} of the isomorphism following from Lemma \ref{lem: Lie re com}$(ii)$, by noting that $(\hat\s^k \circ \hat\t)(\ms X \otimes \varepsilon_x^p) = \omega^{-kp} \sigma^k \tau \ms X \otimes \varepsilon_x^p$.

Consider now $x \in \z_{\rm f}$. In this case we have $\tg^{n_x x} = (\Tak^{n_x}_x \tg)^\Pi = ((\Tak^{n_x}_x \tg)^\Gamma)^{\hat \t}$, which is the real form of $(\Tak^{n_x}_x \tg)^\Gamma$ with respect to the anti-linear involution $\hat\t$. It follows once again from Lemma \ref{lem: Lie re com}$(ii)$ that for such $x$ we have
\begin{equation*}
\tg^{n_x x} \otimes_\RR \CC \cong
(\Tak^{n_x}_x \tg)^\Gamma.
\end{equation*}
The explicit form \eqref{Lie re com iso 0} of the isomorphism follows from Lemma \ref{lem: Lie re com}$(ii)$ using the fact that $\hat\t (\ms X \otimes \varepsilon_x^p) = \tau \ms X \otimes \varepsilon_x^p$. Note also that here $\ms X \otimes \varepsilon_x^p \in (\Tak^{n_x}_x \tg)^\Gamma$ provided $\ms X \in \tg_{(p), \CC}$ when $x = 0$ or provided $\ms X \in \tg_{(-p), \CC}$ when $x = \infty$.
\end{proof}
\end{proposition}

Using the notation of Proposition \ref{prop: complexified Takiff}, a basis for $\tg^\D_\CC = \tg^\D \otimes_\RR \CC$ is given by
\begin{subequations} \label{tgD R C basis}
\begin{equation}
I^{\wt a (x)}_{\ul p} \coloneqq (I^{\wt a})^{(x)}_{\ul p}, \qquad
I^{\wt a (\bar x)}_{\ul p} \coloneqq (I^{\wt a})^{(\bar x)}_{\ul p},
\end{equation}
for $x \in \z_{\rm c}$ with $p = 0, \ldots, n_x - 1$,
\begin{equation} \label{tgD R C basis b}
I^{\wt a (x)}_{k \ul p} \coloneqq (I^{\wt a}_{k, p})^{(x)}_{\ul p} = (I^{\wt a}_{k, p} \otimes \varepsilon_x^p) \otimes 1,
\end{equation}
for $x \in \z^k_{\rm r}$ with $k \in \ZZ_T$, $p = 0, \ldots, n_x - 1$, and
\begin{align}
\label{tgD R C basis c} I^{\wt \alpha (0)}_{\ul p} &\coloneqq \big( I^{(p, \wt \alpha)} \big)^{(0)}_{\ul p} = (I^{(p, \wt \alpha)} \otimes \varepsilon_0^p) \otimes 1,\\
\label{tgD R C basis d} I^{\wt \alpha (\infty)}_{\ul {q+1}} &\coloneqq \big( I^{(-q-1, \wt \alpha)} \big)^{(\infty)}_{\ul{q+1}} = (I^{(-q-1, \wt \alpha)} \otimes \varepsilon_{\infty}^{q+1}) \otimes 1,
\end{align}
\end{subequations}
for $p = 0, \ldots, n_0 - 1$ and $q = 0, \ldots, n_\infty - 2$.

We will also often make use the notation
\begin{subequations} \label{notation Ix}
\begin{equation} \label{notation Ix zcr}
I^{a (x)}_{n, \ul p} \coloneqq (I^a_n)^{(x)}_{\ul p}, \qquad I^{(x)}_{a, n, \ul p} \coloneqq (I_{a, n})^{(x)}_{\ul p}
\end{equation}
for any $x \in \z_{\rm c} \cup \bar \z_{\rm c} \cup \z'_{\rm r}$, $p = 0, \ldots, n_x - 1$, $a = 1, \ldots, \dim \g$ and $n \in \ZZ$. Likewise, for the origin we write
\begin{equation} \label{notation Ix 0}
I^{\alpha (0)}_{n, \ul p} \coloneqq \big( I^{(p, \alpha)}_n \big)^{(0)}_{\ul p}, \qquad
I^{(0)}_{\alpha, n, \ul p} \coloneqq \big( I_{(p, \alpha), n} \big)^{(0)}_{\ul p},
\end{equation}
for any $p = 0, \ldots, n_0 - 1$, $\alpha = 1, \ldots, \dim \g_{(p), \CC}$ and $n \in \ZZ$, and for the point at infinity
\begin{equation} \label{notation Ix inf}
I^{\alpha (\infty)}_{n, \ul{q+1}} \coloneqq \big( I^{(-q-1, \alpha)}_n \big)^{(\infty)}_{\ul{q+1}}, \qquad
I^{(\infty)}_{\alpha, n, \ul{q+1}} \coloneqq \big( I_{(-q-1, \alpha), n} \big)^{(\infty)}_{\ul{q+1}},
\end{equation}
\end{subequations}
for $q = 0, \ldots, n_\infty - 2$, $\alpha = 1, \ldots, \dim \g_{(-q-1), \CC}$ and $n \in \ZZ$.

\section{Classical dihedral affine Gaudin models} \label{sec: affine Gaudin}

Let $\tg$ be an affine Kac-Moody algebra equipped with an action $r : \Pi \to \bAut \tg$ of the dihedral group $\Pi$ by (anti-)linear automorphisms as in \S\ref{sec: KM} and let us fix a divisor $\D = \sum_{x \in \z} n_x \, x \in \text{Div}_{\geq 1}(\z)$, cf. \S\ref{sec: Divisors}. We introduce the set of points
\begin{equation*}
\Z \coloneqq \z_{\rm c} \cup \bar \z_{\rm c} \cup \z_{\rm r},
\end{equation*}
and let $\Z' \coloneqq \Z \setminus \{ 0, \infty \} = \z_{\rm c} \cup \bar \z_{\rm c} \cup \z'_{\rm r}$.
Let $\bm \lvl$ be a tuple of complex numbers
\begin{equation} \label{tuple of k}
\begin{split}
&\lvl^x_0, \ldots, \lvl^x_{n_x - 1}, \qquad \text{for each} \quad x \in \Z',\\
&\lvl^0_0, \ldots, \lvl^0_{n_0-1}, \quad \text{and} \quad \lvl^\infty_1, \ldots, \lvl^\infty_{n_\infty - 1}.
\end{split}
\end{equation}
We require that $\lvl^0_p = \lvl^\infty_p = 0$ for $p \not\equiv 0 \;\text{mod}\, T$. Note that the latter condition on $p$ is never satisfied when $T = 1$, so that in this case none of the levels at the origin and infinity are required to vanish.
Throughout this section we will always assume that
\begin{equation} \label{assumption lvl}
\lvl^x_{n_x - 1} \neq 0
\end{equation}
for all $x \in \Z$. In particular, we assume that $n_0, n_\infty \equiv 1 \;\textup{mod}\, T$. The condition \eqref{assumption lvl} will be important in the discussion of \S\ref{sec: level fix}. However, 
we will show in \S\ref{sec: Toda FT} how this assumption can be relaxed on the example of affine $\tg$-Toda field theory where such a condition does not hold.
By convention, we set $\lvl^x_p = 0$ for every $x \in \Z$ and $p \geq n_x$.
We refer to the tuple $\bm \lvl$ as the \emph{levels} and require them to satisfy the following reality conditions: \begin{subequations} \label{levels real}
\begin{alignat}{2}
\label{levels real zc} \overline{\lvl^x_p} &= \lvl^{\bar x}_p &\qquad &\textup{for}\quad x \in \z_{\rm c} \cup \bar \z_{\rm c},\\
\label{levels real zr} \overline{\lvl^x_p} &= \omega^{-kp} \lvl^x_p &\qquad &\textup{for}\quad x \in \z^k_{\rm r},\\
\label{levels real zf} \lvl^x_p &\in \RR & \qquad &\textup{for}\quad x \in \z_{\rm f}.
\end{alignat}
\end{subequations}

In this section we associate a classical dihedral affine Gaudin model to the datum
\begin{equation*}
(\tg, r, \D, \pi_{\bm \lvl})
\end{equation*}
where $\pi_{\bm \lvl}$ is a homomorphism depending on the tuple of levels $\bm \lvl$, from the complexified algebra of formal observables to the complexified algebra of local observables, both introduced in \S\ref{sec: Alg Obs} below. We will construct an explicit such homomorphism in \S\ref{sec: level fix} under the assumption that the condition \eqref{assumption lvl} is satisfied. In \S\ref{sec: Toda FT} we will also give a definition of $\pi_{\bm \lvl}$ when \eqref{assumption lvl} fails to hold, on the example of affine $\tg$-Toda field theory.

\subsection{Algebra of formal observables} \label{sec: Alg Obs}

Let $S(\tg^\D_\CC)$ denote the symmetric algebra on $\tg^\D_\CC$. The Lie bracket on $\tg^\D_\CC$ uniquely extends to a Poisson bracket on $S(\tg^\D_\CC)$, which we denote
\begin{equation} \label{Sg PB}
\{ \cdot, \cdot \} : S(\tg^\D_\CC) \times S(\tg^\D_\CC) \longrightarrow S(\tg^\D_\CC).
\end{equation}
Explicitly, we require $1 \in S(\tg^\D_\CC)$ to lie in the centre of \eqref{Sg PB} and set $\{ \mathfrak X, \mathfrak Y \} \coloneqq [\mathfrak X, \mathfrak Y]$ for any $\mathfrak X, \mathfrak Y \in \tg^\D_\CC$, then use linearity and the Leibniz rule to uniquely define $\{ f, g \}$ for every $f, g \in S(\tg^\D_\CC)$.

The commutative Poisson algebra $S(\tg^\D_\CC)$ is not large enough for our purposes. For instance it does not contain the quadratic Hamiltonians constructed in \S\ref{sec: quad Ham} below. A suitable completion of it is defined as follows.
Recalling the descending $\ZZ_{\geq 0}$-filtration on $\tg$ defined in \eqref{Fil n tg}, to every $n \in \ZZ_{\geq 0}$ and each finite point $x \in \z \setminus \{ \infty \}$ we associate the subspace
\begin{equation*}
\Fil_n \big( \Tak^{n_x}_x \tg \big) \coloneqq \Fil_n \tg \otimes \CC[\varepsilon_x] / \varepsilon_x^{n_x} \CC[\varepsilon_x] \subset \Tak^{n_x}_x \tg,
\end{equation*}
and for the point at infinity we define the subspaces
\begin{equation*}
\Fil_n \big( \Tak^{n_\infty}_\infty \tg \big) \coloneqq \Fil_n \tg \otimes \varepsilon_\infty \CC[\varepsilon_\infty] / \varepsilon_\infty^{n_\infty} \CC[\varepsilon_\infty] \subset \Tak^{n_\infty}_\infty \tg.
\end{equation*}
This defines a descending $\ZZ_{\geq 0}$-filtration on $\Tak^{n_x}_x \tg$ for each $x \in \z$. At every $x \in \z_{\rm c}$, let us also introduce a `conjugate' descending $\ZZ_{\geq 0}$-filtration on $\Tak^{n_x}_x \tg$, which we denote as $\big( \cFil_n \big( \Tak^{n_x}_x \tg \big) \big)_{n \in \ZZ_{\geq 0}}$, defined by the subspaces, cf. \eqref{Fil- n tg},
\begin{equation} \label{cFil n Tak tg}
\cFil_n \big( \Tak^{n_x}_x \tg \big) \coloneqq \cFil_n \tg \otimes \CC[\varepsilon_x] / \varepsilon_x^{n_x} \CC[\varepsilon_x] \subset \Tak^{n_x}_x \tg.
\end{equation}
For any $x \in \z_{\rm f}$, the action \eqref{Pi action Tg} of $\Gamma \subset \Pi$ on $\Tak^{n_x}_x \tg$ preserves the respective subspaces $\Fil_n( \Tak^{n_x}_x \tg)$, $n \in \ZZ_{\geq 0}$ so we may consider the $\Gamma$-invariant subspaces $(\Fil_n( \Tak^{n_x}_x \tg))^\Gamma$. We can now define a descending $\ZZ_{\geq 0}$-filtration on the complex vector space $\tg^\D_\CC$ introduced in Proposition \ref{prop: complexified Takiff} using the isomorphism \eqref{Takiff complex}. Specifically, we let
\begin{equation*}
\Fil_n(\tg^\D_\CC) \coloneqq \psi \Biggl( \bigoplus_{x \in \z_{\rm f}} \big( \Fil_n(\Tak^{n_x}_x \tg) \big)^\Gamma \oplus \bigoplus_{x \in \z'_{\rm r}} \Fil_n(\Tak^{n_x}_x \tg) \oplus \bigoplus_{x \in \z_{\rm c}} \big( \Fil_n(\Tak^{n_x}_x \tg) \oplus \overline{\cFil_n(\Tak^{n_x}_x \tg)} \big) \Biggr),
\end{equation*}
for each $n \in \ZZ_{\geq 0}$. Recalling the fact that the subspaces \eqref{Fil n tg} define a descending $\ZZ_{\geq 0}$-filtration on $\tg$ as a complex Lie algebra and that the linear map $\psi$ from Proposition \ref{prop: complexified Takiff} is an isomorphism of complex Lie algebras, we have
\begin{equation*}
\big[ \Fil_m(\tg^\D_\CC), \Fil_n(\tg^\D_\CC) \big] \subset \Fil_{m+n}(\tg^\D_\CC)
\end{equation*}
for every $m, n \in \ZZ_{\geq 0}$.
It follows that $(\Fil_n(\tg^\D_\CC) )_{n \in \ZZ_{\geq 0}}$ defines a descending $\ZZ_{\geq 0}$-filtration on $\tg^\D_\CC$ as a complex Lie algebra. In turn, using this we set
\begin{equation} \label{S tgDC Fil 1}
\Fil_n\big( S(\tg^\D_\CC) \big) \coloneqq \Fil_n(\tg^\D_\CC) S(\tg^\D_\CC) \cap c\big( \Fil_n(\tg^\D_\CC) \big) S(\tg^\D_\CC),
\end{equation}
where $c : \tg^\D_\CC \to \tg^\D_\CC$ is the anti-linear map introduced in Proposition \ref{prop: complexified Takiff}.
This defines a descending $\ZZ_{\geq 0}$-filtration on the commutative algebra $S(\tg^\D_\CC)$ by ideals. 
It therefore follows that the corresponding completion, which we denote by
\begin{equation*}
\hat S(\tg^\D_\CC) \coloneqq \varprojlim S(\tg^\D_\CC) \big/ \Fil_n\big( S(\tg^\D_\CC) \big)
\end{equation*}
and call the \emph{complexified algebra of formal observables}, is a commutative $\CC$-algebra. We note that
\begin{equation*}
c\big( \Fil_n(\tg^\D_\CC) \big) = \psi \Biggl( \bigoplus_{x \in \z_{\rm f}} \big( \cFil_n(\Tak^{n_x}_x \tg) \big)^\Gamma \oplus \bigoplus_{x \in \z'_{\rm r}} \cFil_n(\Tak^{n_x}_x \tg) \oplus \bigoplus_{x \in \z_{\rm c}} \big( \cFil_n(\Tak^{n_x}_x \tg) \oplus \overline{\Fil_n(\Tak^{n_x}_x \tg)} \big) \Biggr)
\end{equation*}
for each $n \in \ZZ_{\geq 0}$, which defines the `conjugate' descending $\ZZ_{\geq 0}$-filtration on $\tg^\D_\CC$ as a complex Lie algebra.

Although the $\Fil_n \big( S(\tg^\D_\CC) \big)$ are not Poisson ideals of $S(\tg^\D_\CC)$, the Poisson bracket \eqref{Sg PB} is continuous at the origin with respect to the associated topology, where $S(\tg^\D_\CC) \times S(\tg^\D_\CC)$ is given the product topology, since for all $m, n \in \ZZ_{\geq 0}$ we have
\begin{equation*}
\big\{ \Fil_m\big( S(\tg^\D_\CC) \big), \Fil_n\big( S(\tg^\D_\CC) \big) \big\} \subset \Fil_{\text{min}(m, n)}\big( S(\tg^\D_\CC) \big).
\end{equation*}
By linearity it follows that the Poisson bracket \eqref{Sg PB} is uniformly continuous and hence extends to a Poisson bracket
\begin{equation} \label{wtSg PB}
\{ \cdot, \cdot \} : \hat S(\tg^\D_\CC) \times \hat S(\tg^\D_\CC) \longrightarrow \hat S(\tg^\D_\CC).
\end{equation}
on the completion $\hat S(\tg^\D_\CC)$, which is therefore also a Poisson algebra. We note here that the completion of the Cartesian product with respect to the product topology is the Cartesian product of the completions.

We extend the anti-linear automorphism $c : \tg^\D_\CC \to \tg^\D_\CC$ defined in Proposition \ref{prop: complexified Takiff} to an anti-linear automorphism of the Poisson algebra $S(\tg^\D_\CC)$. Since it preserves each subspace $\Fil_n\big( S(\tg^\D_\CC) \big)$, $n \in \ZZ_{\geq 0}$ of the descending $\ZZ_{\geq 0}$-filtration on $S(\tg^\D_\CC)$ introduced in \eqref{S tgDC Fil 1}, by construction of the latter, it follows that the map $c$ is continuous with respect to the associated topology. It therefore extends to an anti-linear automorphism of the completion $\hat S(\tg^\D_\CC)$ which we still denote $c$.
Hence we can consider the real subalgebra $\hat S(\tg^\D_\CC)^c$ of fixed points under $c$ which we refer to as the \emph{algebra of formal observables}.

In \S\ref{sec: alg of obs} below we use the levels \eqref{tuple of k} to define another Poisson algebra $\hat S_{\bm \lvl}(\hg^\D_\CC)$ and in \S\ref{sec: level fix} we construct a homomorphism of Poisson algebras $\pi_{\bm \lvl} : \hat S(\tg^\D_\CC) \to \hat S_{\bm \lvl}(\hg^\D_\CC)$.

\subsubsection{Algebra of local observables} \label{sec: alg of obs}

Recall the Lie subalgebra $\hg$ of $\tg$ defined by \eqref{hg bg def}.
Let $\hg^\D$ be the real Lie subalgebra of $\tg^\D$ defined in the same way as $\tg^\D$ in \S\ref{sec: real Takiff} with $\hg$ replacing $\tg$. Its complexification $\hg^\D_\CC \coloneqq \hg^\D \otimes_\RR \CC$ is a subalgebra of $\tg^\D_\CC$. In particular, the Poisson bracket \eqref{Sg PB} restricts to the symmetric algebra $S(\hg^\D_\CC)$ on $\hg^\D_\CC$ which is thus a Poisson subalgebra of $S(\tg^\D_\CC)$.

Let $J_{\bm \lvl}$ denote the ideal of $S(\hg^\D_\CC)$ generated by the elements
\begin{equation*}
\cent^{(x)}_{\ul p} - i \lvl^x_p 1, \qquad
\cent^{(y)}_{\ul q} + i \lvl^y_q 1, \qquad
\cent^{(0)}_{\ul r} - i \lvl^0_r 1, \qquad
\cent^{(\infty)}_{\ul{s+1}} - i \lvl^\infty_{s+1} 1
\end{equation*}
for every $x \in \z_{\rm c} \cup \z'_{\rm r}$ with $p = 0,\ldots, n_x - 1$, every $y \in \bar\z_{\rm c}$ with $q = 0, \ldots, n_y - 1$ and $r, s+1 \equiv 0 \;\text{mod}\, T$. Since these elements all lie in the centre of the Poisson bracket \eqref{Sg PB}, $J_{\bm \lvl}$ is also a Poisson ideal of $S(\hg^\D_\CC)$. The quotient $S_{\bm \lvl}(\hg^\D_\CC) \coloneqq S(\hg^\D_\CC)/ J_{\bm \lvl}$ is therefore a Poisson algebra whose induced Poisson bracket we denote
\begin{equation} \label{Skg PB}
\{ \cdot, \cdot \} : S_{\bm \lvl}(\hg^\D_\CC) \times S_{\bm \lvl}(\hg^\D_\CC) \longrightarrow S_{\bm \lvl}(\hg^\D_\CC).
\end{equation}
Let $\hg^\D_{\CC, \bm \lvl}$ denote the image of the Lie subalgebra $\hg^\D_\CC \subset S(\hg^\D_\CC)$ under the quotient map $S(\hg^\D_\CC) \twoheadrightarrow S_{\bm \lvl}(\hg^\D_\CC)$. By a slight abuse of notation we will denote the image in $\hg^\D_{\CC, \bm \lvl}$ of an element $\ms X^{(x)}_{\ul p}
\in \hg^\D_\CC$ with $\ms X \in \Loop\g$ by the same symbol.

The descending $\ZZ_{\geq 0}$-filtration on the subalgebra $\hg^\D_\CC$ inherited from $\tg^\D_\CC$ is simply given by $\Fil_n (\hg^\D_\CC) = \hg^\D_\CC \cap \Fil_n (\tg^\D_\CC) = \Fil_n (\tg^\D_\CC)$ for each $n \in \ZZ_{\geq 0}$. Let $\Fil_n (\hg^\D_{\CC, \bm \lvl})$ denote the subspaces of the induced descending $\ZZ_{\geq 0}$-filtration on $\hg^\D_{\CC, \bm \lvl}$. The ideals
\begin{equation} \label{Fn Sl gDC}
\Fil_n\big( S_{\bm \lvl}(\hg^\D_\CC) \big) \coloneqq \Fil_n(\hg^\D_{\CC, \bm \lvl}) S_{\bm \lvl}(\hg^\D_\CC) \cap c\big(  \Fil_n(\hg^\D_{\CC, \bm \lvl}) \big) S_{\bm \lvl}(\hg^\D_\CC),
\end{equation}
for $n \in \ZZ_{\geq 0}$, define a descending $\ZZ_{\geq 0}$-filtration on the Poisson algebra $S_{\bm \lvl}(\hg^\D_\CC)$. As in the above discussion for $\hat S(\tg^\D_\CC)$, since the Poisson bracket \eqref{Skg PB} is uniformly continuous with respect to the associated topology, it follows that the corresponding completion
\begin{equation} \label{hat S l gDC def}
\hat S_{\bm \lvl}(\hg^\D_\CC) \coloneqq \varprojlim S_{\bm \lvl}(\hg^\D_\CC) \big/ \Fil_n\big( S_{\bm \lvl}(\hg^\D_\CC) \big),
\end{equation}
the \emph{complexified algebra of local observables}, is also a Poisson algebra over $\CC$, whose Poisson bracket we denote
\begin{equation} \label{hat Skg PB}
\{ \cdot, \cdot \} : \hat S_{\bm \lvl}(\hg^\D_\CC) \times \hat S_{\bm \lvl}(\hg^\D_\CC) \longrightarrow \hat S_{\bm \lvl}(\hg^\D_\CC).
\end{equation}

The restriction of $c : \tg^\D_\CC \to \tg^\D_\CC$ to $\hg^\D_\CC \subset \tg^\D_\CC$ extends as an anti-linear automorphism to the Poisson algebra $S(\hg^\D_\CC)$. Using the reality conditions \eqref{levels real} on the tuple of levels $\bm\lvl$ and the properties of the isomorphism $\psi$ with regards to complex conjugation $c$ given in Proposition \ref{prop: complexified Takiff}, it follows that the ideal $J_{\bm \lvl}$ is invariant under $c$. Hence $c$ acts on the quotient $S_{\bm \lvl}(\hg^\D_\CC)$. And since it preserves each of the subspaces in \eqref{Fn Sl gDC}, it is continuous with respect to the associated topology and so extends to the completion $\hat S_{\bm \lvl}(\hg^\D_\CC)$.
We then define the \emph{algebra of local observables} of the classical dihedral affine Gaudin model as the real subalgebra $\hat S_{\bm \lvl}(\hg^\D_\CC)^c$ of fixed points under $c : \hat S_{\bm \lvl}(\hg^\D_\CC) \to \hat S_{\bm \lvl}(\hg^\D_\CC)$.

\subsubsection{Fixing the levels} \label{sec: level fix}

Let $\mathcal C(n)$ denote the set of compositions of $n$ if $n \in \ZZ_{\geq 0}$ or the empty set if $n \in \ZZ_{< 0}$. By convention the empty composition is the only composition of $0$, \emph{i.e.} $\mathcal C(0) = \{ \emptyset \}$. Given a composition $c \in \mathcal C(n)$ for some $n \in \ZZ_{\geq 0}$ we denote by $|c|$ its length and by $c_j \neq 0$, $j=1, \ldots, |c|$ its parts so that $c_1 + \ldots + c_{|c|} = n$.

\begin{lemma} \label{lem: kappa sol}
Let $n \in \ZZ_{\geq 1}$ and $\lvl_p \in \CC$, $p \in \ZZ_{\geq 0}$ be such that $\lvl_{n-1} \neq 0$ and $\lvl_p = 0$ for all $p \geq n$. Consider the system of linear equations for $\kappa_p$, $p = 0, \ldots, 2 n - 2$ given by
\begin{equation} \label{kappa rel lem}
\sum_{p \geq 0} \kappa_{p+q} \lvl_{p + s} = \delta_{q, s}
\end{equation}
with $q, s = 0, \ldots, n - 1$. It has the unique solution
\begin{equation} \label{kappa sol lem}
\kappa_p = \kappa_p(\lvl_0, \ldots, \lvl_{n-1}) \coloneqq \sum_{c \in \mathcal C(p-n+1)} (-1)^{|c|} \frac{\prod_{j=1}^{|c|} \lvl_{n-1-c_j}}{(\lvl_{n-1})^{|c|+1}}.
\end{equation}
Moreover, if $n \equiv 1 \;\textup{mod}\, T$ and $\lvl_p = 0$ for $p \not\equiv 0 \;\textup{mod}\, T$ then $\kappa_p = 0$ for $p \not\equiv 0 \;\textup{mod}\, T$.
\begin{proof}
Note that the sum over $p$ in \eqref{kappa rel lem} truncates to a sum from $0$ to $n-1$ by virtue of the assumption $\lvl_r = 0$ for $r \geq n$.
Performing the change of variable $p \mapsto n - 1 - p$ in \eqref{kappa rel lem} we obtain
\begin{equation*}
\sum_{p=0}^{n - 1} \kappa_{n-1+q-p} \lvl_{n-1+s-p} = \delta_{q, s}.
\end{equation*}
Since $(\lvl_{n-1+s-p})_{p,s=0}^{n-1}$ is a triangular (Toeplitz) matrix with $\lvl_{n-1} \neq 0$ along the diagonal it is invertible, and hence \eqref{kappa rel lem} admits a unique solution.

Setting $q = 0$ in \eqref{kappa rel lem} we get $\sum_{p=0}^{n - 1} \kappa_p \lvl_{p+s} = \delta_{0, s}$. By considering this equation for all values of $s$ from $n-1$ down to $1$, it follows using induction that $\kappa_p = 0$ for all $p = 0, \ldots, n-2$. Finally, setting $q = n-1$ in \eqref{kappa rel lem} gives $\sum_{p=0}^{n - 1} \kappa_{p+n-1} \lvl_{p+s} = \delta_{n-1, s}$, which can be rewritten as the following recurrence relation
\begin{equation*}
\kappa_{n-1} = \frac{1}{\lvl_{n-1}}, \qquad
\kappa_{n-1+r} = - \sum_{s=1}^r \frac{\lvl_{n-1-s}}{\lvl_{n-1}} \kappa_{n-1+r-s}
\end{equation*}
for $r = 1, \ldots, n-1$.
The explicit expression for $\kappa_p$ with $p = 0, \ldots, 2 n - 2$ now follows by induction, and the last statement can be seen directly from the above recurrence relation.
\end{proof}
\end{lemma}

We define a linear map
\begin{subequations} \label{pi D k def}
\begin{equation}
\pi_{\bm \lvl} : \tg^\D_\CC \longrightarrow
\hat S_{\bm \lvl}(\hg^\D_\CC)
\end{equation}
as follows. For any $x \in \z_{\rm c} \cup \z'_{\rm r}$ and $r = 0, \ldots, n_x - 1$ we let
\begin{align} \label{pi l Dzczr def}
\pi_{\bm \lvl}\big(\cocent^{(x)}_{\ul r}\big) &\coloneqq \frac{i}{2} \sum_{p, q \geq 0} \sum_{n \in \ZZ} \kappa^x_{p+q-r} I^{(x)}_{a, -n, \ul p} I^{a (x)}_{n, \ul q},\\
\pi_{\bm \lvl}\big(\cent^{(x)}_{\ul r}\big) &\coloneqq i \lvl^x_r 1, \qquad
\pi_{\bm \lvl}\big(I^{a (x)}_{n, \ul r}\big) \coloneqq I^{a (x)}_{n, \ul r},
\end{align}
where $\kappa^x_p \coloneqq \kappa_p(\lvl^x_0, \ldots, \lvl^x_{n_x-1})$ for $p = 0, \ldots, 2 n_x - 2$ is given by Lemma \ref{lem: kappa sol}. Note that the assumption \eqref{assumption lvl} is used here for satisfying the conditions of the lemma.
For any $y \in \bar\z_{\rm c}$ and $r = 0, \ldots, n_y - 1$ we let
\begin{align} \label{pi l Dbzc def}
\pi_{\bm \lvl}\big(\cocent^{(y)}_{\ul r}\big) &\coloneqq - \frac{i}{2} \sum_{p, q \geq 0} \sum_{n \in \ZZ} \kappa^y_{p+q-r} I^{(y)}_{a, -n, \ul p} I^{a (y)}_{n, \ul q},\\
\pi_{\bm \lvl}\big(\cent^{(y)}_{\ul r}\big) &\coloneqq - i \lvl^y_r 1, \qquad
\pi_{\bm \lvl}\big(I^{a (y)}_{n, \ul r}\big) \coloneqq I^{a (y)}_{n, \ul r},
\end{align}
where $\kappa^y_p \coloneqq \kappa_p(\lvl^y_0, \ldots, \lvl^y_{n_y-1})$ for $p = 0, \ldots, 2 n_y - 2$ is given again by Lemma \ref{lem: kappa sol}. In particular, it follows from the reality conditions \eqref{levels real zc} and the definition \eqref{kappa sol lem} that $\overline{\kappa^x_p} = \kappa^{\bar x}_p$ for any $x \in \z_{\rm c} \cup \bar\z_{\rm c}$. At the origin, for any $r = 0, \ldots, n_0 - 1$ with $r \equiv 0 \;\text{mod}\, T$ and any $s = 0, \ldots, n_0 - 1$ we let
\begin{align} \label{pi l D0 def}
\pi_{\bm \lvl}\big(\cocent^{(0)}_{\ul r}\big) &\coloneqq \frac{i}{2} \sum_{p, q \geq 0} \sum_{n \in \ZZ} \kappa^0_{p+q-r} I^{(0)}_{\alpha, -n, \ul p} I^{\alpha (0)}_{n, \ul q},\\
\pi_{\bm \lvl}\big(\cent^{(0)}_{\ul r}\big) &\coloneqq i \lvl^0_r 1, \qquad
\pi_{\bm \lvl}\big(I^{\alpha (0)}_{n, \ul s}\big) \coloneqq I^{\alpha (0)}_{n, \ul s},
\end{align}
where $\kappa^0_p \coloneqq \kappa_p(\lvl^0_0, \ldots, \lvl^0_{n_0-1})$ for $p = 0, \ldots, 2 n_0 - 2$ is determined by Lemma \ref{lem: kappa sol}. Here we used again the assumption that $\lvl^0_{n_0 - 1} \neq 0$ from \eqref{assumption lvl}. We note that $\kappa^0_p = 0$ whenever $p \not\equiv 0 \;\text{mod}\, T$, using the last part of Lemma \ref{lem: kappa sol}, so that the double sum over $p, q \geq 0$ restricts to $p$ and $q$ satisfying $p+q \equiv 0 \;\text{mod}\, T$ and hence the implicit summation over the repeated index $\alpha = 1, \ldots, \dim \g_{(p), \CC} = \dim \g_{(q), \CC}$ in \eqref{pi l D0 def} makes sense.
Finally, at infinity we set, for $r = 0, \ldots, n_\infty - 2$ with $r+1 \equiv 0 \; \textup{mod}\, T$ and any $s = 0, \ldots, n_\infty - 2$,
\begin{align} \label{pi l Dinf def 1}
\pi_{\bm \lvl}\big(\cocent^{(\infty)}_{\ul{r+1}}\big) &\coloneqq \frac{i}{2} \sum_{p, q \geq 0} \sum_{n \in \ZZ} \kappa^\infty_{p+q-r+1} I^{(\infty)}_{\alpha, -n, \ul{p+1}} I^{\alpha (\infty)}_{n, \ul{q+1}},\\
\label{pi l Dinf def 2} \pi_{\bm \lvl}\big(\cent^{(\infty)}_{\ul{r+1}}\big) &\coloneqq i \lvl^\infty_{r+1} 1, \qquad
\pi_{\bm \lvl}\big(I^{\alpha (\infty)}_{n, \ul{s+1}}\big) \coloneqq I^{\alpha (\infty)}_{n, \ul{s+1}},
\end{align}
\end{subequations}
where $\kappa^\infty_{p+1} \coloneqq \kappa_{p+1}(0, \lvl^\infty_1, \ldots, \lvl^\infty_{n_\infty - 1})$ for each $p = 0, \ldots, 2 n_\infty - 4$. We note here that the expression $\kappa_{p+1}(\ell, \lvl^\infty_1, \ldots, \lvl^\infty_{n_\infty - 1})$ does not depend on $\ell$ whenever $p \leq 2 n_\infty - 4$, as can be seen directly from the explicit formula \eqref{kappa sol lem}. Now from Lemma \ref{lem: kappa sol} we have $\kappa^\infty_{p+q-r+1} = 0$ unless $p+q-r+1 \equiv 0 \; \textup{mod}\, T$, or in other words $p+q+2 \equiv 0 \;\textup{mod}\, T$, so that the implicit sum over $\alpha = 1, \ldots, \dim \g_{(-p-1), \CC} = \dim \g_{(-q-1), \CC}$ in \eqref{pi l Dinf def 1} makes sense.

\begin{remark} \label{rem: infinite sums}
The infinite sums over $n \in \ZZ$ in \eqref{pi l Dzczr def}, \eqref{pi l Dbzc def}, \eqref{pi l D0 def} and \eqref{pi l Dinf def 1} are used to denote elements of the completion $\hat S_{\bm \lvl}(\hg^\D_\CC)$ defined in \S\ref{sec: alg of obs}. For instance, for each $x \in \z_{\rm c} \cup \z'_{\rm r}$ and $p, q, r = 0, \ldots, n_x - 1$, the formal infinite sum
\begin{equation*}
\sum_{n \in \ZZ} \kappa^x_{p+q-r} I^{(x)}_{a, -n, \ul p} I^{a (x)}_{n, \ul q}
\end{equation*}
appearing in \eqref{pi l Dzczr def} represents the element of the inverse limit \eqref{hat S l gDC def} given by
\begin{equation} \label{inv lim example zc}
\Bigg( \sum_{n = -k+1}^{k-1} \kappa^x_{p+q-r} I^{(x)}_{a, -n, \ul p} I^{a (x)}_{n, \ul q} + \Fil_k \big( S_{\bm \lvl}(\hg^\D_\CC) \big) \Bigg)_{k \in \ZZ_{\geq 0}}.
\end{equation}
To see that this defines an element of the inverse limit, let $k \in \ZZ_{\geq 0}$ and consider the corresponding term in the above sequence. Given any $j \in \ZZ_{\geq 0}$ with $j < k$ we note that the terms in the finite sum over $n$ for which $|n| \geq j$ belong to $\Fil_j \big( S_{\bm \lvl}(\hg^\D_\CC) \big)$. Indeed, if $j \leq n < k$ then the factor $I^{a(x)}_{n, \ul q}$ in such a term belongs to the image of the subspace $\psi( \Fil_j(\Tak^{n_x}_x \hg) ) \subset \hg^\D_\CC$ in $\hg^\D_{\CC, \bm \lvl}$ whereas $I^{(x)}_{a, -n, \ul p}$ belongs to the image of $\psi( \cFil_j(\Tak^{n_x}_x \hg) ) \subset \hg^\D_\CC$ in $\hg^\D_{\CC, \bm \lvl}$, and vice versa if $- k < n \leq - j$. Therefore, under the canonical linear map
\begin{equation*}
\pi^k_j : S_{\bm \lvl}(\hg^\D_\CC) / \Fil_k \big( S_{\bm \lvl}(\hg^\D_\CC) \big) \longtwoheadrightarrow S_{\bm \lvl}(\hg^\D_\CC) / \Fil_j \big( S_{\bm \lvl}(\hg^\D_\CC) \big)
\end{equation*}
the $k^{\rm th}$ term in \eqref{inv lim example zc} is sent to the $j^{\rm th}$ one, as required. The same argument applies to the infinite sums over $n \in \ZZ$ in \eqref{pi l Dbzc def}, \eqref{pi l D0 def} and \eqref{pi l Dinf def 1}.
\end{remark}

\begin{example} \label{ex: piD 2x inf}
Let $\D = \sum_{x \in \CP} n_x x \in \textup{Div}_{\geq 1}(\z)$. If $x \in \z \setminus \{ 0, \infty \}$ is such that $n_x = 1$ then
\begin{equation} \label{Seg-Sug standard}
\pi_{\bm \lvl}\big(\cocent^{(x)}_{\ul 0}\big) = \frac{i}{2 \lvl^x_0} \sum_{n \in \ZZ} I^{(x)}_{a, -n, \ul 0} I^{a (x)}_{n, \ul 0}.
\end{equation}
This coincides with the standard expression for the classical Segal-Sugawara operator $L_0$ in terms of the Kac-Moody algebra generators $I^{a (x)}_{n, \ul 0}$.
If $x \in \z \setminus \{ 0, \infty \}$ is such that $n_x = 2$ then instead we have
\begin{align} \label{Seg-Sug Takiff}
\pi_{\bm \lvl}\big(\cocent^{(x)}_{\ul 0}\big) &= \frac{i}{\lvl^x_1} \sum_{n \in \ZZ} I^{(x)}_{a, -n, \ul 0} I^{a (x)}_{n, \ul 1} - \frac{i \lvl^x_0}{2 (\lvl^x_1)^2} \sum_{n \in \ZZ} I^{(x)}_{a, -n, \ul 1} I^{a (x)}_{n, \ul 1},\\
\notag \pi_{\bm \lvl}\big(\cocent^{(x)}_{\ul 1}\big) &= \frac{i}{2 \lvl^x_1} \sum_{n \in \ZZ} I^{(x)}_{a, -n, \ul 1} I^{a (x)}_{n, \ul 1}.
\end{align}
Expression \eqref{Seg-Sug Takiff} is the classical analogue of the generalised Segal-Sugawara operator $L_0$ constructed using the generators $I^{a(x)}_{n, \ul 0}$ and $I^{a(x)}_{n, \ul 1}$ of the generalised Takiff algebra for $\hg$, whose quantum counterpart can be found in \cite[Theorem 1]{Babichenko:2012uq}.
\end{example}

Recall the definition of the anti-linear automorphisms $c$ of $\tg^\D_\CC$ and $\hat S_{\bm \lvl}(\hg^\D_\CC)$ defined in Proposition \ref{prop: complexified Takiff} and \S\ref{sec: alg of obs} respectively, along with the reality conditions \eqref{levels real} on the tuple of levels $\bm \lvl$.

\begin{proposition} \label{prop: pil equiv}
The map \eqref{pi D k def} is $c$-equivariant in the sense that the diagram
\begin{equation*}
\begin{tikzpicture}
\matrix (m) [matrix of math nodes, row sep=3em, column sep=4em,text height=1.5ex, text depth=0.25ex]    
{
\tg^\D_\CC & \hat S_{\bm \lvl}(\hg^\D_\CC)\\
\tg^\D_\CC & \hat S_{\bm \lvl}(\hg^\D_\CC)\\
};
\path[->] (m-1-1) edge node[above]{$\pi_{\bm \lvl}$} (m-1-2);
\path[->] (m-2-1) edge node[below]{$\pi_{\bm \lvl}$} (m-2-2);
\path[->] (m-1-1) edge node[left]{$c$} (m-2-1);
\path[->] (m-1-2) edge node[right]{$c$} (m-2-2);
\end{tikzpicture}
\end{equation*}
is commutative.
\begin{proof}
We wish to show that
\begin{equation*}
(\pi_{\bm \lvl} \circ c)\big( \ms X^{(x)}_{\ul r} \big) = (c \circ \pi_{\bm \lvl})\big( \ms X^{(x)}_{\ul r} \big)
\end{equation*}
for every $x \in \Z$, $\ms X \in \tg$ and $r \geq 0$. This is clear for all $x \in \Z$ and any $\ms X \in \Loop\g$ since $\pi_{\bm \lvl}$ effectively acts as the identity in this case. By linearity it remains to consider the cases when $\ms X = \cent$ and $\ms X = \cocent$.

Consider first the case $\ms X = \cent$. For any $x \in \z_{\rm c}$ and any $r = 0, \ldots, n_x - 1$ we have
\begin{equation*}
(\pi_{\bm \lvl} \circ c)\big( \cent^{(x)}_{\ul r} \big) = \pi_{\bm \lvl}\big( \cent^{(\bar x)}_{\ul r} \big) = - i \lvl^{\bar x}_r \, 1 = \overline{i \lvl^x_r} \, 1 = (c \circ \pi_{\bm \lvl})\big( \cent^{(x)}_{\ul r} \big) 
\end{equation*}
where the first equality uses Proposition \ref{prop: complexified Takiff}$(i)$, the second is by definition of $\pi_{\bm \lvl}$, the third equality uses the reality conditions \eqref{levels real zc} and in the last equality we use once again the definition of $\pi_{\bm \lvl}$. The proof of the statement for $x \in \bar\z_{\rm c}$ now also follows since $c\big( \cent^{(x)}_{\ul r} \big) = \cent^{(\bar x)}_{\ul r}$ from Proposition \ref{prop: complexified Takiff}$(i)$ and $c$ is an involution.
For a point $x \in \z^k_{\rm r}$ with $k \in \ZZ_T$ and $r = 0, \ldots, n_x - 1$ we have
\begin{equation*}
(\pi_{\bm \lvl} \circ c)\big( \cent^{(x)}_{\ul r} \big) = - \omega^{-k r} \pi_{\bm \lvl}\big( \cent^{(x)}_{\ul r} \big) = - i \omega^{-k r} \lvl^x_r \, 1 = \overline{i \lvl^x_r} \, 1 = (c \circ \pi_{\bm \lvl})\big( \cent^{(x)}_{\ul r} \big) 
\end{equation*}
where in the first equality we used Proposition \ref{prop: complexified Takiff}$(ii)$ and the fact that $\sigma \cent = \cent$ and $\tau \cent = - \cent$, in the second equality we used the definition of $\pi_{\bm \lvl}$, in the third equality the reality conditions \eqref{levels real zr} and in last equality the definition of $\pi_{\bm \lvl}$ once again. The cases of the points at the origin and infinity are shown similarly.

Finally, suppose $\ms X = \cocent$ and consider first a point $x \in \z_{\rm c}$. For any $r = 0, \ldots, n_x - 1$ we have
\begin{equation*}
(\pi_{\bm \lvl} \circ c)\big( \cocent^{(x)}_{\ul r} \big) = \pi_{\bm \lvl}\big( \cocent^{(\bar x)}_{\ul r} \big) = - \frac{i}{2} \sum_{p, q \geq 0} \sum_{n \in \ZZ} \kappa^{\bar x}_{p+q-r} I^{(\bar x)}_{a, -n, \ul p} I^{a (\bar x)}_{n, \ul q}.
\end{equation*}
On the other hand,
\begin{align*}
(c \circ \pi_{\bm \lvl})\big( \cocent^{(x)}_{\ul r} \big) &= - \frac{i}{2} \sum_{p, q \geq 0} \sum_{n \in \ZZ} \overline{\kappa^x_{p+q-r}} c \big( I^{(x)}_{a, -n, \ul p} \big) c \big( I^{a (x)}_{n, \ul q} \big)\\
&= - \frac{i}{2} \sum_{p, q \geq 0} \sum_{n \in \ZZ} \overline{\kappa^x_{p+q-r}} I^{(\bar x)}_{a, -n, \ul p} I^{a (\bar x)}_{n, \ul q},
\end{align*}
where in the second line we used Proposition \ref{prop: complexified Takiff}$(i)$. It therefore remains to check that $\overline{\kappa^x_{p+q-r}} = \kappa^{\bar x}_{p+q-r}$, which follows from the definition of $\kappa^x_p$ for $x \in \z_{\rm c} \cup \bar \z_{\rm c}$ and $p = 0, \ldots, 2n_x - 2$ in terms of \eqref{kappa sol lem}, and the reality conditions \eqref{levels real zc}. The statement for $x \in \bar \z_{\rm c}$ now follows immediately using the fact that $c\big( \cocent^{(x)}_{\ul r} \big) = \cocent^{(\bar x)}_{\ul r}$ from Proposition \ref{prop: complexified Takiff}$(i)$ and that $c$ is an involution.

Next, consider a non-fixed real point $x \in \z^k_{\rm r}$. In this case, for any $r = 0, \ldots, n_x - 1$ we have
\begin{equation*}
(\pi_{\bm \lvl} \circ c)\big( \cocent^{(x)}_{\ul r} \big) = - \omega^{-kr} \pi_{\bm \lvl}\big( \cocent^{(x)}_{\ul r} \big) = - \frac{i}{2} \sum_{p, q \geq 0} \sum_{n \in \ZZ} \omega^{-kr} \kappa^x_{p+q-r} I^{(x)}_{a, -n, \ul p} I^{a (x)}_{n, \ul q},
\end{equation*}
where the first equality makes use of Proposition \ref{prop: complexified Takiff}$(ii)$ and the fact that $\sigma \cocent = \cocent$ and $\tau \cocent = - \cocent$. On the other hand, we have
\begin{align*}
(c \circ \pi_{\bm \lvl})\big( \cocent^{(x)}_{\ul r} \big) &= - \frac{i}{2} \sum_{p, q \geq 0} \sum_{n \in \ZZ} \overline{\kappa^x_{p+q-r}} c\big( I^{(x)}_{a, -n, \ul p} \big) c\big( I^{a (x)}_{n, \ul q} \big)\\
&= - \frac{i}{2} \sum_{p, q \geq 0} \sum_{n \in \ZZ} \overline{\kappa^x_{p+q-r}} \omega^{-kp} \big( (\sigma^k I^a)_n \big)^{(x)}_{\ul p} \omega^{-kq} \big( (\sigma^k I_a)_{-n} \big)^{(x)}_{\ul q},\\
&= - \frac{i}{2} \sum_{p, q \geq 0} \sum_{n \in \ZZ} \omega^{-k(p+q)} \overline{\kappa^x_{p+q-r}} I^{(x)}_{a, -n, \ul p} I^{a (x)}_{n, \ul q},
\end{align*}
where in the first line we used the definition of $\pi_{\bm \lvl}$, in the second we used Proposition \ref{prop: complexified Takiff}$(ii)$ and in the last line the $\sigma$-invariance of the canonical element $I_a \otimes I^a$ together with the change variable $n \to -n$ in the sum over $n \in \ZZ$. In order to prove the desired result it remains to check that $\overline{\kappa^x_p} = \omega^{kp} \kappa^x_p$. But from the definition of $\kappa^x_p$ and the reality conditions \eqref{levels real zr} we find
\begin{align*}
\overline{\kappa^x_p} &= \sum_{c \in \mathcal C(p-n_x+1)} (-1)^{|c|} \frac{\prod_{j=1}^{|c|} \overline{\lvl^x_{n_x-1-c_j}}}{(\overline{\lvl^x_{n_x-1}})^{|c|+1}}\\
&= \sum_{c \in \mathcal C(p-n_x+1)} (-1)^{|c|} \frac{\omega^{-k \sum_{j=1}^{|c|} c_j}}{\omega^{-k(n_x - 1)}} \frac{\prod_{j=1}^{|c|} \lvl^x_{n_x-1-c_j}}{(\lvl^x_{n_x-1})^{|c|+1}} = \omega^{-kp} \kappa^x_p,
\end{align*}
where in the last step we have used $\sum_{j=1}^{|c|} c_j = p - n_x + 1$ since $c \in \mathcal C(p-n_x+1)$.

The proof of the statement for $\ms X = \cocent$ when $x$ is the origin or infinity follows from similar considerations.
\end{proof}
\end{proposition}

\begin{proposition} \label{prop: hom pi lvl}
The map \eqref{pi D k def} is a homomorphism of Lie algebras.
\begin{proof}
The only non-trivial relations to check are
\begin{equation*}
\big\{ \pi_{\bm \lvl}\big(\cocent^{(x)}_{\ul r}\big), \pi_{\bm \lvl}\big(\ms X^{(x)}_{\ul s}\big) \big\} = \pi_{\bm \lvl}\big( \big[ \cocent^{(x)}_{\ul r}, \ms X^{(x)}_{\ul s} \big] \big)
\end{equation*}
for any $x \in \Z'$, $\ms X \in \tg$ or $x = 0$, $\ms X \in \tg_{(s)}$ with $r, s \in \ZZ_{\geq 0}$, or for $x = \infty$, $\ms X \in \tg_{(-s)}$ with $r, s \in \ZZ_{\geq 1}$. We consider these three cases $x \in \Z'$, $x = 0$ and $x = \infty$ in turn. In fact, the case $x \in \bar\z_{\rm c}$ will follow from the result with $x \in \z_{\rm c}$ by applying the anti-linear map $c : \tg^\D_\CC \to \tg^\D_\CC$ of Proposition \ref{prop: complexified Takiff}. Therefore we consider only $x \in \z_{\rm c} \cup \z'_{\rm r}$, $x = 0$ and $x = \infty$. In each case, both sides of the above equality clearly vanish when $\ms X = \cent$. We therefore only need to consider the case when $\ms X \in \Loop\g$ and $\ms X = \cocent$.

Let $x \in \z_{\rm c} \cup \z'_{\rm r}$ and suppose first that $\ms X \in \Loop\g$. By linearity it suffices to consider $\ms X = I^b_m$ with $b = 1, \ldots, \dim \g$ and $m \in \ZZ$.
Computing $\big\{ \pi_{\bm \lvl}\big( \cocent^{(x)}_{\ul r} \big), \pi_{\bm \lvl}\big( I^{b (x)}_{m, \ul s} \big) \big\}$ we find
\begin{align*}
&\frac{i}{2} \sum_{p, q \geq 0} \sum_{n \in \ZZ} \kappa^x_{p+q-r} \big( [I_a, I^b]^{(x)}_{m-n, \ul {p+s}} - i n \lvl^x_{p+s} \delta_{m-n, 0} \langle I_a, I^b \rangle \big) I^{a(x)}_{n, \ul q}\\
&\qquad + \frac{i}{2} \sum_{p, q \geq 0} \sum_{n \in \ZZ} \kappa^x_{p+q-r} I^{(x)}_{a, -n, \ul p} \big( [I^a, I^b]^{(x)}_{n+m, \ul {q+s}} + i n \lvl^x_{q+s} \delta_{n+m, 0} \langle I^a, I^b \rangle \big)\\
&= i \sum_{p, q \geq 0} \sum_{n \in \ZZ} \kappa^x_{p+q-r} [I_a, I^b]^{(x)}_{m-n, \ul {p+s}} I^{a(x)}_{n, \ul q} + m \sum_{p, q \geq 0} \kappa^x_{p+q-r} \lvl^x_{p+s} I^{b(x)}_{m, \ul q}.
\end{align*}
We have $q - r \geq n_x - 1 - p \geq s \geq 0$ and $q - r \leq q \leq n_x - 1$ so that $0 \leq q-r \leq n_x - 1$.
The last term on the right then simplifies to $m I^{b(x)}_{m, \ul {r+s}}$ using the identity \eqref{kappa rel lem} from Lemma \ref{lem: kappa sol} satisfied by $\kappa^x_{p+q-r}$. On the other hand, the first term is seen to vanish as follows. We have
\begin{align*}
\sum_{p, q \geq 0} \sum_{n \in \ZZ} &\kappa^x_{p+q-r} [I_a, I^b]^{(x)}_{m-n, \ul {p+s}} I^{a(x)}_{n, \ul q} = - \sum_{p, q \geq 0} \sum_{n \in \ZZ} \kappa^x_{p+q-r} I^{a (x)}_{m-n, \ul {p+s}} [I_a, I^b]^{(x)}_{n, \ul q}\\
&\qquad\qquad = - \sum_{p, q \geq 0} \sum_{n \in \ZZ} \kappa^x_{p+q-r} [I_a, I^b]^{(x)}_{m-n, \ul q} I^{a (x)}_{n, \ul {p+s}}.
\end{align*}
In the first equality we used the adjoint invariance of $I_a \otimes I^a$ together with the linearity of the composition $\g \to \tg \to \tg^\D_\CC$, $\ms x \mapsto \ms x_n \mapsto \ms x^{(x)}_{n, \ul p}$ for each $x \in \z_{\rm c} \cup \z'_{\rm r}$ and $p \in \ZZ_{\geq 0}$, where the second map is from Proposition \ref{prop: complexified Takiff}. In the second equality above we performed the change of variables $n \mapsto m - n$ in the sum over $n$.
Now since $\kappa^x_{p+q-r} = 0$ for all $r = 0, \ldots, n_x - 1$ whenever $p + q < n_x - 1$, it follows that $q \geq n_x - 1 - p \geq s$ in the above sums. Performing the change of variable $p \mapsto q - s$ and $q \mapsto p + s$ in the sums on the right hand side we find that it coincides with the sum on the left hand side up to an extra overall sign, and hence vanishes. We deduce that
\begin{equation*}
\big\{ \pi_{\bm \lvl}\big( \cocent^{(x)}_{\ul r} \big), \pi_{\bm \lvl}\big( I^{b (x)}_{m, \ul s} \big) \big\} = m I^{b(x)}_{m, \ul {r+s}} = m \, \pi_{\bm \lvl} \big(I^{b(x)}_{m, \ul {r+s}}\big) = \pi_{\bm \lvl}\big( \big[ \cocent^{(x)}_{\ul r}, I^{b (x)}_{m, \ul s} \big] \big),
\end{equation*}
the result for arbitrary $\ms X \in \Loop\g$ following by linearity. Let us suppose now that $\ms X = \cocent$. Using the above result and the Leibniz rule we find
\begin{align*}
\big\{ \pi_{\bm \lvl}\big( \cocent^{(x)}_{\ul r} \big), \pi_{\bm \lvl}\big( \cocent^{(x)}_{\ul s} \big) \big\} &= - \frac{i}{2} \sum_{p, q \geq 0} \sum_{n \in \ZZ} n \kappa^x_{p+q-s} I^{(x)}_{a, -n, \ul{p+r}} I^{a (x)}_{n, \ul q}\\
&\qquad\qquad + \frac{i}{2} \sum_{p, q \geq 0} \sum_{n \in \ZZ} n \kappa^x_{p+q-s} I^{(x)}_{a, -n, \ul p} I^{a (x)}_{n, \ul{q+r}}.
\end{align*}
Since $\kappa^x_p = 0$ whenever $p < n_x - 1$, it follows that in the first sum on the right hand side we have $p \geq s$ and $q \geq r+s$ while in the second $p \geq r+s$ and $q \geq s$. Performing the change of variables $p \mapsto p+r$ and $q \mapsto q+r$ in the first and second sum, respectively, we find that they cancel so that
\begin{equation*}
\big\{ \pi_{\bm \lvl}\big( \cocent^{(x)}_{\ul r} \big), \pi_{\bm \lvl}\big( \cocent^{(x)}_{\ul s} \big) \big\} = 0 = \pi_{\bm \lvl}\big( \big[ \cocent^{(x)}_{\ul r}, \cocent^{(x)}_{\ul s} \big] \big)
\end{equation*}
for all $x \in \z_{\rm c} \cup \z'_{\rm r}$ and $r, s \in \ZZ_{\geq 0}$, establishing the result for $\ms X = \cocent$.

Next, consider the origin and let $\ms X = I^{(s, \beta)}_m$ with $\beta = 1, \ldots, \dim \g_{(s), \CC}$ and $m \in \ZZ$.
Computing $\big\{ \pi_{\bm \lvl}\big( \cocent^{(0)}_{\ul r} \big), \pi_{\bm \lvl}\big( I^{\beta (0)}_{m, \ul s} \big) \big\}$ we find
\begin{align*}
&\frac{i}{2} \sum_{p, q \geq 0} \sum_{n \in \ZZ} \kappa^0_{p+q-r} \big( [I_{(p, \alpha)}, I^{(s, \beta)}]^{(0)}_{m-n, \ul {p+s}} - i n \lvl^0_{p+s} \delta_{m-n, 0} \langle I_{(p, \alpha)}, I^{(s, \beta)} \rangle \big) I^{\alpha(0)}_{n, \ul q}\\
&+ \frac{i}{2} \sum_{p, q \geq 0} \sum_{n \in \ZZ} \kappa^0_{p+q-r} I^{(0)}_{\alpha, -n, \ul p} \big( [I^{(q, \alpha)}, I^{(s, \beta)}]^{(0)}_{n+m, \ul {q+s}} + i n \lvl^0_{q+s} \delta_{n+m, 0} \langle I^{(q, \alpha)}, I^{(s, \beta)} \rangle \big)\\
&= i \sum_{p, q \geq 0} \sum_{n \in \ZZ} \kappa^0_{p+q-r} [I_{(p, \alpha)}, I^{(s, \beta)}]^{(0)}_{m-n, \ul {p+s}} I^{\alpha (0)}_{n, \ul q} + m \sum_{p, q \geq 0} \kappa^0_{p+q-r} \lvl^0_{p+s} I^{\beta(0)}_{m, \ul q}.
\end{align*}
As before, the last term simplifies to $m I^{\beta (0)}_{m, [r+s]}$ using the identity \eqref{kappa rel lem} from Lemma \ref{lem: kappa sol} satisfied by $\kappa^0_{p+q-r}$. To see that the first term above vanishes we first note that
\begin{align*}
&\sum_{p, q \geq 0} \sum_{n \in \ZZ} \kappa^0_{p+q-r} [I_{(p, \alpha)}, I^{(s, \beta)}]^{(0)}_{m-n, \ul {p+s}} I^{\alpha (0)}_{n, \ul q}\\
&\qquad\qquad\qquad\qquad = - \sum_{p, q \geq 0} \sum_{n \in \ZZ} \kappa^0_{p+q-r} [I_{(q-s, \alpha)}, I^{(s, \beta)}]^{(0)}_{m-n, \ul q} I^{\alpha (0)}_{n, \ul{p+s}},
\end{align*}
using the identity $[I_{(p,\alpha)}, I^{(s, \beta)}] \otimes I^{(q, \alpha)} = - I_{(p+s, \alpha)} \otimes [I^{(q-s, \alpha)}, I^{(s, \beta)}]$, cf. \eqref{can elem inv j} in the affine case.
Then noting, as before, that $q \geq s$ in the above sums and performing the change of variable $p \mapsto q - s$ and $q \mapsto p+s$ on the right hand side we deduce that both sides must vanish. Therefore
\begin{equation*}
\big\{ \pi_{\bm \lvl}\big( \cocent^{(0)}_{\ul r} \big), \pi_{\bm \lvl}\big( I^{\beta (0)}_{m, \ul s} \big) \big\} = m I^{\beta (0)}_{m, [r+s]} = m \, \pi_{\bm \lvl} \big( I^{\beta (0)}_{m, [r+s]} \big) = \pi_{\bm \lvl} \big( [\cocent^{(0)}_{\ul r}, I^{\beta (0)}_{m, \ul s}] \big).
\end{equation*}
We deduce the result for arbitrary $\ms X \in \Loop\g_{(s)}$ by linearity. Furthermore, the case with $\ms X = \cocent$ follows as before using the above and the Leibniz rule.

Finally, consider the point at infinity and $\ms X = I^{(-s-1, \beta)}_m \in \Loop\g_{(-s-1)}$. We have
\begin{align*}
&\big\{ \pi_{\bm \lvl}\big( \cocent^{(\infty)}_{\ul{r+1}} \big), \pi_{\bm \lvl}\big( I^{\beta (\infty)}_{m, \ul{s+1}} \big) \big\} = m \sum_{p, q \geq 0} \kappa^\infty_{p+q-r+1} \lvl^\infty_{p+s+2} I^{\beta(\infty)}_{m, \ul{q+1}}\\
&\qquad\qquad\qquad + i \sum_{p, q \geq 0} \sum_{n \in \ZZ} \kappa^\infty_{p+q-r+1} [I_{(-p-1, \alpha)}, I^{(-s-1, \beta)}]^{(\infty)}_{m-n, \ul {p+s+2}} I^{\alpha (\infty)}_{n, \ul{q+1}},
\end{align*}
for any $r, s = 0, \ldots, n_\infty - 2$, $\beta = 1, \ldots, \dim \g_{(-s-1), \CC}$ and $m \in \ZZ$. In fact, if $s = n_\infty - 2$ then both sides of the above equality are identically zero, so we may restrict attention to the range of values $0 \leq s \leq n_\infty - 3$.
Moreover, $q - r \geq n_\infty - 2 - p \geq s + 1 \geq 1$ and $q - r \leq q \leq n_\infty - 2$ so that $1 \leq q - r \leq n_\infty - 2$. In other words, $s+2$ and $q-r+1$ both lie between $2$ and $n_\infty - 1$. Using the relation \eqref{kappa rel lem} we then have
\begin{equation*}
\sum_{p \geq 0} \kappa^\infty_{p+q-r+1} \lvl^\infty_{p+s+2} = \delta_{q-r+1, s+2} = \delta_{q, r+s+1},
\end{equation*}
so that the first term above reduces to $m \, I^{\beta (\infty)}_{m, [r+s+2]}$. On the other hand, the second term above can be rewritten as
\begin{align*}
&\sum_{p, q \geq 0} \sum_{n \in \ZZ} \kappa^\infty_{p+q-r+1} [I_{(-p-1, \alpha)}, I^{(-s-1, \beta)}]^{(\infty)}_{m-n, \ul {p+s+2}} I^{\alpha (\infty)}_{n, \ul{q+1}}\\
&\qquad\qquad\qquad = - \sum_{p, q \geq 0} \sum_{n \in \ZZ} \kappa^\infty_{p+q-r+1} [I^{(-q+s, \alpha)}, I^{(-s-1, \beta)}]^{(\infty)}_{m-n, \ul {q+1}} I^{\alpha (\infty)}_{n, \ul{p+s+2}}.
\end{align*}
However, we note that $q \geq s+1$ in these sums since $p+q+1 \geq p+q-r+1 \geq n_\infty - 1$ and $s+2 \leq n_\infty - 1 - p$. Performing the change of variables $p \mapsto q - s - 1$, $q \mapsto p + s + 1$ we deduce that both sides of the above equality vanish. Hence
\begin{equation*}
\big\{ \pi_{\bm \lvl}\big( \cocent^{(\infty)}_{\ul{r+1}} \big), \pi_{\bm \lvl}\big( I^{\beta (\infty)}_{m, \ul{s+1}} \big) \big\} = m \, I^{\beta(\infty)}_{m, \ul{r+s+2}} = m \, \pi_{\bm \lvl} \big( I^{\beta(\infty)}_{m, \ul{r+s+2}} \big) = \pi_{\bm \lvl} \big( \big[ \cocent^{(\infty)}_{\ul{r+1}}, I^{\beta (\infty)}_{m, \ul{s+1}} \big] \big),
\end{equation*}
which implies the result for all $\ms X \in \Loop\g_{(-s-1)}$ by linearity. With $\ms X = \cocent$ we have
\begin{align*}
\big\{ \pi_{\bm \lvl}\big( \cocent^{(\infty)}_{\ul{r+1}} \big), \; &\pi_{\bm \lvl}\big( \cocent^{(\infty)}_{\ul{s+1}} \big) \big\} = - \frac{i}{2} \sum_{p, q \geq 0} \sum_{n \in \ZZ} n \, \kappa^\infty_{p+q-s+1} I^{(\infty)}_{\alpha, -n, \ul{p+r+2}} I^{\alpha (\infty)}_{n, \ul{q+1}}\\
&\qquad\qquad + \frac{i}{2} \sum_{p, q \geq 0} \sum_{n \in \ZZ} n \, \kappa^\infty_{p+q-s+1} I^{(\infty)}_{\alpha, -n, \ul{p+1}} I^{\alpha (\infty)}_{n, \ul{q+r+2}}.
\end{align*}
In the first sum we have $p+q-s+1 \geq n_\infty - 1$, $p+r+2 \leq n_\infty - 1$ and $q + 1 \leq n_\infty - 1$ so that $p \geq s$ and $q \geq r + s + 1$. Likewise, in the second sum $q \geq s$ and $p \geq r + s + 1$. Then performing the change of variables $p \mapsto p + r + 1$ in the first sum and $q \mapsto q + r + 1$ in the second we find that they cancel, as required.
\end{proof}
\end{proposition}

We extend the map \eqref{pi D k def} to a homomorphism $\pi_{\bm \lvl} : S(\tg^\D_\CC) \to \hat S_{\bm \lvl}(\hg^\D_\CC)$ of commutative algebras. It follows from Proposition \ref{prop: hom pi lvl} that the latter is in fact a homomorphism of Poisson algebras. For each $n \in \ZZ_{\geq 0}$, the image of $\Fil_n\big( S(\tg^\D_\CC) \big)$, defined in \eqref{S tgDC Fil 1}, under this homomorphism lies in $\Fil_n \big( \hat S_{\bm \lvl}(\hg^\D_\CC) \big) \coloneqq \ker \big( \hat S_{\bm \lvl}(\hg^\D_\CC) \twoheadrightarrow S_{\bm \lvl}(\hg^\D_\CC) / \Fil_n \big( S_{\bm \lvl}(\hg^\D_\CC) \big) \big)$ so that it extends by continuity to a homomorphism
\begin{equation} \label{hom pi l S to S}
\pi_{\bm \lvl} : \hat S(\tg^\D_\CC) \longrightarrow \hat S_{\bm \lvl}(\hg^\D_\CC).
\end{equation}

\subsection{Fields}

\subsubsection{Formal fields} \label{sec: formal fields}

Let $\F$ denote the completion of the tensor product of complex vector spaces $\tg \otimes \tg^\D_\CC$ with respect to the descending $\ZZ_{\geq 0}$-filtration given by
\begin{equation} \label{Fil formal fields}
\Fil_n (\tg \otimes \tg^\D_\CC) \coloneqq \Fil_n \tg \otimes c\big( \Fil_n \tg^\D_\CC \big) + \cFil_n \tg \otimes \Fil_n(\tg^\D_\CC)
\end{equation}
for all $n \in \ZZ_{\geq 0}$. We refer to the first tensor factor in $\F$ as the \emph{auxiliary space} and the second tensor factor as the \emph{operator space}.

We gather the basis \eqref{tgD R C basis} of $\tg^\D_\CC$ together into a finite collection of elements of $\F$, which we will call the \emph{formal fields} of the dihedral affine Gaudin model, defined as
\begin{subequations} \label{fields A}
\begin{align}
\label{fields A zc} \A^x_{\ul p} \coloneqq I_{\wt a} \otimes I^{\wt a (x)}_{\ul p}, \qquad \A^{\bar x}_{\ul p} \coloneqq \tau I_{\wt a} \otimes I^{\wt a (\bar x)}_{\ul p}, \qquad &\text{for} \quad x \in \z_{\rm c}\\
\label{fields A zr} \A^x_{\ul p} \coloneqq I_{\wt a; k, - p} \otimes I^{\wt a (x)}_{k \ul p}, \qquad\qquad\qquad &\text{for} \quad x \in \z^k_{\rm r}, \; k \in \ZZ_T,
\end{align}
where in each case $p = 0, \ldots, n_x - 1$, and
\begin{equation} \label{fields A zf}
\A^0_{\ul p} \coloneqq I_{(-p, \wt \alpha)} \otimes I^{\wt \alpha (0)}_{\ul p}, \qquad
\A^{\infty}_{\ul{q+1}} \coloneqq I_{(q+1, \wt \alpha)} \otimes I^{\wt \alpha (\infty)}_{\ul{q+1}}.
\end{equation}
\end{subequations}
with $p = 0, \ldots, n_0 - 1$ and $q = 0, \ldots, n_\infty - 2$.
Note in particular that for each $x \in \Z$ we have $\A^x_{\ul p} = 0$ whenever $p \geq n_x$. In order to see that the formal field \eqref{fields A zr} indeed defines an element of $\F$ it is enough to note that by a $\CC$-linear change of basis, recalling the definition \eqref{tg kp bases}, we may rewrite it equivalently as $\A^x_{\ul p} = I_{\wt a} \otimes I^{\wt a (x)}_{\ul p}$ making use of the notation introduced in \eqref{Lie re com iso 2}. Similar considerations apply to the formal fields \eqref{fields A zf} at the origin and infinity.

We equip $\tg^\D_\CC = \tg^\D \otimes_\RR \CC$ with an action of $\Pi$ by letting $\s$ act trivially and letting $\t$ act by complex conjugation on the second tensor factor, cf. Proposition \ref{prop: complexified Takiff}. Combining this with the action \eqref{hom r tg} of $\Pi$ on $\tg$ we obtain an action on the tensor product $\tg \otimes \tg^\D_\CC$, cf. \eqref{Pi act g otimes V}. Now each subspace $\Fil_n (\tg \otimes \tg^\D_\CC)$ given by \eqref{Fil formal fields} is preserved by this action of $\Pi$. In other words, $\Pi$ acts on $\tg \otimes \tg^\D_\CC$ by continuous (anti-)linear maps.
We extend this by continuity to an action of $\Pi$ on $\F$, which we denote by
\begin{equation} \label{Pi act Fields}
r : \Pi \longrightarrow \bAut \F, \qquad \alpha \longmapsto r_\alpha.
\end{equation}
In order to reflect the fact that the original action of $\Pi$ on $\tg$, as given in \eqref{hom r tg}, was defined in terms of the pair of automorphisms $\sigma, \tau \in \bAut \tg$, we will sometimes also write $r_\s$ and $r_\t$ for the above action \eqref{Pi act Fields} of $\Pi$ on $\F$ simply as $\sigma$ and $\tau$.

Let $\tg \hotimes \tg \hotimes \tg^\D_\CC$ denote the completion of the tensor product $\tg \otimes \tg \otimes \tg^\D_\CC$ with respect to the descending $\ZZ_{\geq 0}$-filtration whose subspace $\Fil_n \big( \tg \hotimes \tg \hotimes \tg^\D_\CC \big)$, $n \in \ZZ_{\geq 0}$ is, cf. \S\ref{sec: tensor index},
\begin{align*}
&\Fil_n \tg \otimes \cFil_n \tg \otimes \tg^\D_\CC + \Fil_n \tg \otimes \tg \otimes c \big( \Fil_n (\tg^\D_\CC) \big) + \tg \otimes \Fil_n \tg \otimes c\big( \Fil_n (\tg^\D_\CC) \big)\\
&\qquad\qquad + \cFil_n \tg \otimes \Fil_n \tg \otimes \tg^\D_\CC + \cFil_n \tg \otimes \tg \otimes \Fil_n (\tg^\D_\CC) + \tg \otimes \cFil_n \tg \otimes \Fil_n (\tg^\D_\CC).
\end{align*}
The following proposition describes identities in $\tg \hotimes \tg \hotimes \tg^\D_\CC$ so that in the tensor index notation of \S\ref{sec: tensor index} we have $\a = \b = \tg$, $\c = \tg^\D_\CC$ and $\mathfrak A = \CC$ (hence we do not include the superfluous tensor factor in $\mathfrak A$). Note that although the elements in the third tensor factor all belong to the Lie algebra $\tg^\D_\CC \subset \hat S(\tg^\D_\CC)$, we prefer to use the Poisson bracket notation \eqref{wtSg PB} for their Lie bracket to emphasise that they ought be regarded as `linear' functions in the complexified algebra of formal observables $\hat S(\tg^\D_\CC)$. Moreover, in order to conform to the standard notation, cf. Corollary \ref{cor: Fields PB} below, we shall also drop the tensor index $\null_\3$ corresponding to the operator space. For instance, instead of $\A^x_{{\ul p} \1\3}$ we shall simply write $\A^x_{{\ul p} \1}$.

\begin{proposition} \label{prop: Fields PB}
The collection of non-zero Poisson brackets between the elements of $\F$ defined in \eqref{fields A} reads
\begin{subequations} \label{Fields PB}
\begin{equation} \label{Fields PB a}
\big\{ \A^x_{{\ul p} \1}, \A^x_{{\ul q} \2} \big\} = - \big[ \wt C_{\1\2}, \A^x_{\ul{p+q} \2} \big],
\end{equation}
for each $x \in \Z'$ with $p, q = 0, \ldots, n_x - 1$,
and
\begin{align}
\label{Fields PB b} \big\{ \A^0_{\ul p \1}, \A^0_{\ul q \2} \big\} &= - \big[ \wt C^{(p)}_{\1\2}, \A^0_{\ul{p+q} \2} \big],\\
\label{Fields PB c} \big\{ \A^{\infty}_{\ul{r+1} \1}, \A^{\infty}_{\ul{s+1} \2} \big\} &= - \big[ \wt C^{(-r-1)}_{\1\2}, \A^{\infty}_{\ul{r+s+2} \2} \big],
\end{align}
for $p, q = 0, \ldots, n_0 - 1$ and $r, s = 0, \ldots, n_\infty - 2$.

\end{subequations}
\begin{proof}
From its definition in Proposition \ref{prop: complexified Takiff}, $\tg^\D_\CC$ is a direct sum of complexified Lie algebras $\tg^{n_x x} \otimes_\RR \CC$ attached to each point $x \in \z$. It follows from the definition of the formal fields \eqref{fields A} and using Proposition \ref{prop: complexified Takiff} that $\{ \A^x_{{\ul p} \1}, \A^y_{{\ul q} \2} \} = 0$ for any distinct $x$ and $y$ in $\Z$.

To show \eqref{Fields PB a} for $x \in \z_{\rm c}$, note that for such $x$ we have
\begin{equation} \label{field PB calc}
I_{\wt a} \otimes I_{\wt b} \otimes \Big[ I^{\wt a (x)}_{\ul p}, I^{\wt b (x)}_{\ul q} \Big]
= I_{\wt a} \otimes I_{\wt b} \otimes \big[ I^{\wt a}, I^{\wt b} \big]^{(x)}_{\ul{p+q}}
= - I_{\wt a} \otimes \big[ I^{\wt a}, I_{\wt b} \big] \otimes I^{\wt b (x)}_{\ul{p+q}}.
\end{equation}
In the first equality here we have used the fact that $\ms X \otimes \varepsilon_x^p \mapsto \ms X^{(x)}_{\ul p}$ is a homomorphism of complex Lie algebras since it is a composition of the isomorphism \eqref{Lie re com iso} with the natural embedding $\Tak^{n_x}_x \tg \hookrightarrow \Tak^{n_x}_x \tg \oplus \overline{\Tak^{n_x}_x \tg}$. The second equality uses Lemma \ref{lem: can elem identity}.

The relation \eqref{Fields PB a} for $x \in \bar \z_{\rm c}$ simply follows from applying the anti-linear involution $\tau \otimes \tau \otimes c$ to the equality \eqref{Fields PB a} for $x \in \z_{\rm c}$ and noting that $\tau I_{\wt a} \otimes \tau I^{\wt a} = \wt C_{\1\2}$, by the $\tau$-invariance \eqref{ip on tg sigma tau} of the bilinear form on $\tg$.

In order to prove \eqref{Fields PB a} holds for $x \in \z'_{\rm r}$, recall that the formal field at such a point can be rewritten as $\A^x_{\ul p} = I_{\wt a} \otimes I^{\wt a (x)}_{\ul p}$. The result then follows by the same calculation as above in \eqref{field PB calc}, using the fact that \eqref{Lie re com iso 2} is a homomorphism of complex Lie algebras followed by Lemma \ref{lem: can elem identity}.

The Poisson bracket relations \eqref{Fields PB b} between the formal fields at the origin follow from a similar calculation, namely
\begin{align*}
&I_{(-p, \wt \alpha)} \otimes I_{(-q, \wt \beta)} \otimes \big[ I^{\wt \alpha (0)}_{\ul p}, I^{\wt \beta (0)}_{\ul q} \big]
= I_{(-p, \wt \alpha)} \otimes I_{(-q, \wt \beta)} \otimes [I^{(p, \wt \alpha)}, I^{(q, \wt \beta)}]^{(0)}_{\ul{p+q}}\\
&\qquad = - I_{(-p, \wt \alpha)} \otimes [I^{(p, \wt \alpha)}, I_{(-p-q, \wt \beta)}] \otimes I^{\wt \beta (0)}_{\ul{p+q}}
\end{align*}
where in the first step we used the fact that \eqref{Lie re com iso 0} with $x = 0$ is a homomorphism of complex Lie algebras and in the last line we used the second part of Lemma \ref{lem: can elem identity}. The proof of the final relation \eqref{Fields PB c} between the formal fields at infinity is completely analogous.
\end{proof}
\end{proposition}

Applying $\sigma^k \otimes \sigma^\ell \otimes \id$ for some $k, \ell \in \ZZ_T$ to the relations in \eqref{Fields PB a} and using the identity $\sigma^\ell_{\2} \wt C_{\1\2} = \sigma^{-\ell}_{\1} \wt C_{\1\2}$, where $\sigma_\1 \coloneqq \sigma \otimes \id$ and $\sigma_\2 \coloneqq \id \otimes \sigma$, we obtain
\begin{equation} \label{Fields PB sigma}
\big\{ \sigma^k \A^x_{\ul p \1}, \sigma^\ell \A^x_{\ul q \2} \big\} = - \big[ \sigma^{k-\ell}_{\1} \wt C_{\1\2}, \sigma^\ell \A^x_{\ul{p+q} \2} \big],
\end{equation}
for any $x \in \Z'$ and $p, q = 0, \ldots, n_x - 1$. These relations along with those of Proposition \ref{prop: Fields PB} will be used in the proof of Proposition \ref{prop: Lax algebra} below. Note that the relations \eqref{Fields PB b} and \eqref{Fields PB c} are invariant under the application of $\sigma^k \otimes \sigma^\ell \otimes \id$, $k, \ell \in \ZZ_T$.

\subsubsection{Classical fields} \label{sec: classical fields}

Recall the homomorphisms of Lie algebras $\rep$ and $\pi_{\bm \lvl}$ introduced in \S\ref{sec: Conn on S1} and \S\ref{sec: Alg Obs}, respectively. Consider their tensor product, namely the linear map
$\rep \otimes \pi_{\bm \lvl} : \tg \otimes \tg^\D_\CC \to \textup{Conn}_\g(S^1) \otimes \hat S_{\bm \lvl}(\hg^\D_\CC)$. Given the descending $\ZZ_{\geq 0}$-filtration on its domain with the subspaces \eqref{Fil formal fields}, we endow its codomain with the image descending $\ZZ_{\geq 0}$-filtration whose subspaces are defined by $(\rep \otimes \pi_{\bm \lvl})\big( \Fil_n(\tg \otimes \tg^\D_\CC) \big)$ for each $n \in \ZZ_{\geq 0}$. Let $\textup{Conn}_\g(S^1) \hotimes \hat S_{\bm \lvl}(\hg^\D_\CC)$ denote the corresponding completion.
By construction, the map $\rep \otimes \pi_{\bm \lvl}$ is then continuous at the origin and hence uniformly continuous by linearity so that it extends to the respective completions
\begin{equation*}
\rep \otimes \pi_{\bm \lvl} : \F \longrightarrow \textup{Conn}_\g(S^1) \hotimes \hat S_{\bm \lvl}(\hg^\D_\CC).
\end{equation*}
By applying the latter to the collection of formal fields defined in \eqref{fields A zc} and \eqref{fields A zr} we obtain, using the notation \eqref{notation Ix},
\begin{subequations} \label{rho pi on formal A}
\begin{equation} \label{rho pi on formal A a}
(\rep \otimes \pi_{\bm \lvl}) \A^x_{\ul p} = \lvl^x_p \partial \otimes 1 + \sum_{n \in \ZZ} (I_a \otimes e_{-n}) \otimes I^{a (x)}_{n, \ul p},
\end{equation}
for each $x \in \z_{\rm c} \cup \z'_{\rm r}$ with $p = 0, \ldots, n_x - 1$, whereas
\begin{equation} \label{rho pi on formal A b}
(\rep \otimes \pi_{\bm \lvl}) \A^y_{\ul p} = \lvl^y_p \partial \otimes 1 + \sum_{n \in \ZZ} (I_a \otimes e_n) \otimes I^{a (y)}_{n, \ul p},
\end{equation}
for each $y \in \bar\z_{\rm c}$ with $p = 0, \ldots, n_y - 1$.
We recall here that $I_a$ is by definition a basis of the real form $\g_0$, cf. \S\ref{sec: can elem g}, so that $\tau I_a = I_a$. Similarly, applying $\rep \otimes \pi_{\bm \lvl}$ to the formal fields \eqref{fields A zf} at the origin and infinity we get
\begin{align}
(\rep \otimes \pi_{\bm \lvl}) \A^0_{\ul p} &= \lvl^0_p \partial \otimes 1 + \sum_{n \in \ZZ} (I_{(-p, \alpha)} \otimes e_{-n}) \otimes I^{\alpha (0)}_{n, \ul p},\\
(\rep \otimes \pi_{\bm \lvl}) \A^\infty_{\ul{q+1}} &= \lvl^\infty_{q+1} \partial \otimes 1 + \sum_{n \in \ZZ} (I_{(q+1, \alpha)} \otimes e_{-n}) \otimes I^{\alpha (\infty)}_{n, \ul{q+1}},
\end{align}
\end{subequations}
where we recall that $\lvl^0_p = 0$ if $p \not\equiv 0 \; \text{mod}\, T$ and $\lvl^\infty_{q+1} = 0$ if $q+1 \not\equiv 0 \; \text{mod}\, T$.

To make sense of the above expressions, it is convenient to introduce the following elements of the inverse limit
\begin{equation*}
\mathscr T(S^1) \hotimes \hg^\D_{\CC, \bm \lvl} \coloneqq \varprojlim \big( \mathscr T(S^1) \otimes \hg^\D_{\CC, \bm \lvl} \big) \big/ \Fil_n\big( \mathscr T(S^1) \otimes \hg^\D_{\CC, \bm \lvl} \big)
\end{equation*}
where $\Fil_n\big( \mathscr T(S^1) \otimes \hg^\D_{\CC, \bm \lvl} \big) \coloneqq \Fil_n\big(\mathscr T(S^1)\big) \otimes c\big(\Fil_n(\hg^\D_{\CC, \bm \lvl})\big) + \cFil_n\big(\mathscr T(S^1)\big) \otimes \Fil_n(\hg^\D_{\CC, \bm \lvl})$ for each $n \in \ZZ_{\geq 0}$. For every point $x \in \z_c \cup \z'_{\rm r}$ and $y \in \bar\z_{\rm c}$ we associate to each $p = 0, \ldots, n_x - 1$ and $q = 0, \ldots, n_y - 1$ the finite collection of \emph{classical fields}
\begin{equation*}
A^{a, x}_{\ul p} \coloneqq \sum_{n \in \ZZ} e_{-n} \otimes I^{a (x)}_{n, \ul p}, \qquad
A^{a, y}_{\ul q} \coloneqq \sum_{n \in \ZZ} e_n \otimes I^{a (y)}_{n, \ul q},
\end{equation*}
labelled by $a = 1, \ldots, \dim \g$. Likewise, for the origin and infinity we associate to each $p = 0, \ldots, n_0 - 1$ and $q = 0, \ldots, n_\infty - 2$ the collection of \emph{classical fields}
\begin{equation*}
A^{\alpha, 0}_{\ul p} \coloneqq \sum_{n \in \ZZ} e_{-n} \otimes I^{\alpha (0)}_{n, \ul p}, \qquad
A^{\beta, \infty}_{\ul {q+1}} \coloneqq \sum_{n \in \ZZ} e_{-n} \otimes I^{\beta (\infty)}_{n, \ul {q+1}},
\end{equation*}
labelled by $\alpha = 1, \ldots, \dim \g_{(p), \CC}$ and $\beta = 1, \ldots, \dim \g_{(-q-1), \CC}$ respectively. We then gather all these classical fields at each point in $\Z$ into \emph{$\g$-valued classical fields}, belonging to the space $\mathscr T(S^1, \g) \hotimes \hg^\D_{\CC, \bm \lvl} \coloneqq \g \otimes \mathscr T(S^1) \hotimes \hg^\D_{\CC, \bm \lvl}$ and defined by
\begin{equation} \label{Currents x 0 inf}
A^x_{\ul p} \coloneqq I_a \otimes A^{a, x}_{\ul p}, \qquad
A^0_{\ul q} \coloneqq I_{(-q, \alpha)} \otimes A^{\alpha, 0}_{\ul q}, \qquad
A^\infty_{\ul {r+1}} \coloneqq I_{(r+1, \alpha)} \otimes A^{\alpha, \infty}_{\ul {r+1}},
\end{equation}
for all $x \in \Z'$, $p = 0, \ldots, n_x - 1$, $q = 0, \ldots, n_0 - 1$ and $r = 0, \ldots, n_\infty - 2$.
We may now regard \eqref{rho pi on formal A} as $\g$-valued connections on $S^1$ with components given by these $\g$-valued classical fields, namely
\begin{equation} \label{rho pi Conn}
\begin{split}
(\rep \otimes \pi_{\bm \lvl}) \A^x_{\ul p} &= \lvl^x_p \partial + A^x_{\ul p},\\
(\rep \otimes \pi_{\bm \lvl}) \A^0_{\ul q} = \lvl^0_q \partial + A^0_{\ul q}, \qquad
&(\rep \otimes \pi_{\bm \lvl}) \A^\infty_{\ul {r+1}} = \lvl^\infty_{r+1} \partial + A^\infty_{\ul {r+1}},
\end{split}
\end{equation}
where we suppressed the tensor product with $1 \in \hat S_{\bm \lvl}(\hg^\D_\CC)$ in the derivative terms.

Recall the Dirac $\delta$-distribution $\delta_{\theta\theta'}$ introduced in \S\ref{sec: Conn on S1}. In what follows we shall also need its derivative $\delta' \coloneqq \sum_{n \in \ZZ} e'_n \otimes e_{-n} = \sum_{n \in \ZZ} i n e_n \otimes e_{-n} \in \mathscr T(S^1) \hotimes \mathscr T(S^1)$, with the property that for any $\theta \in S^1$, $\delta'(\theta, \cdot)$ is a distribution on $S^1$ which sends a test function $f \in \mathscr T(S^1)$ to $f'(\theta)$. We use the shorthand notation $\delta'_{\theta \theta'}$ for $\delta'(\theta, \theta')$.

\begin{corollary} \label{cor: Fields PB}
The non-trivial Poisson brackets between the $\g$-valued classical fields \eqref{Currents x 0 inf} read
\begin{equation*}
\big\{ A^x_{{\ul p} \1}(\theta), A^x_{{\ul q} \2}(\theta') \big\} = - \big[ C_{\1\2}, A^x_{\ul{p+q} \2}(\theta) \big] \delta_{\theta \theta'} - \lvl^x_{p+q} C_{\1\2} \delta'_{\theta \theta'},
\end{equation*}
for each $x \in \Z'$ with $p, q = 0, \ldots, n_x - 1$, and
\begin{align*}
\big\{ A^0_{\ul p \1}(\theta), A^0_{\ul q \2}(\theta') \big\} &= - \big[ C^{(p)}_{\1\2}, A^0_{\ul{p+q} \2}(\theta) \big] \delta_{\theta \theta'} - \lvl^0_{p+q} C^{(p)}_{\1\2} \delta'_{\theta \theta'},\\
\big\{ A^{\infty}_{\ul{r+1} \1}(\theta), A^{\infty}_{\ul{s+1} \2}(\theta') \big\} &= - \big[ C^{(-r-1)}_{\1\2}, A^{\infty}_{\ul{r+s+2} \2}(\theta) \big] \delta_{\theta \theta'} - \lvl^\infty_{r+s+2} C^{(-r-1)}_{\1\2} \delta'_{\theta \theta'},
\end{align*}
for $p, q = 0, \ldots, n_0 - 1$ and $r, s = 0, \ldots, n_\infty - 2$.
\begin{proof}
This follows from applying the tensor product of Lie algebra homomorphisms $\rep \otimes \rep \otimes \pi_{\bm \lvl}$ to the identities in Proposition \ref{prop: Fields PB}, and using \eqref{rho pi Conn} together with \eqref{wt C delta} and the analogous relation for \eqref{can elem j}. We also make use of the fact that $\partial_{\theta'} \delta_{\theta \theta'} = - \delta'_{\theta \theta'}$.
\end{proof}
\end{corollary}

\subsection{Canonical element} \label{sec: canonical element}

Recall the dual pair $(\Omega^\Pi_\D(\tg), \tg^\D, \lbf{\cdot, \cdot})$ of Proposition \ref{prop: Takiff dual pair}. Given the dual bases $\{ I^A \}$ of $\tg^\D$ and $\{ I_A \}$ of $\Omega^\Pi_\D(\tg)$ introduced in \eqref{Takiff basis} and \eqref{dual Takiff basis} respectively, we can consider the corresponding canonical element $I_A \otimes I^A$ living in a suitable completion of the tensor product $\Omega^\Pi_\D(\tg) \otimes_\RR \tg^\D$ over $\RR$ defined below. We show in \S\ref{sec: can elem fields} that this canonical element can be naturally rewritten as a linear combination of formal fields from \S\ref{sec: formal fields} with coefficients given by meromorphic differentials on $\CP$. This expression, given in Proposition \ref{prop: Gaudin Lax}, will be directly related to the (formal) Lax matrix of the dihedral affine Gaudin model. In \S\ref{sec: complexified fields} we show that it can be expressed more compactly in terms of \emph{complexified} formal fields.

\subsubsection{Canonical element and formal fields} \label{sec: can elem fields}

By regarding $\tg \otimes_\RR \tg^\D$ as a complex vector space with scalar multiplication acting in the first tensor factor, we get an isomorphism of complex vector spaces $\tg \otimes_\RR \tg^\D \cong_\CC \tg \otimes \tg^\D_\CC$. In turn, taking the tensor product with $\Omega_{\Pi \D}$ induces a linear isomorphism
\begin{equation} \label{pre iso Higgs field}
\zeta : \Omega_{\Pi \D}(\tg) \otimes_\RR \tg^\D \overset{\sim}\longrightarrow \Omega_{\Pi \D}\big( \tg \otimes \tg^\D_\CC \big),
\end{equation}
where we made use of the notation \eqref{Om V notation}. Given any $\ms X \in \tg$, $\varpi \in \Omega_{\Pi \D}$ and $\mathscr Y \in \tg^\D$, the linear map in \eqref{pre iso Higgs field} is given explicitly by $\zeta(\ms X \otimes \varpi \otimes_\RR \mathscr Y) = \ms X \otimes (\mathscr Y \otimes_\RR 1) \otimes \varpi$, where here we denote the tensor products over $\RR$ explicitly by a subscript. Recalling the action of $\Pi$ on the subspace $\Omega_{\Pi \D}(\tg) \subset \Omega_{\Pi \z}(\tg)$ defined by \eqref{Mobius rep Og}, we extend it to an action on $\Omega_{\Pi \D}(\tg) \otimes_\RR \tg^\D$ by letting it act trivially on the second tensor factor. The map \eqref{pre iso Higgs field} can then be made $\Pi$-equivariant if we define the action of $\Pi$ on $\Omega_{\Pi \D}\big( \tg \otimes \tg^\D_\CC \big)$ through the homomorphism
\begin{equation*}
\Pi \longhookrightarrow \bAut \Omega_{\Pi \D}\big( \tg \otimes  \tg^\D_\CC \big), \qquad
\alpha \longmapsto \hat \alpha
\end{equation*}
given explicitly by
$\hat \alpha (\mathcal X \otimes \varpi) \coloneqq r_\alpha \mathcal X \otimes \alpha . \varpi$
for any $\alpha \in \Pi$, $\mathcal X \in \tg \otimes \tg^\D_\CC$ and $\varpi \in \Omega_{\Pi \D}$, where the action of $\Pi$ on $\tg \otimes \tg^\D_\CC$, denoted here $\alpha \mapsto r_\alpha$, was defined in \S\ref{sec: formal fields}.

The linear isomorphism \eqref{pre iso Higgs field} becomes continuous if we equip its codomain with the descending $\ZZ_{\geq 0}$-filtration given by the subspaces $\Omega_{\Pi \D}( \Fil_n(\tg \otimes \tg^\D_\CC))$, cf. \eqref{Fil formal fields}, and its domain with the induced descending $\ZZ_{\geq 0}$-filtration whose subspaces are
\begin{equation} \label{Fil n OmD gD}
\Fil_n \big( \Omega_{\Pi \D}(\tg) \otimes_\RR \tg^\D \big) \coloneqq \zeta^{-1} \big( \Omega_{\Pi \D} ( \Fil_n(\tg \otimes \tg^\D_\CC)) \big),
\end{equation}
for $n \in \ZZ_{\geq 0}$. Let $\Omega_{\Pi \D}(\tg) \hotimes_\RR \tg^\D$ be the completion of the tensor product $\Omega_{\Pi \D}(\tg) \otimes_\RR \tg^\D$ with respect to \eqref{Fil n OmD gD}. We note that the action of $\Pi$ on $\Omega_{\Pi \D}(\tg) \otimes_\RR \tg^\D$ preserves the subspaces \eqref{Fil n OmD gD} and therefore extends to the completion $\Omega_{\Pi \D}(\tg) \hotimes_\RR \tg^\D$.
We obtain a $\Pi$-equivariant linear isomorphism
\begin{equation} \label{pre hom Higgs field}
\zeta : \Omega_{\Pi \D}(\tg) \hotimes_\RR \tg^\D \overset{\sim}\longrightarrow \Omega_{\Pi \D}(\F).
\end{equation}
Here we have used the fact that the completion of the codomain of the isomorphism \eqref{pre iso Higgs field} with respect to the descending $\ZZ_{\geq 0}$-filtration with subspaces $\Omega_{\Pi \D}( \Fil_n(\tg \otimes \tg^\D_\CC))$ is $\Omega_{\Pi \D}(\F)$, where we use again the notation \eqref{Om V notation}. Note that the action of $\Pi$ on the latter is given by the homomorphism
\begin{equation} \label{Pi act OmPiD F}
\Pi \longhookrightarrow \bAut \Omega_{\Pi \D}(\F), \qquad \alpha \longmapsto \hat \alpha
\end{equation}
defined explicitly by
$\hat \alpha (\mathcal X \otimes \varpi) \coloneqq r_\alpha \mathcal X \otimes \alpha . \varpi$
for any $\alpha \in \Pi$, $\mathcal X \in \F$ and $\varpi \in \Omega_{\Pi \D}$, where the action of $\Pi$ on $\F$ is given in \eqref{Pi act Fields}.

Now the $\Pi$-invariant subspace $( \Omega_{\Pi \D}(\tg) \hotimes_\RR \tg^\D )^\Pi$ of the completion $\Omega_{\Pi \D}(\tg) \hotimes_\RR \tg^\D$ coincides with the completion $\Omega^\Pi_\D(\tg) \hotimes_\RR \tg^\D$ of the tensor product $\Omega^\Pi_\D(\tg) \otimes_\RR \tg^\D$ with respect to the descending $\ZZ_{\geq 0}$-filtration defined by the subspaces
\begin{equation} \label{Fil zeta inv def}
\Fil_n \big( \Omega^\Pi_\D(\tg) \otimes_\RR \tg^\D \big) \coloneqq \zeta^{-1} \big( \big(\Omega_{\Pi \D} ( \Fil_n(\tg \otimes \tg^\D_\CC)) \big)^\Pi \big).
\end{equation}
If we define the subspace of $\Pi$-invariants $\Omega^\Pi_\D(\F) \coloneqq \Omega_{\Pi \D}(\F)^\Pi$, then the restriction of \eqref{pre hom Higgs field} to the subspace $\Omega^\Pi_\D(\tg) \hotimes_\RR \tg^\D$ yields a linear isomorphism with $\Omega^\Pi_\D(\F)$ which we denote by the same symbol, namely
\begin{equation} \label{hom Higgs field}
\zeta : \Omega_\D^\Pi(\tg) \hotimes_\RR \tg^\D \overset{\sim}\longrightarrow \Omega_\D^\Pi(\F).
\end{equation}

\begin{proposition} \label{prop: Gaudin Lax}
The canonical element of the dual pair $(\Omega^\Pi_\D(\tg), \tg^\D, \lbf{\cdot, \cdot})$, \emph{i.e.}
\begin{equation} \label{Higgs field def}
\Phi \coloneqq I_A \otimes I^A
\end{equation}
where the tensor product is over $\RR$ and the infinite sum over the repeated multi-index $A$ is implicit, defines an element of the completion $\Omega^\Pi_\D(\tg) \hotimes_\RR \tg^\D$.
Its image under the linear isomorphism \eqref{hom Higgs field} reads
\begin{equation} \label{Gaudin Lax matrix}
\zeta(\Phi) = \frac{1}{T} \sum_{k=0}^{T-1} \hat\s^k \left[ \sum_{x \in \Z'} \sum_{p \geq 0} \frac{\A^x_{\ul p}}{(z - x)^{p+1}} dz \right] + \sum_{p \geq 0} \A^0_{\ul p} z^{-p-1} dz
- \sum_{q \geq 0} \A^{\infty}_{\ul{q+1}} z^q dz.
\end{equation}
\begin{proof}
The proof of both statements, namely that \eqref{Higgs field def} lives in $\Omega^\Pi_\D(\tg) \hotimes_\RR \tg^\D$ and that its image under \eqref{hom Higgs field} is of the form \eqref{Gaudin Lax matrix}, rely on very similar computations. Since our main focus is to prove \eqref{Gaudin Lax matrix}, we choose to begin by proving the second statement assuming that the infinite sum over $A$ in \eqref{Higgs field def} makes sense and then comment on the proof of the first. We work separately on the dual basis elements of $\tg^\D$ and $\Omega^\Pi_\D(\tg)$ associated with points in $\z_{\rm c}$, $\z'_{\rm r}$ and $\z_{\rm f}$.

Consider first basis elements associated with a complex point $x \in \z_{\rm c}$. We may write the elements \eqref{dual Takiff basis a} from the basis \eqref{dual Takiff basis} of $\Omega^\Pi_\D(\tg)$ as
\begin{align*}
\frac{1}{2T} \sum_{\alpha \in \Pi} \hat \alpha \left( \frac{I_{\wt a} dz}{(z - x)^{p+1}} \right) &=
\frac{1}{2T} \sum_{k=0}^{T-1} \frac{\omega^{-k} \sigma^k I_{\wt a} dz}{(\omega^{-k} z - x)^{p+1}} + 
\frac{1}{2T} \sum_{k=0}^{T-1} \frac{\omega^{-k} \sigma^k \tau I_{\wt a} dz}{(\omega^{-k} z - \bar x)^{p+1}},\\
\frac{1}{2T} \sum_{\alpha \in \Pi} \hat \alpha \left( \frac{- i I_{\wt a} dz}{(z - x)^{p+1}} \right) &=
- \frac{1}{2T} \sum_{k=0}^{T-1} \frac{i \omega^{-k} \sigma^k I_{\wt a} dz}{(\omega^{-k} z - x)^{p+1}} +
\frac{1}{2T} \sum_{k=0}^{T-1} \frac{i \omega^{-k} \sigma^k \tau I_{\wt a} dz}{(\omega^{-k} z - \bar x)^{p+1}}.
\end{align*}
Taking their tensor product over $\RR$ with the corresponding dual basis elements of $\tg^\D$ given in \eqref{Takiff basis a} we obtain
\begin{align} \label{Higgs field proof zc}
&\left( \frac{1}{2T} \sum_{k=0}^{T-1} \frac{\omega^{-k} \sigma^k I_{\wt a} dz}{(\omega^{-k} z - x)^{p+1}} + 
\frac{1}{2T} \sum_{k=0}^{T-1} \frac{\omega^{-k} \sigma^k \tau I_{\wt a} dz}{(\omega^{-k} z - \bar x)^{p+1}} \right) \otimes (I^{\wt a} \otimes \varepsilon^p_x) \notag\\
&\qquad + \left( - \frac{1}{2T} \sum_{k=0}^{T-1} \frac{i \omega^{-k} \sigma^k I_{\wt a} dz}{(\omega^{-k} z - x)^{p+1}} +
\frac{1}{2T} \sum_{k=0}^{T-1} \frac{i \omega^{-k} \sigma^k \tau I_{\wt a} dz}{(\omega^{-k} z - \bar x)^{p+1}} \right) \otimes (i I^{\wt a} \otimes \varepsilon^p_x),
\end{align}
where the tensor product between the expressions in brackets is over $\RR$ and as usual the summation over repeated Lie algebra indices $\wt a$ is implicit.
Applying the linear map \eqref{hom Higgs field} to the latter we may rewrite it as
\begin{align*}
&\frac{1}{T} \sum_{k=0}^{T-1} \frac{\omega^{-k} \sigma^k I_{\wt a} dz}{(\omega^{-k} z - x)^{p+1}} \otimes I^{\wt a (x)}_{\ul p} + \frac{1}{T} \sum_{k=0}^{T-1} \frac{\omega^{-k} \sigma^k \tau I_{\wt a} dz}{(\omega^{-k} z - \bar x)^{p+1}} \otimes I^{\wt a (\bar x)}_{\ul p}\\
&\qquad\qquad = \frac{1}{T} \sum_{k=0}^{T-1} \hat \s^k \left[ \left( \frac{\A^x_{\ul p}}{(z - x)^{p+1}} + \frac{\A^{\bar x}_{\ul p}}{(z - \bar x)^{p+1}} \right) dz \right].
\end{align*}
To show that the infinite sum in \eqref{Higgs field proof zc} indeed defines an element of the completion $\Omega^\Pi_\D(\tg) \hotimes_\RR \tg^\D$, consider the same expression but where instead of summing over the index $\wt a$ we replace the dual basis elements $I_{\wt a}$ and $I^{\wt a}$ respectively by $I_{a, -n}$ and $I^a_n$ for some fixed $n \in \ZZ$. Then by applying the linear map in \eqref{pre iso Higgs field} to this we obtain an element of $\big( \Omega_{\Pi \D}( \Fil_{|n|}(\tg \otimes \tg^\D_\CC)) \big)^\Pi$. We therefore deduce from the definition \eqref{Fil zeta inv def} that the original element of $\Omega^\Pi_\D(\tg) \otimes_\RR \tg^\D$ we started with lives in $\Fil_{|n|} \big( \Omega^\Pi_\D(\tg) \otimes_\RR \tg^\D \big)$. In particular, it follows that the infinite sum in \eqref{Higgs field proof zc} defines an element of the inverse limit $\Omega^\Pi_\D(\tg) \hotimes_\RR \tg^\D$, cf. remark \ref{rem: infinite sums}.

Next, for a non-fixed real point $x \in \z^k_{\rm r}$, the elements \eqref{dual Takiff basis b} from the basis \eqref{dual Takiff basis} of $\Omega^\Pi_\D(\tg)$ read
\begin{equation*}
\frac{1}{T} \sum_{\alpha \in \Gamma} \hat \alpha \left( \frac{I_{\wt a; k, -p} dz}{(z - x)^{p+1}} \right) =
\frac{1}{T} \sum_{k=0}^{T-1} \frac{\omega^{-k} \sigma^k I_{\wt a; k, -p} dz}{(\omega^{-k} z - x)^{p+1}}.
\end{equation*}
Their tensor product with the elements $I^{\wt a}_{k, p} \otimes \varepsilon^p_x$ from the dual basis of $\tg^\D$ can then be rewritten, after formally applying the linear map \eqref{hom Higgs field} and recalling the notation \eqref{tgD R C basis b}, as
\begin{equation*}
\frac{1}{T} \sum_{k=0}^{T-1} \frac{\omega^{-k} \sigma^k I_{\wt a; k, -p} dz}{(\omega^{-k} z - x)^{p+1}} \otimes I^{\wt a(x)}_{k \ul p} = \frac{1}{T} \sum_{k = 0}^{T-1} \hat \s^k \left[ \frac{\A^x_{\ul p}}{(z - x)^{p+1}} dz \right].
\end{equation*}
The same argument as above applies to show that the terms in \eqref{Higgs field def} corresponding to a non-fixed real point $x \in \z^k_{\rm r}$ belong to the right completion. Specifically, we consider the tensor product
\begin{equation*}
\frac{1}{T} \sum_{k=0}^{T-1} \frac{\omega^{-k} \sigma^k I_{\wt a; k, -p} dz}{(\omega^{-k} z - x)^{p+1}} \otimes (I^{\wt a}_{k, p} \otimes \epsilon^p_x)
\end{equation*}
over $\RR$, where there is no implicit sum over $\wt a$ but instead $I^{\wt a}_{k, p}$ is a particular basis element from \eqref{tg kp basis} and $I_{\wt a; k, -p}$ is its dual basis element from \eqref{tg kp dual basis}. Applying the linear map \eqref{pre iso Higgs field} we then obtain an element of $\big( \Omega_{\Pi \D}( \Fil_n(\tg \otimes \tg^\D_\CC)) \big)^\Pi$ for some $n \in \ZZ_{\geq 0}$, from which it follows that the above belongs to $\Fil_n \big( \Omega^\Pi_\D(\tg) \otimes_\RR \tg^\D \big)$ as required.

Finally, the tensor product over $\RR$ of the basis elements \eqref{dual Takiff basis c} -- \eqref{dual Takiff basis d} of $\Omega^\Pi_\D(\tg)$ with the basis elements \eqref{Takiff basis c} -- \eqref{Takiff basis d} of $\tg^\D$ gives the last two terms in \eqref{Gaudin Lax matrix} after formally applying the linear map \eqref{hom Higgs field}. The proof that we have in this case as well an element of the completion $\Omega^\Pi_\D(\tg) \hotimes_\RR \tg^\D$ is as above.
\end{proof}
\end{proposition}

\begin{remark} \label{rem: field at origin}
The term involving the formal field at the origin in the expression \eqref{Gaudin Lax matrix} could alternatively be included among the terms in the sum over $x \in \Z'$ as
\begin{align*}
&\frac{1}{T} \sum_{k=0}^{T-1} \hat\s^k \left[ \sum_{p \geq 0} \frac{\A^0_{\ul p}}{z^{p+1}} dz \right] = \frac{1}{T} \sum_{k=0}^{T-1} \sum_{p \geq 0} \frac{\omega^{-k} \sigma^k I_{(-p, \wt \alpha)} \otimes I^{\wt \alpha (0)}_{\ul p}}{(\omega^{-k} z)^{p+1}} dz\\
&\qquad\qquad = \sum_{p \geq 0} \pi_{(-p)} I_{(-p, \wt \alpha)} \otimes I^{\wt \alpha (0)}_{\ul p} z^{-p-1} dz = \sum_{p \geq 0} \A^0_{\ul p} z^{-p-1} dz. \qedhere
\end{align*}
\end{remark}

\subsubsection{Complexified formal fields} \label{sec: complexified fields}

It is possible to treat all the points in $\z$ on a more equal footing by attaching to each of them a `complexified' formal field. In terms of the latter, Proposition \ref{prop: Gaudin Lax 2} below then provides an alternative, more uniform, description of the image of the canonical element \eqref{Higgs field def} under the linear isomorphism \eqref{hom Higgs field}.

Specifically, we start by attaching to every $x \in \z$ the complex Lie algebra $\Tak^{n_x}_x \tg$. By Lemma \ref{lem: Lie re com}$(i)$ we then have, for each $x \in \z$, an isomorphism
\begin{align} \label{Tg bTg to Tg C}
\Tak^{n_x}_x \tg \oplus \overline{\Tak^{n_x}_x \tg} &\overset{\sim}\longrightarrow \Tak^{n_x}_x \tg \otimes_\RR \CC,\\
(\ms X \otimes \varepsilon_x^p, \ms Y \otimes \varepsilon_x^q) &\longmapsto \ms X^{(x)}_{\ul p, \CC} + c \big( \ms Y^{(x)}_{\ul q, \CC} \big) \coloneqq \ha \big( (\ms X \otimes \varepsilon_x^p) \otimes 1 - i (\ms X \otimes \varepsilon_x^p) \otimes i \notag\\
&\qquad\qquad\qquad\qquad\qquad\qquad + (\ms Y \otimes \varepsilon_x^q) \otimes 1 + i (\ms Y \otimes \varepsilon_x^q) \otimes i \big), \notag
\end{align}
where $p, q = 0, \ldots, n_x - 1$ for $x \in \z \setminus \{ \infty \}$ and $p, q = 1, \ldots, n_\infty - 1$ for $x = \infty$. Here $c$ denotes complex conjugation on the second tensor factor of $\Tak^{n_x}_x \tg \otimes_\RR \CC$. Note that for $x \in \z_{\rm c}$ this is precisely the isomorphism of Proposition \ref{prop: complexified Takiff}$(i)$, with the respective notations related as $\ms X^{(x)}_{\ul p} = \ms X^{(x)}_{\ul p, \CC}$ and $\ms X^{(\bar x)}_{\ul p} = c\big( \ms X^{(x)}_{\ul p, \CC} \big)$.

Consider the direct sum of complex Lie algebras
\begin{equation*}
\Tak^\D \tg \coloneqq \bigoplus_{x \in \z} \Tak^{n_x}_x \tg
\end{equation*}
and let $(\Tak^\D \tg)_\CC \coloneqq \Tak^\D \tg \otimes_\RR \CC$ be its complexification.
We gather the above collection of isomorphisms for each point $x \in \z$ into a single isomorphism, cf. \eqref{Takiff complex},
\begin{equation*}
\psi_\CC : \bigoplus_{x \in \z} \big( \Tak^{n_x}_x \tg \oplus \overline{\Tak^{n_x}_x \tg} \big) \overset{\sim}\longrightarrow (\Tak^\D \tg)_\CC.
\end{equation*}
Following the discussion in \S\ref{sec: Alg Obs} we introduce a descending $\ZZ_{\geq 0}$-filtration on $(\Tak^\D \tg)_\CC$ with the subspaces for each $n \in \ZZ_{\geq 0}$ defined as
\begin{equation*}
\Fil_n\big( (\Tak^\D \tg)_\CC \big) \coloneqq \psi_\CC \Biggl( \bigoplus_{x \in \z} \big( \Fil_n(\Tak^{n_x}_x \tg) \oplus \overline{\cFil_n(\Tak^{n_x}_x \tg)} \big) \Biggr).
\end{equation*}
Next, following \S\ref{sec: formal fields} we introduce the completion $\F_\CC$ of the tensor product of complex vector spaces $\tg \otimes (\Tak^\D \tg)_\CC$ with respect to the descending $\ZZ_{\geq 0}$-filtration given by
\begin{equation*}
\Fil_n \big( \tg \otimes (\Tak^\D \tg)_\CC \big) \coloneqq \Fil_n \tg \otimes c\big( \Fil_n (\Tak^\D \tg)_\CC \big) + \cFil_n \tg \otimes \Fil_n (\Tak^\D \tg)_\CC.
\end{equation*}

For every $x \in \z$ we can now introduce a collection of \emph{complexified formal fields} as elements of $\F_\CC$ defined as
\begin{subequations} \label{compl formal fields def}
\begin{equation}
\A^x_{\ul p, \CC} \coloneqq I_{\wt a} \otimes I^{\wt a (x)}_{\ul p, \CC}, \qquad p = 0, \ldots, n_x - 1,
\end{equation}
for $x \in \z \setminus \{ \infty \}$ and for the point at infinity as
\begin{equation}
\A^\infty_{\ul{q+1}, \CC} \coloneqq I_{\wt a} \otimes I^{\wt a (\infty)}_{\ul{q+1}, \CC}, \qquad q = 0, \ldots, n_\infty - 2.
\end{equation}
\end{subequations}
An action $r : \Pi \to \bAut \F_\CC$, $\alpha \mapsto r_\alpha$ of $\Pi$ on the complexified formal fields is defined similarly to \eqref{Pi act Fields}. In particular, we will also occasionally write the maps $r_\s$ and $r_\t$ in $\bAut \F_\CC$ simply as $\sigma$ and $\tau$.

The relation between the formal fields \eqref{fields A} of \S\ref{sec: formal fields} and the complexified formal fields \eqref{compl formal fields def} is given by the following lemma.
\begin{lemma} \label{lem: real complex fields}
For every $x \in \z_{\rm c}$ and $p = 0, \ldots, n_x - 1$ we have
\begin{equation*}
\A^x_{\ul p} = \A^x_{\ul p, \CC}, \qquad
\A^{\bar x}_{\ul p} = r_{\t} \A^x_{\ul p, \CC}.
\end{equation*}
For each $x \in \z_{\rm r}^k$ with $k \in \ZZ_T$ and $p = 0, \ldots, n_x - 1$ we have
\begin{equation*}
\A^x_{\ul p} = \big( \id + \; \omega^{kp} r_{\s^k \t} \big) \A^x_{\ul p, \CC}.
\end{equation*}
For the origin and infinity we have
\begin{equation*}
\A^0_{\ul p} = \frac{1}{T} \sum_{k=0}^{T-1} \omega^{kp} \big( r_{\s^k} + r_{\s^k \t} \big) \A^0_{\ul p, \CC}, \qquad
\A^\infty_{\ul {q+1}} = \frac{1}{T} \sum_{k=0}^{T-1} \omega^{-k(q+1)} \big( r_{\s^k} + r_{\s^k \t} \big) \A^\infty_{\ul {q+1}, \CC}
\end{equation*}
with $p = 0, \ldots, n_0 - 1$ and $q = 0, \ldots, n_\infty - 2$.
\begin{proof}
The statement for a complex point $x \in \z_{\rm c}$ is clear. Next, for any $x \in \z_{\rm r}^k$ with $k \in \ZZ_T$ we can write $\big( \id + \; \omega^{kp} r_{\s^k \t} \big) \A^x_{\ul p, \CC}$ as
\begin{align*}
&I_{\wt a} \otimes I^{\wt a (x)}_{\ul p, \CC} + \omega^{kp} \sigma^k \tau I_{\wt a} \otimes  c \big( I^{\wt a (x)}_{\ul p, \CC} \big) = I_{\wt a} \otimes I^{\wt a (x)}_{\ul p, \CC} + \sigma^k \tau I_{\wt a} \otimes  c \big( \big( \omega^{-kp} I^{\wt a}\big)^{(x)}_{\ul p, \CC} \big)\\
&\qquad = I_{\wt a} \otimes I^{\wt a (x)}_{\ul p, \CC} + I_{\wt a} \otimes  c \big( \big( \omega^{-kp} \sigma^k \tau I^{\wt a} \big)^{(x)}_{\ul p, \CC} \big)\\
&\qquad= I_{\wt a} \otimes \big( (\pi^+_{k, p} I^{\wt a} \otimes \varepsilon_x^p) \otimes 1 - i (\pi^-_{k,p} I^{\wt a} \otimes \varepsilon_x^p) \otimes i \big) = I_{\wt a} \otimes I^{\wt a (x)}_{\ul p} = \A^x_{\ul p},
\end{align*}
as required. In the first equality we have used the anti-linearity of $c$ and the linearity of the map $\ms X \otimes \varepsilon_x^p \mapsto \ms X^{(x)}_{\ul p, \CC}$. In order to see the second equality we write $\sigma^k \tau I_{\wt a} = y^{\;\; \wt c}_{\wt a} I_{\wt c}$ for some $y^{\;\; \wt c}_{\wt a} \in \CC$, where the sum over $\wt c$ is implicit as usual, and it suffices to show that $\overline{y^{\;\; \wt c}_{\wt a}} I^{\wt a} = \sigma^k \tau I^{\wt c}$. But if we let $\sigma^k \tau I^{\wt c} = x^{\wt c}_{\;\; \wt a} I^{\wt a}$ for some $x^{\wt c}_{\;\; \wt a} \in \CC$ then
\begin{equation*}
\overline{y^{\;\; \wt c}_{\wt a}} = \overline{y^{\;\; \wt e}_{\wt a} \big( I_{\wt e} \big| I^{\wt c} \big)} = \overline{\big( \sigma^k \tau I_{\wt a} \big| I^{\wt c} \big)} = \overline{\big( \tau I_{\wt a} \big| \sigma^{-k} I^{\wt c} \big)} = \big( I_{\wt a} \big| \sigma^k \tau I^{\wt c} \big) = x^{\wt c}_{\;\; \wt e} \big( I_{\wt a} \big| I^{\wt e} \big) = x^{\wt c}_{\;\; \wt a}.
\end{equation*}
Finally, in the third line above we used the explicit definitions of the various notations introduced in \eqref{Tg bTg to Tg C} and \eqref{Lie re com iso 2} along with the expression for $\A^x_{\ul p}$ given after \eqref{fields A}.

Similarly, for the complexified formal field at the origin we have
\begin{align*}
&\frac{1}{T} \sum_{k=0}^{T-1} \omega^{kp} \big( r_{\s^k} + r_{\s^k \t} \big) \A^0_{\ul p, \CC} = \frac{1}{T} \sum_{k=0}^{T-1} \omega^{kp} \sigma^k I_{\wt a} \otimes I^{\wt a (0)}_{\ul p, \CC} + \frac{1}{T} \sum_{k=0}^{T-1} \omega^{kp} \sigma^k \tau I_{\wt a} \otimes c \big( I^{\wt a (0)}_{\ul p, \CC} \big)\\
&\quad = \frac{1}{T} \sum_{k=0}^{T-1} I_{\wt a} \otimes \big( \omega^{kp} \sigma^{-k} I^{\wt a} \big)^{(0)}_{\ul p, \CC} + \frac{1}{T} \sum_{k=0}^{T-1} I_{\wt a} \otimes c \big( \big( \omega^{- kp} \sigma^k \tau I^{\wt a} \big)^{(0)}_{\ul p, \CC} \big)\\
&\quad = I_{\wt a} \otimes \big( (\pi^+_0 \pi_{(p)} I^{\wt a} \otimes \varepsilon_0^p) \otimes 1 - i (\pi^-_0 \pi_{(p)} I^{\wt a} \otimes \varepsilon_0^p) \otimes i  \big)
= I_{\wt a} \otimes \big( \pi_{(p)} I^{\wt a} \big)^{(0)}_{\ul p} = \A^0_{\ul p}.
\end{align*}
The proof of the statement for the formal field at infinity is completely analogous.
\end{proof}
\end{lemma}

\begin{proposition} \label{prop: Gaudin Lax 2}
The expression \eqref{Gaudin Lax matrix} for $\zeta(\Phi)$ can be rewritten as
\begin{equation} \label{Gaudin Lax matrix 2}
\zeta(\Phi) = \frac{1}{T} \sum_{\alpha \in \Pi} \hat\alpha \left[ \sum_{x \in \z \setminus \{ \infty \}} \sum_{p \geq 0} \frac{\A^x_{\ul p, \CC}}{(z - x)^{p+1}} dz - \sum_{q \geq 0} \A^\infty_{\ul{q+1}, \CC} z^q dz \right].
\end{equation}
\begin{proof}
We will show that the right hand side of \eqref{Gaudin Lax matrix 2} coincides with the right hand side of \eqref{Gaudin Lax matrix}, working term by term in the sum over $x \in \z$.
The term corresponding to a point $x \in \z_{\rm c}$, for which $\Pi_x = \{ 1 \}$, reads
\begin{align*}
&\frac{1}{T} \sum_{k=0}^{T-1} \hat\s^k \left[ \sum_{p \geq 0} \frac{\A^x_{\ul p, \CC}}{(z - x)^{p+1}} dz + \sum_{p \geq 0} \frac{r_\t \A^x_{\ul p, \CC}}{(z - \bar x)^{p+1}} dz \right],
\end{align*}
which coincides, by Lemma \ref{lem: real complex fields}, with the terms in \eqref{Gaudin Lax matrix} for $x$ and $\bar x$ in $\Z'$.

Consider now a point $x \in \z^k_{\rm r}$ for some $k \in \ZZ_T$ so that $\Pi_x = \langle \s^k \t \rangle$, \emph{i.e.} $\omega^k \bar x = x$. The corresponding term in \eqref{Gaudin Lax matrix 2} reads
\begin{equation*}
\frac{1}{T} \sum_{\ell =0}^{T-1} \hat\s^\ell \left[ \sum_{p \geq 0} \frac{\A^x_{\ul p, \CC}}{(z - x)^{p+1}} dz + \sum_{p \geq 0} \frac{\omega^{kp} r_{\s^k \t} \A^x_{\ul p, \CC}}{( z - \omega^k \bar x)^{p+1}} dz \right]\\
= \frac{1}{T} \sum_{\ell =0}^{T-1} \hat\s^\ell \left[ \sum_{p \geq 0} \frac{\A^x_{\ul p}}{(z - x)^{p+1}} dz \right],
\end{equation*}
where we broke the sum over $\alpha \in \Pi$ into sums over $\s^\ell$ for $\ell = 0, \ldots, T-1$ and $\s^{\ell+k} \t$ for $\ell = 0, \ldots, T-1$. The last step follows from Lemma \ref{lem: real complex fields} and the resulting expression coincides with the corresponding term in the sum over $\Z'$ from \eqref{Gaudin Lax matrix}.

For the point at the origin we have
\begin{align*}
&\frac{1}{T} \sum_{k=0}^{T-1} \left[ \sum_{p \geq 0} \omega^{kp} r_{\s^k} \A^0_{\ul p, \CC} z^{-p-1} dz + \sum_{p \geq 0} \omega^{kp} r_{\s^k \t} \A^0_{\ul p, \CC} z^{-p-1} dz \right] = \sum_{p \geq 0} \A^0_{\ul p} z^{-p-1} dz,
\end{align*}
where we wrote the sum over $\alpha \in \Pi$ as two separate sums over $\s^k$ for $k = 0, \ldots, T-1$ and $\s^k \t$ for $k = 0, \ldots, T-1$, and the equality is by Lemma \ref{lem: real complex fields}. This coincides with the term involving the formal field at the origin from \eqref{Gaudin Lax matrix}.
Likewise, the term in \eqref{Gaudin Lax matrix 2} involving the point at infinity reads
\begin{equation*}
- \frac{1}{T} \sum_{k=0}^{T-1} \left[ \sum_{q \geq 0} \omega^{- k (q+1)} (r_{\s^k} + r_{\s^k \t}) \A^\infty_{\ul{q+1}, \CC} z^q dz \right] = - \sum_{q \geq 0} \A^\infty_{\ul{q+1}} z^q dz,
\end{equation*}
using Lemma \ref{lem: real complex fields}, matching the term in \eqref{Gaudin Lax matrix} for the point at infinity.
\end{proof}
\end{proposition}

\subsection{Classical $\r$-matrix}

Consider the rational function on $\CC^2$ valued in the completed tensor product $\tg \hotimes \tg$, cf. \S\ref{sec: can elem tg}, given by
\begin{equation} \label{r kernel rat}
\r(z, w) \coloneqq \frac{1}{T} \sum_{k=0}^{T-1} \frac{\sigma^k I^{\wt a} \otimes I_{\wt a}}{w - \omega^{-k} z}.
\end{equation}
It is a \emph{non-skew-symmetric} solution of the classical Yang-Baxter equation with spectral parameter. Using the tensor notation of \S\ref{sec: tensor index} with $\a = \b = \c = \tg$ and where $\mathfrak A$ is the algebra of rational functions on $\CC^3$, with coordinates $z, z'$ and $w$, this reads
\begin{equation*}
[\r_{\1\2}(z, z'), \r_{\1\3}(z, w)] + [\r_{\1\2}(z, z'), \r_{\2\3}(z', w)] + [\r_{\3\2}(w, z'), \r_{\1\3}(z, w)] = 0.
\end{equation*}
We will refer to \eqref{r kernel rat} as the \emph{formal $\r$-matrix}.

\subsubsection{Lax matrix algebra} \label{sec: Lax matrix algebra}

We define the \emph{formal Lax matrix} $\L \in \F \otimes R_{\Pi \z}$ by writing $\zeta(\Phi) = \L(z) dz$.
Explicitly, from the expression \eqref{Gaudin Lax matrix} for $\zeta(\Phi)$ we obtain
\begin{equation} \label{Lax simple form}
\L(z) = \frac{1}{T} \sum_{k=0}^{T-1} \sum_{x \in \Z'} \sum_{p \geq 0} \frac{\omega^{-k} \sigma^k \A^x_{\ul p}}{(\omega^{-k} z - x)^{p+1}} + \sum_{p \geq 0} \A^0_{\ul p} z^{-p-1}
- \sum_{q \geq 0} \A^{\infty}_{\ul{q+1}} z^q.
\end{equation}
Since $\Phi \in \Omega^\Pi_\D(\tg) \hotimes_\RR \tg^\D$ by Proposition \ref{prop: Gaudin Lax}, its image $\zeta(\Phi) \in \Omega^\Pi_\D(\F)$ under the isomorphism \eqref{hom Higgs field} is $\Pi$-invariant by construction. Recalling the action of $\Pi$ on $\Omega_{\Pi \D}(\F)$ in \eqref{Pi act OmPiD F}, it follows that the formal Lax matrix is $\Pi$-equivariant in the following sense
\begin{equation} \label{Pi equiv formal Lax}
\sigma \L(z) = \omega \L(\omega z), \qquad \tau \L(z) = \L(\bar z).
\end{equation}

In the following proposition we make use of the tensor index notation from \S\ref{sec: tensor index} with $\a = \b = \tg$, $\c = \tg^\D_\CC$ and where $\mathfrak A$ is the algebra of rational functions on $\CC^2$, with coordinates $z, z'$. Moreover, as in Proposition \ref{prop: Fields PB} we drop the index $\null_\3$ for the operator space and use the Poisson bracket \eqref{wtSg PB} to denote its Lie bracket.

\begin{proposition} \label{prop: Lax algebra}
The bracket of the formal Lax matrix takes the form
\begin{equation} \label{Lax algebra 2}
\{ \L_{\1}(z), \L_{\2}(z') \} = [\r_{\1\2}(z, z'), \L_{\1}(z)] - [\r_{\2\1}(z', z), \L_{\2}(z')].
\end{equation}
\begin{proof}
This is a direct computation making use of Proposition \ref{prop: Fields PB}, its consequence in \eqref{Fields PB sigma} and the identities
\begin{align*}
\sum_{p=0}^r \frac{1}{(z - x)^{p+1}} \frac{1}{(z' - x)^{r-p+1}} &= \frac{1}{z - z'} \left( \frac{1}{(z' - x)^{r+1}} - \frac{1}{(z - x)^{r+1}} \right),\\
\sum_{p=0}^r z^p z'^{r-p} &= \frac{1}{z - z'} \left( z^{r+1} - z'^{r+1} \right),
\end{align*}
valid for any pairwise distinct $x, z, z' \in \CC$ and $r \in \ZZ_{\geq 0}$.

We know from Proposition \ref{prop: Fields PB} that $\{ \A^x_{\ul p \1}, \A^y_{\ul q \2} \} = 0$ whenever $x, y \in \Z$ are distinct. It is therefore sufficient to show that \eqref{Lax algebra 2} holds separately for each summand in $\L(z)$ corresponding to the different points in $\Z$.
For any $x \in \Z'$ we have
\begin{align*}
&\frac{1}{T^2} \sum_{p, q \geq 0} \sum_{k, \ell = 0}^{T-1} \frac{\omega^{-k}}{(\omega^{-k} z - x)^{p+1}} \frac{\omega^{-\ell}}{(\omega^{-\ell} z' - x)^{q+1}} \big\{ \sigma^k \A^x_{\ul p\1}, \sigma^\ell \A^x_{\ul q\2} \big\}\\
&= - \frac{1}{T^2} \sum_{r \geq 0} \sum_{k, \ell = 0}^{T-1} \Bigg( \frac{1}{z - \omega^{k-\ell} z'} \frac{\omega^{-\ell}}{(\omega^{-\ell} z' - x)^{r+1}} \big[ \sigma^{\ell - k}_{\2} \wt C_{\1\2}, \sigma^{\ell} \A^x_{\ul r \2} \big]\\
&\qquad\qquad\qquad\qquad\qquad + \frac{1}{\omega^{\ell-k} z - z'} \frac{\omega^{-k}}{(\omega^{-k} z - x)^{r+1}} \big[ \sigma^{k - \ell}_{\1} \wt C_{\1\2}, \sigma^k \A^x_{\ul r \1} \big] \Bigg)\\
&= \frac{1}{T} \sum_{r \geq 0} \sum_{k = 0}^{T-1} \Bigg( \bigg[ \r_{\1\2}(z, z'), \frac{\omega^{-k} \sigma^k \A^x_{\ul r \1}}{(\omega^{-k} z - x)^{r+1}} \bigg] - \bigg[ \r_{\2\1}(z', z), \frac{\omega^{-k} \sigma^k \A^x_{\ul r \2}}{(\omega^{-k} z' - x)^{r+1}} \bigg] \Bigg),
\end{align*}
which is exactly the right hand side of \eqref{Lax algebra 2} for the summand in $\L(z)$ corresponding to $x \in \Z'$.

For the origin, using the fact that $\wt C^{(p)}_{\1\2} = \pi_{(p) \2} \wt C_{\1\2} = \frac{1}{T} \sum_{k=0}^{T-1} \omega^{-kp} \sigma^k_\2 \wt C_{\1\2}$, we find
\begin{align*}
&\sum_{p, q \geq 0} z^{-p-1} z'^{-q-1} \big\{ \A^0_{\ul p \1}, \A^0_{\ul q \2} \big\} = - \sum_{p, q \geq 0} z^{-p-1} z'^{-q-1} \big[ \wt C^{(p)}_{\1\2}, \A^0_{\ul{p+q}\2} \big]\\
&\quad = - \frac{1}{T} \sum_{k=0}^{T-1} \sum_{r \geq 0} \sum_{p=0}^r \omega^{-(r+1) k} z^{-p-1} (\omega^{-k} z')^{p-r-1} [\sigma^k_\2 \wt C_{\1\2}, \A^0_{\ul r \2}]\\
&\quad = \frac{1}{T} \sum_{k=0}^{T-1} \sum_{r \geq 0} \left( \frac{\omega^{- (r+1) k} z^{-r-1}}{z - \omega^{-k} z'} [\sigma^k_\2 \wt C_{\1\2}, \A^0_{\ul r \2}] - \frac{z'^{-r-1}}{z - \omega^{-k} z'} [\sigma^k_\2 \wt C_{\1\2}, \A^0_{\ul r \2}] \right)\\
&\quad = \sum_{r \geq 0} \big[ \r_{\1\2}(z, z'), \A^0_{\ul r \1} z^{-r-1} \big] - \sum_{r \geq 0} \big[ \r_{\2\1}(z', z), \A^0_{\ul r \2} z'^{-r-1} \big].
\end{align*}
Similarly, for the point at infinity we have
\begin{align*}
&\sum_{p, q \geq 0} z^p z'^q \big\{ \A^\infty_{\ul {p+1}\1}, \A^\infty_{\ul {q+1}\2} \big\} = - \sum_{p, q \geq 0} z^p z'^q \big[ \wt C^{(-p-1)}_{\1\2}, \A^\infty_{\ul{p+q+2}\2} \big]\\
&\quad = - \frac{1}{T} \sum_{k=0}^{T-1} \sum_{r \geq 0} \sum_{p=0}^{r-1} \omega^{r k} z^p (\omega^{-k} z')^{r-1-p} [\sigma^k_\2 \wt C_{\1\2}, \A^\infty_{\ul{r+1} \2}]\\
&\quad = - \frac{1}{T} \sum_{k=0}^{T-1} \sum_{r \geq 0} \left( - \frac{\omega^{-(r+1) k} z^r}{z' - \omega^{-k} z} [\sigma^{-k}_\2 \wt C_{\1\2}, \A^\infty_{\ul{r+1} \2}] - \frac{z'^r}{z - \omega^{-k} z'} [\sigma^k_\2 \wt C_{\1\2}, \A^\infty_{\ul{r+1} \2}] \right)\\
&\quad = - \sum_{r \geq 0} \big[ \r_{\1\2}(z, z'), \A^\infty_{\ul {r+1} \1} z^r \big] + \sum_{r \geq 0} \big[ \r_{\2\1}(z', z), \A^\infty_{\ul {r+1} \2} z'^r \big],
\end{align*}
where in the second line we used $\wt C^{(-p-1)}_{\1\2} = \pi_{(-p-1) \2} \wt C_{\1\2} = \frac{1}{T} \sum_{k=0}^{T-1} \omega^{k(p+1)} \sigma^k_\2 \wt C_{\1\2}$.
\end{proof}
\end{proposition}

\subsubsection{Twist function} \label{sec: twist}

Recall that in Corollary \ref{cor: Fields PB} we applied the tensor product of Lie algebra homomorphisms $\rep \otimes \rep \otimes \pi_{\bm \lvl}$ to the Poisson brackets of formal fields from Proposition \ref{prop: Fields PB} to obtain the Poisson brackets of the $\g$-valued classical fields \eqref{Currents x 0 inf}. In this section we similarly apply the linear map $\rep \otimes \rep \otimes \pi_{\bm \lvl} \otimes \id$ to the algebra of Lax matrices \eqref{Lax algebra 2} from Proposition \ref{prop: Lax algebra}.

Recall the formal Lax matrix $\L \in \F \otimes R_{\Pi \z}$ defined at the start of \S\ref{sec: Lax matrix algebra}. Applying to it the linear map $\rep \otimes \pi_{\bm \lvl} \otimes \id$ it may be written in the form, cf. \S\ref{sec: classical fields},
\begin{equation} \label{Lax twist from formal}
(\rep \otimes \pi_{\bm \lvl} \otimes \id) \L(z) = \varphi(z) \big( \partial + \mathscr L(z) \big) \in \Conn_\g(S^1) \hotimes \hg^\D_{\CC, \bm \lvl} \otimes R_{\Pi \z},
\end{equation}
where $\varphi(z) \in R_{\Pi \z}$ is the \emph{twist function} and is given explicitly by
\begin{equation*}
\varphi(z) = \frac{1}{T} \sum_{k=0}^{T-1} \sum_{x \in \Z'} \sum_{p \geq 0} \frac{\omega^{-k} \lvl^x_p}{(\omega^{-k} z - x)^{p+1}} + \sum_{p \geq 0} \lvl^0_p z^{-p-1}
- \sum_{q \geq 0} \lvl^{\infty}_{q+1} z^q.
\end{equation*}
Here $\varphi(z) \mathscr L(z) \in \mathscr T(S^1, \g) \hotimes \hg^\D_{\CC, \bm \lvl} \otimes R_{\Pi \z}$ is a linear combination of the $\g$-valued classical fields \eqref{Currents x 0 inf} with coefficients given by rational functions in $z$ with poles at the points of the set $\Pi \z$. The element $\mathscr L(z)$ is called the \emph{Lax matrix}. Notice that its poles will typically \emph{not} be at the set of points $\Pi \z$, but rather at the zeroes of the twist function. We deduce from the $\Pi$-equivariance of the homomorphism $\rep$ established in \S\ref{sec: Conn on S1} and the fact that $\pi_{\bm \lvl}$ commutes with $c$ from Proposition \ref{prop: pil equiv}, together with the $\Pi$-equivariance property \eqref{Pi equiv formal Lax} of the formal Lax matrix, that
\begin{alignat*}{2}
\sigma \mathscr L(z) &= \mathscr L(\omega z), &\qquad \tau \mathscr L(z) &= \mathscr L(\bar z),\\
\varphi(z) &= \omega \varphi(\omega z), &\qquad \overline{\varphi(z)} &= \varphi(\bar z).
\end{alignat*}

Next we consider the formal $\r$-matrix defined in \eqref{r kernel rat}. It can be rewritten more explicitly, using the definition of the dual bases $I^{\wt a}$ and $I_{\wt a}$ of $\tg$ in \S\ref{sec: can elem tg}, as
\begin{equation*}
\r(z, z') = \frac{1}{T} \sum_{k=0}^{T-1} \frac{\cent \otimes \cocent + \cocent \otimes \cent}{z' - \omega^{-k} z} + \frac{1}{T} \sum_{n \in \mathbb{Z}} \sum_{k=0}^{T-1} \frac{\sigma^k I_{a, -n} \otimes I^a_n}{z' - \omega^{-k} z} \in \tg \hotimes \tg \otimes \mathfrak A,
\end{equation*}
where $\mathfrak A$ denotes here the algebra of rational functions on $\CC^2$, with coordinates $z, z'$.
Applying to it the linear map $\rep \otimes \rep \otimes \id$ we obtain, cf. \eqref{wt C delta},
\begin{align*}
\big( (\rep \otimes \rep \otimes \id) \r(z, z') \big)(\theta, \theta') &= \frac{1}{T} \sum_{k=0}^{T-1} \frac{\sigma^k I_a \otimes I^a}{z' - \omega^{-k} z} \sum_{n \in \mathbb{Z}} e^{i n (\theta' - \theta)}\\
&= \frac{\sum_{k=0}^{T-1} z^k z'^{T-1-k} C^{(-k)}}{z'^T - z^T} \delta_{\theta \theta'}
\eqqcolon \R^0(z, z') \delta_{\theta \theta'},
\end{align*}
where $C^{(k)} \in \g_{(-k), \CC} \otimes \g_{(k), \CC}$ are the graded components of the canonical element $C$ of $\g$ introduced in \S\ref{sec: can elem g} and $\delta_{\theta \theta'}$ denotes the Dirac $\delta$-distribution introduced in \S\ref{sec: Conn on S1}. Recall also its derivative $\delta'_{\theta \theta'}$ defined before Corollary \ref{cor: Fields PB}.

\begin{corollary} \label{cor: Lax algebra}
The Poisson bracket of the Lax matrix takes the form
\begin{align} \label{Lax algebra}
\{ \mathscr L_{\1}(z), \mathscr L_{\2}(z') \} &= \big[ \R_{\1\2}(z, z'), \mathscr L_{\1}(z) \big] \delta_{\theta \theta'} - \big[ \R_{\2\1}(z', z), \mathscr L_{\2}(z') \big] \delta_{\theta \theta'} \notag\\
&\qquad\qquad\qquad\qquad\qquad\qquad - \big( \R_{\1\2}(z, z') + \R_{\2\1}(z', z) \big) \delta'_{\theta \theta'},
\end{align}
where $\R(z, z') \coloneqq \R^0(z, z') \varphi(z')^{-1}$.
\begin{proof}
Acting with $\rep \otimes \rep \otimes \pi_{\bm \lvl} \otimes \id$ on both sides of \eqref{Lax algebra 2} and using the above we obtain
\begin{equation*}
\{ \mathscr L_{\1}(z), \mathscr L_{\2}(z') \} = \big[ \R_{\1\2}(z, z') \delta_{\theta \theta'}, \partial_\theta + \mathscr L_{\1}(z) \big] - \big[ \R_{\2\1}(z', z) \delta_{\theta \theta'}, \partial_{\theta'} + \mathscr L_{\2}(z') \big],
\end{equation*}
after multiplying through by $\varphi(z)^{-1} \varphi(z')^{-1}$. This is equivalent to \eqref{Lax algebra}.
\end{proof}
\end{corollary}

The Poisson bracket relation \eqref{Lax algebra} can equivalently be rewritten in the form of the non-ultralocal $r/s$-algebra \cite{Maillet1, Maillet2}
\begin{align*}
\{ \mathscr L_{\1}(z), \mathscr L_{\2}(z') \} &= \big[ r_{\1\2}(z, z'), \mathscr L_{\1}(z) + \mathscr L_{\2}(z') \big] \delta_{\theta \theta'}\\
&\qquad\qquad + \big[ s_{\1\2}(z, z'), \mathscr L_{\1}(z) - \mathscr L_{\2}(z') \big] \delta_{\theta \theta'} - 2 s_{\1\2}(z, z') \delta'_{\theta \theta'},
\end{align*}
where the $r$- and $s$-matrices are the skew-symmetric and symmetric parts of $\R(z, z')$ respectively, given explicitly by
\begin{subequations} \label{rs matrices}
\begin{align}
r_{\1\2}(z, z') &\coloneqq \ha \big( \R_{\1\2}(z, z') - \R_{\2\1}(z', z) \big),\\
s_{\1\2}(z, z') &\coloneqq \ha \big( \R_{\1\2}(z, z') + \R_{\2\1}(z', z)\big).
\end{align}
\end{subequations}

\subsection{Formal quadratic Gaudin Hamiltonians} \label{sec: quad Ham}

In this last section we introduce the formal quadratic Hamiltonians and formal momentum of the classical dihedral affine Gaudin model, as elements of the complexified algebra of formal observables $\hat S(\tg^\D_\CC)$ from \S\ref{sec: Alg Obs}. The collection of \emph{local} quadratic Hamiltonians and the momentum, all living in the complexified algebra of local observables $\hat S_{\bm \lvl}(\hg^\D_\CC)$, will be obtained from these in \S\ref{sec: examples of NUL} by applying the homomorphism $\pi_{\bm \lvl}$ constructed in \S\ref{sec: level fix}, or some variant of this for affine $\tg$-Toda field theory in \S\ref{sec: Toda fields}. The Hamiltonian of a classical integrable field theory described by the given dihedral affine Gaudin model is then a certain linear combination of the local quadratic Hamiltonians invariant under complex conjugation, \emph{i.e.} living in the algebra of local observables $\hat S_{\bm \lvl}(\hg^\D_\CC)^c$.

We introduce a bilinear map
\begin{equation} \label{bilinear map F pre}
(\cdot | \cdot) : \tg \otimes \tg^\D_\CC \times \tg \otimes \tg^\D_\CC \longrightarrow S(\tg^\D_\CC),
\end{equation}
defined by applying the bilinear form \eqref{ip on tg} of $\tg$ to the pair of first tensor factors, \emph{i.e.} the auxiliary space in the terminology of \S\ref{sec: formal fields}, and multiplying the second tensor factors in $S(\tg^\D_\CC)$.
Recall the descending $\ZZ_{\geq 0}$-filtrations on $\tg \otimes \tg^\D_\CC$ and $S(\tg^\D_\CC)$ defined by the subspaces \eqref{Fil formal fields} and \eqref{S tgDC Fil 1}, respectively.
The bilinear map \eqref{bilinear map F pre} is continuous at the origin with respect to the associated topology, where $\tg \otimes \tg^\D_\CC \times \tg \otimes \tg^\D_\CC$ is equipped with the product topology. Indeed, for any $m, n \in \ZZ_{\geq 0}$ we have
\begin{equation*}
\big( \Fil_m (\tg \otimes \tg^\D_\CC) \big| \Fil_n (\tg \otimes \tg^\D_\CC) \big) \subset \Fil_{\textup{max}(m, n)}(\tg^\D_\CC) c\big( \Fil_{\textup{max}(m, n)} \tg^\D_\CC \big) \subset \Fil_{\textup{max}(m, n)}\big( S(\tg^\D_\CC) \big),
\end{equation*}
using the fact that $(\Fil_m \tg | \Fil_n \tg) = 0$ and $(\cFil_m \tg | \cFil_n \tg) = 0$. By linearity, \eqref{bilinear map F pre} is therefore uniformly continuous and hence extends to a bilinear map between the completions
\begin{equation} \label{bilinear map F}
(\cdot | \cdot) : \F \times \F \longrightarrow \hat S(\tg^\D_\CC).
\end{equation}
We extend \eqref{bilinear map F} in the obvious way to a bilinear map
\begin{equation*}
(\cdot | \cdot) : \F \otimes R_{\Pi \z} \times \F \otimes R_{\Pi \z} \longrightarrow \hat S(\tg^\D_\CC) \otimes R_{\Pi \z},
\end{equation*}
by sending the pair $(\mathcal X \otimes f, \mathcal Y \otimes g)$ for any $\mathcal X, \mathcal Y \in \F$ and $f, g \in R_{\Pi \z}$ to $(\mathcal X | \mathcal Y) \otimes fg$.

Recall the formal Lax matrix $\L \in \F \otimes R_{\Pi \z}$ from \S\ref{sec: Lax matrix algebra}. The following proposition is to be compared with \cite[Proposition 2.5]{ViY3} corresponding to the case of a quantum cyclotomic Gaudin model associated with a semisimple Lie algebra $\g$.

\begin{proposition} \label{prop: Ham}
We have
\begin{equation*}
\ha \big( \L(z) \big| \L(z) \big) = \frac{1}{T^2} \sum_{k=0}^{T-1} \sum_{x \in \Z'} \sum_{p \geq 0} 
\frac{\omega^{k(p-1)} \mathcal H^x_p}{(z - \omega^k x)^{p+1}}
+ \!\! \sum_{\substack{p \geq 0\\ p \equiv 1 \, \textup{mod}\, T}} \!\!\!\! \mathcal H^0_p z^{-p-1}
+ \!\!\!\! \sum_{\substack{q \geq 0\\ q+2 \equiv 0 \, \textup{mod}\, T}} \!\!\!\!\!\!\! \mathcal H^\infty_q z^q,
\end{equation*}
where the \emph{formal quadratic Gaudin Hamiltonians} in $\hat S(\tg^\D_\CC)$ are given by
\begin{subequations} \label{formal quad Gaud Ham}
\begin{align}
\mathcal H^x_p &\coloneqq \underset{(\ell, y) \neq (0, x)}{\sum_{\ell = 0}^{T-1} \sum_{y \in \Z' \cup \{ 0 \}}} \sum_{q, s \geq 0} \binom{q+s}{s} (-1)^s \frac{\omega^{\ell q} \big( \A^x_{\ul{p + s}} \big| \sigma^\ell \A^y_{\ul q} \big)}{(x - \omega^\ell y)^{q + s + 1}} \notag\\
&\qquad - T \sum_{q, s \geq 0} \binom{q}{s} x^{q - s} \big( \A^x_{\ul{p + s}} \big| \A^\infty_{\ul{q+1}} \big) + \ha \sum_{q=0}^{p-1} \big( \A^x_{\ul q} \big| \A^x_{\ul{p - q - 1}} \big),
\end{align}
for all $x \in \Z'$ and $p = 0, \ldots, 2 n_x - 1$,
\begin{align}
\mathcal H^0_p &= \frac{1}{T} \sum_{x \in \Z'} \sum_{q, s \geq 0} (-1)^{q+1} \binom{q+s}{s} \frac{\big( \A^0_{\ul{p+s}} \big| \A^x_{\ul q} \big)}{x^{q+s+1}} \notag\\
&\qquad\qquad - \sum_{q \geq 0} \big( \A^0_{\ul {p+q}} \big| \A^\infty_{\ul{q+1}} \big) + \sum_{q=0}^{p-1} \ha \big( \A^0_{\ul {p-q-1}} \big| \A^0_{\ul q} \big),
\end{align}
for $p = 0, \ldots, 2 n_0 - 1$ such that $p \equiv 1 \, \textup{mod}\, T$, and
\begin{align}
\mathcal H^\infty_q &= - \sum_{x \in \Z'} \sum_{p, s \geq 0} \binom{s}{p} x^{s - p} \big( \A^x_{\ul p} \big| \A^\infty_{\ul{q+s+2}} \big) \notag\\
&\qquad\qquad - \sum_{p \geq 0} \big( \A^0_{\ul p} \big| \A^\infty_{\ul{q+p+2}} \big) + \sum_{p=0}^q \ha \big( \A^\infty_{\ul{p+1}} \big| \A^\infty_{\ul{q - p +1}} \big),
\end{align}
\end{subequations}
for all $q = 0, \ldots, 2 n_\infty - 4$ such that $q+2 \equiv 0 \, \textup{mod}\, T$.
\begin{proof}
This is a direct computation making use of the identities
\begin{subequations} \label{identities Ham proof}
\begin{align}
\frac{1}{(z - x)^{p+1} (z - y)^{q+1}} &= \sum_{s=0}^p (-1)^s \binom{q+s}{s} \frac{1}{(x - y)^{q+s+1}} \frac{1}{(z - x)^{p-s+1}} \notag\\
\label{identity x y}
&\qquad\quad + \sum_{r=0}^q (-1)^r \binom{p+r}{r} \frac{1}{(y - x)^{p+r+1}} \frac{1}{(z - y)^{q-r+1}},\\
\label{identity x inf}
\frac{z^q}{(z - x)^{p+1}} &= \sum_{s = p}^{q-1} \binom{s}{p} z^{q-s-1} x^{s - p} + \sum_{s=0}^p \binom{q}{s} \frac{x^{q-s}}{(z - x)^{p - s + 1}}.
\end{align}
\end{subequations}
for any distinct $x, y, z \in \CC$ and $p, q \in \ZZ_{\geq 0}$.

We start from the expression for $\L(z)$ in \eqref{Lax simple form} and for brevity we write the three terms on the right hand side of this expression, corresponding to the set $\Z'$, the origin and infinity, as $\L_{\Z'}(z)$, $\L_0(z)$ and $\L_\infty(z)$, respectively. We wish to compute
\begin{align} \label{LL calculate}
\ha \big( \L(z) \big| \L(z) \big) &= \ha \big( \L_{\Z'}(z) \big| \L_{\Z'}(z) \big) + \ha \big( \L_0(z) \big| \L_0(z) \big) + \ha \big( \L_\infty(z) \big| \L_\infty(z) \big) \notag\\
&\qquad + \big( \L_0(z) \big| \L_{\Z'}(z) \big) + \big( \L_{\Z'}(z) \big| \L_\infty(z) \big) + \big( \L_0(z) \big| \L_\infty(z) \big).
\end{align}
After a lengthy computation using the identity \eqref{identity x y} we find that the first term on the right hand side is given by
\begin{align*}
&\ha \big( \L_{\Z'}(z) \big| \L_{\Z'}(z) \big) = \frac{1}{T^2} \sum_{k=0}^{T-1} \sum_{x \in \Z'} \sum_{p \geq 0} \frac{\omega^{k (p-1)}}{(z - \omega^k x)^{p+1}}\\
&\quad \times \left( \underset{(\ell, y) \neq (0, x)}{\sum_{\ell = 0}^{T-1} \sum_{y \in \Z'}} \sum_{q, s \geq 0}
(-1)^s \binom{q+s}{s} \frac{\omega^{\ell q} \big( \A^x_{\ul{p+s}} \big| \sigma^\ell \A^y_{\ul q} \big)}{(x - \omega^\ell y)^{q+s+1}} + \ha \sum_{q=0}^{p-1} \big( \A^x_{\ul q} \big| \A^x_{\ul{p-q-1}} \big) \right).
\end{align*}
The remaining two terms on the first line of the left hand side of \eqref{LL calculate} read
\begin{equation*}
\ha \big( \L_0(z) \big| \L_0(z) \big) = \!\!\! \sum_{\substack{p \geq 0\\ p \equiv 1 \, \textup{mod}\, T}} \!\!\! z^{-p-1} \sum_{q=0}^{p-1} \ha \big( \A^0_{\ul {p-q-1}} \big| \A^0_{\ul q} \big),
\end{equation*}
\begin{equation*}
\ha \big( \L_\infty(z) \big| \L_\infty(z) \big) = \sum_{\substack{q \geq 0\\q+2 \equiv 0 \, \textup{mod}\, T}} \!\!\! z^q \sum_{p=0}^q \ha \big( \A^\infty_{\ul{p+1}} \big| \A^\infty_{\ul{q - p +1}} \big).
\end{equation*}
Similarly, the three cross terms on the second line of \eqref{LL calculate} evaluate to
\begin{align*}
&\big( \L_0(z) \big| \L_{\Z'}(z) \big)
= \frac{1}{T} \!\!\! \sum_{\substack{p \geq 0\\ p \equiv 1 \, \textup{mod}\, T}} \!\!\!
z^{-p-1} \sum_{x \in \Z'} \sum_{q, s \geq 0}
(-1)^{q+1} \binom{q+s}{s} \frac{\big( \A^0_{\ul{p+s}} \big| \A^x_{\ul q} \big)}{x^{q+s+1}}\\
&\qquad + \frac{1}{T} \sum_{k=0}^{T-1} \sum_{x \in \Z'} \sum_{p \geq 0} \frac{\omega^{k(p-1)}}{(z - \omega^k x)^{p+1}} \left( \sum_{q, s \geq 0} (-1)^s \binom{q+s}{s} \frac{\big( \A^x_{\ul{p+s}} \big| \A^0_{\ul q} \big)}{x^{q+s+1}} \right),
\end{align*}
\begin{align*}
\big( \L_{\Z'}(z) \big| \L_\infty(z) \big) 
&= - \frac{1}{T} \sum_{k=0}^{T-1} \sum_{x \in \Z'} \sum_{p \geq 0} \frac{\omega^{k (p-1)}}{(z - \omega^k x)^{p+1}} \sum_{q, s \geq 0} \binom{q}{s} x^{q-s} \big( \A^x_{\ul{p+s}} \big| \A^\infty_{\ul{q+1}} \big)\\
&\qquad - \sum_{\substack{q \geq 0\\ q+2 \equiv 0 \, \textup{mod}\, T}} \!\!\! z^q \sum_{x \in \Z'} \sum_{p, s \geq 0} \binom{s}{p} x^{s - p} \big( \A^x_{\ul p} \big| \A^\infty_{\ul{q+s+2}} \big),
\end{align*}
\begin{align*}
\big( \L_0(z) \big| \L_\infty(z) \big)
&= - \!\!\! \sum_{\substack{q \geq 0\\ q+2 \equiv 0\, \textup{mod}\, T}} \!\!\! z^q \sum_{p \geq 0} \big( \A^0_{\ul p} \big| \A^\infty_{\ul{q+p+2}} \big)\\
&\qquad\qquad\qquad - \sum_{\substack{p \geq 0\\ p \equiv 1 \, \textup{mod}\, T}} \!\!\! z^{-p-1} \sum_{q \geq 0} \big( \A^0_{\ul {p+q}} \big| \A^\infty_{\ul{q+1}} \big).
\end{align*}
The result now follows from combining all the above.
\end{proof}
\end{proposition}

\subsubsection{Lax equations}

\begin{proposition} \label{prop: Lax equations}
The formal quadratic Gaudin Hamiltonians from Proposition \ref{prop: Ham} Poisson commute, \emph{i.e.} \begin{equation} \label{quad Ham commute}
\{ \H^x_p, \H^y_q \} = 0,
\end{equation}
for any pair of points $x, y \in \Z$ and $p , q \in \ZZ_{\geq 0}$. Moreover, we have the Lax equations $\{ \H^x_p, \L(z) \} = [\mathcal M^x_p(z), \L(z)]$ for every $x \in \Z$, where
\begin{gather*}
\mathcal M^x_p(z) \coloneqq \sum_{\ell=0}^{T-1} \sum_{s \geq 0} \frac{\sigma^\ell \A^x_{\ul{p+s}}}{(\omega^{- \ell} z - x)^{s+1}}, \\
\mathcal M^0_q(z) \coloneqq \sum_{s \geq 0} \A^0_{\ul{q+s}} z^{-s-1}, \qquad
\mathcal M^\infty_r(z) \coloneqq \sum_{s \geq 0} \A^\infty_{\ul{r+s+2}} z^s
\end{gather*}
for all $x \in \Z'$ and for $p = 0, \ldots, 2n_x - 1$, $q = 0, \ldots, 2n_0 - 1$ and $r = 0, \ldots, 2n_\infty - 4$ with $q \equiv 1 \, \textup{mod}\, T$ and $r+2 \equiv 0 \, \textup{mod}\, T$.
\begin{proof}
We have
\begin{align} \label{PB L2 with L}
\ha \big\{ (\L(z)|\L(z)), \L(w) \big\} &= \big( \L_\1(z) \big| \{ \L_\1(z), \L_\2(w) \} \big)_\1 \notag\\
&= \big( \L_\1(z) \big| [ \r_{\1\2}(z, w), \L_\1(z)] - [ \r_{\2\1}(w,z), \L_\2(w)] \big)_\1 \notag\\
&= - \big( \L_\1(z) \big| [ \r_{\2\1}(w,z), \L_\2(w)] \big)_\1 = \big[ \mathcal M(z, w), \L(w) \big]
\end{align}
where in the second line we used equation \eqref{Lax algebra 2} from Proposition \ref{prop: Lax algebra} and in the last step defined $\mathcal M(z, w) \coloneqq - \big( \r_{\2\1}(w, z) \big| \L_\1(z) \big)_\1$ which is given by
\begin{align*}
\mathcal M(z, w) &= - \frac{1}{T} \sum_{\ell=0}^{T-1} \frac{\sigma^\ell I^{\wt a}}{z - \omega^{-\ell} w} ( I_{\wt a} | \L(z) ) = - \frac{1}{T} \sum_{\ell=0}^{T-1} \frac{\sigma^\ell \L(z)}{z - \omega^{-\ell} w}\\
&= - \frac{1}{T^2} \sum_{k=0}^{T-1} \sum_{x \in \Z'} \sum_{p \geq 0} \sum_{\ell=0}^{T-1} \frac{1}{z - \omega^{-\ell} w} \frac{\omega^{k p} \sigma^{k+\ell} \A^x_{\ul p}}{(z - \omega^k x)^{p+1}}\\
&\qquad\qquad - \frac{1}{T} \sum_{p \geq 0} \sum_{\ell=0}^{T-1} \frac{\omega^{-\ell p} z^{-p-1}}{z - \omega^{-\ell} w} \A^0_{\ul p} + \frac{1}{T} \sum_{q \geq 0} \sum_{\ell=0}^{T-1} \frac{\omega^{\ell (q+1)} z^q}{z - \omega^{-\ell} w} \A^{\infty}_{\ul{q+1}}.
\end{align*}
In the last line we used the explicit form of the formal Lax matrix in \eqref{Lax simple form}. The first assertion now follows from
\begin{align*}
\qa \big\{ (\L(z)|\L(z)), (\L(w)|\L(w)) \big\} &= \ha \big( \big\{ (\L(z)|\L(z)), \L(w) \big\} \big| \L(w) \big)\\
&= \big( \big[ \mathcal M(z, w), \L(w) \big] \big| \L(w) \big) = 0,
\end{align*}
and taking different residues in $z$ and $w$ to extract the desired formal quadratic Gaudin Hamiltonians.

Next, we obtain the Lax equations from \eqref{PB L2 with L}. The identities \eqref{identities Ham proof} in the proof of Proposition \ref{prop: Ham} can be used to show that
\begin{align*}
\frac{1}{z - \omega^{-\ell} w} \frac{1}{(z - \omega^k x)^{p+1}} &= \frac{\omega^\ell}{z - \omega^{-\ell} w} \frac{\omega^{\ell p}}{(w - \omega^{k+\ell} x)^{p+1}}\\
&\qquad - \sum_{r=0}^p \frac{\omega^{\ell(r+1)}}{(w - \omega^{k+\ell} x)^{r+1}} \frac{1}{(z - \omega^k x)^{p-r+1}},\\
\frac{\omega^{-\ell p} z^{-p-1}}{z - \omega^{-\ell} w} &= \frac{\omega^\ell w^{-p-1}}{z - \omega^{-\ell} w} - \sum_{r=0}^p w^{-r-1} \omega^{\ell(-p+r+1)} z^{-p+r-1},\\
\frac{\omega^{\ell (q+1)} z^q}{z - \omega^{-\ell} w} &= \frac{\omega^\ell w^q}{z - \omega^{-\ell} w} + \sum_{s=0}^{q-1} z^{q-s-1} \omega^{\ell (q-s+1)} w^s.
\end{align*}
Using these we can rewrite the above expression for $\mathcal M(z, w)$ as
\begin{align*}
\mathcal M(z, w) &= \frac{w z^{T-2}}{w^T - z^T} \L(w) + \frac{1}{T^2} \sum_{k=0}^{T-1} \sum_{x \in \Z'} \sum_{p \geq 0} \sum_{\ell=0}^{T-1} \sum_{r=0}^p \frac{\omega^{\ell(r+1)}}{(w - \omega^{k+\ell} x)^{r+1}} \frac{\omega^{k p} \sigma^{k+\ell} \A^x_{\ul p}}{(z - \omega^k x)^{p-r+1}}\\
&\qquad + \underset{p-r \equiv 1 \, \textup{mod}\, T}{\sum_{p \geq 0} \sum_{r=0}^p} \A^0_{\ul p} w^{-r-1} z^{-p+r-1} + \underset{q-s+1 \equiv 0 \, \textup{mod}\, T}{\sum_{q \geq 0} \sum_{s=0}^{q-1}} \A^\infty_{\ul{q+1}} w^s z^{q-s-1}\\
&= \frac{w z^{T-2}}{w^T - z^T} \L(w) + \frac{1}{T^2} \sum_{k=0}^{T-1} \sum_{x \in \Z'} \sum_{p \geq 0} \frac{\omega^{k(p-1)} \mathcal M^x_p(w)}{(z - \omega^k x)^{p+1}}\\
&\qquad\qquad\qquad\qquad + \sum_{\substack{p \geq 0\\p \equiv 1 \, \textup{mod}\, T}} \mathcal M^0_p(w) z^{-p-1} + \sum_{\substack{q \geq 0\\ q+2 \equiv 0 \, \textup{mod}\, T}} \mathcal M^\infty_q(w) z^q.
\end{align*}
The result now follows from equation \eqref{PB L2 with L} together with the above expression for $\mathcal M(z, w)$ and  the expression for $\ha (\L(z)|\L(z))$ given in Proposition \ref{prop: Ham}.
\end{proof}
\end{proposition}

We define the \emph{local quadratic Gaudin Hamiltonians} in $\hat S_{\bm \lvl}(\hg^\D_\CC)$ as
\begin{equation} \label{loc quad Gaud Ham}
H^x_p \coloneqq \pi_{\bm \lvl}(\H^x_p), \qquad
H^\infty_q \coloneqq \pi_{\bm \lvl}(\H^\infty_q)
\end{equation}
for all $x \in \Z \setminus \{ \infty \}$ with $p = 0, \ldots, 2 n_x - 1$ and $q = 0, \ldots, 2 n_\infty - 4$.

Following the discussion in \S\ref{sec: twist}, we apply the linear map $\rep \otimes \pi_{\bm \lvl} \otimes \id$ to the element $\mathcal M^x_p \in \F \otimes R_{\Pi \z}$ from Proposition \ref{prop: Lax equations}. It follows from the assumption \eqref{assumption lvl} made at the start of this section and the explicit form of $\mathcal M^x_p$ in Proposition \ref{prop: Lax equations} that the result takes the form, cf. \eqref{Lax twist from formal},
\begin{equation*}
(\rep \otimes \pi_{\bm \lvl} \otimes \id) \mathcal M^x_p(z) = \varphi^x_p(z) \big( \partial + \mathscr M^x_p(z) \big) \in \Conn_\g(S^1) \hotimes \hg^\D_{\CC, \bm \lvl} \otimes R_{\Pi \z},
\end{equation*}
where $\varphi^x_p(z) \in R_{\Pi \z}$ is not identically zero provided $p \leq n_x - 1$ when $x \in \Z \setminus \{ \infty \}$ and $p \leq n_\infty - 3$ for the point at infinity. Here $\varphi^x_p(z) \mathscr M^x_p(z) \in \mathscr T(S^1, \g) \hotimes \hg^\D_{\CC, \bm \lvl} \otimes R_{\Pi \z}$ is a linear combination of the $\g$-valued classical fields \eqref{Currents x 0 inf} with coefficients given by rational functions in $z$ with poles in $\Pi \z$. In \S\ref{sec: Toda FT Lax Ham} below we will consider also the case where the assumption \eqref{assumption lvl} fails in the context of affine Toda field theory.

\begin{corollary} \label{cor: ZC eq}
The local quadratic Gaudin Hamiltonians \eqref{loc quad Gaud Ham} Poisson commute. Moreover, we have the zero curvature equation
\begin{equation*}
\{ H^x_p, \mathscr L(z) \} = \partial \mathscr N^x_p(z) + [\mathscr L(z), \mathscr N^x_p(z)],
\end{equation*}
where $\mathscr N^x_p(z) \coloneqq \varphi^x_p(z) \big( \mathscr L(z) - \mathscr M^x_p(z) \big)$.
\begin{proof}
The first statement follows from applying the homomorphism \eqref{hom pi l S to S} of Poisson algebras to the relation \eqref{quad Ham commute} in $\hat S(\tg^\D_\CC)$.

Applying the tensor product of homomorphisms $\rep \otimes \pi_{\bm \lvl} \otimes \id$ to the Lax equations from Proposition \ref{prop: Lax equations}, which is a relation in $\tg \hotimes \hat S(\tg^\D_\CC) \otimes R_{\Pi \z}$, we find
\begin{align*}
\{ H^x_p, \varphi(z) \mathscr L(z) \} &= \big[ \varphi^x_p(z) \big(\partial + \mathscr M^x_p(z)\big), \varphi(z) \big(\partial + \mathscr L(z)\big) \big]\\
&= \varphi(z) \varphi^x_p(z) [\partial + \mathscr M^x_p(z), \partial + \mathscr L(z)]\\
&= \varphi(z) \varphi^x_p(z) \big( \partial \mathscr L(z) - \partial \mathscr M^x_p(z) + [\mathscr M^x_p(z), \mathscr L(z)] \big)\\
&= \varphi(z) \big( \partial \mathscr N^x_p(z) + [\mathscr L(z), \mathscr N^x_p(z)] \big).
\end{align*}
The result now follows by dividing through by $\varphi(z)$.
\end{proof}
\end{corollary}

\subsubsection{Hamiltonian and momentum} \label{sec: H and P}

In all the examples of integrable field theories discussed in \S\ref{sec: examples of NUL} below, the Hamiltonian of the model will be related to a specific linear combination of the formal quadratic Gaudin Hamiltonians introduced in Proposition \ref{prop: Ham}. Recall that the latter are associated with the set $\Pi \z$ of poles of the twist function and appear as coefficients in the partial fraction decomposition of $\ha \big( \L(z) \big| \L(z) \big)$. The Hamiltonian of a specific model will instead be naturally associated with \emph{zeroes} of the twist function. Specifically, to each zero $x \in \CP$ of the twist function, \emph{i.e.} such that $\varphi(x) = 0$, we associate a formal quadratic Hamiltonian
\begin{equation} \label{local quad Ham}
\H_x \coloneqq \res_x \ha \big( \L(z) \big| \L(z) \big) \varphi(z)^{-1} dz.
\end{equation}
It is given by a linear combination of the formal quadratic Gaudin Hamiltonians from Proposition \ref{prop: Ham}. We can then associate with \eqref{local quad Ham} a local Hamiltonian by applying to it the homomorphism $\pi_{\bm \lvl}$ from \S\ref{sec: level fix}, namely $H_x \coloneqq \pi_{\bm \lvl}(\H_x) \in \hat S_{\bm \lvl}(\hg^\D_\CC)$.

We shall also need the local momentum of the integrable field theory, which is an element $P \in \hat S_{\bm \lvl}(\hg^\D_\CC)$ whose Poisson bracket with any classical field, cf. \S\ref{sec: classical fields}, is equal to its derivative, \emph{i.e.}
\begin{equation*}
\{ P, A^x_{\ul p} \} = \partial A^x_{\ul p}, \qquad
\{ P, A^\infty_{\ul{q+1}} \} = \partial A^\infty_{\ul{q+1}}
\end{equation*}
for all $x \in \Z \setminus \{ \infty \}$ with $p = 0, \ldots, n_x - 1$ and $q = 0, \ldots, n_\infty - 2$.
For every example of dihedral affine Gaudin model considered in \S\ref{sec: examples of NUL} we always have $n_\infty \leq 2$, cf. \S\ref{sec: Divisors}. If $n_\infty = 1$ then there is no classical field attached to infinity, as will be the case in \S\ref{sec: cycl examples}. When $n_\infty = 2$ the only $\g$-valued classical field at infinity is $A^\infty_{\ul 1}$ which turns out to be a Casimir of the Poisson bracket \eqref{hat Skg PB}, see Corollary \ref{cor: Fields PB}. By a suitable modification of the homomorphism $\pi_{\bm \lvl}$ from \S\ref{sec: level fix} we will set this Casimir to a constant in both \S\ref{sec: non-cycl examples} and \S\ref{sec: Toda FT}. In each of these cases the local momentum can be defined as $P \coloneqq \pi_{\bm \lvl}(\mathcal P)$ with the \emph{formal momentum} given by
\begin{equation} \label{momentum def}
\mathcal P \coloneqq -i \sum_{x \in \z \setminus \{ \infty \}} \cocent^{(x)}_{\ul 0} + i \sum_{x \in \z_{\rm c}} \cocent^{(\bar x)}_{\ul 0}.
\end{equation}
By inspection, cf. also the proof of Proposition \ref{prop: first class} in \S\ref{sec: constraint} below, we see it Poisson commutes with the formal quadratic Gaudin Hamiltonians defined in Proposition \ref{prop: Ham}, \emph{i.e.} we have $\{ \mathcal P, \H^x_p \} = 0$ for all $x \in \Z$ and $p \geq 0$.

\subsubsection{Constraints and Hamiltonian reduction} \label{sec: constraint}

The integrable field theories discussed in \S\ref{sec: cycl examples} below all possess Hamiltonian constraints. To account for such constraints, in this section we introduce a set of first class constraints in the dihedral affine Gaudin model and describe the associated Hamiltonian reduction of the complexified Poisson algebra of formal observables $\hat S(\hg^\D_\CC)$. The reduction remains non-trivial when passing to the complexified Poisson algebra of local observables $\hat S_{\bm \lvl}(\hg^\D_\CC)$ provided the tuple of levels $\bm \lvl$ satisfies an extra condition.

We introduce the \emph{formal constraint} $\mathcal C \in \F$ by
\begin{equation} \label{formal field C}
\mathcal C \coloneqq \res_\infty \zeta(\Phi) = \sum_{x \in \Z \setminus \{ \infty \}} \proj_{(0)} \A^x_{\ul 0},
\end{equation}
where the second equality follows from the explicit form of the formal Lax matrix in \eqref{Lax simple form}. This can be rewritten explicitly as
\begin{equation} \label{formal C expl}
\mathcal C = \cocent \otimes i \mathcal K + \cent \otimes i \P + \sum_{n \in \ZZ} I_{(0, \alpha), -n} \otimes \mathcal C^\alpha_n
\end{equation}
where $\P$ is the formal momentum given by \eqref{momentum def}, we introduced the element
\begin{equation*}
\mathcal K \coloneqq - i \sum_{x \in \z \setminus \{ \infty \}} \cent^{(x)}_{\ul 0} + i \sum_{x \in \z_{\rm c}} \cent^{(\bar x)}_{\ul 0}
\end{equation*}
and the modes $\mathcal C^\alpha_n$ with $\alpha = 1, \ldots, \dim \g_{(0), \CC}$ and $n \in \ZZ$ of the formal constraint $\mathcal C$ are defined by
\begin{equation} \label{C alpha n def}
\mathcal C^\alpha_n \coloneqq \sum_{x \in \z \setminus \{ \infty \}} \big( I^{(0, \alpha)}_n \big)^{(x)}_{\ul 0} + \sum_{x \in \z_{\rm c}} \big( I^{(0, \alpha)}_{-n} \big)^{(\bar x)}_{\ul 0}.
\end{equation}
Note that $\mathcal K$ lies in the centre of the Poisson algebra $\hat S(\tg^\D_\CC)$ and that $\mathcal C^\alpha_n \in \Fil_n (\tg^\D_\CC)$ and $\mathcal C^\alpha_{-n} \in c\big( \Fil_n (\tg^\D_\CC) \big)$ for all $\alpha = 1, \ldots, \dim \g_{(0), \CC}$ and $n \in \ZZ_{\geq 0}$, recalling the definition of the pair of conjugate descending $\ZZ_{\geq 0}$-filtrations on $\tg^\D_\CC$ introduced in \S\ref{sec: Alg Obs}.

Let $J_{\mathcal C}$ denote the ideal of the commutative algebra $S(\tg^\D_\CC)$ generated by $\mathcal K$ and $\mathcal C^\alpha_n$ for all $\alpha = 1, \ldots, \dim \g_{(0), \CC}$ and every $n \in \ZZ$. Let $\hat J_{\mathcal C}$ be the corresponding ideal of the completion $\hat S(\tg^\D_\CC)$ defined by the inverse limit
\begin{equation*}
\hat J_{\mathcal C} \coloneqq \varprojlim J_{\mathcal C} \big/ \big( J_{\mathcal C} \cap \Fil_n\big( S(\tg^\D_\CC) \big) \big).
\end{equation*}
We would like to set $\mathcal K$ and every $\mathcal C^\alpha_n$ for $\alpha = 1, \ldots, \dim \g_{(0), \CC}$ and $n \in \ZZ$ to zero. In other words, we want to impose the set of constraints
\begin{equation} \label{set of constraints}
\mathcal K \approx 0, \qquad
\mathcal C^\alpha_n \approx 0
\end{equation}
for $\alpha = 1, \ldots, \dim \g_{(0), \CC}$ and $n \in \ZZ$. However, simply working in the quotient algebra $\hat S(\tg^\D_\CC) \big/ \hat J_{\mathcal C}$ is not enough since the latter is not a Poisson algebra. Indeed, although $\hat J_{\mathcal C}$ is an ideal of $\hat S(\tg^\D_\CC)$ it is not a \emph{Poisson} ideal, \emph{i.e.} we do not have $\{ \hat S(\tg^\D_\CC), \hat J_{\mathcal C} \} \subset \hat J_{\mathcal C}$, as can be deduced from \eqref{PB L with C} in the proof of the next proposition.

\begin{proposition} \label{prop: constraint ideal}
The ideal $\hat J_{\mathcal C} \subset \hat S(\tg^\D_\CC)$ is a Poisson subalgebra, \emph{i.e.} $\{ \hat J_{\mathcal C}, \hat J_{\mathcal C} \} \subset \hat J_{\mathcal C}$.
In other words, the set of constraints in \eqref{set of constraints} is \emph{first class}.
\begin{proof}
Using the definition \eqref{r kernel rat} of the formal $\r$-matrix we find
\begin{equation*}
\res_\infty \r(z, w) dw = \frac{1}{T} \sum_{k=0}^{T-1} \res_\infty \frac{\sigma^k I^{\wt a} \otimes I_{\wt a}}{w - \omega^{-k} z} dw = - \proj_{(0)} I^{\wt a} \otimes I_{\wt a} = - \wt C^{(0)}.
\end{equation*}
On the other hand, recalling the tensor notation of \S\ref{sec: tensor index} we also have
\begin{align*}
\res_\infty \big[ \r_{\2\1}(w, z), \L_{\2}(w) \big] dw &= \frac{1}{T} \sum_{k=0}^{T-1} \sum_{q, r \geq 0} \res_\infty w^{q-r-1} \omega^{k(r+1)} z^r [ \sigma^k_{\2} \wt C_{\1\2}, \A^\infty_{\ul{q+1} \2}] dw\\
&= - \sum_{q \geq 0} z^q [ \wt C^{(-q-1)}_{\1\2}, \A^\infty_{\ul{q+1} \2}],
\end{align*}
where in the first equality we expanded the $\r$-matrix in the region $|w| > |z|$ and kept only the terms in the formal Lax matrix which contribute to the residue at infinity, namely the last summand on the right hand side of \eqref{Lax simple form}.
Then taking the residue at infinity in $z'$ of the Poisson algebra \eqref{Lax algebra 2} of the formal Lax matrix from Proposition \ref{prop: Lax algebra} we find
\begin{align} \label{PB L with C}
\{ \L_{\1}(z), \mathcal C_{\2} \} &= \big[ \res_\infty \r_{\1\2}(z, z') dz', \L_\1(z) \big] - \res_\infty \big[ \r_{\2\1}(z', z), \L_\2(z') \big] dz' \notag\\
&= - \big[ \wt C^{(0)}_{\1\2}, \L_{\1}(z) \big] + \sum_{q \geq 0} z^q \big[ \wt C^{(-q-1)}_{\1\2}, \A^\infty_{\ul{q+1} \2} \big].
\end{align}
Next, taking the residue of this equation at infinity in $z$, the second term on the right hand side does not contribute and we find that \eqref{formal field C} has the Poisson bracket
\begin{equation} \label{PB CC}
\{ \mathcal C_{\1}, \mathcal C_{\2} \} = - \big[ \wt C^{(0)}_{\1\2}, \mathcal C_{\1} \big].
\end{equation}
Using the explicit form \eqref{formal C expl} of the formal constraint we can write the left hand side of \eqref{PB CC} as
\begin{align*}
&\sum_{m \in \ZZ} I_{(0, \alpha), -m} \otimes \cent \otimes i \{ \mathcal C^\alpha_m, \P \} + \sum_{n \in \ZZ} \cent \otimes I_{(0, \beta), -n} \otimes i \{ \P, \mathcal C^\beta_n \}\\
&\qquad\qquad\qquad\qquad\qquad\qquad + \sum_{m, n \in \ZZ} I_{(0, \alpha), -m} \otimes I_{(0, \beta), -n} \otimes \{ \mathcal C^\alpha_m, \mathcal C^\beta_n \}.
\end{align*}
Likewise, the right hand side of \eqref{PB CC} explicitly reads
\begin{align*}
&\sum_{m \in \ZZ} I_{(0, \alpha), -m} \otimes \cent \otimes m \mathcal C^\alpha_m - \sum_{n \in \ZZ} \cent \otimes I_{(0, \beta), -n} \otimes n \mathcal C^\beta_n\\
&\qquad + \sum_{m, n \in \ZZ} I_{(0, \alpha), -m} \otimes I_{(0, \beta), -n} \otimes \big( - f^{\alpha \beta}_{\quad \gamma} \mathcal C^\gamma_{m+n} - i m \langle I^{(0,\alpha)}, I^{(0, \beta)} \rangle \delta_{m+n, 0} \mathcal K \big),
\end{align*}
where the structure constants $f^{\alpha \beta}_{\quad \gamma}$ are defined by $[I^{(0, \beta)}, I_{(0, \gamma)}] = f^{\alpha \beta}_{\quad \gamma} I_{(0, \alpha)}$. Finally, comparing components of the linearly independent elements $I_{(0, \alpha), -m} \otimes \cent$, $\cent \otimes I_{(0, \beta), -n}$ and $I_{(0, \alpha), -m} \otimes I_{(0, \beta), -n}$ in $\tg \otimes \tg$ with $m, n \in \ZZ$ and $\alpha, \beta = 1, \ldots, \dim \g_{(0), \CC}$ on both sides, we deduce that
\begin{subequations}
\begin{align}
\label{PB P with C} \{ \P, \mathcal C^\alpha_n \} &= i n \mathcal C^\alpha_n\\
\label{PB C with C} \{ \mathcal C^\alpha_m, \mathcal C^\beta_n \} &= - f^{\alpha \beta}_{\quad \gamma} \mathcal C^\gamma_{m+n} - i m \langle I^{(0, \alpha)}, I^{(0, \beta)} \rangle \delta_{m+n, 0} \mathcal K.
\end{align}
\end{subequations}
The result now follows by the Leibniz rule from the relation \eqref{PB C with C} together with the fact that $\mathcal K$ is central in $\hat S(\tg^\D_\CC)$.
\end{proof}
\end{proposition}

Consider the normaliser of $\hat J_{\mathcal C}$ in $\hat S(\tg^\D_\CC)$ defined by
\begin{equation*}
N( \hat J_{\mathcal C} ) \coloneqq \big\{ \mathscr X \in \hat S(\tg^\D_\CC) \,\big|\, \{ \mathscr X, \hat J_{\mathcal C} \} \subset \hat J_{\mathcal C} \big\}.
\end{equation*}
This is the subalgebra of \emph{first class} elements in $\hat S(\tg^\D_\CC)$. By Proposition \ref{prop: constraint ideal} we have that $\hat J_{\mathcal C}$ is contained in $N( \hat J_{\mathcal C} )$, but then by definition of $N( \hat J_{\mathcal C} )$ it is in fact a Poisson ideal. The subquotient
\begin{equation*}
\hat S(\tg^\D_\CC)_{\rm red} \coloneqq N( \hat J_{\mathcal C} ) \big/ \hat J_{\mathcal C}
\end{equation*}
is therefore a Poisson algebra, called the \emph{Hamiltonian reduction} of $\hat S(\tg^\D_\CC)$ with respect to the set of first class constraints given in \eqref{set of constraints}. It consists of equivalence classes of first class observables in $\hat S(\tg^\D_\CC)$, where two such observables are considered equivalent if they differ by an element of $\hat J_{\mathcal C}$, \emph{i.e.} a term proportional to the constraints.

\begin{proposition} \label{prop: first class}
Let $n_\infty = 1$ in the notation of \S\ref{sec: Divisors}. Then the formal momentum $\P$ and the collection of formal quadratic Gaudin Hamiltonians $\H^x_p$ for every $x \in \Z$ and $p \geq 0$ all belong to $N(\hat J_{\mathcal C})$. Hence they descend to the Hamiltonian reduction $\hat S(\tg^\D_\CC)_{\rm red}$.
\begin{proof}
The statement for the formal momentum follows from the fact that $\{ \P, \mathcal K \} = 0$ since $\K$ is central in $\hat S(\tg^\D_\CC)$ and the relation \eqref{PB P with C} obtained in Proposition \ref{prop: constraint ideal}.

For the statement about the formal quadratic Gaudin Hamiltonians we start from the relation \eqref{PB L with C} in the proof of Proposition \ref{prop: constraint ideal}. Since we are assuming $n_\infty = 1$, there are no formal fields attached to infinity and hence the second term on the right hand side of \eqref{PB L with C} is absent. Hence we find
\begin{equation*}
\ha \big\{ (\L(z) | \L(z) ), \mathcal C \big\} = \big( \L_{\1}(z) \big| \{ \L_{\1}(z), \mathcal C_{\2} \} \big)_\1 = - \big( \L_{\1}(z) \big| \big[ \wt C^{(0)}_{\1\2}, \L_{\1}(z) \big] \big)_\1 = 0.
\end{equation*}
We then deduce at once from Proposition \ref{prop: Ham} that $\{ \H^x_p, \mathcal C \} = 0$ for every $x \in \Z$ and $p \geq 0$. In turn, it follows from the explicit form \eqref{formal C expl} of the formal constraint $\mathcal C$ that $\{ \H^x_p, \mathcal K \} = \{ \H^x_p, \mathcal C^\alpha_n \} = 0$ for all $x \in \Z$, $p \geq 0$, $\alpha = 1, \ldots, \dim \g_{(0), \CC}$ and $n \in \ZZ$. Note that we also deduce $\{ \H^x_p, \P \} = 0$ for all $x \in \Z$ and $p \geq 0$ as claimed in \S\ref{sec: H and P}.
\end{proof}
\end{proposition}

The set of elements $\mathcal K$ and $\mathcal C^\alpha_n$, $\alpha = 1, \ldots, \dim \g_{(0), \CC}$, $n \in \ZZ$ is seen to be invariant under the anti-linear automorphism $c : \tg^\D_\CC \to \tg^\D_\CC$ using Proposition \ref{prop: complexified Takiff}. Specifically, $c(\mathcal K) = \mathcal K$ and $c(\mathcal C^\alpha_n) = \mathcal C^\alpha_{-n}$ for all $\alpha = 1, \ldots, \dim \g_{(0), \CC}$ and $n \in \ZZ$. It follows that the ideal $\hat J_{\mathcal C}$ is stable under the action of the extension of the anti-linear automorphism $c$ to the completion $\hat S(\tg^\D_\CC)$ defined in \S\ref{sec: Alg Obs}. In particular, its normaliser $N(\hat J_{\mathcal C})$ is also stable since the anti-linear map $c$ is an automorphism of the Poisson algebra $\hat S(\tg^\D_\CC)$.
We may therefore consider the real subalgebra $\hat S(\tg^\D_\CC)_{\rm red}^c$ of fixed points under $c$ in the Hamiltonian reduction $\hat S(\tg^\D_\CC)_{\rm red}$.

Since the linear map $\pi_{\bm \lvl}$ in \eqref{hom pi l S to S} is a homomorphism of Poisson algebras, it sends the ideal and Poisson subalgebra $\hat J_{\mathcal C}$ of $\hat S(\tg^\D_\CC)$ from Proposition \ref{prop: constraint ideal} to an ideal and Poisson subalgebra $\pi_{\bm \lvl}(\hat J_{\mathcal C})$ of $\hat S_{\bm \lvl}(\hg^\D_\CC)$. Let us describe the latter more explicitly.

Recall from \S\ref{sec: alg of obs} that, given any element $\ms X^{(x)}_{\ul p} \in \hg^\D_\CC$ with $x \in \Z \setminus \{ \infty \}$, $\ms X \in \Loop\g$ and $p = 0, \ldots, n_x - 1$, we denote its image in $\hg^\D_{\CC, \bm \lvl}$ by the same symbol. In particular, we will also keep the same notation for the image in $\hg^\D_{\CC, \bm \lvl}$ of the modes $\mathcal C^\alpha_n$ of the formal constraint defined in \eqref{C alpha n def}, so that $\pi_{\bm \lvl}(\mathcal C^\alpha_n) = \mathcal C^\alpha_n$ for every $\alpha = 1, \ldots, \dim \g_{(0), \CC}$ and $n \in \ZZ$, cf. the definition \eqref{pi D k def}. On the other hand, the central element $\mathcal K$ is sent to
\begin{equation*}
\pi_{\bm\lvl}(\mathcal K) = \Bigg( \sum_{x \in \Z \setminus \{ \infty \}} \lvl^x_0 \Bigg) 1 \in \hat S_{\bm \lvl}(\hg^\D_\CC).
\end{equation*}
In other words, $\pi_{\bm \lvl}(\hat J_{\mathcal C})$ is a proper ideal of $\hat S_{\bm \lvl}(\hg^\D_\CC)$ if and only if the tuple of levels $\bm \lvl$ satisfies the condition
\begin{equation} \label{constraint levels}
\res_\infty \varphi(z) dz = \sum_{x \in \Z \setminus \{ \infty \}} \lvl^x_0 = 0.
\end{equation}
In the remainder of this section we will assume this condition to hold. Then applying the homomorphism $\rep \otimes \pi_{\bm \lvl}$ to the formal constraint \eqref{formal C expl} we obtain the $\g_{(0)}$-valued classical field
\begin{equation*}
(\rep \otimes \pi_{\bm \lvl}) \mathcal C = \sum_{n \in \ZZ} ( I_{(0, \alpha)} \otimes e_{-n} ) \otimes \mathcal C^\alpha_n \in \mathscr T(S^1, \g) \hotimes \hg^\D_{\CC, \bm\lvl},
\end{equation*}
where we note that the derivative term disappeared by virtue of the conditions \eqref{constraint levels}.

Since the linear map $\pi_{\bm \lvl}$ in \eqref{hom pi l S to S} is a homomorphism of Poisson algebras it maps elements of $N(\hat J_{\mathcal C})$ into the normaliser of $\pi_{\bm \lvl}(\hat J_{\mathcal C})$ in $\hat S_{\bm \lvl}(\hg^\D_\CC)$, namely
\begin{equation*}
N\big( \pi_{\bm \lvl}(\hat J_{\mathcal C}) \big) \coloneqq \big\{ \mathscr X \in \hat S_{\bm \lvl}(\hg^\D_\CC) \,\big|\, \{ \mathscr X, \pi_{\bm \lvl}(\hat J_{\mathcal C}) \} \subset \pi_{\bm \lvl}(\hat J_{\mathcal C}) \big\}.
\end{equation*}
In other words, by restricting \eqref{hom pi l S to S} to the normaliser $N(\hat J_{\mathcal C}) \subset \hat S(\tg^\D_\CC)$ we obtain a homomorphism of Poisson algebras
\begin{equation*}
\pi_{\bm \lvl} : N(\hat J_{\mathcal C}) \longrightarrow N\big( \pi_{\bm \lvl}\big( \hat J_{\mathcal C}) \big),
\end{equation*}
which sends the subalgebra of first class formal observables in $\hat S(\tg^\D_\CC)$ to the subalgebra of first class local observables in $\hat S_{\bm \lvl}(\hg^\D_\CC)$.

As in the formal setting, the ideal $\pi_{\bm \lvl}(\hat J_{\mathcal C})$ of $\hat S_{\bm \lvl}(\hg^\D_\CC)$ is contained in $N\big( \pi_{\bm \lvl}(\hat J_{\mathcal C}) \big)$, and by definition of the latter it follows that $\pi_{\bm \lvl}(\hat J_{\mathcal C})$ is in fact a Poisson ideal of $N\big( \pi_{\bm \lvl}(\hat J_{\mathcal C}) \big)$. The corresponding subquotient
\begin{equation*}
\hat S_{\bm \lvl}(\hg^\D_\CC)_{\rm red} \coloneqq N\big( \pi_{\bm \lvl}(\hat J_{\mathcal C}) \big) \big/ \pi_{\bm \lvl}(\hat J_{\mathcal C})
\end{equation*}
is therefore a Poisson algebra. It is the \emph{Hamiltonian reduction} of $\hat S_{\bm \lvl}(\hg^\D_\CC)$ with respect to the set of first class constraints $\mathcal C^\alpha_n \approx 0$ for $\alpha = 1, \ldots, \dim \g_{(0), \CC}$ and $n \in \ZZ$, \emph{i.e.}
\begin{equation*}
(\rep \otimes \pi_{\bm \lvl}) \mathcal C \approx 0.
\end{equation*}
Finally, the $c$-equivariance of $\pi_{\bm \lvl}$, cf. Proposition \ref{prop: pil equiv}, ensures that the ideal $\pi_{\bm \lvl}(\hat J_{\mathcal C})$ of $\hat S_{\bm \lvl}(\hg^\D_\CC)$ and its normaliser $N\big(\pi_{\bm \lvl}(\hat J_{\mathcal C})\big)$ are both stable under the action of $c$ so that we can consider the real subalgebra $\hat S_{\bm \lvl}(\hg^\D_\CC)_{\rm red}^c$ of fixed points under $c$ in $\hat S_{\bm \lvl}(\hg^\D_\CC)_{\rm red}$.

\section{Examples of non-ultralocal field theories} \label{sec: examples of NUL}

In this section we explicitly show how all the classical integrable field theories listed in Table \ref{table intro} of the introduction can be formulated as dihedral affine Gaudin models.

In the case of the principal chiral model and its various multi-parameter integrable deformations, discussed in \S\ref{sec: non-cycl examples}, the affine Gaudin model formulation exactly matches the conventional description of these models for an arbitrary real finite-dimensional Lie group. We relate the standard notation for the Lie algebra valued fields and Lax matrix of these models to the notation introduced in \S\ref{sec: classical fields} and \S\ref{sec: twist}, respectively, in the context of a general classical dihedral affine Gaudin model.

In \S\ref{sec: cycl examples} we turn to the description of $\ZZ_T$-graded coset $\sigma$-models and their integrable deformations. Here as well, the affine Gaudin model formulation of these models exactly matches the standard one. More precisely, for the symmetric space $\sigma$-model, corresponding to the case $T=2$, two different formulations exist in the literature. We show that only one of these formulations admits an affine Gaudin model description and argue, at the end of \S\ref{sec: coset}, that the other one does not. The bi-Yang-Baxter $\sigma$-model is also formulated as an affine Gaudin model in \S\ref{sec: bYB model}. More precisely, we match its description as a two-parameter deformation of a symmetric space $\sigma$-model established in \cite{Delduc:2015xdm} within the Hamiltonian formalism. Its more standard description as a deformation of the principal chiral model, originally given in \cite{Klimcik:2008eq}, was only shown to be integrable within the Lagrangian formalism \cite{Klimcik:2014bta}. This alternative formulation therefore does not lend itself to an affine Gaudin model interpretation.

We give a general treatment of affine Toda field theory in \S\ref{sec: Toda FT}. Whereas the Lie algebra homomorphism $\pi_{\bm \lvl}$ constructed in \S\ref{sec: level fix} could be used in all previous examples, here this is no longer possible because the basic assumption \eqref{assumption lvl} on the levels does not hold in the present case. We thus introduce an alternative homomorphism $\widetilde{\pi}_{\bm\lvl}$ which can be seen as realising the formal fields of the affine Gaudin model in terms of classical Toda fields. In the case of affine Toda field theories there is another substantial difference to the previous cases. The standard formulation of affine Toda field theory is ultralocal, whereas the affine Gaudin model formulation is inherently non-ultralocal, cf. Corollary \ref{cor: Lax algebra}. We show that the Lax matrices in both formulations are related by a gauge transformation. We illustrate the general discussion of affine Toda field theories in detail on the example of the sinh-Gordon model, corresponding to the $\mathfrak{sl}_2$ case, in \S\ref{sec: sinh-Gordon}.

\subsection{Principal chiral model and deformations}
\label{sec: non-cycl examples}

Throughout this subsection we set $T=1$ so that $\Pi = D_2 = \langle \t \,|\, \t^2 = 1 \rangle \simeq \ZZ_2$. In particular, the automorphism $\sigma$ of $\g$ is  the identity. We pick and fix an anti-linear automorphism $\tau \in \bAut_- \g$ and denote by $\g_0$ the corresponding real form.

\subsubsection{Principal chiral model} \label{sec: PCM}

Consider the divisor
\begin{equation} \label{PCM divisor}
\D = 2 \cdot 0 + 2 \cdot \infty.
\end{equation}
In the notation of \S\ref{sec: affine Gaudin} we have $\Z = \{ 0, \infty \}$ and $\Z' = \emptyset$.

There are three formal fields $\A^0_{\ul 0}$, $\A^0_{\ul 1}$ and $\A^\infty_{\ul 1}$. Since $0$ and $\infty$ are both real points and we are in the non-cyclotomic setting, these fields are defined by \eqref{fields A zr}. In terms of them, the formal Lax matrix reads, cf. Proposition \ref{prop: Gaudin Lax},
\begin{equation} \label{PCM formal Lax}
\L(z) = \frac{\A^0_{\ul 1}}{z^2} + \frac{\A^0_{\ul 0}}{z} - \A^\infty_{\ul 1}.
\end{equation}
We fix the levels 
\begin{equation} \label{PCM levels}
\lvl^0_0 = 0, \qquad
\lvl^0_1 = 1, \qquad
\lvl^\infty_1 = 1.
\end{equation}
Consider the $\g_0$-valued classical fields $j_p \coloneqq A^0_{\ul p}$ for $p = 0,1$ associated with the origin and $\xi \coloneqq A^\infty_{\ul 1}$ associated with infinity, defined by \eqref{rho pi Conn}. By Corollary \ref{cor: Fields PB} these satisfy the Poisson brackets
\begin{subequations} \label{PB PCM}
\begin{align}
\big\{ j_{0 \1}(\theta), j_{0 \2}(\theta') \big\} &= - \big[ C_{\1\2}, j_{0 \2}(\theta) \big] \delta_{\theta \theta'},\\
\big\{ j_{0 \1}(\theta), j_{1 \2}(\theta') \big\} &= - \big[ C_{\1\2}, j_{1 \2}(\theta) \big] \delta_{\theta \theta'} - C_{\1\2} \delta'_{\theta \theta'},\\
\big\{ j_{1 \1}(\theta), j_{1 \2}(\theta') \big\} &= 0,
\end{align}
\end{subequations}
and $\{ \xi_\1(\theta), \xi_\2(\theta') \} = 0$, with all other brackets being zero.
In particular, the $\g_0$-valued classical field $\xi$ is a Casimir of the Poisson bracket. We therefore choose to set it to zero from now on. This is formally achieved by altering slightly the definition of the homomorphism $\pi_{\bm \lvl}$ in \eqref{pi D k def}. Specifically, we replace the definitions \eqref{pi l Dinf def 1}-\eqref{pi l Dinf def 2} of $\pi_{\bm \lvl}$ on the elements of $\tg^{n_\infty \infty} \otimes_\RR \CC$ attached to infinity by
\begin{equation} \label{pi l Dinf modif}
\pi_{\bm \lvl}\big(\cocent^{(\infty)}_{\ul 1}\big) \coloneqq 0, \qquad
\pi_{\bm \lvl}\big(\cent^{(\infty)}_{\ul 1}\big) \coloneqq i, \qquad
\pi_{\bm \lvl}\big(I^{\alpha (\infty)}_{n, \ul 1}\big) \coloneqq 0.
\end{equation}
The resulting linear map $\pi_{\bm \lvl} : \tg^\D_\CC \to \hat S_{\bm \lvl}(\hg^D_\CC)$ is still seen to be a homomorphism of Lie algebras, cf. Proposition \ref{prop: hom pi lvl}, using the fact that $n_\infty = 2$. In what follows we use this altered definition of $\pi_{\bm \lvl}$. We recognise \eqref{PB PCM} as the Poisson brackets of the principal chiral model written in terms of the components of the current $1$-form $j = - dg g^{-1}$, where $g$ denotes the principal chiral field, see for instance \cite[\S I.5 of Part 2]{FTbook}.

Applying the representation $\rep \otimes \pi_{\bm \lvl} \otimes \id$ to the formal Lax matrix \eqref{PCM formal Lax}, as in \S\ref{sec: twist}, we obtain
\begin{align*}
(\rep \otimes \pi_{\bm \lvl} \otimes \id)\L(z) &= \frac{\partial + j_1}{z^2} + \frac{j_0}{z} - \partial = \left( \frac{1}{z^2} - 1 \right) \partial + \frac{1}{z^2} (j_1 + z j_0)\\
&= \frac{1 - z^2}{z^2} \left( \partial + \frac{1}{1 - z^2} (j_1 + z j_0) \right).
\end{align*}
By comparing with \eqref{Lax twist from formal} we read off the Lax matrix and twist function to be
\begin{equation} \label{PCM L phi}
\mathscr L(z) = \frac{1}{1 - z^2} (j_1 + z j_0), \qquad \varphi(z) = \frac{1}{z^2} - 1,
\end{equation}
which coincide with those of the principal chiral model. See \emph{e.g.} \cite[\S I.3 of Part 2]{FTbook} for the Lax matrix and \cite{Sevostyanov, Delduc:2012qb} for the twist function.

It remains to be checked that the Hamiltonian of the principal chiral model can be obtained from the formal quadratic Gaudin Hamiltonians of Proposition \ref{prop: Ham}. These are given by
\begin{alignat*}{2}
\H^0_0 &= - \big( \A^0_{\ul 0} \big| \A^\infty_{\ul 1} \big), &\qquad\qquad
\H^0_1 &= - \big( \A^0_{\ul 1} \big| \A^\infty_{\ul 1} \big) + \ha \big( \A^0_{\ul 0} \big| \A^0_{\ul 0} \big), \\
\H^0_2 &= \big( \A^0_{\ul 0} \big| \A^0_{\ul 1} \big), &\qquad\qquad
\H^0_3 &= \ha \big( \A^0_{\ul 1} \big| \A^0_{\ul 1} \big), \qquad\qquad \H^\infty_0 = \ha \big( \A^\infty_{\ul 1} \big| \A^\infty_{\ul 1} \big).
\end{alignat*}
The twist function defined in \eqref{PCM L phi} has a pair of simple zeroes located at $\pm 1$. We then find that the associated formal quadratic Hamiltonians \eqref{local quad Ham} read
\begin{equation*}
\H_{\pm 1} = \mp \qa \big( \A^0_{\ul 0} \pm \A^0_{\ul 1} \mp \A^\infty_{\ul 1} \big| \A^0_{\ul 0} \pm \A^0_{\ul 1} \mp \A^\infty_{\ul 1} \big).
\end{equation*}
The terms involving $\cocent$ and $\cent$ on the right hand side take the form
\begin{equation*}
\mp \ha \left( \cocent^{(0)}_{\ul 0} \pm \cocent^{(0)}_{\ul 1} \mp \cocent^{(\infty)}_{\ul 1} \right) \left( \cent^{(0)}_{\ul 0} \pm \cent^{(0)}_{\ul 1} \mp \cent^{(\infty)}_{\ul 1} \right).
\end{equation*}
However, these disappear upon applying the homomorphism $\pi_{\bm \lvl}$ since $\lvl^0_1 \pm \lvl^0_0 - \lvl^\infty_1 = 0$, which is related to the fact that $\pm 1$ are both simple zeroes of the twist function $\varphi(z)$ in \eqref{PCM L phi}. Recalling that we have set $\xi$ to zero, cf. \eqref{pi l Dinf modif}, we therefore find
\begin{equation*}
H_{\pm 1} = \pi_{\bm \lvl} \big( \H_{\pm 1} \big) = \mp \qa (j_0 \pm j_1| j_0 \pm j_1) = \mp \frac{1}{8 \pi} \int_{S^1} d\theta \big\langle j_0(\theta) \pm j_1(\theta), j_0(\theta) \pm j_1(\theta) \big\rangle,
\end{equation*}
where in the last equality we used Lemma \ref{lem: eval rep}.
The difference of these local quadratic Hamiltonians is
\begin{equation} \label{PCM Ham}
H \coloneqq H_{-1} - H_1 = \frac{1}{4\pi} \int_{S^1} d\theta \big( \langle j_0(\theta), j_0(\theta) \rangle + \langle j_1(\theta), j_1(\theta) \rangle \big),
\end{equation}
which coincides, up to an overall factor, with the Hamiltonian of the principal chiral model, see \cite[\S I.5 of Part 2]{FTbook}. Using \eqref{Seg-Sug Takiff} from example \ref{ex: piD 2x inf} we find the momentum $P = \pi_{\bm \lvl}(\P)$ to be
\begin{equation} \label{PCM Momentum}
P = \frac{1}{2\pi} \int_{S^1} d\theta \langle j_0(\theta), j_1(\theta) \rangle.
\end{equation}
We note that $\H_{-1} - \H_1 = \H^0_1 + \H^0_3 + \H^\infty_0$. Moreover, $\pi_{\bm \lvl}(\H^0_3) = \pi_{\bm \lvl}(\H^\infty_0) = 0$ so that the local quadratic Hamiltonian can also be obtained as $H = \pi_{\bm \lvl}(\H^0_1)$.
In particular, applying Corollary \ref{cor: ZC eq} we obtain the zero curvature equation
\begin{equation*}
\{ H, \mathscr L(z) \} = \partial \mathscr M(z) + [\mathscr L(z), \mathscr M(z)],
\end{equation*}
where, noting that $\varphi^0_1(z) = z^{-1}$ and $\mathscr M^0_1(z) = j_1$, we have
\begin{equation*}
\mathscr M(z) \coloneqq \mathscr N^0_1(z) = \frac{1}{z} \left( \frac{1}{1-z^2}(j_1 + z j_0) - j_1 \right) = \frac{1}{1-z^2}(j_0 + z j_1).
\end{equation*}

We deduce from the above analysis that the classical dihedral affine Gaudin model associated with the divisor \eqref{PCM divisor} and the corresponding choice of levels \eqref{PCM levels} coincides with the principal chiral model described in terms of the current $1$-form $j = - dg g^{-1}$. Recall, however, that the actual phase space of the principal chiral model is given by the cotangent bundle $T^\ast \Loop G_0$ of the loop group $\Loop G_0$ of the real Lie group $G_0$ with Lie algebra $\g_0$. The above formulation in terms of the $1$-form current $j$ therefore only describes the principal chiral model dynamics on the quotient $(T^\ast \Loop G_0) / G_0$ of the cotangent bundle by the right action of the subgroup of constant loops. Obtaining a complete description of the principal chiral model requires introducing a $G_0$-valued field $g$ satisfying $j_1 = - \partial g g^{-1}$ so that together with the $\g$-valued field $X \coloneqq - g^{-1} j_0 g$ they parametrise the global (left) trivialisation of $T^\ast \Loop G_0$. We shall not discuss this issue further here, and refer to \cite{Vicedo:2015pna} for further details.

\subsubsection{Two-parameter deformations}
\label{sec: YB}

Integrable deformations of the principal chiral model may be constructed by altering the data of the affine Gaudin model of \S\ref{sec: PCM} in various ways. For instance, one could keep the divisor \eqref{PCM divisor} unchanged and simply modify the value of the levels $\lvl^0_0$, $\lvl^0_1$ and $\lvl^\infty_1$. In what follows we will only be concerned with deformations which alter the divisor \eqref{PCM divisor} itself. More precisely, we shall deform the divisor $\Pi \D$, cf. \S\ref{sec: Divisors}, while preserving its $\Pi$-invariance. One possible way of doing so is to split the double pole at the origin into a pair of simple poles, either both real or complex conjugate of one another. We refer to these as the real and complex branches respectively \cite{Vicedo:2015pna}.

To describe the deformation in the complex branch we use the divisor
\begin{equation} \label{two param C divisor}
\D = x_+ + 2 \cdot \infty.
\end{equation}
for arbtirary $x_+ \in \CC$ such that $\Im x_+ > 0$. Writing $x_+ = k + i A$ we can regard $k \in \RR$ and $A \in \RR_{> 0}$ as two \emph{real} deformation parameters. We shall treat the limiting case when $A \to 0$ with $k \neq 0$ in \S\ref{sec: PCM with WZ} below. Similarly, to construct the deformation in the real branch we would start from the divisor
\begin{equation} \label{two param R divisor}
\D = x_+ + x_- + 2 \cdot \infty,
\end{equation}
with distinct $x_\pm \in \RR$ being the two real deformation parameters.
To treat the two branches in a unified way we note that in both cases the divisor $\Pi \D$ takes the form
\begin{equation*}
\Pi \D = x_+ + x_- + 2 \cdot \infty,
\end{equation*}
with $x_+ \neq x_-$, where $x_\pm \in \RR$ in the real branch and $x_- = \bar x_+$ in the complex branch. In particular, we then have $\Z = \{ x_+, x_-, \infty \}$ in the notation of \S\ref{sec: affine Gaudin}.

There are three formal fields $\A^{x_+}_{\ul 0}$, $\A^{x_-}_{\ul 0}$ and $\A^\infty_{\ul 1}$, which are respectively attached to the points $x_+$, $x_-$ and infinity.
By Proposition \ref{prop: Gaudin Lax}, the formal Lax matrix is expressed in terms of these as
\begin{equation} \label{YB formal Lax}
\L(z) = \frac{\A^{x_+}_{\ul 0}}{z - x_+} + \frac{\A^{x_-}_{\ul 0}}{z - x_-} - \A^\infty_{\ul 1}.
\end{equation}
To fix the levels, recall that the twist function \eqref{PCM L phi} of the principal chiral model has a double pole at the origin. The requirement that this double pole is recovered in the limit $x_\pm \to 0$ uniquely fixes the singular behaviour of the levels associated with the points $x_+$ and $x_-$. We shall also require that the zeroes of the twist function remain fixed at $\pm 1$ under the deformation, which leads us to set
\begin{equation} \label{two param levels}
\lvl^{x_\pm}_0 = \pm \frac{1-x_\pm^2}{x_+ - x_-}, \qquad
\lvl^\infty_1 = 1.
\end{equation}
We introduce the $\g$-valued classical fields $\mathcal J_\pm \coloneqq A^{x_\pm}_{\ul 0}$ and $\xi \coloneqq A^\infty_{\ul 1}$, which according to Corollary \ref{cor: Fields PB} satisfy the Poisson brackets
\begin{equation*}
\big\{ \mathcal J_{\pm \1}(\theta), \mathcal J_{\pm \2}(\theta') \big\} = - \big[ C_{\1\2}, \mathcal J_{\pm \2}(\theta) \big] \delta_{\theta \theta'} - \lvl^{x_\pm}_0 C_{\1\2} \delta'_{\theta \theta'},
\end{equation*}
with all other brackets being zero. As in \S\ref{sec: PCM} we observe that $\xi$ is a Casimir so in what follows we will set it to zero by suitably modifying the definition of $\pi_{\bm \lvl}$ on the elements of $\tg^{n_\infty \infty} \otimes_\RR \CC$ given in \eqref{pi l Dinf def 1}-\eqref{pi l Dinf def 2}, replacing it by \eqref{pi l Dinf modif}.

To obtain the Lax matrix and twist function we apply the linear map $\rep \otimes \pi_{\bm \lvl} \otimes \id$ to the formal Lax matrix in \eqref{YB formal Lax}. We find
\begin{align*}
(\rep \otimes \pi_{\bm \lvl} \otimes \id)\L(z) &= \frac{\lvl^{x_+}_0 \partial + \mathcal J_+}{z - x_+} + \frac{\lvl^{x_-}_0 \partial + \mathcal J_-}{z - x_-} - \partial\\
&= \frac{1 - z^2}{(z - x_+)(z - x_-)} \left( \partial + \frac{1}{1 - z^2} \big( j_1 + z j_0 \big) \right),
\end{align*}
where in the second line we introduced the linear combinations $j_1 \coloneqq - x_- \mathcal J_+ - x_+ \mathcal J_-$ and $j_0 \coloneqq \mathcal J_+ + \mathcal J_-$ of the $\g$-valued classical fields. In terms of these, the Lax matrix takes the same form as for the principal chiral model in \eqref{PCM L phi}. However, these fields satisfy a two parameter deformation of the Poisson bracket \eqref{PB PCM} which reads
\begin{subequations} \label{two param PB}
\begin{align}
\big\{ j_{0 \1}(\theta), j_{0\2}(\theta') \big\} &= - \big[ C_{\1\2}, j_{0 \2}(\theta) \big] \delta_{\theta \theta'} + (x_+ + x_-) C_{\1\2} \delta'_{\theta \theta'},\\
\big\{ j_{0 \1}(\theta), j_{1\2}(\theta') \big\} &= - \big[ C_{\1\2}, j_{1 \2}(\theta) \big] \delta_{\theta \theta'} - (1 + x_+ x_-) C_{\1\2} \delta'_{\theta \theta'},\\
\big\{ j_{1 \1}(\theta), j_{1\2}(\theta') \big\} &= \big[ C_{\1\2}, (x_+ + x_-) j_{1 \2} + x_+ x_- j_{0 \2}(\theta) \big] \delta_{\theta \theta'} + (x_+ + x_-) C_{\1\2} \delta'_{\theta \theta'}.
\end{align}
\end{subequations}
Such a deformation of the Poisson brackets of the principal chiral model was considered in \cite{Balog} for $\g = \mathfrak{su}_2$, but the same brackets immediately extend to any Lie algebra (see \emph{e.g.} \cite{Itsios:2014vfa, Delduc:2014uaa}). By construction, the Poisson brackets \eqref{two param PB} for the $\g$-valued classical fields $j_0$ and $j_1$ are equivalent to the Lax matrix \eqref{PCM L phi} satisfying the Poisson bracket from Corollary \ref{cor: Lax algebra} with the twist function
\begin{equation} \label{two param twist}
\varphi(z) = \frac{1 - z^2}{(z - x_+)(z - x_-)}.
\end{equation}

Consider now the collection of formal quadratic Gaudin Hamiltonians introduced in Proposition \ref{prop: Ham}, given here by
\begin{align*}
\H^{x_\pm}_0 &= \pm \frac{\big( \A^{x_+}_{\ul 0} \big| \A^{x_-}_{\ul 0} \big)}{x_+ - x_-} - \big( \A^{x_\pm}_{\ul 0} \big| \A^\infty_{\ul 1} \big),\\
\H^{x_\pm}_1 &= \ha \big( \A^{x_\pm}_{\ul 0} \big| \A^{x_\pm}_{\ul 0} \big), \qquad
\H^\infty_0 = \ha \big( \A^\infty_{\ul 1} \big| \A^\infty_{\ul 1} \big).
\end{align*}
Since the twist function \eqref{two param twist} still has simple zeroes at the points $\pm 1$, the specific linear combinations \eqref{local quad Ham} of the above quadratic Gaudin Hamiltonians read
\begin{equation*}
\H_{\pm 1} = \mp \qa (1 \mp x_+)(1 \mp x_-) \big( \L(\pm 1) \big| \L(\pm 1) \big).
\end{equation*}
Just as in \S\ref{sec: PCM}, the terms on the right hand side involving $\cocent$ and $\cent$ disappear once we apply the homomorphism $\pi_{\bm \lvl}$ using the fact that $\varphi(\pm 1) = 0$. We then find
\begin{equation} \label{Hpm two param}
H_{\pm 1} = \pi_{\bm \lvl}\big( \H_{\pm 1} \big) = \mp \frac{( j_0 \pm j_1 | j_0 \pm j_1)}{4 (1 \mp x_+)(1 \mp x_-)}.
\end{equation}
Their difference $H \coloneqq H_{-1} - H_1$ is then given by
\begin{equation*}
H = \int_{S^1} d\theta \frac{ (1 + x_+ x_-) \big( \langle j_0(\theta), j_0(\theta) \rangle + \langle j_1(\theta), j_1(\theta) \rangle \big) + 2 (x_+ + x_-) \langle j_0(\theta), j_1(\theta) \rangle}{4 \pi (1 - x_+^2)(1 - x_-^2)},
\end{equation*}
which coincides, in the complex branch, with the Hamiltonian for the two-parameter integrable deformation of the principal chiral model constructed in \cite{Delduc:2013fga}. The latter model has two interesting limits. Writing $x_+ = k + i A$, the limit $k \to 0$ corresponds to the Yang-Baxter $\sigma$-model first introduced by Klimcik in \cite{Klimcik}. Alternatively, taking the limit $A \to 0$ yields the principal chiral model with a Wess-Zumino term which we describe in more detail in \S\ref{sec: PCM with WZ} below. Note also that the momentum $P = \pi_{\bm \lvl}(\P)$ is found, using \eqref{Seg-Sug standard} from example \ref{ex: piD 2x inf}, to be
\begin{equation*}
P = \int_{S^1} d\theta \frac{ (x_+ + x_-) \big( \langle j_0(\theta), j_0(\theta) \rangle + \langle j_1(\theta), j_1(\theta) \rangle \big) + 2 (1 + x_+ x_-) \langle j_0(\theta), j_1(\theta) \rangle}{4 \pi (1 - x_+^2)(1 - x_-^2)}.
\end{equation*}

The classical integrable structure of the real branch was studied in \cite{Itsios:2014vfa} but the action of the corresponding model has not yet been identified. The above analysis shows that at the Hamiltonian level this model is described by a dihedral affine Gaudin model associated with the divisor \eqref{two param R divisor}, where $x_\pm \in \RR$, together with the choice of levels \eqref{two param levels}.
When $x_+ = - x_-$ we obtain a one-parameter deformation described by an integrable gauged WZW-type theory introduced by Sfetsos \cite{Sfetsos:2013wia}, which interpolates between the WZW model and the non-abelian $T$-dual of the principal chiral model.

The remark made about the principal chiral field at the end of \S\ref{sec: PCM} also applies to its two-parameter deformation. Consider for simplicity the one-parameter deformation where $x_+ = - x_-$. In this case, defining a pair of fields parametrising the global (left) trivialisation of the cotangent bundle $T^\ast \Loop G_0$ of the loop group of the real Lie group $G_0$ requires the introduction of a solution $R \in \End \g_0$ of the modified classical Yang-Baxter equation on $\g_0$. We refer to \cite{Vicedo:2015pna} for more details.

\subsubsection{Principal chiral model with WZ-term} \label{sec: PCM with WZ}

Another way to deform the divisor $\Pi \D$, where $\D$ is given by \eqref{PCM divisor}, besides splitting its double pole at the origin into a pair of simple poles as in \S\ref{sec: YB}, is to simply shift this double pole along the real axis. That is, we define the divisor
\begin{equation} \label{divisor PCM WZ}
\D = 2 \cdot k + 2 \cdot \infty,
\end{equation}
where $k \in \RR \setminus \{ 0 \}$ plays the role of the deformation parameter.
For a suitable choice of levels, cf. \eqref{WZ levels} below, this can also be seen as a limit of the setup in \S\ref{sec: YB} where the pair of simple poles $x_\pm$ collide at $k$.

As in the principal chiral model case we have three formal fields $\A^k_{\ul 0}$, $\A^k_{\ul 1}$ and $\A^\infty_{\ul 1}$. By Proposition \ref{prop: Gaudin Lax}, the corresponding formal Lax matrix reads
\begin{equation} \label{PCM WZ formal Lax}
\L(z) = \frac{\A^k_{\ul 1}}{(z-k)^2} + \frac{\A^k_{\ul 0}}{z-k} - \A^\infty_{\ul 1}.
\end{equation}
We require the deformation not to alter the position of the zeroes of the twist function at $\pm 1$, as in \S\ref{sec: YB}. This will ensure that the Lax matrix takes the same form as in the principal chiral model \eqref{PCM L phi} for suitably defined linear combinations $j_0$ and $j_1$ of the classical fields. We therefore fix the levels to be
\begin{equation} \label{WZ levels}
\lvl^k_0 = - 2 k, \qquad
\lvl^k_1 = 1 - k^2, \qquad
\lvl^\infty_1 = 1.
\end{equation}

If we define the $\g$-valued classical fields $\tilde\jmath_p \coloneqq A^k_{\ul p}$ for $p = 0, 1$ and $\xi \coloneqq A^\infty_{\ul 1}$, then according to Corollary \ref{cor: Fields PB} their Poisson brackets read
\begin{subequations} \label{PCM WZ PB tilde j}
\begin{align}
\big\{ \tilde\jmath_{0 \1}(\theta), \tilde\jmath_{0\2}(\theta') \big\} &= - \big[ C_{\1\2}, \tilde\jmath_{0 \1}(\theta) \big] \delta_{\theta \theta'} + 2k C_{\1\2} \delta'_{\theta \theta'},\\
\big\{ \tilde\jmath_{0 \1}(\theta), \tilde\jmath_{1\2}(\theta') \big\} &= - \big[ C_{\1\2}, \tilde\jmath_{1 \2}(\theta) \big] \delta_{\theta \theta'} - (1-k^2) C_{\1\2} \delta'_{\theta \theta'},\\
\big\{ \tilde\jmath_{1 \1}(\theta), \tilde\jmath_{1\2}(\theta') \big\} &= 0,
\end{align}
\end{subequations}
and $\{ \xi_\1(\theta), \xi_\2(\theta') \} = 0$, the remaining brackets all being zero. As usual we shall set $\xi$ to zero since it is a Casimir, cf. \S\ref{sec: PCM} and \S\ref{sec: YB}.

The Lax matrix and twist function are then constructed by applying the linear map $\rep \otimes \pi_{\bm \lvl} \otimes \id$ to the formal Lax matrix $\L(z)$. We have
\begin{align*}
(\rep \otimes \pi_{\bm \lvl} \otimes \id)\L(z) &=  \frac{(1-k^2) \partial + \tilde\jmath_1}{(z-k)^2} + \frac{- 2 k \partial + \tilde\jmath_0}{z-k} - \partial\\
&= \frac{1 - z^2}{(z - k)^2} \left( \partial + \frac{1}{1 - z^2} \big( j_1 + z j_0 \big) \right),
\end{align*}
where we introduced the linear combinations $j_1 \coloneqq \tilde\jmath_1 - k \tilde\jmath_0$ and $j_0 \coloneqq \tilde\jmath_0$. In terms of these, the expression for the Lax matrix is exactly the same as in the principal chiral model \eqref{PCM L phi}. However, the Poisson brackets between the classical fields $j_0$ and $j_1$ are a deformation of \eqref{PB PCM}. They follow from \eqref{PCM WZ PB tilde j} and take the form
\begin{subequations} \label{WZW PB}
\begin{align}
\big\{ j_{0 \1}(\theta), j_{0\2}(\theta') \big\} &= - \big[ C_{\1\2}, j_{0 \1}(\theta) \big] \delta_{\theta \theta'} + 2k C_{\1\2} \delta'_{\theta \theta'},\\
\big\{ j_{0 \1}(\theta), j_{1\2}(\theta') \big\} &= - \big[ C_{\1\2}, j_{1 \2}(\theta) \big] \delta_{\theta \theta'} - (1+k^2) C_{\1\2} \delta'_{\theta \theta'},\\
\big\{ j_{1 \1}(\theta), j_{1\2}(\theta') \big\} &= \big[ C_{\1\2}, 2 k\, j_{1 \2}(\theta) + k^2 j_{0 \2}(\theta) \big] \delta_{\theta \theta'} + 2k C_{\1\2} \delta'_{\theta \theta'}.
\end{align}
\end{subequations}
This corresponds precisely to the coinciding point limit $x_\pm \to k$ of the Poisson brackets \eqref{two param PB} of \S\ref{sec: YB}. We note that the twist function
\begin{equation} \label{WZ twist}
\varphi(z) = \frac{1 - z^2}{(z - k)^2}
\end{equation}
can also be obtained from the two-parameter twist function \eqref{two param twist} by taking the same coinciding point limit.

The formal quadratic Gaudin Hamiltonians from Proposition \ref{prop: Ham} are of exactly the same form as those of the principal chiral model, specifically
\begin{alignat*}{2}
\mathcal H^k_0 &= - \big( \A^k_{\ul 0} \big| \A^\infty_{\ul 1} \big), &\qquad\qquad
\mathcal H^k_1 &= - \big( \A^k_{\ul 1} \big| \A^\infty_{\ul 1} \big) + \ha \big( \A^k_{\ul 0} \big| \A^k_{\ul 0} \big),\\
\mathcal H^k_2 &= \big( \A^k_{\ul 0} \big| \A^k_{\ul 1} \big), &\qquad\qquad
\mathcal H^k_3 &= \ha \big( \A^k_{\ul 1} \big| \A^k_{\ul 1} \big), \qquad\qquad
\H^\infty_0 = \ha \big( \A^\infty_{\ul 1} \big| \A^\infty_{\ul 1} \big).
\end{alignat*}
Since the twist function \eqref{WZ twist} still has a pair of simple zeroes at $\pm 1$ by construction, we may associate to these formal quadratic Hamiltonians \eqref{local quad Ham} which read
\begin{equation*}
\H_{\pm 1} = \mp \qa (1 \mp k)^2 \big( \L(\pm 1) \big| \L(\pm 1) \big).
\end{equation*}
Applying the homomorphism $\pi_{\bm \lvl}$ we find that the terms involving $\cocent$ and $\cent$ drop out because $\varphi(\pm 1) = 0$. The resulting Hamiltonians are
\begin{equation*}
H_{\pm 1} = \pi_{\bm \lvl}(\H_{\pm 1}) = \mp \frac{(j_0 \pm j_1 | j_0 \pm j_1)}{4 (1 \mp k)^2},
\end{equation*}
which coincide with the limit $x_\pm \to k$ of the Hamiltonians \eqref{Hpm two param} of the two-parameter deformation. Finally, we have
\begin{equation*}
H \coloneqq H_{-1} - H_1 = \int_{S^1} d\theta \frac{ (1 + k^2) \big( \langle j_0(\theta), j_0(\theta) \rangle + \langle j_1(\theta), j_1(\theta) \rangle \big) + 4 k \langle j_0(\theta), j_1(\theta) \rangle}{4 \pi (1 - k^2)^2}.
\end{equation*}
This Hamiltonian and the Poisson brackets \eqref{WZW PB} agree with those resulting from a Hamiltonian analysis of the principal chiral model action with a Wess-Zumino term added, see \emph{e.g.} \cite{Delduc:2014uaa} and setting $\eta = 0$ in the notation used there.

\subsection{$\ZZ_T$-graded coset $\sigma$-model and deformations}
\label{sec: cycl examples}

Let $\g$ be a finite-dimensional complex Lie algebra and $\sigma \in \Aut \g$ an automorphism of order $T \in \ZZ_{\geq 2}$. We also fix an anti-linear involution $\tau \in \bAut_- \g$ and denote by $\g_0$ the corresponding real form of $\g$. In the present case we have $\Pi = D_{2T} \simeq \ZZ_T \rtimes \ZZ_2$. Throughout this section we will impose the set of first class constraints \eqref{set of constraints} as described in \S\ref{sec: constraint}.

\subsubsection{$\ZZ_T$-graded coset $\sigma$-model} \label{sec: coset}

Consider the divisor
\begin{equation} \label{sym space divisor}
\D = 2 \cdot 1 + \infty.
\end{equation}
In the notation of \S\ref{sec: affine Gaudin} we have $\Z = \{ 1, \infty \}$ and $\Z' = \{ 1 \}$.
There are two formal fields $\A^1_{\ul p}$, $p = 0, 1$ associated with the point $1$. However, since $n_\infty = 1$ there is no formal field associated with the point at infinity. It follows from Proposition \ref{prop: Gaudin Lax} that the formal Lax matrix is given by
\begin{equation} \label{sym space formal Lax}
\L(z) = \frac{1}{T} \sum_{k=0}^{T-1} \left( \frac{\omega^k \sigma^k \A^1_{\ul 1}}{(z - \omega^k)^2} + \frac{\sigma^k \A^1_{\ul 0}}{z - \omega^k} \right),
\end{equation}
where $\omega$ is a primitive $T^{\rm th}$-root of unity. We shall fix the levels to be
\begin{equation} \label{sym space levels}
\lvl^1_0 = 0, \qquad \lvl^1_1 = 1,
\end{equation}
and define the $\g$-valued classical fields $A \coloneqq A^1_{\ul 1}$, $\Pi \coloneqq A^1_{\ul 0}$. Using Corollary \ref{cor: Fields PB} we find their Poisson brackets to be
\begin{subequations} \label{sym space PB}
\begin{align}
\label{sym space PB a} \big\{ A_\1(\theta), A_\2(\theta') \big\} &= 0,\\
\label{sym space PB b} \big\{ A_\1(\theta), \Pi_\2(\theta') \big\} &= - \big[ C_{\1\2}, A_\2(\theta) \big] \delta_{\theta \theta'} - C_{\1\2} \delta'_{\theta \theta'},\\
\label{sym space PB c} \big\{ \Pi_\1(\theta), \Pi_\2(\theta') \big\} &= - \big[ C_{\1\2}, \Pi_\2(\theta) \big] \delta_{\theta \theta'}.
\end{align}
\end{subequations}
These coincide with the Poisson brackets of the $\ZZ_2$-graded coset $\sigma$-model (for which $T=2$),  also known as the symmetric space $\sigma$-model, see \emph{e.g.} \cite{Delduc:2012qb}. If $\g$ is the Grassmann envelope of a Lie superalgebra then these Poisson brackets are also those of the $\ZZ_4$-graded supercoset $\sigma$-model, \emph{i.e.} semi-symmetric space $\sigma$-model, for which $\sigma \in \Aut \g$ is of order $T=4$ \cite{Magro, Vicedo:2009sn}, or more generally those of the $\ZZ_{2n}$-graded supercoset $\sigma$-model for which $\sigma \in \Aut \g$ is of order $T=2n$ with $n \in \ZZ_{\geq 2}$ \cite{Ke:2011zzb}. 

To compute the Lax matrix and twist function we apply the linear map $\rep \otimes \pi_{\bm \lvl} \otimes \id$ to the formal Lax matrix \eqref{sym space formal Lax} which gives
\begin{align} \label{Lax from formal sym space}
&(\rep \otimes \pi_{\bm \lvl} \otimes \id)\L(z) = \frac{1}{T} \sum_{k=0}^{T-1} \left( \frac{\omega^k \partial + \omega^k \sigma^k A}{(z - \omega^k)^2} + \frac{\sigma^k \Pi}{z - \omega^k} \right) \notag\\
&\qquad\qquad = \frac{T z^{T-1}}{(1-z^T)^2} \left( \partial + \sum_{j=1}^T \frac{T - j + j \, z^{-T}}{T} z^j A^{(j)} + \sum_{j=1}^T \frac{1 - z^{-T}}{T} z^j \Pi^{(j)} \right).
\end{align}
Here we have introduced the projections of the $\g_0$-valued classical fields $A$ and $\Pi$ onto the graded subspaces $\g_{(j)}$ for $j \in \ZZ_T$ relative to the direct sum decomposition \eqref{g0 eigen decomp}. Specifically, we set
\begin{equation*}
A^{(j)} \coloneqq \proj_{(j)} A, \qquad
\Pi^{(j)} \coloneqq \proj_{(j)} \Pi.
\end{equation*}
Comparing the above expression with the general form in \eqref{Lax twist from formal} we read off the Lax matrix to be
\begin{equation} \label{sym space Lax}
\mathscr L(z) = \sum_{j=1}^T \frac{T - j + j \, z^{-T}}{T} z^j A^{(j)} + \sum_{j=1}^T \frac{1 - z^{-T}}{T} z^j \Pi^{(j)},
\end{equation}
and the corresponding twist function
\begin{equation} \label{sym space twist}
\varphi(z) = \frac{T z^{T-1}}{(1-z^T)^2}.
\end{equation}
When $T=2$ these are precisely the expressions for the Lax matrix and corresponding twist function of the $\ZZ_2$-graded coset $\sigma$-model, see \emph{e.g.} \cite{Delduc:2012qb}. If $\g$ is taken to be the Grassmann envelope of a Lie superalgebra and $T = 4$ then \eqref{sym space Lax} is the Hamiltonian Lax matrix of a semi-symmetric space $\sigma$-model, \emph{i.e.} a $\ZZ_4$-graded supercoset $\sigma$-model \cite{Magro, Vicedo:2010qd}, the Green-Schwarz superstring on $AdS_5 \times S^5$ being a particular example of the latter. More generally, when $\g$ is the Grassmann envelope of a Lie superalgebra and $T=2n$ with $n \in \ZZ_{\geq 2}$, the expressions for the Lax matrix \eqref{sym space Lax} and the twist function \eqref{sym space twist} agree up to differences in conventions with those obtained from the Hamiltonian analysis of the $\ZZ_{2n}$-graded supercoset $\sigma$-model \cite{Young:2005jv} performed in \cite{Ke:2011zzb}. Moreover, for a semisimple Lie algebra $\g$ and general $T \in \ZZ_{\geq 2}$, the above expressions \eqref{sym space Lax} and \eqref{sym space twist} reproduce, again up to conventions, those of the bosonic truncation of the $\ZZ_{2T}$-graded supercoset $\sigma$-model from \cite{Ke:2011zzb}.

The formal constraint defined in \eqref{formal field C} is given by
\begin{equation*}
\mathcal C = \proj_{(0)} \A^1_{\ul 0}.
\end{equation*}
We note that the levels defined in \eqref{sym space levels} trivially satisfy the condition \eqref{constraint levels} of \S\ref{sec: constraint}. Applying to it the linear map $\rep \otimes \pi_{\bm \lvl}$ and using the $\Pi$-equivariance of $\rep$ we find
\begin{equation} \label{Pi0 constraint}
(\rep \otimes \pi_{\bm \lvl}) \mathcal C = \Pi^{(0)}.
\end{equation}
This coincides with the constraint of the $\ZZ_2$-graded coset $\sigma$-model, see \emph{e.g.} \cite{Delduc:2012qb}, or more generally with that of the $\ZZ_T$-graded coset $\sigma$-model for $T \in \ZZ_{\geq 2}$ \cite{Ke:2011zzb}.

Finally, we turn to the definition of the Hamiltonian. The twist function \eqref{sym space twist} has zeroes at the origin and infinity. Consider the associated quadratic Hamiltonians defined as in \eqref{local quad Ham}, namely
\begin{equation} \label{loc Ham sym space}
\H_x \coloneqq \res_x \ha \big( \L(z) \big| \L(z) \big) \varphi(z)^{-1} dz,
\end{equation}
for $x \in \{ 0, \infty \}$.

\begin{lemma} \label{lem: loc Ham sym space}
The quadratic Hamiltonians \eqref{loc Ham sym space} are given by
\begin{align*}
\H_0 &= \frac{1}{2 T} \sum_{j=1}^{T-1} \Big( j \, \A_{\ul 1}^{(j)} - \A_{\ul 0}^{(j)} \Big| (T-j) \A_{\ul 1}^{(-j)} - \A_{\ul 0}^{(-j)} \Big),\\
\H_\infty &= - \frac{1}{2 T} \sum_{j=1}^{T-1} \Big( (T-j) \A_{\ul 1}^{(j)} + \A_{\ul 0}^{(j)} \Big| j\, \A_{\ul 1}^{(-j)} + \A_{\ul 0}^{(-j)} \Big) - \Big( \A_{\ul 1}^{(0)} \Big| \A_{\ul 0}^{(0)} \Big),
\end{align*}
where $\A_{\ul p}^{(j)} \coloneqq \proj_{(j)} \A_{\ul p}^1$ for $p = 0, 1$.
\begin{proof}
Although these can be determined with the help of Proposition \ref{prop: Ham} as we did in the previous examples of \S\ref{sec: non-cycl examples}, here we will compute these quadratic Hamiltonians more directly using the following observations. Firstly, by comparing the right hand side of the first line in \eqref{Lax from formal sym space} with the formal Lax matrix \eqref{sym space formal Lax}, one can show by following the same steps leading to the second line in \eqref{Lax from formal sym space} that
\begin{align} \label{sym space tilde L}
\widetilde\L(z) &\coloneqq \varphi(z)^{-1} \L(z) = \sum_{j=1}^T \frac{T - j + j \, z^{-T}}{T} z^j \A_{\ul 1}^{(j)} + \sum_{j=1}^T \frac{1 - z^{-T}}{T} z^j \A_{\ul 0}^{(j)} \notag\\
&\, = \frac{1}{T} \sum_{j=1}^T z^j \Big( (T - j) \A_{\ul 1}^{(j)} + \A_{\ul 0}^{(j)} \Big) + \frac{1}{T} \sum_{j=1}^T z^{-T + j} \Big( j \, \A_{\ul 1}^{(j)} - \A_{\ul 0}^{(j)} \Big).
\end{align}
The above quadratic Hamiltonians \eqref{loc Ham sym space} can now be rewritten in terms of the Laurent polynomial expression \eqref{sym space tilde L} as
\begin{equation} \label{Hx coset alternative}
\H_x = \res_x \ha \big( \widetilde\L(z) \big| \widetilde\L(z) \big) \varphi(z) dz
\end{equation}
for $x \in \{ 0, \infty \}$.
Secondly, we note that with the twist function $\varphi(z)$ given by \eqref{sym space twist}, the $1$-form $\varphi(z) dz$ has a zero of order $T-1$ both at the origin and at infinity.

We are now in a position to compute \eqref{loc Ham sym space} explicitly. For the quadratic Hamiltonian associated with the origin we have
\begin{align*}
\H_0 &= \frac{T}{2} \res_0 \big( \widetilde\L(z) \big| \widetilde\L(z) \big) z^{T-1} dz\\
&= \frac{1}{2T} \res_0 \sum_{j,k=1}^T \Big( j \, \A_{\ul 1}^{(j)} - \A_{\ul 0}^{(j)} \Big| k \, \A_{\ul 1}^{(k)} - \A_{\ul 0}^{(k)} \Big) z^{-T-1+j+k} dz\\
&= \frac{1}{2 T} \sum_{j=1}^{T-1} \Big( j \, \A_{\ul 1}^{(j)} - \A_{\ul 0}^{(j)} \Big| (T-j) \A_{\ul 1}^{(-j)} - \A_{\ul 0}^{(-j)} \Big),
\end{align*}
where in the first equality we have used the behaviour $\varphi(z) \sim T z^{T-1}$ as $z \to 0$ of the twist function and the fact that the most singular terms at the origin in the Laurent polynomial expression \eqref{sym space tilde L} are of the order $z^{-T+1}$. In the second line we used the invariance property of the bilinear form on $\tg$ under $\sigma$ and the observation that terms from the first sum over $j$ in \eqref{sym space tilde L} cannot contribute to the residue.

Similarly, for the quadratic Hamiltonian associated with infinity we have
\begin{align*}
\H_\infty &= \frac{T}{2} \res_\infty \big( \widetilde\L(z) \big| \widetilde\L(z) \big) \left( \frac{1}{z^{T+1}} + \frac{2}{z^{2T+1}} \right) dz\\
&= \frac{1}{2T} \res_\infty \Bigg( \sum_{j,k=1}^T \Big( (T-j) \A_{\ul 1}^{(j)} + \A_{\ul 0}^{(j)} \Big| (T-k) \A_{\ul 1}^{(k)} + \A_{\ul 0}^{(k)} \Big) z^{j+k}\\
&\qquad\qquad\qquad\qquad + 2 \Big( T \A_{\ul 1}^{(0)} - \A_{\ul 0}^{(0)} \Big| \A_{\ul 0}^{(0)} \Big) z^T \Bigg) \left( \frac{1}{z^{T+1}} + \frac{2}{z^{2T+1}} \right) dz\\
&= - \frac{1}{2 T} \sum_{j=1}^{T-1} \Big( (T-j) \A_{\ul 1}^{(j)} + \A_{\ul 0}^{(j)} \Big| j\, \A_{\ul 1}^{(-j)} + \A_{\ul 0}^{(-j)} \Big) - \Big( \A_{\ul 1}^{(0)} \Big| \A_{\ul 0}^{(0)} \Big),
\end{align*}
where in the first equality we kept only the first two terms in the Laurent expansion of $\varphi(z)$ at infinity since the highest power in the Laurent polynomial expression \eqref{sym space tilde L} is $z^T$. The second line follows from the $\sigma$-invariance of the bilinear form on $\tg$ and is a result of having kept only the terms which can contribute to the residue.
\end{proof}
\end{lemma}

Note that the expression for $\H_0$ in Lemma \ref{lem: loc Ham sym space} does not involve $\cocent$ and $\cent$ since these are in the grade $0$ component $\tg_{(0)}$ of $\tg$, cf. \S\ref{sec: tg action of Pi}. Therefore its image under $\pi_{\bm \lvl}$ reads
\begin{equation} \label{H0 coset}
H_0 \coloneqq \pi_{\bm \lvl}(\H_0) = \frac{1}{2 T} \sum_{j=1}^{T-1} \big( j \, A^{(j)} - \Pi^{(j)} \big| (T-j) A^{(-j)} - \Pi^{(-j)} \big).
\end{equation}
On the other hand, the image under $\pi_{\bm \lvl}$ of the momentum defined in \eqref{momentum def} is given by, see example \ref{ex: piD 2x inf},
\begin{equation} \label{P coset}
P \coloneqq \pi_{\bm \lvl}(\P) = - i \pi_{\bm \lvl}\big( \cocent^{(1)}_{\ul 0} \big) = (A| \Pi) = \sum_{j=0}^{T-1} \big( A^{(j)} \big| \Pi^{(-j)} \big).
\end{equation}
We define the Hamiltonian as $H \coloneqq 2 H_0 + P$, which reads
\begin{align} \label{Ham ZT model}
H &= \frac{1}{2 \pi T} \int_{S^1} d\theta \Bigg( \sum_{j=0}^{T-1} \Big( j (T-j) \big\langle A^{(j)}(\theta), A^{(-j)}(\theta) \big\rangle + (T - 2 j) \big\langle A^{(j)}(\theta), \Pi^{(-j)}(\theta) \big\rangle \Big) \notag\\
&\qquad\qquad\qquad\qquad\qquad\qquad\qquad\qquad + \sum_{j=1}^{T-1} \big\langle \Pi^{(j)}(\theta), \Pi^{(-j)}(\theta) \big\rangle \Bigg).
\end{align}
As discussed in \S\ref{sec: constraint}, in the reduced theory the Hamiltonian is given by this same expression but up to the addition of a term proportional to the constraint \eqref{Pi0 constraint}.
In the simplest case when $T = 2$ the expression \eqref{Ham ZT model} coincides with the Hamiltonian of the symmetric space $\sigma$-model, see \emph{e.g.} \cite{Vicedo:2015pna}.
If, however, we take $\g$ to be the Grassmann envelope of a Lie superalgebra and $T=2n$ then \eqref{Ham ZT model} coincides, up to conventions, with the Hamiltonian of the $\ZZ_{2n}$-graded supercoset $\sigma$-model \cite{Ke:2011zzb}. Moreover, for semisimple $\g$ and $T \in \ZZ_{\geq 2}$ arbitrary, the Hamiltonian \eqref{Ham ZT model} reproduces that of the bosonic truncation of the $\ZZ_{2T}$-graded supercoset $\sigma$-model \cite{Ke:2011zzb}.

The twist function \eqref{sym space twist} in the case $T=2$, namely
\begin{equation} \label{sym space twist T2}
\varphi(z) = \frac{2 z}{(1-z^2)^2},
\end{equation}
coincides with the twist function of the symmetric space $\sigma$-model in both the formalism of \cite{Sevostyanov} and \cite{Vicedo:2015pna}. More precisely, due to a difference in conventions, the twist function in \cite{Sevostyanov} which is denoted as $\phi(z^2)$ there, is related to the twist function \eqref{sym space twist T2} as $\phi(z^2) = - 2z \varphi(z)$.
However, it is worth pointing out here that the Lax matrix \eqref{sym space Lax}, which coincides with that given in \cite[(3.12)]{Vicedo:2015pna} for the case $T=2$, namely
\begin{equation*}
\mathscr L(z) = A^{(0)} + \ha (1 - z^{-2}) \Pi^{(0)} + \ha (z + z^{-1}) A^{(1)} + \ha (z - z^{-1}) \Pi^{(1)},
\end{equation*}
differs from the expression in \cite[(61)]{Sevostyanov} for the same model, which in the present notation reads
\begin{equation*}
\widetilde{\mathscr L}(z) = - A^{(0)} - \ha (z + z^{-1}) A^{(1)} - \ha (z - z^{-1}) \Pi^{(1)}.
\end{equation*}
Aside from a trivial overall sign, we see that the difference between these two Lax matrices is a term proportional to the Hamiltonian constraint $\Pi^{(0)}$. However, the presence of the latter is crucial. Indeed, whereas the Lax matrix of \cite[(3.12)]{Vicedo:2015pna} admits an affine Gaudin model description, as shown above, the Lax matrix of \cite[(61)]{Sevostyanov} does not. This can be seen explicitly from the form of the $r$- and $s$-matrices in \cite[Theorem 3.1]{Sevostyanov}. Indeed, the $r$- and $s$-matrices of \cite{Sevostyanov} which we denote here by $\widetilde{r}_{\1\2}(z, z')$ and $\widetilde{s}_{\1\2}(z, z')$ respectively, take the form
\begin{equation} \label{rs matrix Sevostyanov}
\widetilde{r}_{\1\2}(z, z') = r_{\1\2}(z, z') + \frac{z^2 -z'^2}{4} C^{(0)}_{\1\2}, \qquad
\widetilde{s}_{\1\2}(z, z') = s_{\1\2}(z, z') - \frac{2 - z^2 - z'^2}{4} C^{(0)}_{\1\2},
\end{equation}
in terms of the $r$- and $s$-matrices in \eqref{rs matrices} which in the present case with $T = 2$ explicitly read
\begin{align*}
r_{\1\2}(z, z') &= \frac{2 - 2 z^2 - 2 z'^2 + z^4 + z'^4}{4 (z^2 - z'^2)} C^{(0)}_{\1\2} + \frac{z^2 + z'^2 - 4 z^2 z'^2 + z^4 z'^2 + z^2 z'^4}{4 z z' (z^2 - z'^2)} C^{(1)}_{\1\2},\\
s_{\1\2}(z, z') &= \frac{2 - z^2 - z'^2}{4} C^{(0)}_{\1\2} + \frac{(1 - z^2 z'^2)}{4 z z'} C^{(1)}_{\1\2}.
\end{align*}
We refer the reader to \cite[\S3]{Vicedo:2009sn} for a more detailed discussion of such a difference in the $r$- and $s$-matrices coming from the omission of the Hamiltonian constraint $\Pi^{(0)}$ in the Lax matrix, in the context of semi-symmetric space $\sigma$-models, \emph{i.e.} the case $T=4$. Since the $r$- and $s$-matrices \eqref{rs matrix Sevostyanov} of \cite{Sevostyanov} differ from those of a general affine Gaudin model derived in Corollary \ref{cor: Lax algebra}, it is therefore clear that the formulation of the symmetric space $\sigma$-model given in \cite{Sevostyanov} cannot admit a Gaudin model interpretation.

\subsubsection{One-parameter deformations} \label{sec: Z2 one-param}

Let $\g$ be a finite-dimensional complex Lie algebra equipped with (anti-)automorphisms $\sigma \in \Aut \g$ and $\tau \in \bAut \g$ as in \S\ref{sec: coset}.

To construct integrable deformations of the $\ZZ_T$-graded coset $\sigma$-model we proceed as we did in \S\ref{sec: YB} for deforming the principal chiral model. Namely we will deform the divisor $\langle \t \rangle \D$, with $\D$ the divisor \eqref{sym space divisor} underlying the affine Gaudin model from \S\ref{sec: coset}, while preserving its $\langle \t \rangle$-invariance. There are two ways of splitting the double pole at $1$, either into two real points or into a pair of complex conjugate points, which we will also refer to as the real and complex branches respectively.
We note, however, that moving the double pole in the divisor \eqref{sym space divisor} from $1$ to $r > 0$ amounts to a trivial rescaling $z \mapsto z' \coloneqq r^{-1} z$ of the coordinate on $\CP$ along with a rescaling of the formal field $\A^1_{\ul 1}$ appearing in \eqref{sym space formal Lax} by $r$, namely $\A^r_{\ul 1} = r \A^1_{\ul 1}$, cf. \cite{Delduc:2014kha}. It follows that the Poisson brackets \eqref{sym space PB} of the $\g$-valued classical fields remain undeformed except for a possible change in the level in front of the $\delta'_{\theta\theta'}$-term in \eqref{sym space PB b}. We shall therefore only consider one-parameter deformations. To build the deformation in the complex branch we start from the divisor
\begin{subequations}
\begin{equation} \label{divisor eta def}
\D = e^{i \vartheta} + \infty
\end{equation}
where $\vartheta \in \, ]0, \frac{\pi}{T}[$ plays the role of the real deformation parameter.
Similarly, in the real branch the deformation is constructed using the divisor
\begin{equation} \label{divisor lambda def}
\D = \lambda^{1/2} + \lambda^{-1/2} + \infty
\end{equation}
\end{subequations}
where $\lambda \in \RR_{> 0}$ is the deformation parameter. In order to treat both branches at once we note that in both cases the divisor $\langle \t \rangle \D$ is of the form
\begin{equation*}
\langle \t \rangle \D = x_+ + x_- + \infty
\end{equation*}
where $x_- \coloneqq x_+^{-1}$. When $x_+ \in \RR$ this corresponds to the real branch whereas $|x_+| = 1$ corresponds to the complex branch. In the notation of \S\ref{sec: affine Gaudin} we have $\Z = \{ x_+, x_-, \infty \}$ and $\Z' = \{ x_+, x_- \}$.

By virtue of Proposition \ref{prop: Gaudin Lax}, the formal Lax matrix can be written in terms of the pair of formal fields $\A^{x_\pm}_{\ul 0}$ associated with the pair of points $x_\pm$ as
\begin{equation} \label{sym def formal Lax}
\L(z) = \frac{1}{T} \sum_{k=0}^{T-1} \left( \frac{\sigma^k \A^{x_+}_{\ul 0}}{z - \omega^k x_+} + \frac{\sigma^k \A^{x_-}_{\ul 0}}{z - \omega^k x_-} \right).
\end{equation}
We fix the levels to be
\begin{equation} \label{sym def levels}
\lvl^{x_\pm}_0 = \pm \frac{T}{x_+^T - x_-^T}.
\end{equation}
Define the pair of $\g$-valued classical fields $\mathcal J_\pm \coloneqq A^{x_\pm}_{\ul 0}$. If the deformation is in the real branch then these are in fact both $\g_0$-valued and moreover $\lvl^{x_\pm} \in \RR$. By contrast, in the complex branch we have $\tau \mathcal J_+ = \mathcal J_-$, recalling the action of $\Pi$ on $\F$ from \S\ref{sec: formal fields}, cf. also \eqref{rho pi on formal A a} and \eqref{rho pi on formal A b}. Moreover, the levels \eqref{sym def levels} satisfy $\overline{\lvl^{x_+}} = \lvl^{x_-}$. It follows from Corollary \ref{cor: Fields PB} that the non-trivial Poisson brackets of the classical fields are given by
\begin{equation} \label{sym def KM}
\big\{ \mathcal J_{\pm \1}(\theta), \mathcal J_{\pm \2}(\theta') \big\} = - \big[ C_{\1\2}, \mathcal J_{\pm \2}(\theta) \big] \delta_{\theta \theta'} - \lvl^{x_\pm}_0 C_{\1\2} \delta'_{\theta \theta'}.
\end{equation}

The Lax matrix and twist function are then obtained from the formal Lax matrix \eqref{sym def formal Lax} by the procedure described in \S\ref{sec: twist}. We find
\begin{align} \label{sym def formal to Lax}
&(\rep \otimes \pi_{\bm \lvl} \otimes \id)\L(z) = \frac{1}{T} \sum_{k=0}^{T-1} \left( \frac{\lvl^{x_+}_0 \partial + \sigma^k \mathcal J_+}{z - \omega^k x_+} + \frac{\lvl^{x_-}_0 \partial + \sigma^k \mathcal J_-}{z - \omega^k x_-} \right)\\
&\quad = \frac{T z^{T-1}}{(z^T - x_+^T)(z^T - x_-^T)} \left( \partial + \sum_{j=1}^T \frac{T - j + j \, z^{-T}}{T} z^j A^{(j)} + \sum_{j=1}^T \frac{1 - z^{-T}}{T} z^j \Pi^{(j)} \right). \notag
\end{align}
Here we defined the following linear combination of the $\g$-valued classical fields,
\begin{subequations} \label{sym def A Pi}
\begin{align}
\label{sym def A}
A^{(j)} &\coloneqq \frac{1 - x_-^T}{T} x_+^{T-j} \mathcal J_+^{(j)} + \frac{1 - x_+^T}{T} x_-^{T-j} \mathcal J_-^{(j)},\\
\label{sym def Pi} \Pi^{(j)} &\coloneqq \frac{j + (T-j) x_-^T}{T} x_+^{T-j} \mathcal J_+^{(j)} + \frac{j + (T-j) x_+^T}{T} x_-^{T-j} \mathcal J_-^{(j)},
\end{align}
\end{subequations}
for $j = 1, \ldots, T$, and where $\mathcal J_\pm^{(j)} \coloneqq \proj_{(j)} \mathcal J_\pm$. By comparing \eqref{sym def formal to Lax} with the general form in \eqref{Lax twist from formal} we deduce that, when written in terms of the classical fields \eqref{sym def A Pi}, the Lax matrix is of exactly the same form as it was in \S\ref{sec: coset} for the undeformed case, cf. \eqref{sym space Lax}. On the other hand, the twist function \eqref{sym space twist} gets deformed to
\begin{equation} \label{sym def twist}
\varphi(z) = \frac{T z^{T-1}}{(z^T - x_+^T)(z^T - x_-^T)}.
\end{equation}
As expected, we recover the original twist function \eqref{sym space twist} in the limit $x_\pm \to 1$. Now by Corollary \ref{cor: Lax algebra}, the twist function being deformed corresponds to the fact that the fields \eqref{sym def A Pi} satisfy a deformation of the Poisson brackets \eqref{sym space PB}. These can be derived using the expressions \eqref{sym def A Pi} and the Poisson brackets \eqref{sym def KM}. However, since they are equivalent to \eqref{sym def KM} we will not give their explicit form. See \cite[Appendix D]{Delduc:2013fga} for explicit expressions in the case $T = 2$.

In the present case, the formal constraint defined in \eqref{formal field C} reads
\begin{equation*}
\mathcal C = \proj_{(0)} \A^{x_+}_{\ul 0} + \proj_{(0)} \A^{x_-}_{\ul 0}.
\end{equation*}
We note that the levels \eqref{sym def levels} still satisfy the condition \eqref{constraint levels} of \S\ref{sec: constraint}.
Upon applying the linear map $\rep \otimes \pi_{\bm \lvl}$ we find
\begin{equation} \label{Pi0 constraint deform}
(\rep \otimes \pi_{\bm \lvl}) \mathcal C = \mathcal J_+^{(0)} + \mathcal J_-^{(0)} = \Pi^{(0)},
\end{equation}
where the first equality uses the $\Pi$-equivariance of $\rep$ and the second equality follows from the definition \eqref{sym def Pi} with $j = T$.
When $T=2$ this corresponds to the constraint of the deformed $\ZZ_2$-graded coset $\sigma$-model \cite{Delduc:2013fga}.

Consider, finally, the quadratic Hamiltonians defined as in \eqref{local quad Ham}, namely
\begin{equation} \label{loc Ham sym def}
\hat\H_x \coloneqq \res_x \ha \big( \L(z) \big| \L(z) \big) \varphi(z)^{-1} dz,
\end{equation}
for $x \in \{ 0, \infty \}$, where $\varphi(z)$ is the twist function given in \eqref{sym def twist} whose zeroes are at the origin and infinity. By comparing the expression for the formal Lax matrix \eqref{sym def formal Lax} to the right hand side of the first line in \eqref{sym def formal to Lax}, it follows by the same reasoning leading to the second line in \eqref{sym def formal to Lax} that we can write, cf. \eqref{sym space tilde L} in the undeformed case,
\begin{equation} \label{tilde Lax deformed coset}
\wt \L(z) \coloneqq \varphi(z)^{-1} \L(z) = \sum_{j=1}^T \frac{T - j + j z^{-T}}{T} z^j \A^{(j)}_{\ul 1} + \sum_{j=1}^T \frac{1 - z^{-T}}{T} z^j \A^{(j)}_{\ul 0},
\end{equation}
where here we defined the formal fields $\A^{(j)}_{\ul p}$ for $j \in \ZZ_T$ and $p = 0, 1$, by analogy with the definition of the $\g$-valued classical fields \eqref{sym def A Pi}, as
\begin{subequations} \label{formal Ajp def coset}
\begin{align}
\A^{(j)}_{\ul 1} &\coloneqq \frac{1 - x_-^T}{T} x_+^{T-j} \pi_{(j)} \A^{x_+}_{\ul 0} + \frac{1 - x_+^T}{T} x_-^{T-j} \pi_{(j)} \A^{x_-}_{\ul 0},\\
\A^{(j)}_{\ul 0} &\coloneqq \frac{j + (T-j) x_-^T}{T} x_+^{T-j} \pi_{(j)} \A^{x_+}_{\ul 0} + \frac{j + (T-j) x_+^T}{T} x_-^{T-j} \pi_{(j)} \A^{x_-}_{\ul 0},
\end{align}
\end{subequations}
for $j = 1, \ldots, T$.

\begin{lemma} \label{lem: loc Ham sym def}
The expression for the quadratic Hamiltonians \eqref{loc Ham sym def} in terms of \eqref{formal Ajp def coset} are related to those of the quadratic Hamiltonians from Lemma \ref{lem: loc Ham sym space} as
\begin{equation*}
\hat\H_0 = \H_0, \qquad
\hat\H_\infty = \H_\infty + \frac{2 - x_+^T - x_-^T}{2T} \Big( \A_{\ul 0}^{(0)} \Big| \A_{\ul 0}^{(0)} \Big).
\end{equation*}
\begin{proof}
This follows from the same argument as used in the proof of Lemma \ref{lem: loc Ham sym space}. Specifically, using the expression for the deformed twist function \eqref{sym def twist} we find that
\begin{equation*}
\hat\H_0 = \frac{T}{2} \res_0 \big( \widetilde\L(z) \big| \widetilde\L(z) \big) z^{T-1} dz = \H_0,
\end{equation*}
where we recall that $x_+ x_- = 1$ and the second step follows the derivation of $\H_0$ in the proof of Lemma \ref{lem: loc Ham sym space}.
Likewise, for the quadratic Hamiltonian associated to infinity we have
\begin{equation*}
\hat\H_\infty = \frac{T}{2} \res_\infty \big( \widetilde\L(z) \big| \widetilde\L(z) \big) \left( \frac{1}{z^{T+1}} + \frac{x_+^T + x_-^T}{z^{2T+1}} \right) dz,
\end{equation*}
from which the result easily follows by comparison with the rest of the calculation of $\H_\infty$ in the proof of Lemma \ref{lem: loc Ham sym space}.
\end{proof}
\end{lemma}

Applying $\pi_{\bm \lvl}$ to $\hat \H_0$ we obtain $\hat H_0 \coloneqq \pi_{\bm \lvl}(\hat \H_0) = \pi_{\bm \lvl}(\H_0) = H_0$ using Lemma \ref{lem: loc Ham sym def}, where the latter is given by \eqref{H0 coset}. Likewise, the momentum is obtained by applying $\pi_{\bm \lvl}$ to \eqref{momentum def}, which gives
\begin{align*}
\hat P &\coloneqq \pi_{\bm \lvl}(\P) = - i \pi_{\bm \lvl}\Big( D^{(x_+)}_{\ul 0} - D^{(x_-)}_{\ul 0} \Big) = \frac{1}{2 \lvl^{x_+}_0} \big( \mathcal J_+ \big| \mathcal J_+ \big) + \frac{1}{2 \lvl^{x_-}_0} \big( \mathcal J_- \big| \mathcal J_- \big)\\
&\, = (A | \Pi) - \frac{2 - x_+^T - x_-^T}{2 T} \big( \Pi^{(0)} \big| \Pi^{(0)} \big) = P - \frac{2 - x_+^T - x_-^T}{2 T} \big( \Pi^{(0)} \big| \Pi^{(0)} \big).
\end{align*}
We define the Hamiltonian of the deformed $\ZZ_T$-graded coset $\sigma$-model as, cf. \eqref{Ham ZT model},
\begin{equation} \label{Ham ZT def}
\hat H \coloneqq 2 \hat H_0 + \hat P = H - \frac{2 - x_+^T - x_-^T}{4 \pi T} \int_{S^1} d\theta \big\langle \Pi^{(0)}(\theta), \Pi^{(0)}(\theta) \big\rangle.
\end{equation}
Note that since $H$ and $\hat H$ differ by a term proportional to the constraint \eqref{Pi0 constraint deform}, they define equivalent Hamiltonians in the reduced theory, cf. \S\ref{sec: constraint}.
When $T=2$ the expression \eqref{Ham ZT def} coincides with the Hamiltonian of the deformed symmetric space $\sigma$-model in \cite[\S 4.3]{Vicedo:2015pna} for the complex branch and \cite[(5.5)]{Hollowood:2014rla} for the real branch.
For $T=4$ and $\g$ the Grassmann envelope of a Lie superalgebra, the expression \eqref{Ham ZT def} corresponds, in the complex branch, to the Hamiltonian of the so called $\eta$-deformed semi-symmetric space $\sigma$-model \cite{Delduc:2013qra, Delduc:2014kha} where $\eta = \tan \vartheta$ with $x_+ = e^{i \vartheta}$. Likewise, in the real branch with $x_+ = \lambda^{1/2} \in \mathbb{R}$ it corresponds to the Hamiltonian of the so called $\lambda$-deformation \cite{Hollowood:2014qma}. More precisely, the $\eta$- and $\lambda$-deformations can both be defined as certain `lifts' to the cotangent bundle $T^\ast \Loop G_0$ of the dihedral affine Gaudin model with divisor \eqref{divisor eta def} and \eqref{divisor lambda def} respectively, and levels \eqref{sym def levels}, cf. discussions at the end of \S\ref{sec: PCM} and \S\ref{sec: YB}.

The maximal deformation limit in the complex branch is $\vartheta \to \frac{\pi}{T}$. It corresponds to the point $x_+$ coinciding with $\omega x_-$. In this limit the divisor becomes
\begin{equation*}
\D = 2 \cdot x_+ + \infty
\end{equation*}
with $x_+ = \omega^{1/2} \in \z_{\rm r}^1$ in the notation of \S\ref{sec: Double loop}. In particular, the $\g$-valued fields $\omega^{-p/2} A^{x_+}_{\ul p}$ for $p = 0, 1$, associated with the double point at $x_+$, take value in the real form $\g_1$, see \emph{e.g.} Lemma \ref{lem: real complex fields}. This is by contrast with the $\g$-valued classical fields $A^1_{\ul p}$ associated with $1$ in the undeformed model of \S\ref{sec: coset} which take value in the real form $\g_0$.

\subsubsection{bi-Yang-Baxter $\sigma$-model} \label{sec: bYB model}

The bi-Yang-Baxter $\sigma$-model \cite{Klimcik:2008eq} is yet another two-parameter deformation of the principal chiral model different from the one presented in \S\ref{sec: YB}. Its integrability within the Hamiltonian formalism was studied in \cite{Delduc:2015xdm}. It was found to correspond to a non-cyclotomic one-parameter deformation of the deformed $\ZZ_2$-graded coset $\sigma$-model discussed in \S\ref{sec: Z2 one-param}. In other words, the $\Pi = D_4$ dihedral symmetry of the latter is broken down to a $\langle \t \rangle = D_2 = \ZZ_2$ symmetry. This is achieved by deforming the divisor $\Pi \D = e^{i \vartheta} + e^{-i \vartheta} + e^{i(\vartheta + \pi)} + e^{-i(\vartheta + \pi)} + \infty$, with $\D$ given in \eqref{divisor eta def}, to the new divisor $\langle \t \rangle \D'$ where
\begin{equation*}
\D' \coloneqq e^{i \vartheta} + e^{i (\psi + \pi)} + \infty
\end{equation*}
and $\vartheta, \psi \in \, ]0, \frac{\pi}{2}[$. In the notation of \S\ref{sec: affine Gaudin} we have $\Z = \{ x_+, x_-, y_+, y_-, \infty \}$, where we defined $x_\pm \coloneqq e^{\pm i \vartheta}$ and $y_\pm \coloneqq e^{\pm i (\psi + \pi)}$, and $\Z' = \{ x_+, x_-, y_+, y_- \}$.

There are four formal fields $\A^{x_\pm}_{\ul 0}$ and $\A^{y_\pm}_{\ul 0}$ associated to the points of $\Z \setminus \{ \infty \}$, in terms of which the formal Lax matrix takes the form
\begin{equation} \label{biYB formal Lax}
\L(z) = \frac{\A^{x_+}_{\ul 0}}{z - x_+} + \frac{\A^{x_-}_{\ul 0}}{z - x_-} + \frac{\A^{y_+}_{\ul 0}}{z - y_+} + \frac{\A^{y_-}_{\ul 0}}{z - y_-}.
\end{equation}
The levels are chosen as
\begin{equation*}
\lvl^{x_\pm}_0 = \pm \frac{2}{(x_+ + x_- - y_+ - y_-)(x_+ - x_-)}, \quad \lvl^{y_\pm}_0 = \mp \frac{2}{(x_+ + x_- - y_+ - y_-)(y_+ - y_-)}.
\end{equation*}
They satisfy the condition \eqref{constraint levels} from \S\ref{sec: constraint}. Let $\mathcal J_\pm \coloneqq A^{x_\pm}_{\ul 0}$ and $\tilde{\mathcal J}_\pm \coloneqq A^{y_\pm}_{\ul 0}$ denote the four $\g$-valued classical fields. Their non-trivial Poisson brackets read
\begin{subequations} \label{biYB KM}
\begin{align}
\big\{ \mathcal J_{\pm \1}(\theta), \mathcal J_{\pm \2}(\theta') \big\} &= - \big[ C_{\1\2}, \mathcal J_{\pm \2}(\theta) \big] \delta_{\theta \theta'} - \lvl^{x_\pm}_0 C_{\1\2} \delta'_{\theta \theta'},\\
\big\{ \tilde{\mathcal J}_{\pm \1}(\theta), \tilde{\mathcal J}_{\pm \2}(\theta') \big\} &= - \big[ C_{\1\2}, \tilde{\mathcal J}_{\pm \2}(\theta) \big] \delta_{\theta \theta'} - \lvl^{y_\pm}_0 C_{\1\2} \delta'_{\theta \theta'},
\end{align}
\end{subequations}
from Corollary \ref{cor: Fields PB}. These coincide with the Poisson brackets from \cite[\S 5.1]{Delduc:2015xdm}.

As usual, the Lax matrix and twist function of the model are obtained by applying the linear map $\rep \otimes \pi_{\bm \lvl} \otimes \id$ to the formal Lax matrix \eqref{biYB formal Lax}, as in \S\ref{sec: twist}. We find
\begin{align*}
\mathscr L(z) &= \varphi(z)^{-1} \bigg( \frac{\mathcal J_+}{z - x_+} + \frac{\mathcal J_-}{z - x_-} + \frac{\tilde{\mathcal J}_+}{z - y_+} + \frac{\tilde{\mathcal J}_-}{z - y_-} \bigg),\\
\varphi(z) &= \frac{2 z}{(z - x_+)(z - x_-)(z - y_+)(z - y_-)}.
\end{align*}
Up to an overall normalisation, these are the Lax matrix and twist function of the bi-Yang-Baxter $\sigma$-model in the Hamiltonian formalism obtained in \cite{Delduc:2015xdm}.
We also note that in the limit when $\psi = \vartheta$, \emph{i.e.} $x_\pm = - y_\pm$, we recover the twist function of the deformed $\ZZ_2$-graded coset $\sigma$-model, namely \eqref{sym def twist} with $T=2$.

The formal constraint defined in \eqref{formal field C} is given by the sum of all the formal fields
\begin{equation*}
\mathcal C = \A^{x_+}_{\ul 0} + \A^{x_-}_{\ul 0} + \A^{y_+}_{\ul 0} + \A^{y_-}_{\ul 0}.
\end{equation*}
Note that $\pi_{(0)} = \id$ since in the present case $T=1$. Applying the linear map $\rep \otimes \pi_{\bm \lvl}$ we find that the $\g$-valued classical fields attached to the four points in $\Z \setminus \{ \infty \}$ satisfy the constraint
\begin{equation*}
(\rep \otimes \pi_{\bm \lvl}) \mathcal C = \mathcal J_+ + \mathcal J_- + \tilde{\mathcal J}_+ + \tilde{\mathcal J}_- \approx 0.
\end{equation*}
The latter is equivalent to the Hamiltonian constraint $X + \wt X \approx 0$ of \cite{Delduc:2015xdm}, if in the notation of that paper, cf. equations (5.1) there, we replace $X$ in (5.1b) by $\wt X$ to get the expressions for $\widetilde{\mathcal J}_\pm$ before imposing the constraint.

\subsection{Affine Toda field theory} \label{sec: Toda FT}

Let $\g$ be a finite-dimensional simple Lie algebra over $\CC$ of rank $\ell \coloneqq \rk \g$. Let $\h$ be a Cartan subalgebra and denote by $\Delta \subset \h^\ast$ the root system of $(\g , \h)$. We have the root space decomposition
\begin{equation*}
\g = \h \oplus \bigoplus_{\alpha \in \Delta} \CC E_\alpha.
\end{equation*}
Fix a basis of simple roots $\{ \alpha_i \}_{i=1}^\ell$ and let $\theta \in \Delta$ be the corresponding maximal root. The Coxeter number of $\g$ is defined as $h \coloneqq \hgt \theta + 1$, where $\hgt \alpha$ denotes the height of a root $\alpha \in \Delta$ relative to the system of simple roots $\{ \alpha_i \}_{i=1}^\ell$. The extended system of simple roots is defined by adjoining $\alpha_0 \coloneqq - \theta$ to $\{ \alpha_i \}_{i=1}^\ell$. We will make use of the shorthands $E_i \coloneqq E_{\alpha_i}$ and $F_i \coloneqq E_{-\alpha_i}$ for $i = 0, \ldots, \ell$.

We fix a non-degenerate invariant bilinear form $\langle \cdot, \cdot \rangle : \g \times \g \to \CC$. To any $\alpha \in \Delta$ we associate the Cartan element $H_\alpha \in \h$ defined by $\langle H_\alpha, H \rangle = \alpha(H)$ for all $H \in \h$. We will use the shorthand $H_i \coloneqq H_{\alpha_i}$ for $i = 0, \ldots, \ell$. Restricting $\langle \cdot, \cdot \rangle$ to the Cartan subalgebra $\h$, the bijection $\h^\ast \SimTo \h$, $\alpha \mapsto H_\alpha$ induces a bilinear form on $\h^\ast$ which we also denote $\langle \cdot, \cdot \rangle$. We normalise the bilinear form on $\g$ such that $\langle \theta, \theta \rangle = 2$. Introduce the coroot $\co\alpha \coloneqq \frac{2}{\langle \alpha, \alpha \rangle} H_\alpha \in \h$ for each $\alpha \in \Delta$. Letting $\epsilon_i \coloneqq \frac{2}{\langle \alpha_i, \alpha_i \rangle}$ for all $i = 0, \ldots, \ell$ we then have $\co\alpha_i = \epsilon_i H_i$. Note that $\epsilon_0 = 1$ from the choice of normalisation of the bilinear form $\langle \cdot, \cdot \rangle$. The root vectors $E_{-\alpha}$, $\alpha \in \Delta^+$ can be normalised such that $[E_\alpha, E_{-\alpha}] = \co\alpha$. We then have
\begin{equation*}
[\co\alpha_i, E_j] = a_{ij} E_j, \qquad
[\co\alpha_i, F_j] = - a_{ij} F_j, \qquad
[E_i, F_j] = \delta_{ij} \co\alpha_i,
\end{equation*}
for $i, j = 0, \ldots, \ell$, where $a_{ij} \coloneqq \alpha_j(\co\alpha_i)$ are the components of the Cartan matrix of the untwisted affine Kac-Moody algebra $\tg$ associated with $\g$. Writing the maximal root $\theta$ in the basis of simple roots as $\theta = \sum_{i=1}^\ell a_i \alpha_i \in \Delta$ with $a_i \in \ZZ_{> 0}$ and defining $a_0 \coloneqq 1$, we have $\sum_{j=1}^\ell a_{ij} a_j = 0$. It then also follows that $\co\theta = \sum_{i=1}^\ell \co a_i \co\alpha_i$ where $\co a_i \coloneqq a_i \epsilon_i^{-1}$. And since $\co a_0 = 1$ we deduce that $\sum_{j=1}^\ell a_{ij} \co a_i = 0$. We note that $a_{ij} \epsilon_j = \langle \co\alpha_j, \co\alpha_i \rangle \eqqcolon c_{ij}$ and $a_{ij} \epsilon_i^{-1} = \langle \alpha_j, \alpha_i \rangle \eqqcolon b_{ij}$ are both symmetric. Since $\langle E_i, F_i \rangle = \epsilon_i$, it is convenient to define $E^i \coloneqq \epsilon_i^{-1} E_i$ and $F^i \coloneqq \epsilon_i^{-1} F_i$ for all $i = 0, \ldots, \ell$. We also let $\{ H^i \}_{i=1}^\ell$ be a dual basis of $\{ H_i \}_{i=1}^\ell$ in $\h$, \emph{i.e.} $\langle H^i, H_j \rangle = \delta^i_j$ for every $i, j = 1, \ldots, \ell$. Note that for any $j = 0, \ldots, \ell$ we have the relation $\sum_{k=1}^\ell \alpha_j(H_k) H^k = H_j$.

Let $\omega \coloneqq e^{2 \pi i / h}$ and define an automorphism $\sigma \in \Aut \g$ by letting
\begin{equation*}
\sigma(E_i) = \omega E_i, \qquad
\sigma(\co\alpha_i) = \co\alpha_i, \qquad
\sigma(F_i) = \omega^{-1} F_i,
\end{equation*}
for $i = 1, \ldots, \ell$. This is a Coxeter automorphism in the outer automorphism class of the identity, \emph{i.e.} $[\sigma] = [\text{id}] \in \Aut \g/ \text{Inn}\, \g$. That is, the invariant subalgebra $\g_{(0), \CC} = \h$ is abelian and $\sigma$ has minimum order among all automorphisms $\upsilon \in \Aut \g$ such that $[\upsilon] = [\id] \in \Aut \g / \text{Inn}\, \g$ and $\g^\upsilon$ is abelian, see \emph{e.g.} \cite{BelavinDrinfeld2}. Because $\sigma$ lies in the outer automorphism class of the identity, its action on any root vector $E_\alpha$, $\alpha \in \Delta$ reads
\begin{equation} \label{Coxeter Aut}
\sigma(E_\alpha) = \omega^{\hgt \alpha} E_\alpha.
\end{equation}
We consider the anti-linear automorphism $\tau \in \bAut \g$ defined by
\begin{equation} \label{reality Toda}
\tau(\co\alpha_i) = \co\alpha_i, \qquad
\tau(E_\alpha) = E_\alpha,
\end{equation}
for all $i = 1, \ldots, \ell$ and $\alpha \in \Delta$. The corresponding real Lie algebra $\g_0$ is the split real form of $\g$. Recall the subspaces $\g_{(j)} \subset \g_0$ for $j \in \ZZ_h$ defined in \S\ref{sec: fin dim g}. We have
\begin{equation*}
\g_{(0)} = \h_0 \coloneqq \textup{span}_\RR \{ H_i \}_{i=1}^\ell.
\end{equation*}
Moreover, it follows from \eqref{Coxeter Aut} that a basis of the subspace $\g_{(1)}$ (resp. $\g_{(-1)}$) is given by $E_i$ (resp. $F_i$) for $i = 0, \ldots, \ell$. In other words,
\begin{equation*}
\g_{(1)} =\textup{span}_\RR \{ E_i \}_{i=0}^\ell, \qquad
\g_{(-1)} =\textup{span}_\RR \{ F_i \}_{i=0}^\ell.
\end{equation*}

\subsubsection{Divisor and levels}
Consider the divisor
\begin{equation} \label{aff Toda divisor}
\D = 2 \cdot 0 + 2 \cdot \infty.
\end{equation}
We have $\Z = \{ 0, \infty \}$ so that $\Z' = \emptyset$.
By Proposition \ref{prop: Gaudin Lax} we can then write the formal Lax matrix as
\begin{equation} \label{aff Toda formal Lax}
\L(z) = \A^0_{\ul 0} z^{-1} + \A^0_{\ul 1} z^{-2} - \A^{\infty}_{\ul 1}.
\end{equation}
The formal fields $\A^0_{\ul p}$ for $p = 0, 1$ and $\A^\infty_{\ul 1}$ are defined in \eqref{fields A zf}. Recalling the notation \eqref{tgD R C basis c}, they are given explicitly by
\begin{subequations} \label{formal Gaudin fields Toda}
\begin{align}
\A^0_{\ul 0} &= \cocent \otimes \cent^{(0)}_{\ul 0} + \cent \otimes \cocent^{(0)}_{\ul 0} + \sum_{n \in \ZZ} \sum_{j=1}^\ell H_{j, -n} \otimes H^{j (0)}_{n, \ul 0},\\
\A^0_{\ul 1} &= \sum_{n \in \ZZ} \sum_{j=0}^\ell F_{j, -n} \otimes E^{j (0)}_{n, \ul 1}, \qquad
\A^\infty_{\ul 1} = \sum_{n \in \ZZ} \sum_{j=0}^\ell E_{j, -n} \otimes F^{j (\infty)}_{n, \ul 1}.
\end{align}
\end{subequations}
It follows from the definition of the action \eqref{Pi act Fields} of $\t \in \Pi$ on formal fields in $\tg^\D_\CC$, the reality conditions \eqref{reality Toda} and those on the generators $\cent$, $\cocent$ defined in \S\ref{sec: tg action of Pi} that the formal fields in \eqref{formal Gaudin fields Toda} are all real in the sense that $r_\t \A^0_{\ul p} = \A^0_{\ul p}$ for $p = 0,1$ and $r_\t \A^\infty_{\ul 1} = \A^\infty_{\ul 1}$, cf. \eqref{reality Toda HEF} below. The Poisson brackets among the formal fields \eqref{formal Gaudin fields Toda} are determined from Proposition \ref{prop: Fields PB} to be
\begin{gather*}
\big\{ \A^0_{\ul 0 \1}, \A^0_{\ul 0 \2} \big\} = - \big[ \widetilde C^{(0)}_{\1\2}, \A^0_{\ul 0 \2} \big], \qquad
\big\{ \A^0_{\ul 0 \1}, \A^0_{\ul 1 \2} \big\} = - \big[ \widetilde C^{(0)}_{\1\2}, \A^0_{\ul 1 \2} \big],
\end{gather*}
with all other brackets vanishing. Recall from the proof of Proposition \ref{prop: Fields PB} that these Poisson brackets are equivalent to the Lie brackets on $\tg^\D_\CC$. The latter is spanned by $\cocent^{(0)}_{\ul 0}$, $\cent^{(0)}_{\ul 0}$ together with $H^{j (0)}_{n, \ul 0}$ for $j = 1, \ldots, \ell$, $n \in \ZZ$ and $E^{j (0)}_{n, \ul 1}$, $F^{j (\infty)}_{n, \ul 1}$ for $j = 0, \ldots, \ell$, $n \in \ZZ$, in terms of which its non-trivial Lie brackets read
\begin{subequations} \label{non-trivial PB Toda}
\begin{alignat}{2}
\big[ H^{i (0)}_{m, \ul 0}, H^{j (0)}_{n, \ul 0} \big] &= m \delta_{m+n, 0} \langle H^i, H^j \rangle \cent^{(0)}_{\ul 0}, &\qquad
\big[ \cocent^{(0)}_{\ul 0}, H^{j (0)}_{n, \ul 0} \big] &= n H^{j (0)}_{n, \ul 0},\\
\big[ H^{i (0)}_{m, \ul 0}, E^{j (0)}_{n, \ul 1} \big] &= \alpha_j(H^i) E^{j (0)}_{m+n, \ul 1}, &\qquad
\big[ \cocent^{(0)}_{\ul 0}, E^{j (0)}_{n, \ul 1} \big] &= n E^{j (0)}_{n, \ul 1}.
\end{alignat}
\end{subequations}
It follows from Proposition \ref{prop: complexified Takiff}$(iii)$ that under the anti-linear map $c : \tg^\D_\CC \to \tg^\D_\CC$ we have
\begin{subequations} \label{reality Toda HEF}
\begin{gather}
c \big( H^{i (0)}_{n, \ul 0} \big) = H^{i (0)}_{-n, \ul 0}, \qquad
c \big( E^{j (0)}_{n, \ul 1} \big) = E^{j (0)}_{-n, \ul 1}, \qquad
c \big( F^{j (\infty)}_{n, \ul 1} \big) = F^{j (\infty)}_{-n, \ul 1},\\
c \big( \cent^{(0)}_{\ul 0} \big) = - \cent^{(0)}_{\ul 0}, \qquad c \big( \cocent^{(0)}_{\ul 0} \big) = - \cocent^{(0)}_{\ul 0}
\end{gather}
\end{subequations}
for any $i = 1, \ldots, \ell$, $j = 0, \ldots, \ell$ and $n \in \ZZ$.

We fix the levels to be
\begin{equation} \label{aff Toda levels}
\lvl^0_0 = 1, \qquad \lvl^0_1 = 0, \qquad \lvl^\infty_1 = 0.
\end{equation}
Note that by contrast with the examples discussed in \S\ref{sec: non-cycl examples} and \S\ref{sec: cycl examples}, since $\lvl^0_1 = \lvl^\infty_1 = 0$ the assumption \eqref{assumption lvl} made throughout \S\ref{sec: affine Gaudin} fails to hold. In particular, we cannot use the homomorphism $\pi_{\bm \lvl}$ as defined in \eqref{pi D k def}. Instead we will define a homomorphism $\widetilde{\pi}_{\bm \lvl}$ from the complex Lie algebra $\tg^\D_\CC$ of Proposition \ref{prop: complexified Takiff} to a completion of the symmetric algebra on a certain Heisenberg-type algebra.

\subsubsection{Toda fields} \label{sec: Toda fields}

Let $\Loop \h \coloneqq \h \otimes \CC[t, t^{-1}]$ and $\Loop \g_{(1), \CC} \coloneqq \g_{(1), \CC} \otimes \CC[t, t^{-1}]$. We consider the subspaces $\mathscr H^{(+)} \coloneqq \Loop \h \dotplus \Loop \g_{(1), \CC}$ and $\mathscr H^{(-)} \coloneqq \Loop \h$ of the loop algebra $\Loop \g$, cf. \S\ref{sec: aff KM}. Let
\begin{equation*}
(\cdot | \cdot) : \mathscr H^{(+)} \times \mathscr H^{(-)} \longrightarrow \CC
\end{equation*}
denote the restriction of the bilinear form on $\Loop \g$ to this pair of subspaces. We endow $\mathscr H^{(-)}$ with its natural abelian Lie algebra structure coming from $\Loop \g$. The Lie algebra structure on $\mathscr H^{(+)}$ is defined by letting $\Loop \h$ and $\Loop \g_{(1), \CC}$ be abelian subalgebras and letting $[\cdot, \cdot] : \Loop \h \times \Loop \g_{(1), \CC} \to \Loop \g_{(1), \CC}$ be given by the restriction of the Lie bracket on $\Loop \g$. Consider the direct sum of complex vector spaces $\mathscr H \coloneqq \mathscr H^{(+)} \oplus \mathscr H^{(-)} \oplus \CC {\bf 1}$. For any $\ms X \in \mathscr H^{(+)}$ we use the notation $\ms X^{(+)} \coloneqq (\ms X, 0, 0) \in \mathscr H$ to represent $\ms X$ regarded as an element of $\mathscr H$, and similarly $\ms Y^{(-)} \coloneqq (0, \ms Y, 0) \in \mathscr H$ for any $\ms Y \in \mathscr H^{(-)}$.
We define a Lie bracket on $\mathscr H$ by setting
\begin{equation*}
\big[ \ms X^{(+)} + \ms X'^{(-)} + \alpha {\bf 1}, \ms Y^{(+)} + \ms Y'^{(-)} + \beta {\bf 1} \big] \coloneqq [\ms X, \ms Y]^{(+)} + \big( (\ms X | \ms Y') - (\ms Y | \ms X') \big) {\bf 1}.
\end{equation*}
for any $\ms X, \ms Y \in \mathscr H^{(+)}$, $\ms X', \ms Y' \in \mathscr H^{(-)}$ and $\alpha, \beta \in \CC$.
A basis of $\mathscr H$ is given by $E^{i(+)}_n$ and $H^{j(\pm)}_n$ for $i=0, \ldots, \ell$, $j = 1, \ldots, \ell$ and $n \in \ZZ$, together with ${\bf 1}$. In terms of these the non-trivial Lie brackets on $\mathscr H$ read
\begin{equation} \label{com Hpm Ep}
\big[ H^{i(+)}_m, H^{j(-)}_n \big] = \delta_{m+n, 0} \langle H^i, H^j \rangle {\bf 1}, \qquad
\big[ H^{i(+)}_m, E^{j(+)}_n \big] = \alpha_j(H^i) E^{j(+)}_{m+n}.
\end{equation}
We introduce an anti-linear automorphism $c : \mathscr H \to \mathscr H$ simply by letting
\begin{equation} \label{def c on H}
c \big( H^{i(\pm)}_n \big) = H^{i(\pm)}_{-n}, \qquad
c \big( E^{j(+)}_n \big) = E^{j(+)}_{-n}, \qquad
c ({\bf 1}) = {\bf 1},
\end{equation}
for all $i = 1, \ldots, \ell$, $j = 0, \ldots, \ell$ and $n \in \ZZ$.

The Lie bracket on $\mathscr H$ extends by the Leibniz rule and linearity to a Poisson bracket on the symmetric algebra $S(\mathscr H)$, cf. \S\ref{sec: Alg Obs},
\begin{equation} \label{PB SH}
\{ \cdot, \cdot \} : S(\mathscr H) \times S(\mathscr H) \longrightarrow S(\mathscr H).
\end{equation}
Recall the descending $\ZZ_{\geq 0}$-filtrations \eqref{Fil n tg} and \eqref{Fil- n tg} on $\tg$, or equivalently on the loop algebra $\Loop \g$. We similarly define a pair of descending $\ZZ_{\geq 0}$-filtrations $(\Fil_n \mathscr H)_{n \in \ZZ_{\geq 0}}$ and $(\cFil_n \mathscr H)_{n \in \ZZ_{\geq 0}}$ on the vector space $\mathscr H$ as
\begin{subequations} \label{Fil cFil H}
\begin{align}
\Fil_n \mathscr H &\coloneqq (\h \dotplus \g_{(1), \CC}) \otimes t^n \CC[t] \oplus \h \otimes t^n \CC[t],\\
\cFil_n \mathscr H &\coloneqq (\h \dotplus \g_{(1), \CC}) \otimes t^{-n} \CC[t^{-1}] \oplus \h \otimes t^{-n} \CC[t^{-1}],
\end{align}
\end{subequations}
for $n \in \ZZ_{\geq 0}$. In turn, we use this to define a descending $\ZZ_{\geq 0}$-filtration on the commutative algebra $S(\mathscr H)$ by ideals $\Fil_n \big( S(\mathscr H) \big) \coloneqq (\Fil_n \mathscr H) S(\mathscr H) \cap (\cFil_n \mathscr H) S(\mathscr H)$, cf. \eqref{S tgDC Fil 1}. The corresponding completion is a commutative algebra over $\CC$ which we denote by
\begin{equation*}
\hat S(\mathscr H) \coloneqq \varprojlim S(\mathscr H) \big/ \Fil_n \big( S(\mathscr H) \big). 
\end{equation*}
In the same way as in \S\ref{sec: Alg Obs}, the Poisson bracket \eqref{PB SH} extends to one on the completion which we also denote
\begin{equation*}
\{ \cdot, \cdot \} : \hat S(\mathscr H) \times \hat S(\mathscr H) \longrightarrow \hat S(\mathscr H).
\end{equation*}

\begin{lemma} \label{lem: Toda exp}
The set of elements $\wh E^{j(+)}_n \coloneqq \sum_{p \in \ZZ} p E^{j(+)}_{n-p} H^{(-)}_{j, p} - n E^{j(+)}_n$ for all $n \in \ZZ$, $j = 0, \ldots, \ell$ and ${\bf 1} - 1$, generate a Poisson ideal of $\hat S(\mathscr H)$. We denote the corresponding quotient by $\hat S_{\bm \lvl}(\mathscr H)$.
\begin{proof}
The element ${\bf 1} - 1$ is clearly central in $\hat S(\mathscr H)$. For every $i = 1, \ldots, \ell$, $j = 0, \ldots, \ell$ and $m, n \in \ZZ$ we have
\begin{equation*}
\big\{ H^{i(+)}_m, n E^{j(+)}_n \big\} = \alpha_j(H^i) n E^{j(+)}_{m+n} = \alpha_j(H^i) (m+n) E^{j(+)}_{m+n} - \alpha_j(H^i) m E^{j(+)}_{m+n}.
\end{equation*}
On the other hand we find that
\begin{align*}
&\sum_{p \in \ZZ} \big\{ H^{i(+)}_m, p E^{j(+)}_{n-p} H^{(-)}_{j, p} \big\} = \alpha_j(H^i) \sum_{p \in \ZZ} p E^{j(+)}_{m+n-p} H^{(-)}_{j, p} + \sum_{p \in \ZZ} p E^{j(+)}_{n-p} \langle H^i, H_j \rangle \delta_{m+p, 0} {\bf 1}\\
&\qquad = \alpha_j(H^i) \sum_{p \in \ZZ} p E^{j(+)}_{m+n-p} H^{(-)}_{j, p} - \alpha_j(H^i) m E^{j(+)}_{m+n} - \alpha_j(H^i) m E^{j(+)}_{m+n} ({\bf 1} - 1).
\end{align*}
Combining the above yields
\begin{equation*}
\big\{ H^{i(+)}_m, \wh E^{j(+)}_n \big\} = \alpha_j(H^i) \wh E^{j(+)}_{m+n} - \alpha_j(H^i) m E^{j(+)}_{m+n} ({\bf 1} - 1).
\end{equation*}
Moreover, the elements $\wh E^{j(+)}_n$ have vanishing Lie bracket with both $H^{i(-)}_m$ and $E^{i(+)}_m$. The result now follows.
\end{proof}
\end{lemma}

The map $c : \mathscr H \to \mathscr H$ extends as an anti-linear automorphism to $S(\mathscr H)$. And since it preserves the subspaces $\Fil_n\big( S(\mathscr H) \big)$ for each $n \in \ZZ_{\geq 0}$ by sending the pair of subspaces in \eqref{Fil cFil H} to one another, it extends by continuity to an anti-linear automorphism of the completion $\hat S(\mathscr H)$. Moreover, since $c(\wh E^{j(+)}_n) = - \wh E^{j(+)}_{-n}$ and $c({\bf 1} - 1) = {\bf 1} - 1$, the corresponding Poisson ideal of $\hat S(\mathscr H)$ from Lemma \ref{lem: Toda exp} is invariant under it. Thus we obtain an anti-linear automorphism
\begin{equation} \label{def c SH}
c : \hat S_{\bm\lvl}(\mathscr H) \longrightarrow \hat S_{\bm\lvl}(\mathscr H)
\end{equation}
of the quotient defined in Lemma \ref{lem: Toda exp}.

We are now in a position to define an analogue of the homomorphism $\pi_{\bm \lvl}$ in the present case, replacing the definition \eqref{pi D k def}. Pick real parameters $\mu_i, \nu_i \in \RR$ for each $i = 0, \ldots, \ell$ such that $\mu_i \nu_i \geq 0$. Consider the linear map
\begin{subequations} \label{pi D l Toda}
\begin{equation}
\widetilde{\pi}_{\bm \lvl} : \tg^\D_\CC \longrightarrow \hat S_{\bm \lvl}(\mathscr H),
\end{equation}
defined on the basis elements of $\tg^\D_\CC$ as
\begin{gather}
\label{pi D l Toda b}
\widetilde{\pi}_{\bm \lvl} \big( E^{i (0)}_{n, \ul 1} \big) \coloneqq \frac{\mu_i}{\sqrt{2}} E^{i(+)}_n, \qquad
\widetilde{\pi}_{\bm \lvl} \big( F^{i (\infty)}_{n, \ul 1} \big) \coloneqq - \frac{\nu_i}{\sqrt{2}} \delta_{n, 0}, \qquad
\widetilde{\pi}_{\bm \lvl} \big( \cent^{(0)}_{\ul 0} \big) \coloneqq i,\\
\label{pi D l Toda c}
\widetilde{\pi}_{\bm \lvl} \big( H^{j (0)}_{n, \ul 0} \big) \coloneqq H^{j(+)}_n - \frac{i n}{2} H^{j(-)}_n, \qquad
\widetilde{\pi}_{\bm \lvl} \big( \cocent^{(0)}_{\ul 0} \big) \coloneqq \sum_{n \in \ZZ} \sum_{k=1}^\ell n H_{k, -n}^{(+)} H^{k(-)}_n,
\end{gather}
\end{subequations}
for $i = 0, \ldots, \ell$, $j = 1, \ldots, \ell$ and $n \in \ZZ$. This map is seen to be equivariant with respect to the pair of anti-linear automorphisms $c$ on $\tg^\D_\CC$ and $\hat S_{\bm \lvl}(\mathscr H)$ defined respectively in Proposition \ref{prop: complexified Takiff} and \eqref{def c SH}, using \eqref{reality Toda HEF} and \eqref{def c on H}.

\begin{lemma} \label{lem: tilde pi lvl hom}
The map \eqref{pi D l Toda} is a homomorphism of Lie algebras.
\begin{proof}
For all $j = 0, \ldots, \ell$ and $n \in \ZZ$ we have
\begin{align*}
\sum_{m \in \ZZ} \sum_{i=1}^\ell \big\{ m H_{i, -m}^{(+)} H^{i(-)}_m, E^{j(+)}_n \big\} &= \sum_{m \in \ZZ} \sum_{i=1}^\ell m \, \alpha_j(H_i) E^{j(+)}_{n-m} H^{i(-)}_m\\
&= \sum_{m \in \ZZ} m E^{j(+)}_{n-m} H^{(-)}_{j, m} = n E^{j(+)}_n,
\end{align*}
where the last equality uses the definition of $\hat S_{\bm \lvl}(\mathscr H)$ given in Lemma \ref{lem: Toda exp}. Similarly, for $j = 1, \ldots, \ell$ and $n \in \ZZ$ we find
\begin{equation*}
\sum_{m \in \ZZ} \sum_{i=1}^\ell \big\{ m H_{i, -m}^{(+)} H^{i (-)}_m, H^{j(\pm)}_n \big\} = n H^{j(\pm)}_n.
\end{equation*}
We deduce at once from the above that
\begin{align*}
\widetilde{\pi}_{\bm \lvl} \big( \big[ \cocent^{(0)}_{\ul 0}, E^{i(0)}_{n \ul 1} \big] \big) &= \big\{ \widetilde{\pi}_{\bm \lvl}\big( \cocent^{(0)}_{\ul 0} \big), \widetilde{\pi}_{\bm \lvl}\big( E^{i(0)}_{n \ul 1} \big) \big\},\\
\widetilde{\pi}_{\bm \lvl} \big( \big[ \cocent^{(0)}_{\ul 0}, H^{j(0)}_{n \ul 0} \big] \big) &= \big\{ \widetilde{\pi}_{\bm \lvl}\big( \cocent^{(0)}_{\ul 0} \big), \widetilde{\pi}_{\bm \lvl}\big( H^{j(0)}_{n \ul 0} \big) \big\},
\end{align*}
for all $i = 0, \ldots, \ell$, $j = 1, \ldots, \ell$ and $n \in \ZZ$. Next, we have
\begin{align*}
\big\{ \widetilde{\pi}_{\bm \lvl} \big( H^{j (0)}_{m, \ul 0} \big), \widetilde{\pi}_{\bm \lvl} \big( H^{k (0)}_{n, \ul 0} \big) \big\} &= - \frac{i n}{2} \big[ H^{j(+)}_m, H^{k(-)}_n \big] - \frac{i m}{2} \big[ H^{j(-)}_m, H^{k(+)}_n \big]\\
&= - \frac{i n}{2} \delta_{m+n, 0} \langle H^j, H^k \rangle + \frac{i m}{2} \delta_{m+n, 0} \langle H^k, H^j \rangle\\
&= i m \delta_{m+n, 0} \langle H^j, H^k \rangle = \widetilde{\pi}_{\bm \lvl} \big( \big[ H^{j(0)}_{m, \ul 0}, H^{k(0)}_{n, \ul 0} \big] \big),
\end{align*}
for every $j, k = 1, \ldots, \ell$ and $m, n \in \ZZ$.
We also have that
\begin{align*}
\big\{ \widetilde{\pi}_{\bm \lvl} \big( H^{j (0)}_{m, \ul 0} \big), \widetilde{\pi}_{\bm \lvl} \big( E^{k(0)}_{n, \ul 1} \big) \big\}
&= \frac{\mu_k}{\sqrt{2}} \Big[ H^{j(+)}_m - \frac{i m}{2} H^{j(-)}_m, E^{k(+)}_n \Big]
= \frac{\mu_k}{\sqrt{2}} \big[ H^{j(+)}_m, E^{k(+)}_n \big]\\
&= \frac{\mu_k}{\sqrt{2}} \alpha_k(H^j) E^{k(+)}_{m+n}
= \widetilde{\pi}_{\bm \lvl} \big( \big[ H^{j (0)}_{m, \ul 0}, E^{k (0)}_{n, \ul 1} \big] \big),
\end{align*}
for $j = 1, \ldots, \ell$, $k = 0, \ldots, \ell$ and $m, n \in \ZZ$. This completes the proof that the linear map \eqref{pi D l Toda} preserves all the non-trivial Lie brackets \eqref{non-trivial PB Toda} on $\tg^\D_\CC$. Finally, one checks that all the trivial Lie brackets are also preserved by \eqref{pi D l Toda}, as required.
\end{proof}
\end{lemma}

We define the \emph{formal Toda fields} as
\begin{equation*}
\A^{(\pm)} \coloneqq \sum_{n \in \ZZ} \sum_{j=1}^\ell H_{j, -n} \otimes H^{j(\pm)}_n, \qquad
\mathcal E = \sum_{n \in \ZZ} \sum_{j=0}^\ell F_{j, -n} \otimes E^{j(+)}_n.
\end{equation*}
These are elements of the completed tensor product $\Loop (\h \dotplus \g_{(-1), \CC}) \hotimes \hat S_{\bm \lvl}(\mathscr H)$, and both are real in the sense that they are invariant under the action of $\tau \otimes c$.

\begin{lemma} \label{lem: formal Toda PB}
The collection of non-trivial Poisson brackets between the formal Toda fields read
\begin{equation*}
\big\{ \A^{(+)}_\1, \A^{(-)}_\2 \big\} = \wh C^{(0)}_{\1\2}, \qquad
\big\{ \A^{(+)}_\1, \mathcal E_\2 \big\} = - \big[ \wh C^{(0)}_{\1\2}, \mathcal E_\2 \big],
\end{equation*}
where $\wh C^{(0)}_{\1\2} \coloneqq \sum_{n \in \ZZ} \sum_{j=1}^\ell H_{j, -n} \otimes H^j_n$.
\begin{proof}
The first relation follows from
\begin{equation*}
\big\{ \A^{(+)}_\1, \A^{(-)}_\2 \big\} = \sum_{m, n \in \ZZ} \sum_{i, j=1}^\ell H_{i, -n} \otimes H^j_m \otimes \big[ H^{i(+)}_n, H^{(-)}_{j, -m} \big] = \wh C^{(0)}_{\1\2} \otimes {\bf 1} = \wh C^{(0)}_{\1\2} \otimes 1,
\end{equation*}
where in the last equality we used the fact that ${\bf 1} = 1$ in $\hat S_{\bm \lvl}(\mathscr H)$.
For the second relation we have
\begin{align*}
\{ \A^{(+)}_\1, \mathcal E_\2 \} &= \sum_{m, n \in \ZZ} \sum_{i, j=1}^\ell H_{i, - n} \otimes F_{j, -m} \otimes \big[ H^{i(+)}_n, E^{j(+)}_m \big]\\
&= \sum_{m, n \in \ZZ} \sum_{i, j=1}^\ell H_{i, - n} \otimes \alpha_j(H^i) F_{j, -m} \otimes E^{j(+)}_{m+n}\\
&= - \sum_{m, n \in \ZZ} \sum_{i, j=1}^\ell H_{i, - n} \otimes [ H^i_n, F_{j, -m-n} ] \otimes E^{j(+)}_{m+n} = - \big[ \wh C^{(0)}_{\1\2}, \mathcal E_\2 \big],
\end{align*}
where in the last step we shifted the summation over $m$ by $-n$.
\end{proof}
\end{lemma}

The \emph{Toda field} $\Phi$ and its conjugate momentum $\Pi$ are the $\h_0$-valued classical fields defined as
\begin{subequations} \label{Toda fields def}
\begin{align}
\Phi &\coloneqq (\rep \otimes \id) \A^{(-)} = \sum_{j=1}^\ell H_j \otimes \sum_{n \in \ZZ} e_{-n} \otimes H^{j(-)}_n,\\
\Pi &\coloneqq (\rep \otimes \id) \A^{(+)} = \sum_{j=1}^\ell H_j \otimes \sum_{n \in \ZZ} e_{-n} \otimes H^{j(+)}_n.
\end{align}
\end{subequations}
To see that $\Pi$ is the conjugate momentum we apply $\rep \otimes \rep \otimes \id$ to the first relation in Lemma \ref{lem: formal Toda PB}. This yields the canonical Poisson brackets
\begin{equation*}
\big\{ \Pi_\1(\theta), \Phi_\2(\theta') \big\} = C^{(0)}_{\1\2} \delta_{\theta \theta'}.
\end{equation*}

Applying the root $\alpha_j$ to the Toda field $\Phi$ we obtain the classical field
\begin{equation} \label{alpha j Phi}
\alpha_j(\Phi) = \sum_{n \in \ZZ} e_{-n} \otimes \sum_{k=1}^\ell \alpha_j(H_k) H^{k(-)}_n = \sum_{n \in \ZZ} e_{-n} \otimes \sum_{k=1}^\ell H^{(-)}_{j, n}.
\end{equation}
We also define the exponential of $\alpha_j(\Phi)$ as the classical field
\begin{equation} \label{exp Toda def}
e^{\alpha_j(\Phi)} \coloneqq \sum_{n \in \ZZ} e_{-n} \otimes E^{j(+)}_n.
\end{equation}
The justification for this definition comes from Lemma \ref{lem: exp justification} below. First we note that it is related to the formal field $\mathcal E$ defined above as $(\rep \otimes \id) \mathcal E = \sum_{j=0}^\ell F_j \otimes e^{\alpha_j(\Phi)}$.
Then applying the linear map $\rep \otimes \rep \otimes \id$ to the second relation in Lemma \ref{lem: formal Toda PB} gives
\begin{align*}
\sum_{j=0}^\ell &\big\{ \Pi_\1(\theta), \big( F_j \otimes e^{\alpha_j(\Phi)} \big)_\2(\theta') \big\} = - \sum_{j=0}^\ell \sum_{i=1}^\ell H_i \otimes [H^i, F_j] \otimes e^{\alpha_j(\Phi(\theta))} \delta_{\theta \theta'}\\
&= \sum_{j=0}^\ell \sum_{i=1}^\ell \alpha_j(H^i) H_i \otimes F_j \otimes e^{\alpha_j(\Phi(\theta))} \delta_{\theta \theta'}
= \sum_{j=0}^\ell H_j \otimes F_j \otimes e^{\alpha_j(\Phi(\theta))} \delta_{\theta \theta'},
\end{align*}
where we introduced the notation $e^{\alpha_j(\Phi(\theta))} \coloneqq e^{\alpha_j(\Phi)}(\theta)$ for $j = 0, \ldots, \ell$.
Equivalently, by writing the conjugate momentum $\Pi$ in components as $\Pi = \sum_{i=1}^\ell H_i \otimes \Pi^i$, that is $\Pi^i \coloneqq \sum_{n \in \ZZ} e_{-n} \otimes H^{i(+)}_n$, we can express the above relation in components as
\begin{align*}
\big\{ \Pi^i(\theta), e^{\alpha_j(\Phi(\theta'))} \big\} = e^{\alpha_j(\Phi(\theta))} \delta_{\theta \theta'} \delta_{ij}, \qquad \big\{ \Pi^i(\theta), e^{\alpha_0(\Phi(\theta'))} \big\} = - a_i e^{\alpha_0(\Phi(\theta))} \delta_{\theta \theta'},
\end{align*}
for all $i, j = 1, \ldots, \ell$. Here we have used the fact that $H_0 = - \sum_{i=1}^\ell a_i H_i$.

\begin{lemma} \label{lem: exp justification}
For all $j = 0, \ldots, \ell$, we have $\partial \big( e^{\alpha_j(\Phi)} \big) = \alpha_j(\partial \Phi) e^{\alpha_j(\Phi)}$.
\begin{proof}
Taking the derivative of the field \eqref{exp Toda def} we find
\begin{equation*}
\partial \big( e^{\alpha_j(\Phi)} \big) = - i \sum_{n \in \ZZ} e_{-n} \otimes n E^{j(+)}_n.
\end{equation*}
On the other hand, the derivative $\partial \alpha_j(\Phi) = \alpha_j(\partial \Phi)$ of the classical field \eqref{alpha j Phi} takes the form $\alpha_j(\partial \Phi) = - i \sum_{p \in \ZZ} e_{-p} \otimes p H^{(-)}_{j, p}$. Multiplying it by \eqref{exp Toda def} we obtain
\begin{equation*}
\alpha_j(\partial \Phi) e^{\alpha_j(\Phi)} = - i \sum_{m, p \in \ZZ} e_{-m-p} \otimes p E^{j(+)}_m H^{(-)}_{j, p} = - i \sum_{n \in \ZZ} e_{-n} \otimes \sum_{p \in \ZZ} p E^{j(+)}_{n-p} H^{(-)}_{j, p},
\end{equation*}
where in the first equality we used the relation $e_{-m} e_{-p} = e_{-m-p}$, cf. \S\ref{sec: Conn on S1}, and in the last step we performed the change of variables $m \to n - p$. The result now follows from Lemma \ref{lem: Toda exp}.
\end{proof}
\end{lemma}

\subsubsection{Lax matrix and Hamiltonian} \label{sec: Toda FT Lax Ham}

We define the Lax matrix and twist function as in \S\ref{sec: twist}, through the relation, cf. \eqref{Lax twist from formal},
\begin{equation*}
( \rep \otimes \widetilde{\pi}_{\bm \lvl} \otimes \id ) \L(z) = \varphi(z) \big( \partial + \mathscr L(z) \big),
\end{equation*}
where the formal Lax matrix $\L(z)$ is given in \eqref{aff Toda formal Lax}.
To compute these we begin by applying the linear map $\rep \otimes \widetilde{\pi}_{\bm \lvl}$ to the expressions \eqref{formal Gaudin fields Toda} for the formal fields, with $\rep$ defined as in \S\ref{sec: Conn on S1} and $\widetilde{\pi}_{\bm \lvl}$ in \eqref{pi D l Toda}.

For the pair of formal fields at the origin we have
\begin{align*}
\big( \rep \otimes \widetilde{\pi}_{\bm \lvl} \big) \A^0_{\ul 0} &= \partial + \sum_{j=1}^\ell H_j \otimes \sum_{n \in \ZZ} e_{-n} \otimes \widetilde{\pi}_{\bm \lvl} \big( H^{j (0)}_{n, \ul 0} \big)\\
&= \partial + \sum_{j=1}^\ell H_j \otimes \sum_{n \in \ZZ} e_{-n} \otimes \Big( H^{j(+)}_n - \frac{i n}{2} H^{j(-)}_n \Big) = \partial + \Pi + \ha \partial \Phi,\\
\big( \rep \otimes \widetilde{\pi}_{\bm \lvl} \big) \A^0_{\ul 1} &= \sum_{j=0}^\ell F_j \otimes \sum_{n \in \ZZ} e_{-n} \otimes \widetilde{\pi}_{\bm \lvl} \big( E^{j (0)}_{n, \ul 1} \big) = \frac{1}{\sqrt{2}} \sum_{j=0}^\ell \mu_j F_j \otimes e^{\alpha_j(\Phi)}.
\end{align*}
Likewise, for the field at infinity we find
\begin{equation*}
\big( \rep \otimes \widetilde{\pi}_{\bm \lvl} \big) \A^\infty_{\ul 1} = \sum_{j=0}^\ell E_j \otimes \sum_{n \in \ZZ} e_{-n} \otimes \widetilde{\pi}_{\bm \lvl} \big( F^{j (\infty)}_{n, \ul 1} \big) = - \frac{1}{\sqrt{2}} \sum_{j=0}^\ell \nu_j E_j.
\end{equation*}
Applying the linear map $\rep \otimes \widetilde{\pi}_{\bm \lvl} \otimes \id$ to the formal Lax matrix \eqref{aff Toda formal Lax} and using the above we find the twist function $\varphi(z) = z^{-1}$ and the Lax matrix
\begin{equation} \label{Lax aff Toda gauge}
\mathscr L(z) = \Pi + \ha \partial \Phi + \frac{1}{\sqrt{2}} \sum_{j=0}^\ell \Big( e^{\alpha_j(\Phi)} z^{-1} \mu_j F_j + z \, \nu_j E_j \Big),
\end{equation}
where we drop some tensor products to conform to standard notations.
By Corollary \ref{cor: Lax algebra} it satisfies the non-ultralocal Poisson algebra \eqref{Lax algebra} with $\R$-matrix given by
\begin{equation*}
\mathcal R(z, w) \coloneqq \frac{1}{h} \sum_{n \in \ZZ} \sum_{r = 0}^{h-1} \frac{\sigma^r I_{a, -n} \otimes I^a_n}{w - \omega^{-r} z} \varphi(w)^{-1} = \frac{1}{w^h - z^h} \sum_{r = 0}^{h-1} z^r w^{h - r} C^{(-r)},
\end{equation*}
whose skew-symmetric and symmetric parts, cf. \eqref{rs matrices}, read
\begin{subequations} \label{skew r-matrix}
\begin{align}
\label{skew r-matrix a} r_{\1\2}(z, w) &= - \ha C^{(0)}_{\1\2} + \frac{1}{w^h - z^h} \sum_{r = 0}^{h-1} z^r w^{h - r} C^{(-r)}_{\1\2},\\
\label{skew r-matrix b} s_{\1\2}(z, w) &= \ha C^{(0)}_{\1\2}.
\end{align}
\end{subequations}

The expression \eqref{Lax aff Toda gauge} coincides with the Lax matrix of affine Toda field theory in its non-ultralocal formulation. In order to see this, note that if we formally apply a gauge transformation by $e^{\frac{1}{2} \Phi}$  to \eqref{Lax aff Toda gauge} then we obtain
\begin{equation} \label{Lax aff Toda}
\mathscr L'(z) \coloneqq - \ha \partial \Phi + e^{\frac{1}{2} \ad \Phi} \mathscr L(z) = \Pi + \frac{1}{\sqrt{2}} \sum_{j=0}^\ell e^{\frac{1}{2} \alpha_j(\Phi)} \big( z^{-1} \mu_j F_j + z \, \nu_j E_j \big).
\end{equation}
This is the usual Lax matrix used in the standard treatment of affine Toda field theory, see \emph{e.g.} \cite[ch. 12]{BBT}. It satisfies the \emph{ultralocal} Poisson bracket
\begin{equation*}
\{ \mathscr L'_\1(z), \mathscr L'_\2(w) \} = [r_{\1\2}(z, w), \mathscr L'_\1(z) + \mathscr L'_\2(w)] \delta_{\theta \theta'},
\end{equation*}
with the $r$-matrix given in \eqref{skew r-matrix a}.

Finally, we show that the Hamiltonian of affine Toda field theory can be obtained from the local quadratic Gaudin Hamiltonians. It follows from Proposition \ref{prop: Ham} that the only non-trivial formal quadratic Gaudin Hamiltonian is
\begin{equation*}
\H^0_1 = - \big( \A^0_{\ul 1} \big| \A^\infty_{\ul 1} \big) + \ha \big( \A^0_{\ul 0} \big| \A^0_{\ul 0} \big).
\end{equation*}
Note that this can also be obtained from the formal quadratic Hamiltonian \eqref{local quad Ham} associated with infinity, the only zero of the twist function $\varphi(z) = z^{-1}$, since this reads $\H_\infty = \res_\infty \ha \big( \L(z) \big| \L(z) \big) \varphi(z)^{-1} dz = - \H^0_1$.
When applying the homomorphism $\widetilde{\pi}_{\bm \lvl}$ defined in \S\ref{sec: Toda fields}, the terms not involving $\cocent$ and $\cent$ give
\begin{align} \label{Ham Toda compute}
&\sum_{i=0}^\ell \frac{m^2_i}{4 \pi} \epsilon_i \int_{S^1} d\theta \, e^{\alpha_i(\Phi(\theta))} + \ha \big( \Pi + \ha \partial \Phi \big| \Pi + \ha \partial \Phi \big) \notag\\
&\quad = \sum_{i=0}^\ell \frac{m^2_i}{4 \pi} \epsilon_i \int_{S^1} d\theta \, e^{\alpha_i(\Phi(\theta))} + \ha (\Pi| \Pi) + \mbox{\small $\frac{1}{8}$} (\partial \Phi|\partial \Phi) + \ha ( \Pi | \partial \Phi ).
\end{align}
Here we introduced the Toda masses $m_i \in \RR_{\geq 0}$ by $m_i^2 \coloneqq \mu_i \nu_i$ and used the relation $\langle E_i, F_j \rangle = \delta_{ij} \epsilon_i$ for all $i, j = 0, \ldots, \ell$.
The first three terms on the right hand side of \eqref{Ham Toda compute} are exactly (half) the Hamiltonian of affine Toda field theory
\begin{equation} \label{Toda Ham def}
H \coloneqq (\Pi| \Pi) + \qa (\partial \Phi|\partial \Phi) + \sum_{i=0}^\ell \frac{m^2_i}{2 \pi} \epsilon_i \int_{S^1} d\theta \, e^{\alpha_i(\Phi(\theta))}.
\end{equation}
Moreover, the last term on the right hand side of \eqref{Ham Toda compute} is (half) the momentum since
\begin{equation*}
P \coloneqq \widetilde{\pi}_{\bm \lvl}(\P) = -i \widetilde{\pi}_{\bm \lvl}(\cocent^{(0)}_{\ul 0}) = -i \sum_{n \in \ZZ} \sum_{k=1}^\ell n H_{k, -n}^{(+)} H^{k(-)}_n = (\Pi | \partial \Phi)
\end{equation*}
where the second equality is from the definition \eqref{momentum def} of the formal momentum, the third equality from \eqref{pi D l Toda c} and the last equality uses the definition \eqref{Toda fields def} of the Toda field and its conjugate momentum.

Consider now the terms in $\ha \big( \mathcal A^0_{\ul 0} | \mathcal A^0_{\ul 0} \big)$ containing $\cocent$ and $\cent$. They are of the form
\begin{equation} \label{DK terms Toda}
\ha \Big(\cocent^{(0)}_{\ul 0} \cent^{(0)}_{\ul 0} + \cent^{(0)}_{\ul 0} \cocent^{(0)}_{\ul 0} \Big) = \cocent^{(0)}_{\ul 0} \cent^{(0)}_{\ul 0}.
\end{equation}
We have $\widetilde{\pi}_{\bm \lvl} \big( \cent^{(0)}_{\ul 0} \big) = i$ and $\widetilde{\pi}_{\bm \lvl} \big( \cocent^{(0)}_{\ul 0} \big) = i (\Pi | \partial \Phi)$, hence combining the above we find
\begin{equation*}
H_0 \coloneqq \widetilde{\pi}_{\bm \lvl}(\H^0_1) = \ha H - \ha (\Pi | \partial \Phi) = \ha (H - P).
\end{equation*}
In particular, as was the case for the $\ZZ_T$-graded coset $\sigma$-model and its deformation in \S\ref{sec: cycl examples}, the Hamiltonian \eqref{Toda Ham def} of affine Toda field theory is given by $H = 2 H_0 + P$.

Applying the homomorphism $\rep \otimes \widetilde{\pi}_{\bm \lvl}$ to the Lax equation $\{ \H^0_1, \L(z) \} = [\mathcal M^0_1(z), \L(z)]$ from Proposition \ref{prop: Lax equations} and dividing through by $\varphi(z)$ we find
\begin{equation*}
\{ H_0, \mathscr L(z) \} = z^{-1} \big[ (\rep \otimes \widetilde{\pi}_{\bm \lvl})\A^0_{\ul 1}, \partial + \mathscr L(z) \big]
\end{equation*}
Combining this with the fact that
$\{ P, \mathscr L(z) \} = \partial \mathscr L(z)$ and using the definition of $H$ we find the zero curvature equation
\begin{equation*}
\{ H, \mathscr L(z) \} = \partial \mathscr M(z) + \big[ \mathscr L(z), \mathscr M(z) \big],
\end{equation*}
where we defined $\mathscr M(z) \coloneqq \mathscr L(z) - 2 (\rep \otimes \widetilde{\pi}_{\bm \lvl})\A^0_{\ul 1}$. The explicit form of the latter reads
\begin{equation*}
\mathscr M(z) = \Pi + \ha \partial \Phi + \frac{1}{\sqrt{2}} \sum_{j=0}^\ell \Big( - e^{\alpha_j(\Phi)} z^{-1} \mu_j F_j + z \, \nu_j E_j \Big).
\end{equation*}
In terms of the gauge transformed Lax matrix \eqref{Lax aff Toda} we have
\begin{equation*}
\{ H, \mathscr L'(z) \} = \partial \mathscr M'(z) + \big[ \mathscr L'(z), \mathscr M'(z) \big],
\end{equation*}
where $\mathscr M'(z) \coloneqq - \Pi + e^{\frac{1}{2} \ad \Phi} \mathscr M(z)$ is given explicitly by
\begin{equation*}
\mathscr M'(z) = \ha \partial \Phi + \frac{1}{\sqrt{2}} \sum_{j=0}^\ell e^{\frac{1}{2} \alpha_j(\Phi)} \big( - z^{-1} \mu_j F_j + z \, \nu_j E_j \big).
\end{equation*}
Up to differences in conventions, this agrees with the temporal component of the Lax connection in the standard treatment of affine Toda field theory \cite[ch. 12]{BBT}.

\subsubsection{The $\g = \mathfrak{sl}_2$ case: sinh-Gordon model} \label{sec: sinh-Gordon}

It is helpful to illustrate the above general procedure for obtaining the Lax matrix \eqref{Lax aff Toda gauge} of affine Toda field theory from the formal Lax matrix
\begin{equation} \label{Gaudin Lax for sinh-Gordon}
\L(z) = \A^0_{\ul 0} z^{-1} + \A^0_{\ul 1} z^{-2} - \A^{\infty}_{\ul 1}
\end{equation}
of the dihedral affine Gaudin model 
explicitly in the case $\g = \mathfrak{sl}_2$, which corresponds to the sinh-Gordon model.

Let $\{ E, H, F \}$ be the standard basis of $\g = \mathfrak{sl}_2$. A basis of the corresponding untwisted affine Kac-Moody algebra $\tg = \widetilde{\mathfrak{sl}}_2 = \mathfrak{sl}_2[t,t^{-1}] \oplus \CC \cent \oplus \CC \cocent$ then consists of $E_n = E \otimes t^n$, $H_n = H \otimes t^n$ and $F_n = F \otimes t^n$ for all $n \in \ZZ$ as well as $\cent$ and $\cocent$. Its dual basis, with respect to the standard bilinear form on $\widetilde{\mathfrak{sl}}_2$, cf. \eqref{ip on tg}, induced from the non-degenerate invariant bilinear form $\langle \cdot, \cdot \rangle$ on $\mathfrak{sl}_2$ normalised such that $\langle H, H \rangle = 2$, then consists of the elements $F_{-n}$, $\ha H_{-n}$ and $E_{-n}$ for all $n \in \ZZ$ as well as $\cocent$ and $\cent$. In the notation introduced at the start of \S\ref{sec: Toda FT} we have $E_1 = E^1 = F_0 = F^0 = E$, $F_1 = F^1 = E_0 = E^0 = F$, $H_1 = H$ and $H^1 = \ha H$. As we shall not use this notation here it should not lead to confusion with the above notation $E_n$, $F_n$ and $H_n$ for $n \in \ZZ$.

We extend the Coxeter automorphism $\sigma \in \Aut \mathfrak{sl}_2$ defined by $\sigma(E) = - E$, $\sigma(H) = H$ and $\sigma(F) = - F$ to the affine Kac-Moody algebra, cf. \S\ref{sec: tg action of Pi}, by letting $\sigma(E_n) = - E_n$, $\sigma(H_n) = H_n$ and $\sigma(F_n) = - F_n$ for any $n \in \ZZ$ as well as $\sigma(\cent) = \cent$ and $\sigma(\cocent) = \cocent$. 
A basis for the $1$-eigenspace $\tg_{(0), \CC} \subset \tg$ of $\sigma \in \Aut \widetilde{\mathfrak{sl}}_2$ consists of $H_n$, $n \in \ZZ$ together with $\cent$ and $\cocent$, while the $(-1)$-eigenspace $\tg_{(1), \CC} \subset \tg$ has basis elements $E_n$ and $F_n$ for $n \in \ZZ$.

The affine Gaudin model we are considering has divisor $\D = 2 \cdot 0 + 2 \cdot \infty$. This means that to the sites at the origin and infinity we attach certain copies of the basis elements of the affine Kac-Moody algebra $\widetilde{\mathfrak{sl}}_2$, namely 
\begin{align*}
\textup{origin}: \qquad & H^{(0)}_{n, \ul 0}, E^{(0)}_{n, \ul 1} \textup{ and } F^{(0)}_{n, \ul 1} \textup{ for each } n \in \ZZ, \textup{ and } \cent^{(0)}_{\ul 0}, \cocent^{(0)}_{\ul 0},\\
\textup{infinity}: \qquad & E^{(\infty)}_{n, \ul 1} \textup{ and } F^{(\infty)}_{n, \ul 1} \textup{ for each } n \in \ZZ.
\end{align*}
In order to avoid overburdening the present section with technicalities, we shall not discuss the precise definitions of these objects further here. 
We refer to \S\ref{sec: real Takiff} for the definition of the Lie algebra $\tg^\D_\CC$ to which they belong. We shall simply note that their Lie brackets read, cf. \eqref{non-trivial PB Toda},
\begin{subequations} \label{com H0 E0inf F0inf}
\begin{alignat}{2}
\big[ H^{(0)}_{n, \ul 0}, H^{(0)}_{-n, \ul 0} \big] &= 2 n \cent^{(0)}_{\ul 0}, &\qquad
\big[ H^{(0)}_{m, \ul 0}, E^{(0)}_{n, \ul 1} \big] &= 2 E^{(0)}_{m+n, \ul 1},\\
\big[ H^{(0)}_{m, \ul 0}, F^{(0)}_{n, \ul 1} \big] &= - 2 F^{(0)}_{m+n, \ul 1}, &\qquad
\big[ \cocent^{(0)}_{\ul 0}, H^{(0)}_{n, \ul 0} \big] &= n H^{(0)}_{n, \ul 0},\\
\big[ \cocent^{(0)}_{\ul 0}, E^{(0)}_{n, \ul 1} \big] &= n E^{(0)}_{n, \ul 1}, &\qquad
\big[ \cocent^{(0)}_{\ul 0}, F^{(0)}_{n, \ul 1} \big] &= n F^{(0)}_{n, \ul 1},
\end{alignat}
\end{subequations}
for all $m, n \in \ZZ$, with all other Lie brackets being trivial.

The three formal fields $\A^0_{\ul 0}$, $\A^0_{\ul 1}$ and $\A^\infty_{\ul 1}$ of the dihedral affine Gaudin model under consideration may now be written in the present $\mathfrak{sl}_2$ case as, cf. \eqref{formal Gaudin fields Toda},
\begin{subequations} \label{formal fields sinh-Gordon}
\begin{align}
\label{formal fields sinh-Gordon a} \A^0_{\ul 0} &= \cocent \otimes \cent^{(0)}_{\ul 0} + \cent \otimes \cocent^{(0)}_{\ul 0} + \sum_{n \in \ZZ} \ha H_{-n} \otimes H^{(0)}_{n, \ul 0}, \\
\label{formal fields sinh-Gordon b} \A^0_{\ul 1} &= \sum_{n \in \ZZ} \big( E_{-n} \otimes F^{(0)}_{n, \ul 1} + F_{-n} \otimes E^{(0)}_{n, \ul 1} \big), \\
\label{formal fields sinh-Gordon c} \A^\infty_{\ul 1} &= \sum_{n \in \ZZ}  \big( E_{-n} \otimes F^{(\infty)}_{n, \ul 1} + F_{-n} \otimes E^{(\infty)}_{n, \ul 1} \big).
\end{align}
\end{subequations}
Following the nomenclature introduced at the start of \S\ref{sec: formal fields}, we refer to the first tensor factor in each of these expressions as the \emph{auxiliary factor}. 
In order to obtain more familiar looking $\mathfrak{sl}_2$-valued fields on the circle $S^1$ from the above formal fields, we should represent the affine Kac-Moody algebra elements in the auxiliary factor in terms of $\mathfrak{sl}_2$-valued connections on $S^1$ with trigonometric polynomial coefficients, cf. \S\ref{sec: classical fields}. Specifically, we apply the representation $\rep$ defined in Lemma \ref{lem: eval rep}, which in the present case explicitly reads
\begin{gather*}
\rep(\cocent) = - i \partial_\theta, \qquad
\rep(\cent) = 0, \qquad
\rep(E_n) = E e^{i n \theta}, \\
\rep(H_n) = H e^{i n \theta}, \qquad
\rep(F_n) = F e^{i n \theta},
\end{gather*}
in terms of the coordinate $\theta \in \RR / 2 \pi \ZZ = S^1$ on the circle. To be even more concrete, we shall modify the above representation so that the basis elements of $\mathfrak{sl}_2$ are further represented by $2 \times 2$ matrices. That is, instead of applying $\rep$ to the auxiliary factor of the formal fields and the formal Lax matrix, we shall use a representation $\widetilde{\rep}$ of $\tg$ into $2 \times 2$ matrix connections on the circle $S^1$, defined by
\begin{gather*}
\widetilde{\rep}(\cocent) = - i \partial_\theta, \qquad
\widetilde{\rep}(\cent) = 0, \qquad
\widetilde{\rep}(E_n) =
\left( \begin{matrix}
0 & e^{i n \theta}\\
0 & 0
\end{matrix} \right), \\
\widetilde{\rep}(H_n) = 
\left( \begin{matrix}
e^{i n \theta} & 0\\
0 & - e^{i n \theta}
\end{matrix} \right), \qquad
\widetilde{\rep}(F_n) = 
\left( \begin{matrix}
0 & 0\\
e^{i n \theta} & 0
\end{matrix} \right).
\end{gather*}
For instance, in the case of the formal field $\A^0_{\ul 1}$ defined in \eqref{formal fields sinh-Gordon b} we obtain
\begin{equation*}
(\widetilde{\rep} \otimes \id) \A^0_{\ul 1} = \left( \begin{matrix}
0 & \sum_{n \in \ZZ} e^{-in\theta} F^{(0)}_{n, \ul 1}\\
\sum_{n \in \ZZ} e^{-in\theta} E^{(0)}_{n, \ul 1} & 0
\end{matrix} \right).
\end{equation*}
The two off-diagonal components of the matrix on the right hand side are formal distributions on $S^1$ with Fourier coefficients $F^{(0)}_{n, \ul 1}$ and $E^{(0)}_{n, \ul 1}$, respectively. That is, their integral pairing with any trigonometric polynomial yields an element of the Lie algebra $\tg^\D_\CC$.

To formulate the sinh-Gordon model as an affine Gaudin model we should realise all the fields of the latter in terms of sinh-Gordon fields. This is the purpose of the homomorphism \eqref{pi D l Toda} defined in \S\ref{sec: Toda fields}, expressing the Fourier modes of the affine Gaudin model fields in terms of those of the classical Toda fields. Let us spell out its definition in the present $\mathfrak{sl}_2$ case. Let $\mathscr H$ denote the Lie algebra spanned by elements $E^{(+)}_n$, $F^{(+)}_n$ and $H^{(\pm)}_n$ for all $n \in \ZZ$ subject to the relations, cf. \eqref{com Hpm Ep},
\begin{equation} \label{com Hpm Ep sinh}
\big[ H^{(+)}_n, H^{(-)}_{-n} \big] = 2, \qquad
\big[ H^{(+)}_m, E^{(+)}_n \big] = 2 E^{(+)}_{m+n}, \qquad
\big[ H^{(+)}_m, F^{(+)}_n \big] = -2 F^{(+)}_{m+n},
\end{equation}
with all other Lie brackets being trivial.
In terms of these, let us introduce the sinh-Gordon field and its conjugate momentum as
\begin{equation*}
\phi(\theta) \coloneqq \ha \sum_{n \in \ZZ} e^{-in \theta} H^{(-)}_n, \qquad
\pi(\theta) \coloneqq \ha \sum_{n \in \ZZ} e^{-in \theta} H^{(+)}_n.
\end{equation*}
Note that their Poisson bracket is normalised as $\{ \pi(\theta), \phi(\theta') \} = \ha \delta_{\theta \theta'}$. This can be seen directly from the first relation in \eqref{com Hpm Ep sinh} or, alternatively, from the Poisson bracket $\big\{ \Pi_\1(\theta), \Phi_\2(\theta') \big\} = C^{(0)}_{\1\2} \delta_{\theta \theta'}$ by noting that we have $C^{(0)} = \ha H \otimes H$ and observing that the scalar fields $\phi$ and $\pi$ are related to the $\h_0$-valued fields $\Phi$ and $\Pi$ defined in \eqref{Toda fields def} simply as $\Phi = \phi \, H$ and $\Pi = \pi \, H$. Moreover, in the notation \eqref{exp Toda def}, which was motivated by Lemma \ref{lem: exp justification}, the exponential of (twice) the sinh-Gordon field $\phi$ is then given by $e^{2 \phi(\theta)} = \sum_{n \in \ZZ} e^{-in\theta} E^{(+)}_n$ where we used the fact that $\alpha_1(H) = 2$ and also $e^{- 2 \phi(\theta)} = \sum_{n \in \ZZ} e^{-in\theta} F^{(+)}_n$ since we have $\alpha_0(H) = - \alpha_1(H) = - 2$.

The desired homomorphism \eqref{pi D l Toda} in the present case takes the form
\begin{subequations} \label{Gaudin to sinh-Gordon fields}
\begin{gather}
\label{Gaudin to sinh-Gordon fields a}
\widetilde{\pi}_{\bm \lvl} \big( E^{(0)}_{n, \ul 1} \big) = \frac{m}{\sqrt{2}} E^{(+)}_n, \qquad
\widetilde{\pi}_{\bm \lvl} \big( F^{(0)}_{n, \ul 1} \big) = \frac{m}{\sqrt{2}} F^{(+)}_n, \\
\label{Gaudin to sinh-Gordon fields b}
\widetilde{\pi}_{\bm \lvl} \big( E^{(\infty)}_{n, \ul 1} \big) = \widetilde{\pi}_{\bm \lvl} \big( F^{(\infty)}_{n, \ul 1} \big) = - \frac{m}{\sqrt{2}} \delta_{n, 0}, \qquad
\widetilde{\pi}_{\bm \lvl} \big( \cent^{(0)}_{\ul 0} \big) = i, \\
\label{Gaudin to sinh-Gordon fields c} \widetilde{\pi}_{\bm \lvl} \big( H^{(0)}_{n, \ul 0} \big) = H^{(+)}_n - \frac{i n}{2} H^{(-)}_n, \qquad
\widetilde{\pi}_{\bm \lvl} \big( \cocent^{(0)}_{\ul 0} \big) = \ha \sum_{n \in \ZZ} \sum_{k=1}^\ell n H_{-n}^{(+)} H^{(-)}_n,
\end{gather}
\end{subequations}
for all $n \in \ZZ$. Here we have fixed all the parameters $\mu_0 = \mu_1 = \nu_0 = \nu_1 = m$ in terms of a single real parameter $m \in \RR_{\geq 0}$ which will turn out to be the mass of the sinh-Gordon field. Note that to make sense of the infinite sum on the right hand side of the last expression in \eqref{Gaudin to sinh-Gordon fields c}, the map $\widetilde{\pi}_{\bm \lvl}$ takes value in a suitable completion of the symmetric algebra $S(\mathscr H)$, see \S\ref{sec: Toda fields} for details. More precisely, we work in its quotient by the relation $\sum_{p \in \ZZ} p E^{(+)}_{n-p} H^{(-)}_{p} = n E^{(+)}_n$, cf. Lemma \ref{lem: Toda exp}. According to Lemma \ref{lem: tilde pi lvl hom}, the map $\widetilde{\pi}_{\bm \lvl}$ defined by \eqref{Gaudin to sinh-Gordon fields} then provides a realisation of the Lie algebra $\tg^\D_\CC$ with relations \eqref{com H0 E0inf F0inf}.

We may now proceed to construct the Lax matrix of the sinh-Gordon model from the formal Lax matrix \eqref{Gaudin Lax for sinh-Gordon} of the affine Gaudin model. Substituting the expressions \eqref{formal fields sinh-Gordon} for the formal fields of the affine Gaudin model into $\L(z)$, it explicitly reads
\begin{align*}
\L(z) &= \bigg( \cocent \otimes \cent^{(0)}_{\ul 0} + \cent \otimes \cocent^{(0)}_{\ul 0} + \sum_{n \in \ZZ} \ha H_{-n} \otimes H^{(0)}_{n, \ul 0} \bigg) z^{-1}\\
&\quad + \sum_{n \in \ZZ} \big( E_{-n} \otimes F^{(0)}_{n, \ul 1} + F_{-n} \otimes E^{(0)}_{n, \ul 1} \big) z^{-2}
- \sum_{n \in \ZZ} \big( E_{-n} \otimes F^{(\infty)}_{n, \ul 1} + F_{-n} \otimes E^{(\infty)}_{n, \ul 1} \big).
\end{align*}
To begin with, applying to it just the representation $\widetilde{\rep} \otimes \id \otimes \id$, \emph{i.e.} applying $\widetilde{\rep}$ defined above to the auxiliary factor, we find
\begin{align*}
&(\widetilde{\rep} \otimes \id \otimes \id) \L(z) =
- \cent^{(0)}_{\ul 0} z^{-1} i \partial_\theta + \sum_{n \in \ZZ} \left( \begin{matrix}
\ha e^{- i n \theta} H^{(0)}_{n, \ul 0} & 0\\
0 & - \ha e^{- i n \theta} H^{(0)}_{n, \ul 0}
\end{matrix} \right) z^{-1}\\
&\qquad\qquad + \sum_{n \in \ZZ} \left( \begin{matrix}
0 & e^{- i n \theta} F^{(0)}_{n, \ul 1}\\
e^{- i n \theta} E^{(0)}_{n, \ul 1} & 0
\end{matrix} \right) z^{-2}
- \sum_{n \in \ZZ} \left( \begin{matrix}
0 & e^{- i n \theta} F^{(\infty)}_{n, \ul 1}\\
e^{- i n \theta} E^{(\infty)}_{n, \ul 1} & 0
\end{matrix} \right).
\end{align*}
Next, applying the further representation $\id \otimes \, \widetilde{\pi}_{\bm \lvl} \otimes \id$, with $\widetilde{\pi}_{\bm \lvl}$ given in \eqref{Gaudin to sinh-Gordon fields}, to the above expression we find
\begin{align*}
(\widetilde{\rep} \otimes \widetilde{\pi}_{\bm \lvl} \otimes \id) \L(z) &=
\frac{\partial_\theta}{z} + \frac{1}{z} \left( \begin{matrix}
\pi(\theta) + \ha \phi'(\theta) & 0\\
0 & - \pi(\theta) - \ha \phi'(\theta)
\end{matrix} \right)\\
&\qquad\qquad\qquad + \frac{m}{\sqrt{2} z^2} \left( \begin{matrix}
0 & e^{-2 \phi(\theta)}\\
e^{2 \phi(\theta)} & 0
\end{matrix} \right)
+ \frac{m}{\sqrt{2}} \left( \begin{matrix}
0 & 1\\
1 & 0
\end{matrix} \right).
\end{align*}
Comparing this with the general expression \eqref{Lax twist from formal}, we read off the twist function as the coefficient of the $\partial_\theta$-term to be $\varphi(z) = z^{-1}$ and then the Lax matrix is
\begin{equation} \label{Sinh-Gordon NUL}
\mathscr L(z) =
\left( \begin{matrix}
\pi(\theta) + \ha \phi'(\theta) & \frac{m}{\sqrt{2}} \big( z^{-1} e^{-2 \phi(\theta)} + z \big)\\
\frac{m}{\sqrt{2}} \big( z^{-1} e^{2 \phi(\theta)} + z \big) & - \pi(\theta) - \ha \phi'(\theta)
\end{matrix} \right).
\end{equation}
This expression can equally be obtained directly from the general result \eqref{Lax aff Toda gauge}, specialised to the case $\g = \mathfrak{sl}_2$, after representing the auxiliary factor of this $\mathfrak{sl}_2$-valued Lax matrix in terms of $2 \times 2$ matrices. 

The Lax matrix \eqref{Sinh-Gordon NUL} is a non-ultralocal version of the Lax matrix of the sinh-Gordon model. But it is gauge equivalent to the standard ultralocal Lax matrix.
Indeed, if we perform a gauge transformation by $\textup{diag}(e^{\frac{1}{2} \phi(\theta)}, e^{-\frac{1}{2} \phi(\theta)})$ on it, or even more directly if we represent the auxiliary factor of the Lax matrix \eqref{Lax aff Toda} in the present case by $2 \times 2$ matrices, then we arrive at
\begin{equation} \label{sinh-Gordon UL}
\mathscr L'(z) =
\left( \begin{matrix}
\pi(\theta) & \frac{m}{\sqrt{2}} \big( z^{-1} e^{-\phi(\theta)} + z e^{\phi(\theta)} \big)\\
\frac{m}{\sqrt{2}} \big( z^{-1} e^{\phi(\theta)} + z e^{-\phi(\theta)} \big) & - \pi(\theta)
\end{matrix} \right).
\end{equation}
This is the standard ultralocal Lax matrix of the sinh-Gordon model in the Hamiltonian formalism, see for instance \cite[\S12.7]{BBT} where the momentum is written as $\pi = \ha \partial_t \phi$ so that the canonical Poisson bracket reads $\{ \partial_t \phi(\theta), \phi(\theta') \} = \delta_{\theta\theta'}$.

Unlike its non-ultralocal counterpart \eqref{Sinh-Gordon NUL}, however, and this is the crucial point, the Lax matrix \eqref{sinh-Gordon UL} cannot be obtained as a representation of the formal Lax matrix of an affine Gaudin model in any natural way. One can see this directly from the form of the Hamiltonian \eqref{Toda Ham def} of the affine Gaudin model obtained in \S\ref{sec: Toda FT Lax Ham}. Indeed, in the present case with $\g = \mathfrak{sl}_2$, the Hamiltonian and momentum take on the simple form
\begin{align*}
H &= \frac{1}{2 \pi} \int_{S^1} d\theta \big( 2 \, \pi(\theta)^2 + \ha \phi'(\theta)^2 + m^2 e^{2 \phi(\theta)} + m^2 e^{- 2 \phi(\theta)} \big),\\
P &= \frac{1}{\pi} \int_{S^1} d\theta \; \pi(\theta) \phi'(\theta).
\end{align*}
Here we have used the identification $\Phi = \phi \, H$ and $\Pi = \pi \, H$ made above, together with the fact that $\langle H, H \rangle = 2$. Recall here also the definition \eqref{ip Conn g} which according to Lemma \ref{lem: eval rep} is identified with the bilinear form $(\cdot | \cdot)$ on $\Loop \g$ defined in \eqref{sec: aff KM}. Since the ultralocal Lax matrix \eqref{sinh-Gordon UL} does not explicitly depend on the derivative $\phi'(\theta)$ of the sinh-Gordon field, the term $\phi'(\theta)^2$ in the density of the Hamiltonian cannot arise from a bilinear expression in the Lax matrix $\mathscr L'(z)$, as it should do if the ultralocal formulation of sinh-Gordon could be given an affine Gaudin model interpretation. One of the effects of the gauge transformation from \eqref{sinh-Gordon UL} to \eqref{Sinh-Gordon NUL} is precisely to introduce such a dependence on $\phi'(\theta)$ in the Lax matrix.

\appendix

\section{Notations and some useful lemmas}

\subsection{Dual pairs} \label{app: dual pairs}

Let $V$ and $W$ be a pair of real vector spaces and $\langle \cdot, \cdot \rangle : V \times W \to \RR$ be a bilinear form. For any two vectors $x \in V$ and $y \in W$ satisfying $\langle x, y \rangle = 0$ we will write $x \perp y$. More generally, for any subspace $E \subset V$ and $y \in W$ we write $E \perp y$ if $\langle x, y \rangle = 0$ for all $x \in E$. Similarly, we write $x \perp F$ if $\langle x, y \rangle = 0$ for any $y \in F$. Also, for any two subspaces $E \subset V$ and $F \subset W$, we write $E \perp F$ if $\langle x, y \rangle = 0$ for all $x \in E$ and $y \in F$. We define the orthogonal complement in $W$ of a subspace $E \subset V$ by
\begin{equation*}
E^\perp \coloneqq \{ y \in W \,|\, E \perp y \}.
\end{equation*}
Likewise, for any subspace $F \subset W$ we define its orthogonal complement in $V$ by
\begin{equation*}
F^\perp \coloneqq \{ x \in V \,|\, x \perp F \}.
\end{equation*}
We will call the triple $(V, W, \langle \cdot, \cdot \rangle)$ a \emph{dual pair} if the bilinear form $\langle \cdot, \cdot \rangle$ is non-degenerate both on the left, \emph{i.e.} $W^\perp = \{ 0 \}$, and on the right, \emph{i.e.} $V^\perp = \{ 0 \}$. In other words, we say $(V, W, \langle \cdot, \cdot \rangle)$ is a dual pair if $x \perp W$ implies $x = 0$ and $V \perp y$ implies $y = 0$.

\begin{lemma} \label{lem: ABC}
Let $V$ be a vector space. For any subspaces $A$, $B$ and $C$ such that $C \subset A$ and $B \cap C = \{ 0 \}$ we have $A \cap (B \dotplus C) = (A \cap B) \dotplus C$. In particular, if $V = B \dotplus C$ and $A \cap B = \{ 0 \}$ then $A = C$.
\begin{proof}
The inclusion $(A \cap B) \dotplus C \subset A \cap (B \dotplus C)$ is obvious since $C \subset A$, $C \subset B \dotplus C$ and $A \cap B \subset A \cap (B \dotplus C)$. For the reverse inclusion, let $x \in A \cap (B \dotplus C)$. We can then write $x = b + c$ for some $b \in B$ and $c \in C$. Then $b = x - c \in A$ since $x \in A$ and $c \in C \subset A$. Therefore $b \in A \cap B$ so that $x = b + c \in (A \cap B) \dotplus C$, as required.
\end{proof}
\end{lemma}

\begin{lemma} \label{lem: dual pairs}
Let $(V, W, \langle \cdot, \cdot \rangle)$ be a dual pair of real vector spaces. Suppose that there are direct sum decompositions $V = V_+ \dotplus V_-$ and $W = W_+ \dotplus W_-$ such that $V_{\pm} \perp W_{\pm}$. Then $V_\pm^\perp = W_\pm$ and $W_\pm^\perp = V_\pm$. In particular, $(V_\pm, W_\mp, \langle \cdot, \cdot \rangle|_{V_\pm \times W_\mp})$ are dual pairs.
\begin{proof}
We will show that $V_+^\perp = W_+$, the proof of the other statements $V_-^\perp = W_-$ and $W_\pm^\perp = V_\pm$ being very similar. Since $V_{\pm} \perp W_{\pm}$ we have $W_\pm \subset V_\pm^\perp$. Now applying Lemma \ref{lem: ABC} with $A = V_+^\perp$, $B = W_-$ and $C = W_+$ of $W$ which satisfy $B \dotplus C = W$, and noting that $A \cap B = V_+^\perp \cap W_- \subset V_+^\perp \cap V_-^\perp \subset V^\perp = \{ 0 \}$, we conclude $V_+^\perp = W_+$.

To see that $(V_\pm, W_\mp, \langle \cdot, \cdot \rangle|_{V_\pm \times W_\pm})$ are dual pairs, let $x \in V_\pm$ and suppose $x \perp W_\mp$. Then $x \in V_\pm \cap W_\mp^\perp = V_\pm \cap V_\mp = \{ 0 \}$ so that $x = 0$. Similarly, if $y \in W_\mp$ is such that $V_\pm \perp y$ then $y \in W_\mp \cap V_\pm^\perp = W_\mp \cap W_\pm = \{ 0 \}$ and hence $y = 0$.
\end{proof}
\end{lemma}

Let $(V, W, \langle \cdot, \cdot \rangle)$ be a dual pair. We can endow any pair of subspaces $E \subset V$ and $F \subset W$ with the restricted bilinear form $\langle \cdot, \cdot \rangle|_{E \times F} : E \times F \to \RR$. In general this will be degenerate, with left kernel given by $E \cap F^\perp = \{ x \in E \,|\, x \perp F \}$ and right kernel given by $F \cap E^\perp = \{ y \in F \,|\, E \perp y \}$.

\begin{lemma} \label{lem: induced bilinear}
The triple $\big( E / (E \cap F^\perp), F / (F \cap E^\perp), \langle \cdot, \cdot \rangle \big)$, with bilinear form
\begin{equation} \label{induced bilinear E F}
\langle \cdot, \cdot \rangle : E / (E \cap F^\perp) \times F / (F \cap E^\perp) \longrightarrow \RR,
\end{equation}
induced from the restriction $\langle \cdot, \cdot \rangle|_{E \times F} : E \times F \to \RR$, is a dual pair.
\begin{proof}
The bilinear form \eqref{induced bilinear E F} is explicitly given by
\begin{equation*}
\langle x + E \cap F^\perp, y + F \cap E^\perp \rangle = \langle x, y \rangle
\end{equation*}
for any $x \in E$ and $y \in F$. Note that this is clearly well defined since $E \perp (F \cap E^\perp)$ and $(E \cap F^\perp) \perp F$. It remains to show non-degeneracy on both sides. So suppose that $x \in E$ satisfies $\langle x + E \cap F^\perp, y + F \cap E^\perp \rangle = 0$ for all $y \in F$. This means $\langle x, y \rangle = 0$ for all $y \in F$ and hence $x \in E \cap F^\perp$. Thus $x + E \cap F^\perp = 0$ in $E / (E \cap F^\perp)$, as required. The proof of non-degeneracy on the right is completely analogous.
\end{proof}
\end{lemma}

\subsection{Tensor index notation} \label{sec: tensor index}

Let $\a$, $\b$ and $\c$ be complex Lie algebras, each equipped with an anti-linear involutive automorphism which we commonly denote by $\tau$. Let us suppose further that these are equipped with descending $\ZZ_{\geq 0}$-filtrations $(\Fil_n \a)_{n \in \ZZ_{\geq 0}}$, $(\Fil_n \b)_{n \in \ZZ_{\geq 0}}$ and $(\Fil_n \c)_{n \in \ZZ_{\geq 0}}$ as complex Lie algebras, respectively. So in particular, for every $m, n \in \ZZ_{\geq 0}$ we have $[\Fil_m \a, \Fil_n \a] \subset \Fil_{m+n} \a$ and similarly for $\b$ and $\c$. Let $\mathfrak A$ be a commutative algebra over $\CC$. Consider elements
\begin{gather*}
\ms x = \sum_i \ms x^1_i \otimes \ms x^2_i \otimes f_i \in \a \hotimes \b \otimes \mathfrak A, \qquad
\ms y = \sum_i \ms y^1_i \otimes \ms y^2_i \otimes g_i \in \a \hotimes \c \otimes \mathfrak A,\\
\ms z = \sum_i \ms z^1_i \otimes \ms z^2_i \otimes h_i \in \b \hotimes \c \otimes \mathfrak A,
\end{gather*}
where $\a \hotimes \b$ is the completion of the tensor product $\a \otimes \b$ with respect to the descending $\ZZ_{\geq 0}$-filtration with subspaces
\begin{equation*}
\Fil_n (\a \otimes \b) = \Fil_n \a \otimes \tau(\Fil_n \b) + \tau(\Fil_n \a) \otimes \Fil_n \b
\end{equation*}
for each $n \in \ZZ_{\geq 0}$ and similarly for $\a \hotimes \c$ and $\b \hotimes \c$. In particular, the sums over $i$ in the elements $\ms x$, $\ms y$ and $\ms z$ above may be infinite. Given any $\ms u \in \a$ and $\ms v \in \b$ we use the tensor index notation
\begin{equation*}
[\ms u_\1, \ms x_{\1\2}] \coloneqq \sum_i [\ms u, \ms x^1_i] \otimes \ms x^2_i \otimes f_i, \qquad
[\ms v_\2, \ms x_{\1\2}] \coloneqq \sum_i \ms x^1_i \otimes [\ms v, \ms x^2_i] \otimes f_i.
\end{equation*}
Similarly, following standard conventions we define the elements
\begin{gather*}
[ \ms x_{\1\2}, \ms y_{\1\3} ] \coloneqq \sum_{i,j} [\ms x^1_i, \ms y^1_j] \otimes \ms x^2_i \otimes \ms y^2_j \otimes f_i g_j, \quad
[ \ms y_{\1\3}, \ms z_{\2\3} ] \coloneqq \sum_{i,j} \ms y^1_i \otimes \ms z^1_j \otimes [ \ms y^2_i, \ms z^2_j ] \otimes g_i h_j,\\
[ \ms x_{\1\2}, \ms z_{\2\3} ] \coloneqq \sum_{i,j} \ms x^1_i \otimes [ \ms x^2_i, \ms z^1_j ] \otimes \ms z^2_j \otimes f_i h_j,
\end{gather*}
of $\a \hotimes \b \hotimes \c \otimes \mathfrak A$ where $\a \hotimes \b \hotimes \c$ denotes the completion of $\a \otimes \b \otimes \c$ with respect to the descending $\ZZ_{\geq 0}$-filtration defined by
\begin{align*}
\Fil_n (\a \otimes \b \otimes \c) &= \Fil_n \a \otimes \tau(\Fil_n \b) \otimes \c + \Fil_n \a \otimes \b \otimes \tau(\Fil_n \c) + \a \otimes \Fil_n \b \otimes \tau(\Fil_n \c)\\
&\qquad\quad + \tau(\Fil_n \a) \otimes \Fil_n \b \otimes \c + \tau(\Fil_n \a) \otimes \b \otimes \Fil_n \c + \a \otimes \tau(\Fil_n \b) \otimes \Fil_n \c
\end{align*}
for all $n \in \ZZ_{\geq 0}$.
If tensor indices appear in decreasing order then one should permute the tensor factors. For instance, if
$\ms w = \sum_i \ms w^1_i \otimes \ms w^2_i \otimes k_i \in \b \hotimes \a \otimes \mathfrak A$ then we let
\begin{equation*}
[ \ms w_{\2\1}, \ms y_{\1\3} ] \coloneqq \sum_{i,j} [ \ms w^2_i, \ms y^1_j ] \otimes \ms w^1_i \otimes \ms y^2_j \otimes k_i g_j,
\end{equation*}
and similarly for other possible permutations of the tensor indices.


\begin{thebibliography}{99}

\bibitem[BBT]{BBT}
  Babelon, O., Bernard, D. and Talon, M.:
  \emph{Introduction to classical integrable systems},
  Cambridge University Press (2003).

\bibitem[BR]{Babichenko:2012uq}
  A.~Babichenko and D.~Ridout,
  \emph{Takiff superalgebras and Conformal Field Theory},
  J.\ Phys.\ A {\bf 46} (2013) 125204.

\bibitem[BFHP]{Balog}
  J.~Balog, P.~Forgacs, Z.~Horvath and L.~Palla,
  \emph{A New family of $SU(2)$ symmetric integrable sigma models},
  Phys. Lett. B {\bf 324} (1994) 403--408.

\bibitem[BHK]{Bazhanov:2001xm}
  V.~V.~Bazhanov, A.~N.~Hibberd and S.~M.~Khoroshkin,
  \emph{Integrable structure of $W(3)$ conformal field theory, quantum Boussinesq theory and boundary affine Toda theory},
  Nucl.\ Phys.\ B {\bf 622} (2002) 475.

\bibitem[BKL]{Bazhanov:2014joa}
  V.~V.~Bazhanov, G.~A.~Kotousov and S.~L.~Lukyanov,
  \emph{Winding vacuum energies in a deformed O(4) sigma model},
  Nucl.\ Phys.\ B {\bf 889} (2014) 817.

\bibitem[BL]{Bazhanov:2013cua}
  V.~V.~Bazhanov and S.~L.~Lukyanov,
  \emph{Integrable structure of Quantum Field Theory: Classical flat connections versus quantum stationary states},
  JHEP {\bf 1409} (2014) 147.
  
\bibitem[BLZ1]{Bazhanov:1994ft}
  V.~V.~Bazhanov, S.~L.~Lukyanov and A.~B.~Zamolodchikov,
  \emph{Integrable structure of conformal field theory, quantum KdV theory and thermodynamic Bethe ansatz},
  Commun.\ Math.\ Phys.\  {\bf 177} (1996) 381.

\bibitem[BLZ2]{Bazhanov:1996dr}
  V.~V.~Bazhanov, S.~L.~Lukyanov and A.~B.~Zamolodchikov,
  \emph{Integrable structure of conformal field theory. II. Q operator and DDV equation},
  Commun.\ Math.\ Phys.\  {\bf 190} (1997) 247.

\bibitem[BLZ3]{Bazhanov:1998dq}
  V.~V.~Bazhanov, S.~L.~Lukyanov and A.~B.~Zamolodchikov,
  \emph{Integrable structure of conformal field theory. III. The Yang-Baxter relation},
  Commun.\ Math.\ Phys.\  {\bf 200} (1999) 297.

\bibitem[BLZ4]{Bazhanov:1998wj}
  V.~V.~Bazhanov, S.~L.~Lukyanov and A.~B.~Zamolodchikov,
  \emph{Spectral determinants for Schrodinger equation and Q operators of conformal field theory},
  J.\ Statist.\ Phys.\  {\bf 102} (2001) 567.

\bibitem[BLZ5]{Bazhanov:2003ni}
  V.~V.~Bazhanov, S.~L.~Lukyanov and A.~B.~Zamolodchikov,
  \emph{Higher level eigenvalues of Q operators and Schroedinger equation},
  Adv.\ Theor.\ Math.\ Phys.\  {\bf 7} (2003) no.4, 711.

\bibitem[BD2]{BelavinDrinfeld2}
  Belavin, A. A. and Drinfeld, V. G.:
  \emph{Triangle equations and simple Lie algebras},
  Classic Reviews in Mathematics and Mathematical Physics. 1.
  Amsterdam: Harwood Academic Publishers. vii, 91 p. (1998).

\bibitem[DLMV]{Delduc:2015xdm}
  F.~Delduc, S.~Lacroix, M.~Magro and B.~Vicedo,
  \emph{On the Hamiltonian integrability of the bi-Yang-Baxter sigma-model},
  JHEP {\bf 1603} (2016) 104.

\bibitem[DMV1]{Delduc:2012qb}
  F.~Delduc, M.~Magro and B.~Vicedo,
  \emph{Alleviating the non-ultralocality of coset sigma models through a generalized Faddeev-Reshetikhin procedure},
  JHEP {\bf 1208} (2012) 019.

\bibitem[DMV2]{Delduc:2012mk}
  F.~Delduc, M.~Magro and B.~Vicedo,
  \emph{A lattice Poisson algebra for the Pohlmeyer reduction of the $AdS_5 \times S^5$ superstring},
  Phys.Lett. B {\bf 713} (2012) 347--349.

\bibitem[DMV3]{Delduc:2013fga}
  F.~Delduc, M.~Magro and B.~Vicedo,
  \emph{On classical $q$-deformations of integrable sigma-models},
  JHEP {\bf 1311} (2013) 192.

\bibitem[DMV4]{Delduc:2013qra}
  F.~Delduc, M.~Magro and B.~Vicedo,
  \emph{An integrable deformation of the $AdS_5 \times S^5$ superstring action},
  Phys.\ Rev.\ Lett.\  {\bf 112} (2014) no.5,  051601.

\bibitem[DMV5]{Delduc:2014kha}
  F.~Delduc, M.~Magro and B.~Vicedo,
  \emph{Derivation of the action and symmetries of the $q$-deformed $AdS_{5} \times S^{5}$ superstring},
  JHEP {\bf 1410} (2014) 132.

\bibitem[DMV6]{Delduc:2014uaa}
  F.~Delduc, M.~Magro and B.~Vicedo,
  \emph{Integrable double deformation of the principal chiral model},
  Nucl. Phys. B {\bf 891} (2015) 312--321.

\bibitem[DDMST]{Dorey:2006an}
  P.~Dorey, C.~Dunning, D.~Masoero, J.~Suzuki and R.~Tateo,
  \emph{Pseudo-differential equations, and the Bethe ansatz for the classical Lie algebras},
  Nucl.\ Phys.\ B {\bf 772} (2007) 249.

\bibitem[DDT1]{Dorey:2000ma}
  P.~Dorey, C.~Dunning and R.~Tateo,
  \emph{Differential equations for general $SU(n)$ Bethe ansatz systems},
  J.\ Phys.\ A {\bf 33} (2000) 8427.

\bibitem[DDT2]{Dorey:2007zx}
  P.~Dorey, C.~Dunning and R.~Tateo,
  \emph{The ODE/IM Correspondence},
  J.\ Phys.\ A {\bf 40} (2007) R205.

\bibitem[DFNT]{Dorey:2012bx}
  P.~Dorey, S.~Faldella, S.~Negro and R.~Tateo,
  \emph{The Bethe Ansatz and the Tzitzeica-Bullough-Dodd equation},
  Phil.\ Trans.\ Roy.\ Soc.\ Lond.\ A {\bf 371} (2013) 20120052.

\bibitem[DT1]{Dorey:1998pt}
  P.~Dorey and R.~Tateo,
  \emph{Anharmonic oscillators, the thermodynamic Bethe ansatz, and nonlinear integral equations},
  J.\ Phys.\ A {\bf 32} (1999) L419.

\bibitem[DT2]{Dorey:1999pv}
  P.~Dorey and R.~Tateo,
  \emph{Differential equations and integrable models: The SU(3) case},
  Nucl.\ Phys.\ B {\bf 571} (2000) 583,
   Erratum: [Nucl.\ Phys.\ B {\bf 603} (2001) 582].

\bibitem[AD]{Adamopoulou:2014fca}
  P.~Adamopoulou and C.~Dunning,
  \emph{Bethe Ansatz equations for the classical $A_n^{(1)}$ affine Toda field theories},
  J.\ Phys.\ A {\bf 47} (2014) 205205.

\bibitem[ISST]{Itsios:2014vfa}
  G.~Itsios, K.~Sfetsos, K.~Siampos and A.~Torrielli,
  \emph{The classical Yang-Baxter equation and the associated Yangian symmetry of gauged WZW-type theories},
  Nucl. Phys. B {\bf 889} (2014) 64--86.
  
\bibitem[FR]{Faddeev:1985qu}
  L.~D.~Faddeev and N.~Y.~Reshetikhin,
  \emph{Integrability of the Principal Chiral Field Model in (1+1)-dimension},
  Annals Phys.\  {\bf 167} (1986) 227.

\bibitem[FT1]{FT}
  L.~Faddeev and L.~Takhtajan,
  \emph{The quantum method of the inverse problem and the Heisenberg XYZ-model},
  Russ. Math. Surveys \textbf{34:5} (1979) 1168.
  
\bibitem[FT2]{FTbook}
  L.~D.~Faddeev and L.~A.~Takhtajan,
  \emph{Hamiltonian Methods in the Theory of Solitons},
  Springer, Berlin Heidelberg New York (1987).

\bibitem[FF1]{FeiginFrenkel}
  B.~Feigin and E.~Frenkel,
  \emph{Affine Kac-Moody algebras at the critical level and Gelfand-Dikii algebras},
  Int. Jour. Mod. Phys. \textbf{A7} (1992), 197--215, Supplement 1A.

\bibitem[FF2]{Feigin:2007mr}
  B.~Feigin and E.~Frenkel,
  \emph{Quantization of soliton systems and Langlands duality},
  Adv. Stud. Pure. Math. 61, Math. Soc. Japan, Tokyo, 2011.

\bibitem[FFR]{FFR}
  B.~Feigin, E.~Frenkel, and N.~Reshetikhin,
  \emph{Gaudin model, Bethe ansatz and correlation functions at the critical level},
  Commun. Math. Phys. \textbf{166} (1994), 27--62.

\bibitem[FFRy]{FFRy}
  B.~Feigin, E.~Frenkel, and L.~Rybnikov,
  \emph{Opers with irregular singularity and spectra of the shift of argument subalgebra},
  Duke Math. J. {\bf 155} (2010), no. 2, 337--363.

\bibitem[FFT]{FFT}
  B.~Feigin, E.~Frenkel and V.~Toledano Laredo,
  \emph{Gaudin models with irregular singularities},
  Adv. Math. {\bf 223} (2010) 873.

\bibitem[F1]{Frenkel}
  E.~Frenkel,
  \emph{Opers on the projective line, flag manifolds and Bethe Ansatz},
  Mosc. Math. J. \textbf{4} (2004), no.~3, 655--705.

\bibitem[F2]{Frenkel2}
  E.~Frenkel, 
  \emph{Gaudin model and opers},
  Progress in Mathematics \textbf{237} (2005), 1--58.

\bibitem[F3]{Frenkel3}
  E.~Frenkel,
  \emph{Wakimoto modules, opers and the center at the critical level},
  Advances in Math. \textbf{195} (2005) 297--404.

\bibitem[F4]{Langlands}
  E.~Frenkel,
  \emph{Langlands Correspondence for Loop Groups},
  Cambridge University Press, 2007.

\bibitem[EHMM1]{Evans:1999mj}
  J.~M.~Evans, M.~Hassan, N.~MacKay and A.~Mountain,
  \emph{Local conserved charges in principal chiral models},
  Nucl. Phys. B {\bf 561} (1999) 385.

\bibitem[EHMM2]{Evans:2000hx}
  J.~M.~Evans, M.~Hassan, N.~MacKay and A.~Mountain,
  \emph{Conserved charges and supersymmetry in principal chiral and WZW models},
  Nucl. Phys. B {\bf 580} (2000) 605.

\bibitem[EM]{Evans:2000qx}
  J.~M.~Evans and A.~Mountain,
  \emph{Commuting charges and symmetric spaces},
  Phys. Lett. B {\bf 483} (2000) 290.

\bibitem[EY]{Evans:2005zd}
  J.~M. Evans and C.~A.~S. Young,
  \emph{Higher-spin conserved currents in supersymmetric sigma models on symmetric spaces},
  Nucl. Phys. B {\bf 717} (2005) 327.

\bibitem[Eva]{Evans:2001sz}
  J.~M.~Evans,
  \emph{Integrable sigma models and Drinfeld-Sokolov hierarchies},
  Nucl. Phys. B {\bf 608} (2001) 591.

\bibitem[FH]{Frenkel:2016gxg}
  E.~Frenkel and D.~Hernandez,
  \emph{Spectra of quantum KdV Hamiltonians, Langlands duality, and affine opers},
  arXiv:1606.05301 [math.QA].
  
\bibitem[Fa]{Fateev:1996ea}
  V.~A.~Fateev,
  \emph{The sigma model (dual) representation for a two-parameter family of integrable quantum field theories},
  Nucl.\ Phys.\ B {\bf 473} (1996) 509.

\bibitem[FM]{Freidel-Maillet}
  L.~Freidel and J.-M.~Maillet,
  \emph{Quadratic algebras and integrable systems},
  Phys. Lett. B {\bf 262}, 2-3 (1991) 278--284.

\bibitem[G]{Gaudin}
  M.~Gaudin,
  \emph{Diagonalisation d'une classe d'Hamiltoniens de spins},
  J. Physique \textbf{37} (1976), 1087--1098.

\bibitem[HK]{Hlavaty:1994vh}
  L.~Hlavaty and A.~Kundu,
  \emph{Quantum integrability of nonultralocal models through Baxterization of quantized braided algebra},
  Int.\ J.\ Mod.\ Phys.\ A {\bf 11} (1996) 2143.

\bibitem[HRT]{Hoare:2014pna}
  B.~Hoare, R.~Roiban and A.~A.~Tseytlin,
  \emph{On deformations of $AdS_n$ x $S^n$ supercosets},
  JHEP {\bf 1406} (2014) 002.

\bibitem[HMS1]{Hollowood:2014rla}
  T.~J.~Hollowood, J.~L.~Miramontes and D.~M.~Schmidtt,
  \emph{Integrable Deformations of Strings on Symmetric Spaces},
  JHEP {\bf 1411} (2014) 009.

\bibitem[HMS2]{Hollowood:2014qma}
  T.~J.~Hollowood, J.~L.~Miramontes and D.~M.~Schmidtt,
  \emph{An Integrable Deformation of the $AdS_5 \times S^5$ Superstring},
  J.\ Phys.\ A {\bf 47} (2014) no.49, 495402.

\bibitem[IL1]{Ito:2013aea}
  K.~Ito and C.~Locke,
  \emph{ODE/IM correspondence and modified affine Toda field equations},
  Nucl.\ Phys.\ B {\bf 885} (2014) 600.

\bibitem[IL2]{Ito:2015nla}
  K.~Ito and C.~Locke,
  \emph{ODE/IM correspondence and Bethe ansatz for affine Toda field equations},
  Nucl.\ Phys.\ B {\bf 896} (2015) 763.

\bibitem[IS]{Ito:2016qzt}
  K.~Ito and H.~Shu,
  \emph{ODE/IM correspondence for modified $B_2^{(1)}$ affine Toda field equation},
  arXiv:1605.04668 [hep-th].

\bibitem[KLWY]{Ke:2011zzb}
  S.~M.~Ke, X.~Y.~Li, C.~Wang and R.~H.~Yue,
  \emph{Classical exchange algebra of the nonlinear sigma model on a supercoset target with Z(2n) grading},
  Chin.\ Phys.\ Lett.\  {\bf 28} (2011) 101101.

\bibitem[K1]{Klimcik}
  C.~Klimcik,
  \emph{Yang-Baxter sigma models and dS/AdS T duality},
  JHEP {\bf 0212} (2002) 051.

\bibitem[K2]{Klimcik:2008eq}
  C.~Klimcik,
  \emph{On integrability of the Yang-Baxter sigma-model},
  J.\ Math.\ Phys.\  {\bf 50} (2009) 043508.

\bibitem[K3]{Klimcik:2014bta}
  C.~Klimcik,
  \emph{Integrability of the bi-Yang-Baxter sigma-model},
  Lett.\ Math.\ Phys.\  {\bf 104} (2014) 1095.

\bibitem[KS]{KS}
  P.~Kulish and E.~Sklyanin,
  \emph{Quantum inverse scattering method and the Heisenberg ferromagnet},
  Phys. Lett. \textbf{A70} (1979) 461--463.
  
\bibitem[LMV]{Lacroix:2017isl}
  S.~Lacroix, M.~Magro and B.~Vicedo,
  \emph{Local charges in involution and hierarchies in integrable sigma-models},
  J. High Energy Phys. {\bf 1709} (2017) 117.

\bibitem[LV]{Lacroix:2016mpg}
  S.~Lacroix and B.~Vicedo,
  \emph{Cyclotomic Gaudin models, Miura opers and flag varieties},
  arXiv:1607.07397 [math.QA].

\bibitem[Lu]{Lukyanov:2013wra}
  S.~L.~Lukyanov,
  \emph{ODE/IM correspondence for the Fateev model},
  JHEP {\bf 1312} (2013) 012.

\bibitem[LZ]{Lukyanov:2010rn}
  S.~L.~Lukyanov and A.~B.~Zamolodchikov,
  \emph{Quantum Sine(h)-Gordon Model and Classical Integrable Equations},
  JHEP {\bf 1007} (2010) 008.

\bibitem[Mai]{Maillet1}  
  J.~M.~Maillet,
  \emph{Kac-Moody algebra and extended Yang-Baxter relations in the O(N) non-linear sigma model},
  Phys. Lett. B {\bf 162} (1985) 137.

\bibitem[Mai2]{Maillet2}
  J.~M.~Maillet,
  \emph{New integrable canonical structures in two-dimensional models},
  Nucl. Phys. B {\bf 269} (1986) 54.

\bibitem[Mag]{Magro}
  M.~Magro,
  \emph{The Classical Exchange Algebra of $AdS_5 \times S^5$},
  JHEP {\bf 0901} (2009) 021.

\bibitem[MRV1]{Masoero:2015lga}
  D.~Masoero, A.~Raimondo and D.~Valeri,
  \emph{Bethe Ansatz and the Spectral Theory of Affine Lie Algebra-Valued Connections I. The simply-laced Case},
  Commun.\ Math.\ Phys.\  {\bf 344} (2016) no.3, 719.

\bibitem[MRV2]{Masoero:2015rcz}
  D.~Masoero, A.~Raimondo and D.~Valeri,
  \emph{Bethe Ansatz and the Spectral Theory of affine Lie algebra--valued connections II. The non simply--laced case},
  arXiv:1511.00895 [math-ph].
  
\bibitem[MW]{Melikyan:2016gkd}
  A.~Melikyan and G.~Weber,
  \emph{On the quantization of continuous non-ultralocal integrable systems},
  Nucl.\ Phys.\ B {\bf 913} (2016) 716.

\bibitem[Mik]{Mikhailov}
  A.~V.~Mikhailov,
  \emph{The reduction problem in the inverse scattering method},
  Phys. D {\bf 3} (1981) 73.
  
\bibitem[MV]{MV}
  E.~Mukhin and A.~Varchenko,
  \emph{Critical points of master functions and flag varieties},
  Commun. Contemp. Math. 6 (2004), no. 1, 111--163.

\bibitem[RS]{RS}
  A.~G.~Reiman and M.~A.~Semenov-Tian-Shansky,
  \emph{Current algebras and nonlinear partial differential equations},
  Dokl. Akad. Nauk SSSR \textbf{251} No. 6 (1980) 1310--1314.

\bibitem[RT]{Ridout:2011wx}
  D.~Ridout and J.~Teschner,
  \emph{Integrability of a family of quantum field theories related to sigma models},
  Nucl.\ Phys.\ B {\bf 853} (2011) 327.

\bibitem[Ry]{Rybnikov:2016}
  L.~Rybnikov,
  \emph{A proof of the Gaudin Bethe ansatz conjecture},
  [arXiv:1608.04625 [math.QA]].

\bibitem[Sch]{Schmidtt:2017ngw}
  D.~M.~Schmidtt,
  \emph{Integrable Lambda Models And Chern-Simons Theories},
  arXiv:1701.04138 [hep-th].

\bibitem[S1]{STS1}
  M.~A.~Semenov-Tian-Shansky,
  \emph{What is a classical r-matrix?},
  Funct. Anal. Appl. \textbf{17} (1983) 259 [Funkt. Anal. Pril. \textbf{17}N4 (1983) 17]

\bibitem[S2]{STS2}
  M.~A.~Semenov-Tian-Shansky,
  \emph{Monodromy map and classical r-matrices},
  J. Math. Sci. {\bf 77}, 3 (1995) 3236--3242.

\bibitem[SS]{SemenovTianShansky:1995ha}
  M.~A.~Semenov-Tian-Shansky and A.~Sevostyanov,
  \emph{Classical and quantum nonultralocal systems on the lattice},
  [hep-th/9509029].

\bibitem[Sev]{Sevostyanov}
  A.~Sevostyanov,
  \emph{The Classical R matrix method for nonlinear sigma model},
  Int.\ J.\ Mod.\ Phys.\  A {\bf 11} (1996) 4241.

\bibitem[Sfe]{Sfetsos:2013wia}
  K.~Sfetsos,
  \emph{Integrable interpolations: From exact CFTs to non-Abelian T-duals},
  Nucl.\ Phys.\ B {\bf 880} (2014) 225.

\bibitem[Su]{Sun:2012xw}
  J.~Sun,
  \emph{Polynomial relations for $q$-characters via the ODE/IM correspondence},
  SIGMA {\bf 8} (2012) 028.
  
\bibitem[Tak]{Takhtajan}
  Takhtajan, L. A.:
  \emph{Quantum field theories on algebraic curves. I. Additive bosons},
  2013 Russian Academy of Sciences, (DoM) and London Mathematical Society, Turpion Ltd -- Izvestiya: Mathematics, Volume 77, Number 2.

\bibitem[Vi1]{Vicedo:2009sn}
  B.~Vicedo,
  \emph{Hamiltonian dynamics and the hidden symmetries of the $AdS_5 \times S^5$ superstring},
  JHEP {\bf 1001} (2010) 102.

\bibitem[Vi2]{Vicedo:2010qd}
  B.~Vicedo,
  \emph{The classical R-matrix of AdS/CFT and its Lie dialgebra structure},
  Lett.\ Math.\ Phys.\  {\bf 95} (2011) 249.

\bibitem[Vi3]{Vicedo:2015pna}
  B.~Vicedo,
  \emph{Deformed integrable $\sigma$-models, classical $R$-matrices and classical exchange algebra on Drinfel'd doubles},
  J.\ Phys.\ A {\bf 48} (2015) no.35, 355203.

\bibitem[ViY1]{Vicedo:2014zza}
  B.~Vicedo and C.~A.~S.~Young,
  \emph{Cyclotomic Gaudin models: construction and Bethe ansatz},
  Comm. Math. Phys. {\bf 343} (2016), no. 3, 971--1024.

\bibitem[ViY2]{ViY2}
  B.~Vicedo, C.~A.~S.~Young,
  \emph{Vertex Lie algebras and cyclotomic coinvariants},
  to appear in Commun. Contemp. Math.

\bibitem[ViY3]{ViY3}
  B.~Vicedo and C.~A.~S.~Young,
  \emph{Cyclotomic Gaudin models with irregular singularities},
  arXiv:1611.09059 [math.QA].

\bibitem[Y]{Young:2005jv}
  C.~A.~S.~Young,
  \emph{Non-local charges, Z(m) gradings and coset space actions},
  Phys.\ Lett.\ B {\bf 632} (2006) 559.
  
\end{thebibliography}
\end{document}